\documentclass[onecolumn,numbers]{els-mrw}
\usepackage{amsmath,amssymb,amsfonts,amsthm,makeidx,graphicx}
\usepackage{txfonts}
\usepackage{helvet}
\usepackage[dvipsnames]{xcolor}
\usepackage{multirow, slashed}
\usepackage{longtable}
\usepackage{amsmath}
\usepackage{mathtools}
\usepackage{commath}
\usepackage{subcaption}
\usepackage{bm}
\usepackage[font=small,labelfont=bf]{caption}
\usepackage{adjustbox}
\usepackage{tikz}
\usepackage[abs]{overpic}
\usepackage{xcolor,varwidth}

\begin{document}

%%%%%%%%%%%%%%%%%%%%%%%%%%%%%%%%%%%%%%%%%%%%%%%%%%%%%%%%%%%%%%%%

\chapter{Charm physics}\label{chap1}

\author[1]{David Friday}%
\author[2]{Evelina Gersabeck}%
\author[3]{Alexander Lenz}%
\author[4,5]{Maria Laura Piscopo}%

\address[1]{
\orgname{University of Manchester}, 
\orgdiv{School of Physics and Astronomy}, \orgaddress{Oxford Rd, Manchester, M13 9PL, United Kingdom}
}\address[2]{
\orgname{Albert-Ludwigs-Universität Freiburg}, 
\orgdiv{Physikalisches Institut}, \orgaddress{Gustav-Mie-Haus 02 025, 
Hermann-Herder-Straße 3b, 
79085 Freiburg im Breisgau, Germany}
}
\address[3]{
\orgname{Universität Siegen}, 
\orgdiv{Physik Department}, 
\orgaddress{ Walter-Flex-Str. 3, 57068 Siegen, Germany}
}
\address[4]{
\orgname{Nikhef}, 
\orgaddress{Science Park 105, NL-1098 XG Amsterdam, Netherlands}
}
\address[5]{
\orgname{Vrije Universiteit Amsterdam}, 
\orgdiv{Department of Physics and Astronomy},
\orgaddress{NL-1081 HV Amsterdam, Netherlands}
}

\articletag{Chapter Article tagline: update of previous edition, reprint.}

\maketitle

%%%%%%%%%%%%%%%%%%%%%%%%%%%%%%%%%%%%%%%%%%%%%%%%%%%%%%%%%%%%%%%%
\begin{abstract}[Abstract]
50 years after the discovery of the first charmed particle, charm physics continues to be an extremely lively field of research and a cornerstone in particle physics. The study of charm, with its unique properties, is characterised by many challenging but also exciting peculiarities, making it an ideal testing ground for Standard Model (SM) predictions and a very sensitive probe of new physics. This chapter is intended to provide a pedagogical introduction to the physics of the charm quark and to its current theoretical and experimental status. Specifically, it discusses the main features of the charm sector of the SM, the theoretical and experimental challenges that arise when dealing with the charm quark, and the methods used to study it. An overview, both from a theoretical and experimental perspective, of fundamental observables such as lifetimes of charm hadrons, $D^0$-meson mixing, charm charge-parity violation~(CPV) and rare charm decays is also presented.

\end{abstract}

\begin{keywords}
 	charm quark \sep heavy quark flavour physics \sep charm hadron decays \sep charm mixing \sep charm CPV. 
\end{keywords}

%%%%%%%%%%%%%%%%%%%%%%%%%%%%%%%%%%%%%%%%%%%%%%%%%%%%%%%%%%%%%%%%
\section*{Objectives}
\begin{itemize}
        \item History of the discovery of the charm quark  and subsequent developments.
        \item Description of the properties of the charm quark and of its weak decays. 
        \item Introduction to the theoretical and experimental methods used in charm physics. 
        \item Theoretical and experimental status for charm observables including lifetimes, CPV, mixing and rare decays.
\end{itemize}

%%%%%%%%%%%%%%%%%%%%%%%%%%%%%%%%%%%%%%%%%%%%%
%%%%%%%%%%%%%%%%%%%%%%%%%%%%%%%%%%%%%%%%%%%%%%%%%%%%%%%%%%%%%%%%%%%%%%%%%%%%%%%%%%%%%%%%%%%%%%%%%%%%%%%%%%%%%%%%%%%%%%%%%%%%%%%%
\section{Introduction}
\label{sec:intro}
\begin{figure}[h]
\centering
\includegraphics[scale=0.56]{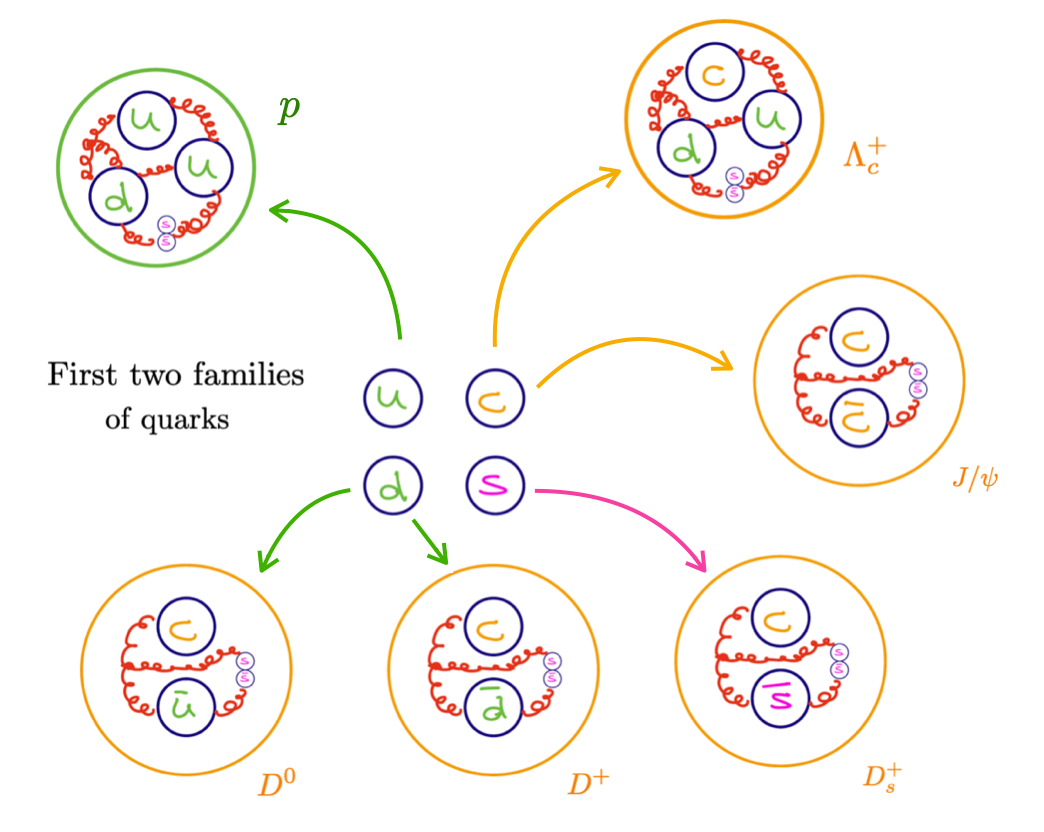}
\caption{Quarks are bound via the strong interaction into hadrons. The most well-known hadron is probably the proton, consisting of two up quarks and one down quarks, as well as sea-quarks, here indicated by the $s \bar{s}$ pair, and gluons, indicated by the curly red lines. The charm quark is the main constituent of the $J/\psi$ resonance, the $D$ mesons and the  $\Lambda_c^+$ baryon.}
\label{sec1:fig1}
\end{figure}

November 2024 marked the
50th anniversary of the discovery of the charm quark, also known as the
{\it November Revolution} in particle
physics~\footnote{See e.g. 
https://cerncourier.com/a/the-new-particles/
and
https://cerncourier.com/a/charming-clues-for-existence/.}. 
In 1974, the groups of Samuel Ting at the Brookhaven National Laboratory (BNL)~\cite{E598:1974sol}
and Burton Richter at the Stanford Linear Accelerator Center (SLAC)~\cite{SLAC-SP-017:1974ind} discovered, more or less
simultaneously, the $J/ \psi$ meson, a narrow resonance consisting of 
a charm quark and a charm anti-quark.
The $J/ \psi$ has a mass of
$3097$ MeV, suggesting a charm-quark mass of the order of $1500$ MeV - close to current values - and a width of $92.6$ keV,
corresponding to a short lifetime of
about $7.1 \cdot 10^{-21}$s. 
The charm quark, in fact, is not one of the main constituents of the ordinary matter i.e.\ protons or neutrons,  
but forms heavier particles that decay after some time, see Fig.~\ref{sec1:fig1}. These bound states, which are glued together by the strong interaction, are conventionally classified into charmed mesons, made of a quark and an anti-quark, just like the $J/\psi$, and  charmed baryons (anti-baryons), made of three quarks (anti-quarks). 
Experimentally, charmed particles are distinguished according to the value of their {\it charm quantum number} $C$.
States with $C = 0$ composed of a charm quark and a charm anti-quark
such as the $J/\psi$, $\psi(2S)$, $\psi(3770)$, etc., 
 are referred to as {\bf hidden charm} states. They are also commonly referred to as charmonium---analogous to positronium---states or charmonia. Particles with $C \neq 0$ containing either charm {\it or} anti-charm quarks such as the $D^0$, $D^+$, $D^+_s$, $\Lambda_c^+$, etc. are referred to as {\bf open charm} particles, see Table~\ref{sec1:tab1} and Fig.~\ref{sec1:fig1p}. In particular, 
the term ``$D$ meson" commonly refers to a particle containing a charm
quark and a light anti-quark, either a up, down or strange, that is the $D^0$, $D^+$, or $D_s^+$, as well as the corresponding excited states.~\footnote{It is interesting to note that prior to 1986 the $D_s^+$ mesons were named $F$ mesons from ``funny", and the name of the $D$ mesons came from ``doublet"~\cite{Rosner}.}.
\\

\noindent
This simultaneous discovery of the $J/\psi$ opened the gates for a new era of particle physics.
The existence of other conventional charm particles was predicted and the second charm particle to be discovered was the  excited $\psi(2S)$ meson~\cite{Abrams:1974yy}~\footnote{Here (2S) indicates that $\psi(2S)$ is the first radial excitation of the ground state $J/\psi$.} at SLAC. 
Several other charmonia were quickly discovered and studied at the SLAC-LBL (Lawrence Berkeley National Laboratory) magnetic detector at the SPEAR electron-positron storage ring, also known as the MARK-I detector~\cite{Augustin:1974xq} and at the DORIS storage ring, also electron-positron, at DESY (DASP and PLUTO detectors)~\cite{Wiik:1975ee}. 
Some of these early studies can be found in~\cite{Augustin:1975yq, Boyarski:1975ci, Rapidis:1977cv, Luth:1975bh, Siegrist:1976br, DASP:1975gzk, Criegee:1975uc, DASP:1977eel}. 
Of particular interest is the state $\psi(3770)$ discovered with the MARK-I detector~\cite{Rapidis:1977cv}, as its  mass is just above the energy threshold for the production of a pair of charmed mesons thus enabling the decays $\psi(3770) \to D^+ D^-$, and $\psi(3770) \to  D^0 \bar{D}^0$, see Table~\ref{sec1:tab1} and Fig.~\ref{sec1:fig1p}. Now, 50 years after the discovery of the first charmed particle, charm physics continues to be an extremely lively field of research and a cornerstone
of modern particle physics. The study of charm decays is characterised by many challenging but also exciting
peculiarities, which make the charm sector an ideal testing ground for Quantum Chromodynamics~(QCD) based frameworks and a very sensitive probe of new physics (NP). In particular, 
it might even reveal us some crucial insights on the phenomenon of the violation of the charge-conjugation and parity~(CP) symmetry in the Standard Model of particle physics~(SM), which is a necessary ingredient
in order to explain the existence of matter in the Universe~\cite{Sakharov:1967dj}. \\
\\

\begin{minipage}{\textwidth}
\begin{minipage}{0.3\textwidth}
\centering
\renewcommand*{\arraystretch}{1.8}
\begin{tabular}[t]{|c||c|}
 \hline
   & $m$[MeV] \\
  \hline
  \hline
   $D^0$ & 1864.84(5) \\
   \hline
   $D^+$ & 1869.66(5)  \\
   \hline
   $D^+_s$ & 1968.35(7) \\
   \hline
   $D^{0*}$ & 2006.85(5) \\
   \hline
   $D^{+*}$ & 2010.26(5) \\
   \hline
   $D^{+*}_s$ & 2112.2(4) \\
   \hline
   $\Lambda_c^+$ & 2286.46(14)\\
   \hline
   $J/\psi$ & 3096.900(6) \\
   \hline
   $\psi(2S)$ & 3686.097(11)\\
   \hline
   $\psi(3770)$ & 3773.7(7)\\
   \hline
\end{tabular}
    \captionof{table}{Masses of the most common charmonia and open charm particles; both ground and excited  states~\cite{ParticleDataGroup:2024cfk}.}
\label{sec1:tab1}
\end{minipage}
\hspace*{5mm}
\begin{minipage}{0.6\textwidth}
 \centering
\includegraphics[scale=0.52]{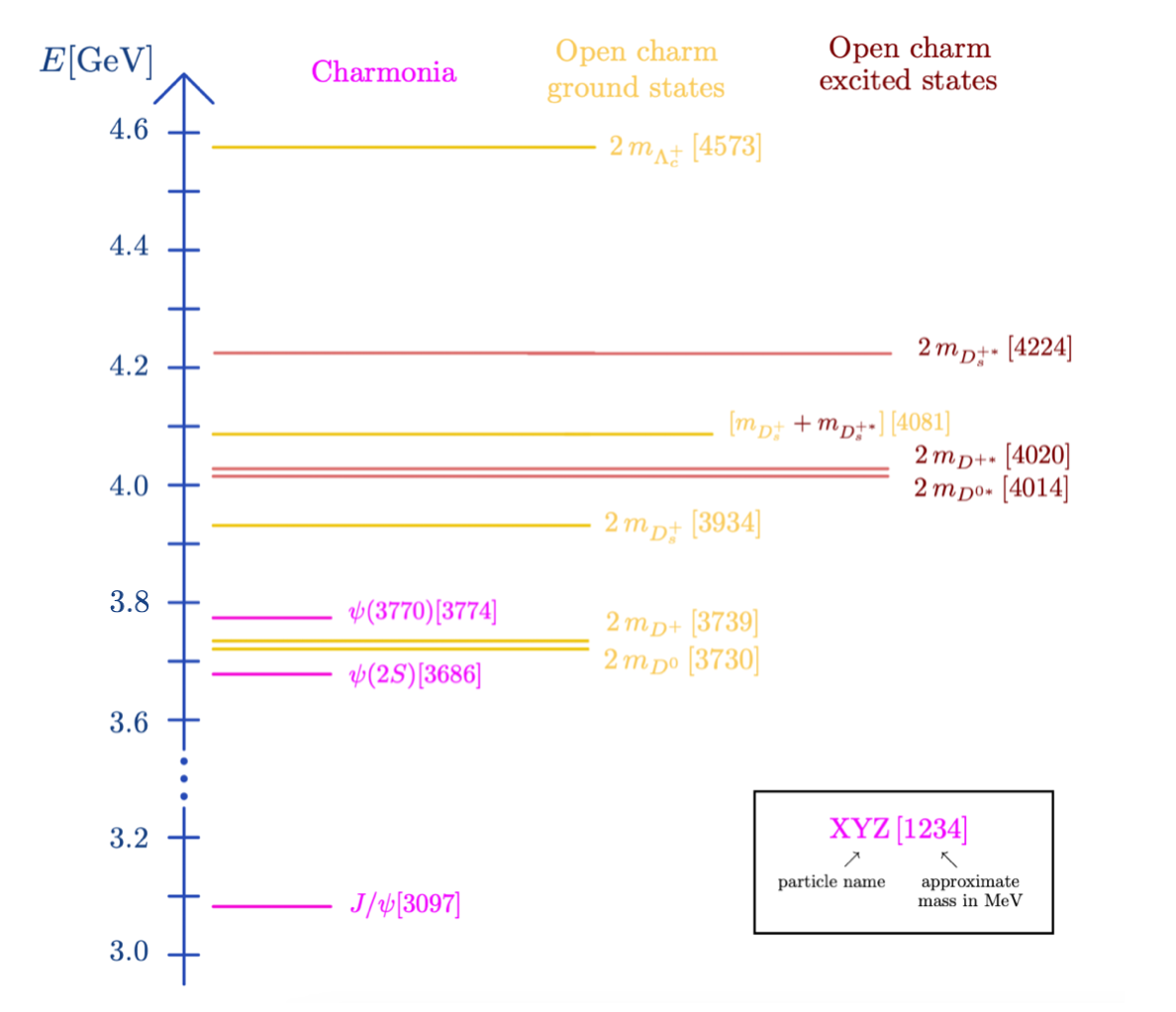}
\captionof{figure}{Energy hierarchy of selected charmonia masses and of the most common thresholds for open charm production channels---both ground and excited states.}
\label{sec1:fig1p}
\end{minipage}
\end{minipage}\\
\\

\begin{figure}
    \centering
    \includegraphics[width=0.37\linewidth]{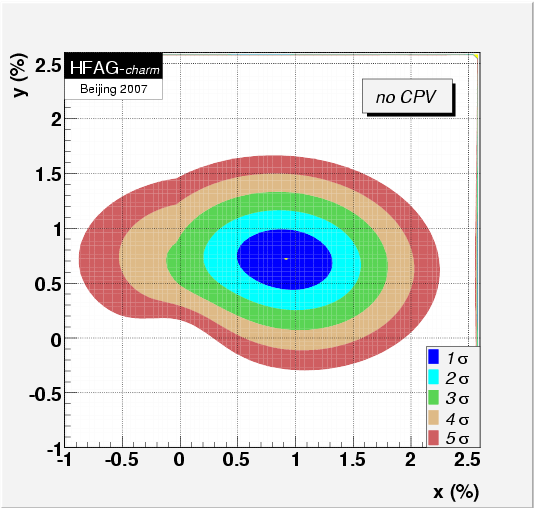}    
    \caption{First combination of the mixing parameters $x$ and $y$ excluding the no-mixing hypothesis (0,0) with more than $5\sigma$. Taken from~\cite{Schwartz:2008wa}. }
    \label{sec1:fig2}
\end{figure}

\noindent
In recent years, significant progress has been made, and the following phenomena have been experimentally established---all of which will be discussed in more detail in the next sections:\\
\begin{enumerate}
\item {\it 2003 onwards: charm spectroscopy---} 
For a long time only baryons and mesons were experimentally known as colour-singlet bound states. 
From a QCD point of view also tetraquarks (bound states of 
two quarks and two anti-quarks) and pentaquarks (bound states of four quarks and one anti-quark) could be colour neutral particles. By now, these so-called exotic hadrons have been observed and many of them contain charm quarks~\footnote{For a nice summary of current results see e.g.~https://cerncourier.com/a/a-bestiary-of-exotic-hadrons/.}, see Section~\ref{subsec:spectroscopy}.
\item {\it 2007/2012/2021: $D^0$-meson mixing---} Through the weak interaction, the neutral $D^0$ meson can transform into its antiparticle, the $\bar{D}^0$ meson, leading to the phenomenon of $D^0$-$\bar D^0$ mixing.
The two fundamental observables that describe the mixing process are the mass and width difference between the corresponding mass eigenstates, encoded in the dimensionless parameters $x$ and $y$, respectively, see Sections~\ref{subsec:basics_mixing},~\ref{subsec:mixing}. Although the existence of $D^0$-meson mixing could already be inferred in 2007 from a combination of measurements performed by Belle~\cite{BELLE:2007dgh}, BaBar~\cite{BaBar:2007kib} and CLEO-c~\cite{Sun:2007fh},
see the original HFLAV (Heavy Flavour Averaging Group, earlier known as HFAG) plot in Fig.~\ref{sec1:fig2}---taken from~\cite{Schwartz:2008wa}--- 
it was only in 2012 that charm mixing was observed by a single experiment. The observation was performed by the LHCb
collaboration~\cite{LHCb:2012zll} and soon confirmed by CDF~\cite{CDF:2013gvz} and Belle~\cite{Belle:2014yoi}. Almost 10 years later, in 2021 the LHCb collaboration also performed the first single measurement of $x$ with a statistical significance of more than five standard deviations~\cite{LHCb:2021ykz}. The parameter $y$ was determined with a statistical significance of more than five standard deviations in 2010 through a HFLAV combination~\cite{HFLAV:2010pgm} of results from Belle~\cite{Belle:2008qhk}, BaBar~\cite{BaBar:2007kib} and CDF~\cite{CDF:2007bdz}, but to this day it has not been determined with more than 5$\sigma$ by a single experiment.
\item {\it 2019: CP violation in the charm sector---} 
In 2019, the 
LHCb collaboration discovered CP violation (CPV) in the charm system~\cite{LHCb:2019hro},
by measuring the difference of the CP asymmetries in the decays 
$D^0 \to K^+ K^-$ and $D^0 \to \pi^+ \pi^-$---a quantity denoted by $\Delta A_{\rm CP}$, see Sections~\ref{subsec:basics_CPV}, \ref{subsec:CPV}. The measured value of $\Delta A_{\rm CP}$ is of the order of $10^{-3}$ and this is 
about a factor of ten larger than the  
naive expectation obtained within the SM. Currently, it is debated whether the naive SM expectation could be enhanced by a factor of ten due to poorly known non-perturbative effects, or whether the measurement of $\Delta A_{\rm CP}$ might be a first glimpse of physics beyond the SM~(BSM).\\
\end{enumerate}

\noindent
The previous points illustrate how the physics of charm is a fast-advancing field of research with many interesting open questions. The remaining sections of this chapter aim, therefore, to provide a pedagogical guide to the physics of the charm quark and its current theoretical and experimental status. In the following, we discuss the main features of the charm sector of the SM and why, from a certain point of view, it represents a system with unique properties. We also describe what are the main theoretical and experimental challenges that arise when dealing with the charm quark, as well as the methods used to study it. Finally, we provide an overview, from both a theoretical and experimental perspective, of fundamental observables such as lifetimes of charm hadrons, $D^0$ mixing, charm CPV and rare charm decays.

%%%%%%%%%%%%%%%%%%%%%%%%%%%%%%%%%%%%%%%%%%%%%%%%%%%%%%%%%%%%%%%%%%%%%%%%%%%%%%%%%%%%%%%%%%%%%%%%%%%%%%%%%%%%%%%%%%%%%%%%%%%%%%%%
\section{Creation of charm}
\label{sec:creation}
\subsection{History of the charm quark }
\label{subsec:discovery}
Before the $J/\psi $ resonance was discovered, only the three light quarks, up, down and strange were known.
Up and down form the first family of quarks and they make up the dominant part of the ordinary matter i.e.\ protons and
neutrons. The detection of muons
in cosmic rays in 1936 by Anderson and Neddermeyer from Caltech~\cite{PhysRev.50.263} gave the evidence for a second generation
of fermions, triggering the famous outcry of the Nobel laureate I.\ I.\ Rabi  {\it ``Who ordered that?"}. Strange particles were found in 1947,
and by now we also know that there is a significant virtual strange-quark content in the proton~\cite{Faura:2020oom}. Over the coming years many new particles were found---{\it particle zoo}---and it took up to 1964 until Gell-Mann and Zweig introduced the concept of quarks, as an ordering principle in the subatomic world~\cite{Gell-Mann:1964ewy, Zweig:1964ruk}.\\

\begin{figure}[h]
\centering
\includegraphics[scale=0.45]{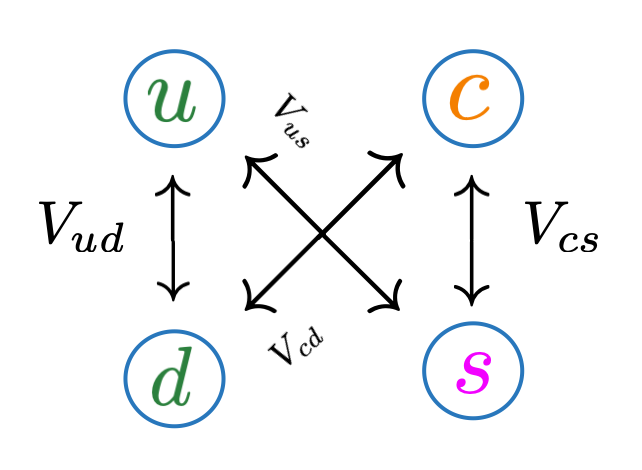}
\caption{Visualisation of the Cabibbo matrix, which describes the coupling strength of the $W$ boson to the first two generations of quarks.}
\label{sec2.1:fig1}
\end{figure}  

\begin{figure}[t]  
\centering
\includegraphics[scale=0.55]{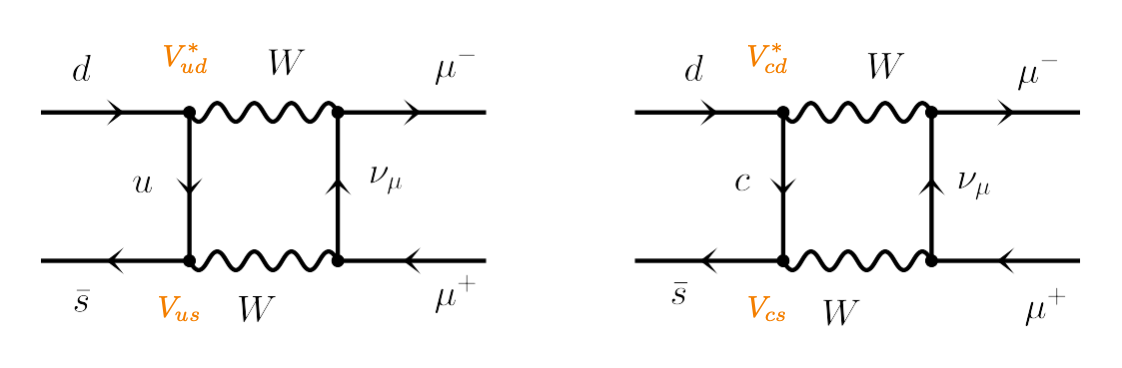}
	\caption{Leading Feynman diagrams with internal up (left) and charm (right) quarks describing the decay $K^0 \to \mu^+ \mu^-$. The contribution of the up quark alone would lead to a much higher decay probability than actually observed. The contribution from the charm quark largely compensates the one of the up quark and the sum of the two diagrams results in a suppression of this decay, in agreement with experiments.}
	\label{sec2.1:fig2}
\end{figure}

\begin{figure}[h]
	\centering     \includegraphics[width=0.9\textwidth]{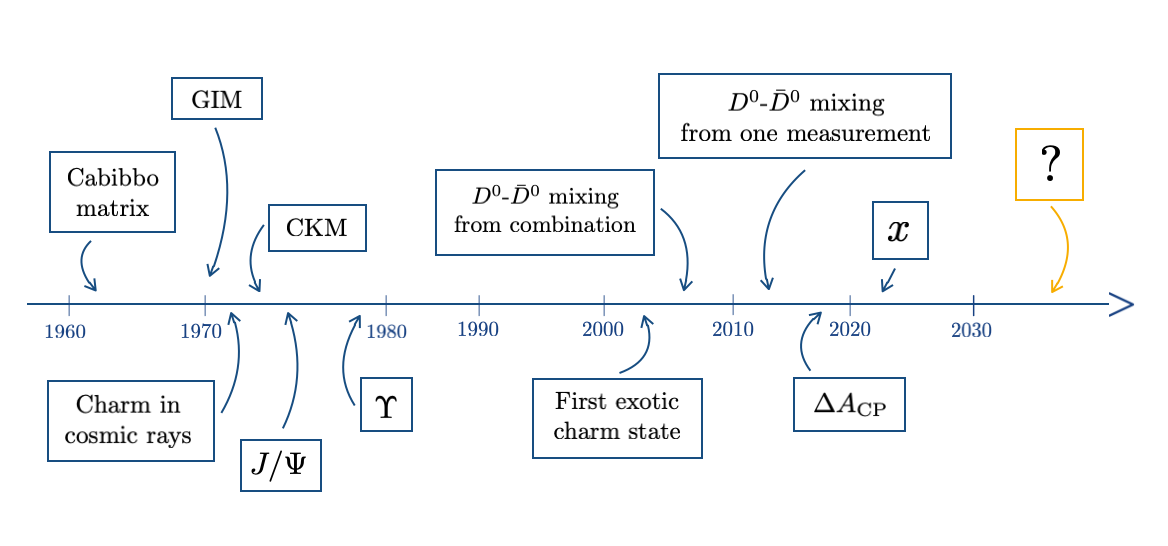}
	\caption{Timeline of key theoretical and experimental milestones in charm physics.}
	\label{sec2.1:fig3}
\end{figure}

\noindent
The existence of a fourth quark---the charm---to complete the second generation of quarks, was first theorised by Bjorken and Glashow in 
1964~\cite{Bjorken:1964gz}. In this scenario, the coupling strength of the weak charged interaction, i.e.\ the $W^\pm$ exchange, between the two corresponding up- and down-type quarks is encoded into a $2 \times 2$ rotation matrix, commonly known as the Cabibbo matrix~\footnote{
In the modern language of particle physics, Nicola Cabibbo in 1963 proposed, based on symmetry arguments for the weak currents, that the mass and weak eigenstates of the down-type quarks would not coincide but that the states weakly coupled would be a linear combination of the mass eigenstates, with coefficients depending on the the Cabibbo angle $\theta_c$~\cite{Cabibbo:1963yz}. This work paved the way for the concept of 'probability' for a certain weak transition to happen.}, that is
\begin{equation}
V_{\rm Cabibbo} = 
\begin{pmatrix}
V_{ud} & V_{cd}\\
V_{us} & V_{cs}
\end{pmatrix}
= 
\begin{pmatrix}
\cos \theta_c &  -\sin \theta_c\\
\sin \theta_c & \cos \theta_c
\end{pmatrix}\,,
\end{equation}
where e.g.\ the matrix element $V_{cd}$ denotes the coupling strength of the weak interaction between the charm and down quark, see Fig.~\ref{sec2.1:fig1},
and $\theta_c$ is the Cabibbo angle, the only independent parameter that characterises this matrix.
In 1970, Glashow, Iliopoulos, and Maiani showed that the existence of the charm quark could explain the observed suppression of the leptonic decay of a neutral $K^0$ meson---the bound state of a down quark and strange anti-quark---into two muons, \ $K^0 \to \mu^+ \mu^-$.
In fact, assuming the existence of two generations, this process would proceed via loop diagrams with internal up and charm quarks, as sketched in Fig.~\ref{sec2.1:fig2}, including all four elements of the Cabibbo matrix. The decay amplitude would be written as a sum of the up- and charm-quark contributions, and using the parametrisation of the Cabibbo matrix in terms of the Cabibbo angle would yield, schematically
\begin{equation}
{\cal A} (K^0 \to \mu^+ \mu^-)  
\propto    
V_{ud}^* V_{us} A(m_u) 
+
V^*_{cd} V_{cs} A(m_c) 
=
\cos \theta_c \sin \theta_c 
\left[  A (m_c) -  A (m_u)\right]\,,
\end{equation}
where $A(m_q)$ is a function of the mass ratio $m_q^2/m_W^2$.
While the individual contributions $A(m_u)$ and $A(m_c)$ can be quite sizable, they cancel in the difference almost exactly, as $m_u^2/m_W^2 \approx m_c^2/m_W^2 \ll 1$, thus leading to a strong suppression of the decay amplitude. 
This illustrates the basic principle of the Glashow-Iliopoulos-Maiani~{\bf (GIM) mechanism}~\cite{Glashow:1970gm}, which states that the existence of the charm quark is
necessary to suppress the \ $K^0 \to \mu^+ \mu^-$ decay to the level observed experimentally.
The measurement of $K^0 \to \mu^+ \mu^-$
in combination with the GIM mechanism is therefore considered to be the {\it indirect 
observation of the charm quark}. 
The direct observation of the first particle containing a charm quark was performed, as introduced in the previous section, four years later by Ting and Richter~\footnote{We note that there was
an earlier direct detection - unfortunately only one event and no estimation of the
background - for a charmed meson in a cosmic ray experiment
performed in 1971 in Japan~\cite{Niu:1971xu}. The latter, however, was not noticed by Western scientists.
Moreover, it is also interesting to note that,
theoretically, the expected properties of charmed particles were quite well known before their discovery, see the seminal review~\cite{Gaillard:1974mw} whose preprint appeared three months before the $J/ \psi$ discovery.}. The quark-mixing formalism was later extended in 1973 by Kobayashi and Maskawa to three quark families, leading to the Cabibbo-Kobayashi-Maskawa~(CKM) matrix~\cite{Kobayashi:1973fv}. 
This 3 $\times$ 3 matrix would now depend on four parameters, three angles and one phase, which could introduce CPV into the SM via complex couplings. The amount of CPV contained in the CKM matrix, however, seems not to be sufficient to explain the observed matter-antimatter asymmetry in the Universe, see e.g.~\cite{Gavela:1993ts}. A timeline of key moments in charm physics
is sketched in Fig.~\ref{sec2.1:fig3}.\\

\noindent
To conclude the excursus through the SM, the bottom quark was experimentally established in 1977 with the creation of the
$\Upsilon$ resonance---bottom and anti-bottom bound state~\footnote{The $\Upsilon$ resonance plays also an important role for current charm-production experiments, see Section~\ref{subsec:experiments}.}. In 1986 the
measurement of the oscillations of neutral $B$ mesons indicated a large value
for the top-quark mass. The diagrams describing $B^0$-meson mixing are similar to those sketched for $K^0 \to \mu^+ \mu^-$ in Fig.~\ref{sec2.1:fig2}, but they present internal up, charm, and top quarks; due to both the different size of the CKM matrix elements involved and the very different hierarchies $m_u^2/m_W^2 \approx m_c^2/m_W^2 \ll 1$ and $m_t^2/m_W^2 >1 $, the corresponding diagrams with internal top quarks largely dominates, strongly breaking the GIM mechanism.
The top quark was directly observed in proton anti-proton collisions in  1995 at the Fermilab Tevatron~\cite{D0:1995jca} and a third generation of leptons was established with the discovery of the $\tau$ lepton in 1975
at SLAC~\cite{Perl:1975bf}. The mediator of the strong interaction, the gluon, was discovered at the PETRA collider at DESY in 1979 by the
TASSO~\cite{TASSO:1979zyf}, MARK-J~\cite{Barber:1979yr}~\footnote{Note that the MARK-J experiment at DESY is different from the MARK-I-III experiments at SLAC.} and PLUTO~\cite{PLUTO:1979dxn} experiments,
while the gauge bosons of 
the weak interaction, that is the $W^{\pm}$ and $Z$, were discovered in  1983 at the UA1 experiment at the 
Super Proton Synchrotron (SPS) - proton anti-proton collider - at 
CERN~\cite{UA1:1983crd}.
With the discovery of the Higgs boson in proton proton collisions in 2012 at the Large Hadron Collider (LHC) by the ATLAS~\cite{ATLAS:2012yve} and CMS~\cite{CMS:2012qbp} collaborations, the complete spectrum of the SM particles has been experimentally confirmed. 
%%%%%%%%%%%%%%%%%%%%%%%%%%%%%%%%%%%%%%%%%%%%%%%%%%%%%%%%%%%%%%%%%%%%%%%%%%%%%%%%%%%%%%%%%%%%%%%%%%%%%%%%%%%%%%%%%%%%%%%%%%%%%%%%
\subsection{Charm experiments: production and cross-sections}
\label{subsec:experiments}

\begin{figure}[b]
    \centering
    \includegraphics[width=0.35\linewidth]{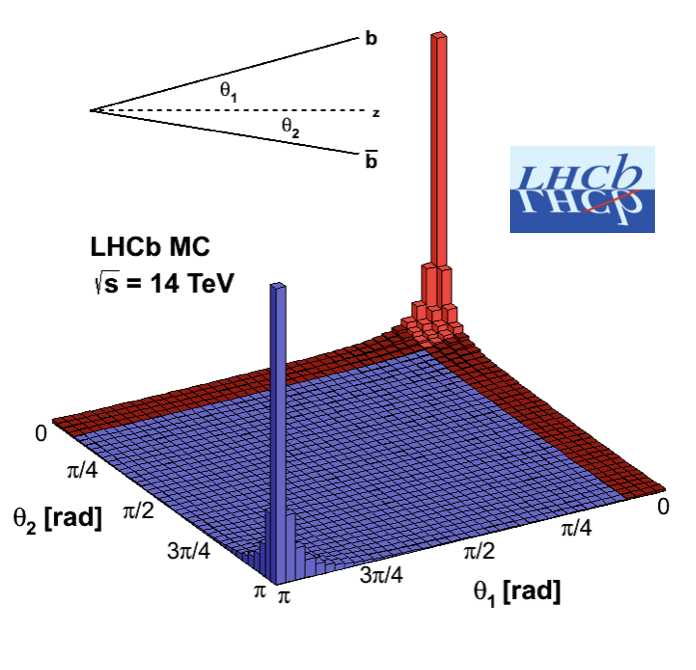}
    \qquad
    \includegraphics[width=0.5\linewidth]{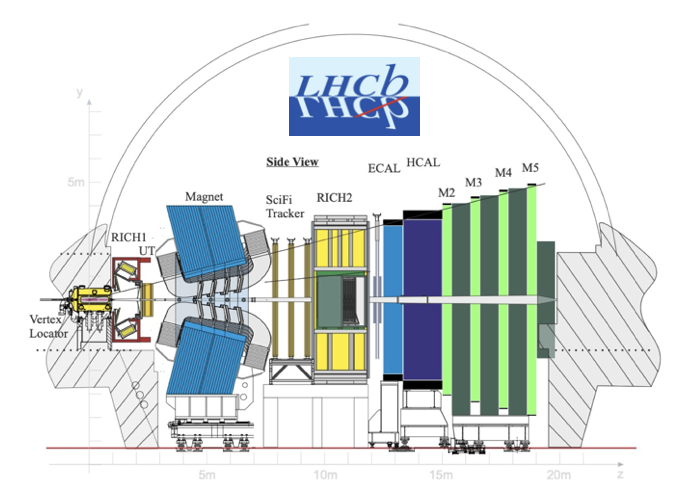}
    \caption{At the LHC, heavy-flavour quark pairs such as $b\bar{b}$, explicitly shown, and $c\bar{c}$ pairs are produced in the forward - or backward - direction~(left). Therefore, the LHCb experiment, and its upgrade (right) are designed as a forward spectrometer. Taken from~\cite{bbangles, LHCb:2023hlw}.}
    \label{sec2.2:fig1}
\end{figure}

After the discovery of the $J/\psi$ particle, many new experiments were built at several different laboratories and a rich charm physics programme was exploited.
More information about the early charm physics experiments can be found at the end of this subsection. The more recent charm physics experiments, namely BaBar at PEPII, SLAC, USA~\cite{BabarWeb, BaBar:2001yhh, BaBar:2013byz}, Belle at KEKB, Tsukuba, Japan~\cite{BelleWeb, Belle:2000cnh, Belle:2012iwr}, together with CDF at the 
Tevatron\footnote{At CDF, charm particles were produced in proton anti-proton collisions at a centre-of-mass energy of 1.96 TeV.}, Fermilab, USA~\cite{CDFWeb, CDF:2013bqv}, and CLEO/CLEO-c\footnote{CLEO-c was the final upgrade of the CLEO experiment, optimised for studies of the charmed hadrons.} at the CESR, Cornell University, Ithaca, USA~\cite{CLEOWeb, CLEO:1991qyy, CLEO:2001bjm} have provided a plethora of measurements. Their aim was to test the SM and search for BSM effects via precision data. This included, for example, high-precision measurements of CPV and mixing parameters, of CKM parameters and of rare and forbidden decays. 
\\

\begin{figure}
    \centering    \includegraphics[width=0.4\linewidth]{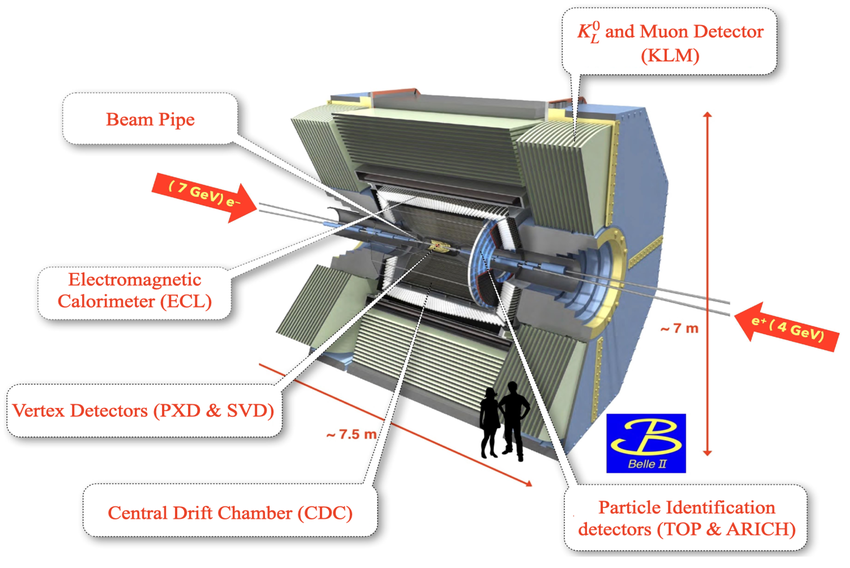}
    \qquad \qquad
    \includegraphics[width=0.4\linewidth]{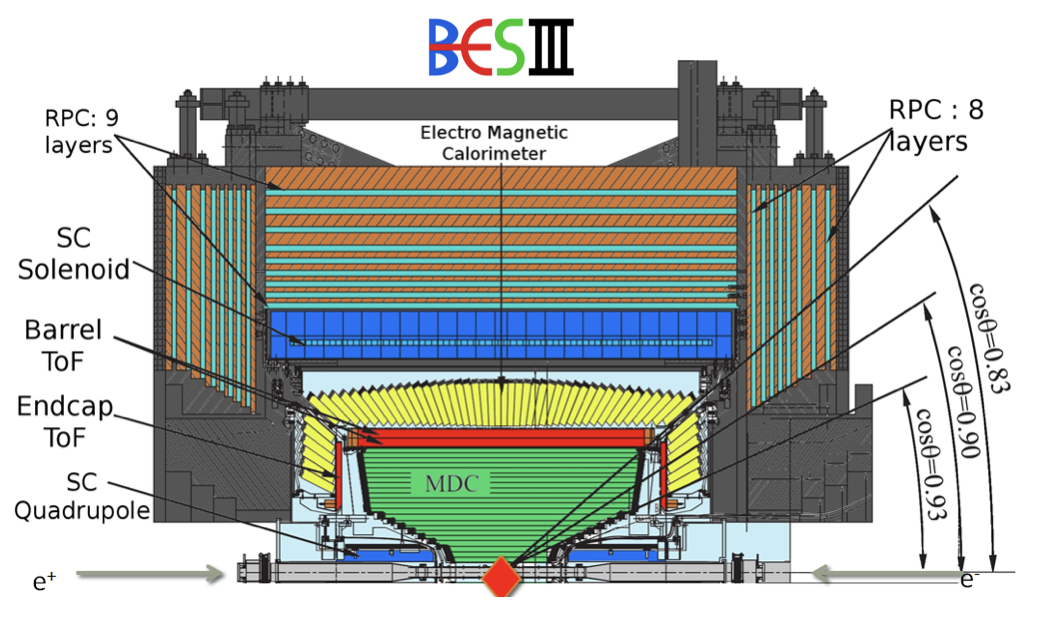}
    \caption{Experiments at electron-positron colliders: Belle II (left) and BESIII (right) - note that only the upper half of the latter is shown. Both detectors are hermetic with a $4\pi$ geometry. At Belle II the beam energies are asymmetric and particles are boosted in contrast to those at BESIII. Taken from~\cite{Banerjee:2022vdd, BESIIIWeb}.
    }
    \label{sec2.2:fig2}
\end{figure}

\begin{figure}[t]
	\centering
       \includegraphics[scale=0.2]{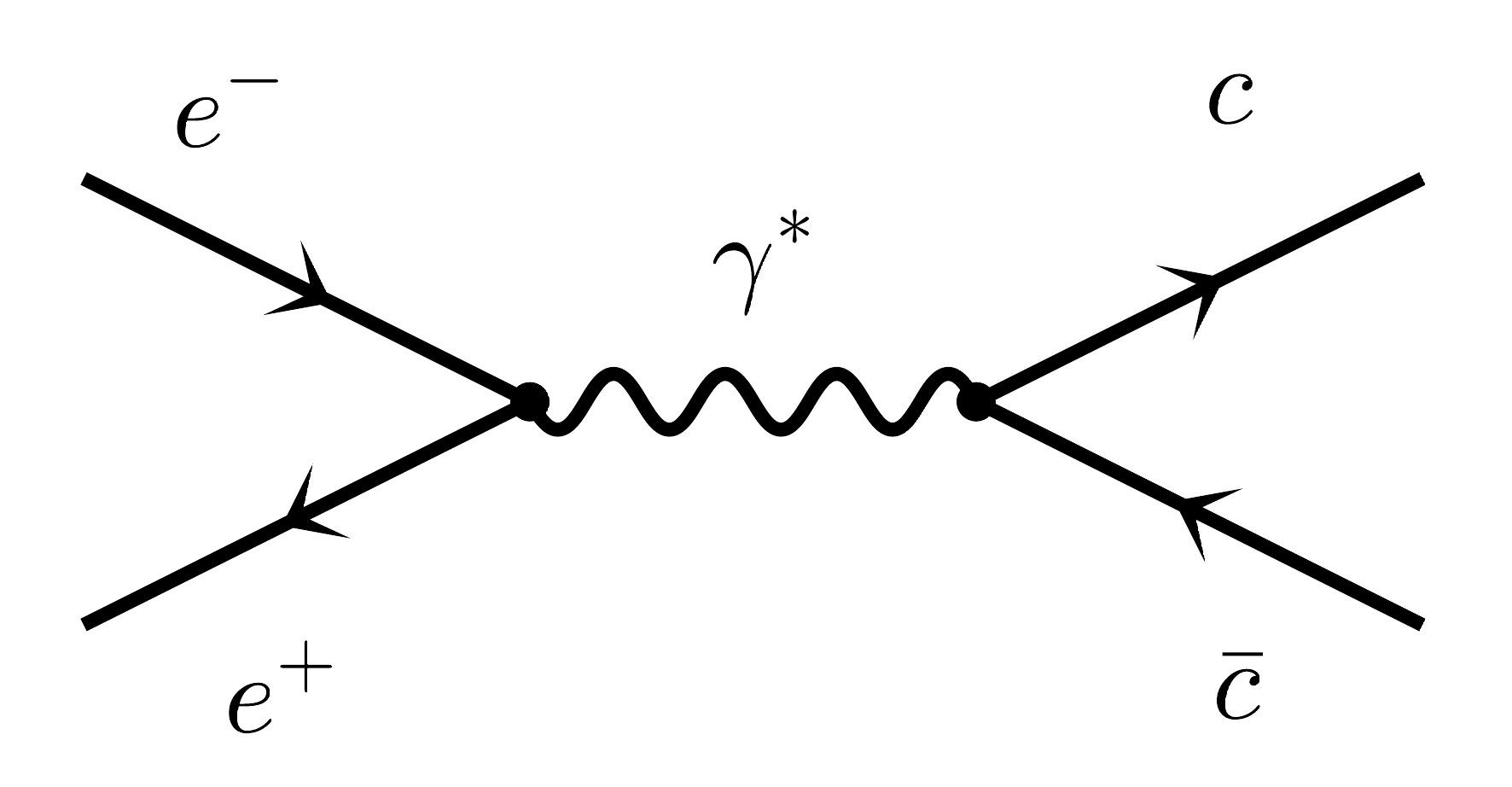}
	\caption{Example of charm-quark continuum production processes - prompt charm production - at electron-positron colliders. The charm anti-charm quarks will hadronise into QCD bound states. Examples are ground-state mesons like $D^0 (\bar{D}^0)$, $D^+ (D^-)$ and $D_s^+ (D_s^-)$, baryons like $\Lambda_c^+ (\Lambda_c^-$) or excited mesons $D^{*0} (\bar{D}^{*0})$, $D^{*+} (D^{*-})$ and $D_s^{*+}(D_s^{*-}$). Unless the collision energy is at threshold for pair production, both sides can hadronise into different charm particle species.
    }
	\label{sec2.2:fig3}
\end{figure}

\begin{figure}
    \centering
    \includegraphics[scale = 0.65]{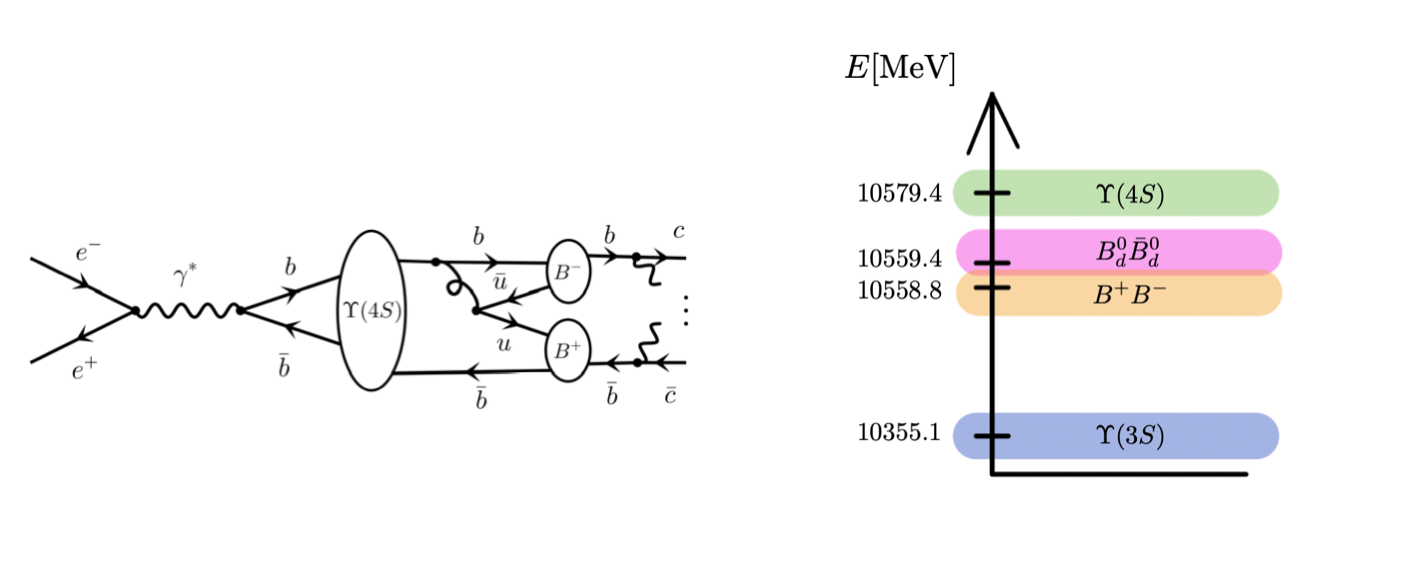}
    \caption{(Left) Secondary charm production at $b$-factories operating at the $\Upsilon(4S)$ resonance: the $\Upsilon(4S)$ decays into $B^+ B^-$ (and also $B_d^0 \bar{B}_d^0$) mesons, which subsequently decay into charmed hadrons. (Right) Illustration of the most common energy thresholds: the mass of the $B_d \bar{B}_d$ and $B^+ B^-$ meson pairs is just below the mass of the $\Upsilon(4S)$ resonance.}
    \label{sec2.2:fig3p}
\end{figure}

\noindent
Currently, charm physics is intensively studied at several experiments, at the LHCb experiment at the LHC, CERN, 
Switzerland~\cite{LHCbWeb, LHCb:2008vvz, LHCb:2014set}, see Fig.~\ref{sec2.2:fig1}, at the Belle~II experiment at the SuperKEKB collider in Tsukuba, Japan~\cite{Belle2Web, Belle-II:2010dht}, and at the BESIII experiment at the BEPCII accelerator in Beijing, China~\cite{BESIIIWeb, BESIII:2009fln}, both shown in Fig.~\ref{sec2.2:fig2}. The other three big experiments at the LHC i.e.\ ATLAS, CMS and ALICE, have also contributed to charm physics with measurements complementary to the LHCb's programme.  
At all these experiments, charm particles are produced in 
different processes and at different centre-of-mass energies, with the charm {\it cross-sections} giving the measure of 
the corresponding production rates. 
Charm particles can be produced as a direct result of the beam collision, referred to as {\bf prompt production}, see Fig.~\ref{sec2.2:fig3}, or as a 
result of the decay of another heavier particle,
e.g.\ of a $B$ meson, 
that was produced in the collisions. The latter case is referred to as {\bf secondary charm production}, see Fig.~\ref{sec2.2:fig3p} (left). 
\\

\noindent
{\bf Charm at the asymmetric $\mathbf b$-factories:} The $b$-factories, Belle and BaBar, were primarily designed to study CPV in $B$-meson decays,
but have also produced numerous charm physics results. 
Electron-positron annihilation at a centre-of-mass energy corresponding to the mass of the $\Upsilon(4S)$ resonance 
is the main mechanism for producing entangled pairs of $B$ mesons, as the mass of $\Upsilon(4S)$ is sufficient to produce only a $B^+B^-$  or $B^0_d \bar{B}^0_d$ pair.
A significant number of charm particles can be created through secondary production
via decays of $B$ mesons produced at the $\Upsilon(4S)$ threshold, see Fig.~\ref{sec2.2:fig3p} (right). However, the slightly more dominant production mechanism at the $b$-factories is prompt production via the continuum process $e^+e^-\rightarrow c\bar{c}$ (non-resonant $e^+e^-$ annihilations), sketched in Fig.~\ref{sec2.2:fig3}.
At the $b$-factories 
the electron-positron collisions are asymmetric in energy in order to ensure that the $\Upsilon(4S)$ is not created at rest~\cite{Oddone:1987up}. This energy asymmetry ensures that subsequent $B$-meson decays are boosted in the laboratory frame. In this way the decay length of the $B$ mesons could be measured and the time-dependent CP-violating effects in the $B$-meson system would not vanish. The beam energies at PEPII for BaBar were, respectively, 9.0 GeV for the $e^-$, and 3.1 GeV for $e^+$; and at KEKB for Belle they were 8.0 GeV for the $e^-$ and 3.5 GeV for the $e^+$. The upgrade of Belle to Belle II was made with the aim of collecting 50 times more data than its predecessor. This will be achieved by significantly improving the accelerator complex, as summarised in~\cite{Aihara:2024zds}. SuperKEKB has already reached the world record instantaneous luminosity of $5.1\times10^{34} cm^{-2}s^{-1}$ in December 2024, with an ultimate target of $6\times10^{35} cm^{-2}s^{-1}$. The beam energies of Belle~II are respectively 7.0 GeV for the $e^-$, and 4.0 GeV for the $e^+$. The positron beam energy was changed to 4.0 GeV in order to increase the beam lifetime while accepting a smaller boost~\cite{BelleIIbeam}.\\

\noindent
{\bf Charm at threshold:}
BESIII is an experiment at the symmetric electron-positron collider BEPCII in Beijing,
where data is collected at a range of different centre-of-mass energies of about several GeVs. For each specific dataset collected, the beam energies are changed and tuned to the collision energy point of interest. Each of the samples can then be used for unique studies, above, at, and below the open charm thresholds, cf.\ Fig.~\ref{sec1:fig1p}.
\begin{figure}
	\centering
       \includegraphics[width=0.6\textwidth]{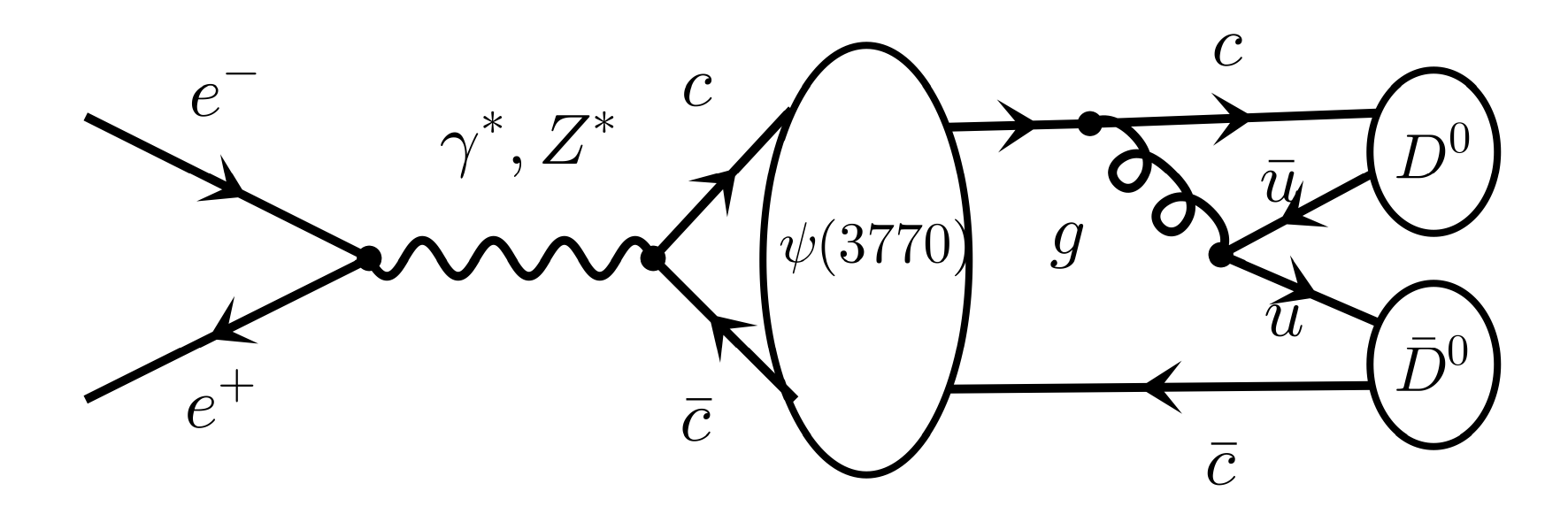}
	\caption{Production mechanism for quantum entangled charm meson pairs at the $\psi(3770)$ threshold at electron positron colliders. The mass of the  $\psi(3770)$ resonance is just slightly higher than the mass of the $D^0 \bar{D}^0$ pair.}
	\label{sec2.5:fig2}
\end{figure}
\noindent
An example is the production of $D^0\bar{D}^0$/$D^+D^-$ at the $\psi(3770)$ threshold via the process $e^+e^-\rightarrow c\bar{c}$, see Fig.~\ref{sec2.5:fig2}, as the resonance $\psi(3770)$ is the lightest state which can decay into pairs of neutral or charged charmed mesons. We note already here that at the $\psi(3770)$ threshold, neutral charm mesons are produced in a {\it quantum entangled state} which provides unique access to information about hadronic decay parameters, as discussed in Section~\ref{subsec:entanglement}.
At BESIII, the data sample for $D_s$ meson studies is collected in the energy range [4.128 - 4.230] GeV, slightly above the $D_s^+D_s^{*-}$ threshold, 
and a small sample has been collected at 4.008 GeV, slightly above the $D_s^+D_s^{-}$ threshold, cf.\ Fig.~\ref{sec1:fig1p}. 
In general, $D_s$ mesons can be produced at the threshold energy for $D_s^+D_s^-$ pairs,
however, as it was shown in~\cite{CLEO:2008ojp}, the cross-section for the $D_s^{*+}D_s^-$ production is greater than that of $D_s^{+}D_s^-$, namely $\sigma(e^+e^-\rightarrow D_s^{*+}D_s^-) > \sigma(e^+e^-\rightarrow D_s^{+}D_s^-)$, see Fig.~\ref{sec2.2:fig4}~(left). 
Specifically, in~\cite{CLEO:2008ojp} it was determined that the largest cross-section for $D_s$ mesons is at 4.17 GeV, as indicated by the green data points in Fig.~\ref{sec2.2:fig4}~(left), 
so that using the threshold for $D_s^{*+}D_s^-$ pair production\footnote{ $m_{D_s^{*+}} + m_{D_s^{-}}$ = 4080.55(50) MeV and $2 m_{D_s^{*+}}$ = 4224.4(8) MeV.}, where the $D^{*+}$ meson decays as $D^{*+}_s\rightarrow \gamma / \pi^0 D_s^+$, results in the most abundant source of charm-strange meson pairs.
\begin{figure}[t]
	\centering       \includegraphics[width=0.52\textwidth]{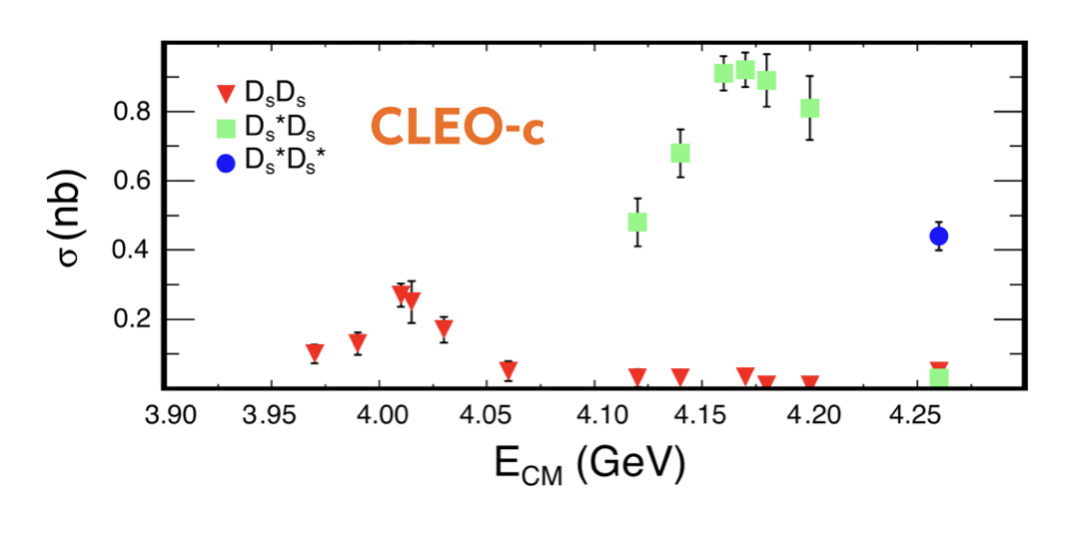}
 \begin{overpic}[width=0.45\linewidth]{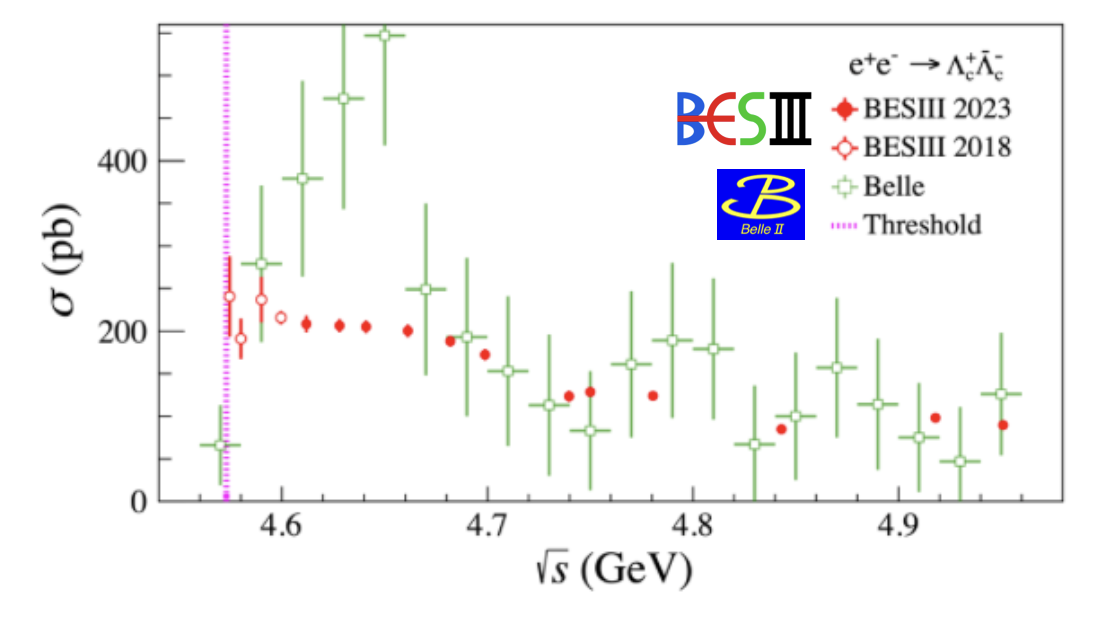} \put(137,73){\includegraphics[width=0.04\linewidth]{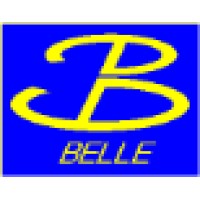}}
  \end{overpic}      	
  \caption{Cross-sections 
    at electron-positron colliders as a function of the centre of mass energy $E_{\rm CM}$ or $\sqrt{s}$ for: (left) strange charm-meson pairs measured by the CLEO collaboration and reprinted from~\cite{CLEO:2008ojp} with permission - $D_sD_s$ (red), $D_s^*D_s$ (green) and $D_s^*D_s^*$ (blue); (right) charm baryon anti-baryon pairs as measured by the BESIII experiment~\cite{BESIII:2023rwv} (red), compared to the measurements from the Belle collaboration~\cite{Belle:2008xmh} (green). The non-flat structure of the Belle data in green corresponds to the resonant structure Y(4630), not confirmed by BESIII. Taken from~\cite{BESIII:2023rwv}.}
    
 \label{sec2.2:fig4}
\end{figure}
Other important data samples produced at BESIII are used for studies of the lowest lying charmed baryons $\Lambda_c^+$. Pairs of $\Lambda_c^+ \Lambda_c^-$ are produced at threshold
in the energy region [4.6 - 4.95] GeV. The $\Lambda_c^+$ is the lightest charm baryon, and until 2014, when BESIII started taking data at the 4.6 GeV threshold, the information about its branching fractions was limited. Since it is likely that $\Lambda_c^+$ baryons appear as an intermediate state in heavier baryon decays, the knowledge of their branching fractions is particularly important. The production cross sections of charmed baryon $\Lambda_c^\pm$ pairs in electron-positron collisions has been measured with a high precision by the BESIII experiment~\cite{BESIII:2023rwv}, see Fig.~\ref{sec2.2:fig4}~(right). Other key samples collected with the BESIII detector include about 10 billion $J/\psi$ events, 2.7 billion $\psi(2S)$ events, and a scan of about 150 different centre-of-mass energy points in the range [3.8-5.0] GeV used for searches of exotic hadrons with hidden charm.  \\

\begin{figure}[t]
	\centering
    \includegraphics[scale = 0.5]{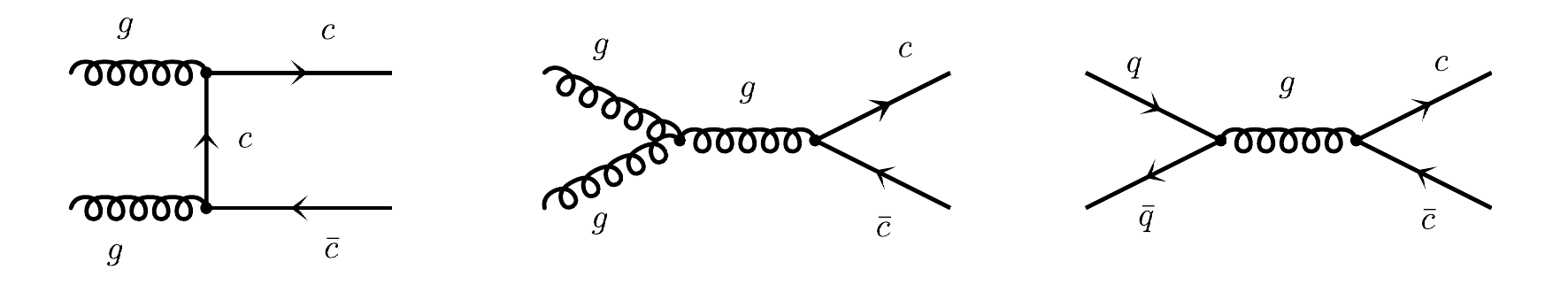}
	\caption{Examples of charm-quark production processes at hadron colliders.  The charm anti-charm pair subsequently  hadronises.}
	\label{sec2.2:fig5}
\end{figure}

\noindent
{\bf Charm at hadron colliders:} 
Charm quarks can also be produced at proton (anti-)proton colliders, see Fig.~\ref{sec2.2:fig5}. Here, the heavy-quark flavour production occurs primarily due to gluon-gluon fusion, shown in the left and the middle diagrams of Fig.~\ref{sec2.2:fig5}. These experiments present the significant advantage of yielding production cross-sections several orders of magnitude greater than those achievable in electron-positron machines, at the price of generating  higher backgrounds and events complexity. 
In this case, in fact, having fast and efficient triggers becomes essential~\footnote{Because of the high data rates and of other limitations like available storage, computing power etc., not all data can be recorded. Modern particle physics experiments use trigger systems to make fast decisions about which events to record and which not. These can be implemented on several levels, both hardware- and software- based.}. Moreover, due to the high collision energies, all different species of charmed hadrons can be created.\\

\noindent
The major experiment for flavour physics at hadron colliders is LHCb, which is 
 a dedicated experiment for $b$- and $c$-physics. At LHCb, pairs of heavy $c\bar{c}$ quarks with large cross-sections are produced at unprecedented rates in inelastic proton proton collisions. Data was collected at centre-of-mass energies of 7 TeV in 2011, 8 TeV in 2012 - both periods are referred to as Run 1 - and at 13 TeV in 2015-2018 - Run 2, followed by the Run 3 period which started in 2022 and is currently planned until 2026. In the long shutdown period before Run~3, the LHCb experiment underwent a major upgrade \cite{LHCb:2023hlw}.  
 The production of charmed mesons can also be studied at the ATLAS, CMS and ALICE experiments at CERN~\footnote{Charm meson production is also studied in nucleus-nucleus collisions at SPS energies by the NA61/SHINE experiment at CERN~\cite{Rybicki:2024mfy}. The experiments at the heavy ion collider RHIC at BNL: STAR and PHENIX also contribute to charm production studies~\cite{Zhou:2017ikn, PHENIX:2022wim, STAR:2019ank}. }, in either proton proton collisions or heavy ion collisions, with ALICE being dedicated to heavy ion and quark-gluon plasma physics, and ATLAS and CMS being general performance detectors. For the latter two, the lack of hadron particle identification limits the type of charm physics measurements they can perform.\\

\noindent
The four big LHC experiments have provided open charm production cross-section energies
in the range of [2.96 - 13] TeV~\cite{ALICE:2012inj, LHCb:2016ikn, CMS:2017qjw, CMS:2019uws, ALICE:2019nxm, ALICE:2020wfu, ALICE:2020wla, ALICE:2021psx, ALICE:2021mgk, ALICE:2022exq, ALICE:2012mhc, ALICE:2012gkr, ATLAS:2015igt, ALICE:2017olh, ALICE:2017thy, LHCb:2015swx, ALICE:2021bli, ALICE:2021rzj, CMS:2021lab, ALICE:2022cop, LHCb:2023kqs, LHCb:2022qvj, LHCb:2023cwu, LHCb:2022cul, LHCb:2022ddg}. 
Among these, \cite{CMS:2019uws, ALICE:2020wfu, ALICE:2020wla, ALICE:2022exq, ALICE:2017thy, ALICE:2021psx, ALICE:2021bli, ALICE:2021rzj, ALICE:2022cop, LHCb:2023cwu, LHCb:2022ddg} reported measurements of the charm baryons $\Lambda_c^+$, $\Xi_c^0$, $\Xi_c^+$, $\Xi_c^{++}$, $\Omega_c^0$ cross-sections. 
At ATLAS, CMS and ALICE, open charm cross-section measurements were taken in the central rapidity region $|y|< 2.5$~\footnote{The rapidity of a particle is defined as $y=(1/2) \ln\big((E+p_z)/(E-p_z)\big)$, where $E$ is the energy of the particle, and $p_z$ is its longitudinal momentum.} 
whereas at LHCb, the cross-sections were measured in the forward rapidity region $2 < y < 4.5$, hence providing a complementary coverage compared to the other three LHC experiments.
This is particularly valuable since cross-sections of open charm at LHCb correspond to measurements at very low values of the fraction
of the total hadron momentum carried by the interacting parton, i.e.\ at $x < 10^{-6}$, where $x$ denotes the Bjorken scaling variable, and these are, for instance, 
important inputs for fits/constraints of parton density functions~\footnote{The parton density functions describe the probability
of finding a parton, i.e.\ quark or gluon, with momentum fraction $x$ inside
a hadron. They
encode intrinsic properties of the partons and cannot be calculated in perturbation theory.}. The combination and QCD analysis of charm production cross-sections is performed by the PROSA collaboration~\cite{PROSA:2015yid}. Note that measurements of charm cross-sections not only allow to test perturbative QCD calculations and to evaluate rates of processes at colliders, but they are also important for high-energy astroparticle physics e.g.\ for evaluating background for atmospheric neutrino experiments from atmospheric charm production~\cite{Garzelli:2016xmx, Zenaiev:2019ktw}.\\

\noindent
In heavy ion collisions, the charm production can be studied via e.g.\ proton lead and lead lead collisions, and in ``fixed target mode". These experiments allow to look for signatures of quark gluon plasma - a state the Universe was in shortly after the Big Bang, where quarks and gluons were not confined. If quark gluon plasma was produced in heavy ion collisions, the formation of $c\bar{c}$ bound states would be suppressed by a mechanism called ``colour screening".
The ``fixed-target mode" is specific for LHCb, and this technique uses the experiment's internal gas target: the SMOG (system for measuring the overlap
with gas)~\cite{LHCb:2014vhh}. Open charm cross-section measurements in such an environment allow to probe the nuclear partonic structure and study the hadronisation in a different regime, while charmonia production (suppression) studies can test the presence of nuclear effects, as well as of a hot nuclear medium~\footnote{A comprehensive study of the effect of medium-induced gluon radiation on the parton splitting function $g \to \bar c c$ in the QCD plasma was performed in~\cite{Attems:2022ubu}.}. 
The SMOG system was initially designed for precise luminosity measurements but was re-designed to allow LHCb to collect data in a fixed-target mode. 
SMOG allows the injection of noble gases such as neon, argon, helium which then are used as a gaseous target system at LHCb.
Results for charm production in lead neon and proton neon collisions with SMOG can be found in~\cite{LHCb:2022qvj, LHCb:2022cul}. \\

\begin{figure}
    \centering
   \begin{minipage}{0.45\textwidth}
    \centering
    \includegraphics[width=1\textwidth]{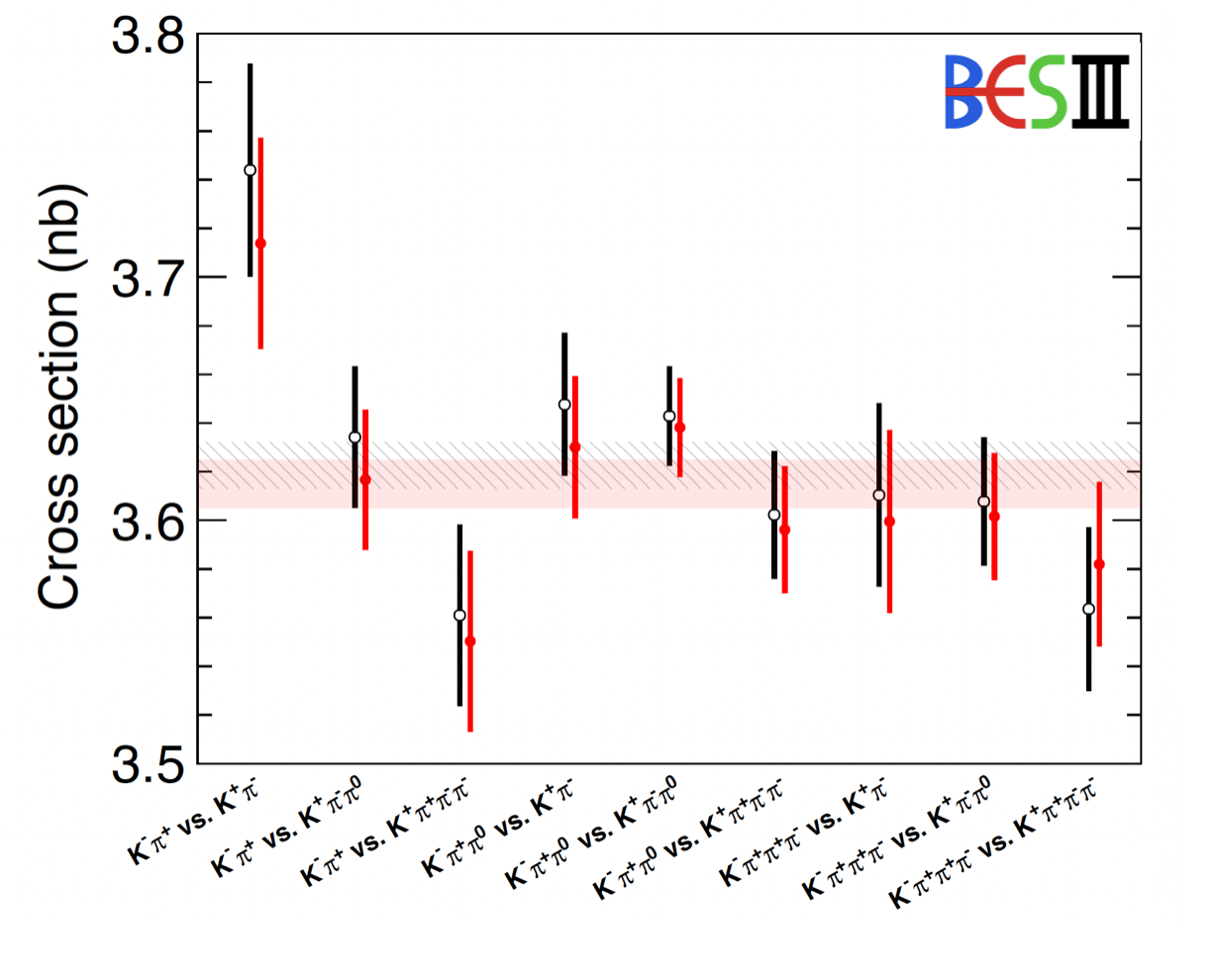} 
    \includegraphics[width=1\textwidth]{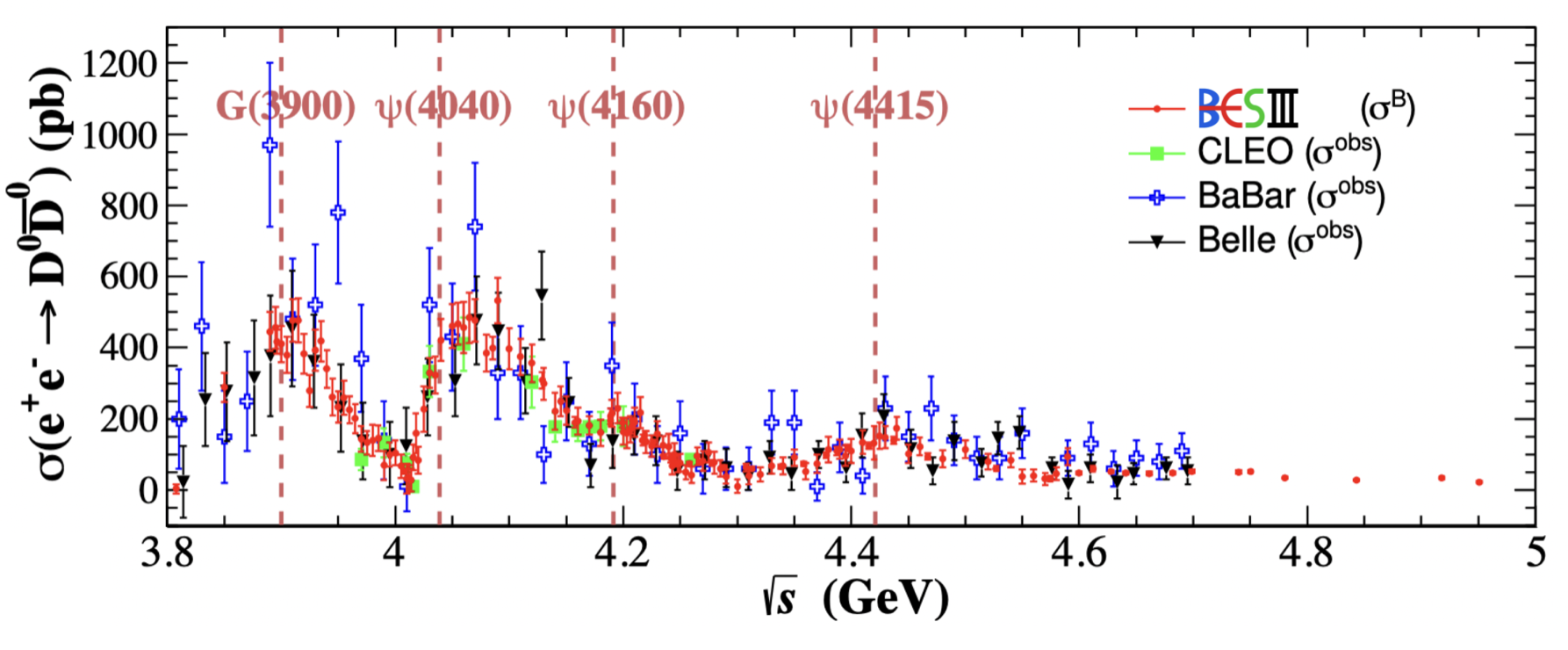}
\caption{(Top) Production cross-sections of $D^0\bar{D}^0$ pairs at the $\psi(3770)$ resonance  as a function of the tag mode at BESIII~\cite{BESIII:2018iev}, with and without quantum correlation corrections, shown in red and black respectively; the horizontal bands show the average values for all tag modes, as explained in Sections~\ref{subsec:tagging}. (Bottom) The production cross-sections at higher energies are measured in the centre-of-mass energies range from 3.8 to 5 GeV at BESIII, and the results are compared with those of CLEO, BaBar and Belle experiments. The vertical red dashed lines correspond to the masses of the charmonia states shown in the plot~\cite{BESIII:2024ths}.}
\label{sec2.2:fig7}
\end{minipage}
\qquad
\begin{minipage}{0.5\textwidth}
    \centering
\begin{overpic}[width=1.0\linewidth]{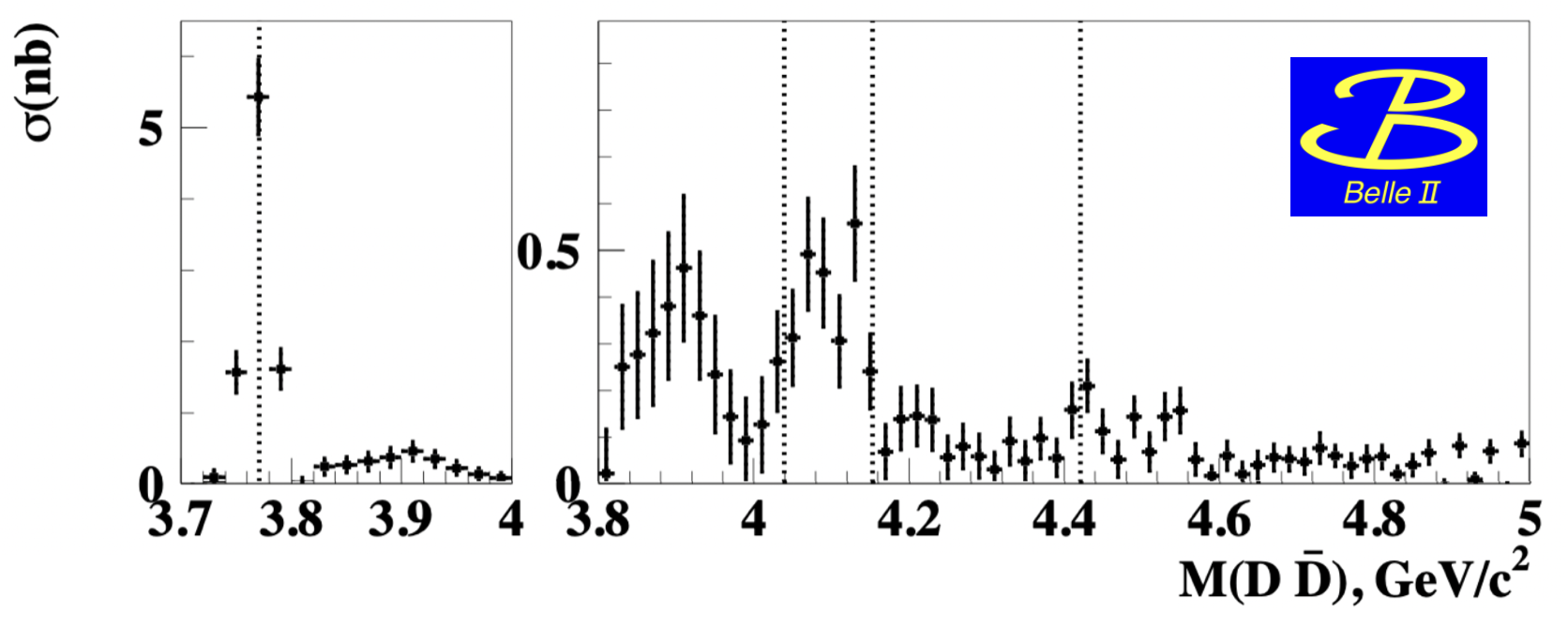} \put(192,55){\includegraphics[width=0.15\linewidth]{belle-logo.jpeg}}
  \end{overpic}    
    \includegraphics[width=1\textwidth]{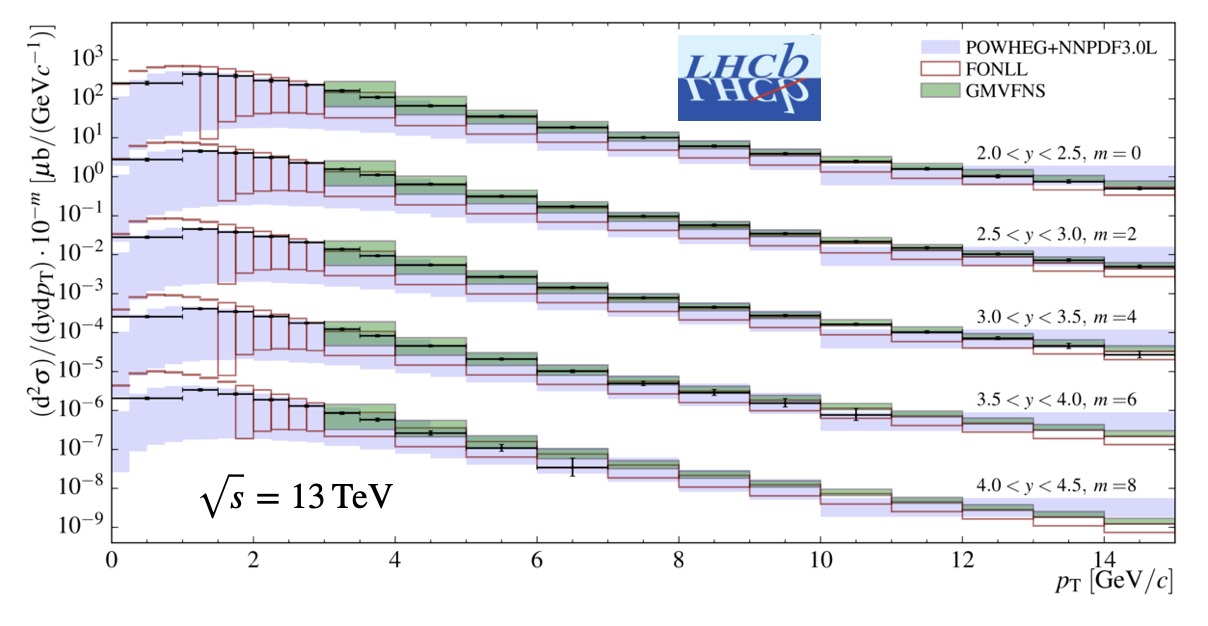}
   \caption{(Top) Production cross-sections of promptly produced $D^0\bar{D}^0$ pairs at Belle, at centre-of-mass energies ranging from the $D^0\bar{D}^0$ threshold to 5~GeV, with the vertical lines corresponding to the $\psi(3770)$, $\psi(4040)$, $\psi(4160)$ and $\psi(4415)$ resonances (reprinted from ~\cite{Belle:2007qxm} with permission). (Bottom) Differential cross sections of promptly produced $D^0$ mesons as a function of the $D^0$ transverse momentum at centre-of-mass energies of 13 TeV at the LHCb experiment. Each curve corresponds to one rapidity interval (bin), where five different bins have been considered. The experimental data points are shown in black and are compared with different theoretical predictions, shown respectively in violet, red and green (and labelled in the legend of the figure)~\cite{LHCb:2015swx}. }
\label{sec2.2:fig8}
\end{minipage}
\end{figure}

\noindent
{\textbf{Comparison between the main active flavour physics experiments:}} 
Several important features are required to ensure the success of a flavour physics experiment. First, an excellent {\it vertex reconstruction}, which allows to identify the heavy flavour particles from the prompt collision, as well as to distinguish the secondary from promptly produced particles. This is characterised by the vertex resolution
 and by the impact parameter resolution \footnote{The performance numbers quoted below are sourced to give the reader an appreciation of scale. There is enough variation between detectors that numbers quoted here should be used only as a guide. The primary vertex resolution at LHCb in the transverse plane is 13 $\mu m$, while the resolution in the longitudinal direction $z$ is 71 $\mu m$ (for a vertex of about 25 tracks)~\cite{LHCb:2018zdd}. The impact parameter is usually of $\mathcal{O}(10-100)\,\mu m$ and measures the transverse distance of closest approach of a track to the primary vertex. This parameter scales with $1/P_T$, meaning tracks with the highest transverse momentum have the best impact parameter resolution. The boosted direction has a much larger resolution across all flavour detectors.}. By constraining the flight distance of heavy flavour particles, experiments can more precisely access their lifetime.
Second, an excellent {\it momentum resolution}, translating into mass resolution, which allows to separate the different heavy flavour particle species e.g.\ $B^0$ from $B^0_s$. \footnote{The momentum resolution is commonly around 0.3-0.5\,\% of the total momentum for low momentum systems (1\,GeV), and up to 1\,\% for higher momentum across Belle II, BESIII and LHCb.}
Third, an excellent {\it particle identification}, i.e.\ $>$ 90\%, with a few percent mis-identification rates, which allows to distinguish different species of particles and suppress background levels. Hadron separation such as kaon - pion separation is particularly important for CPV studies. Fourth, {\it high trigger and reconstruction efficiencies}, which are essential to maximise the number of signal yields \footnote{The trigger efficiency characterises the ability of the trigger system to quickly identify correctly physics events of interest, and to record them for subsequent analysis offline (as not all events can be recorded due to storage, computing limitations and rates). The reconstruction efficiency measures how well the detector identifies and reconstructs the trajectories of the particles interacting with the material of the detector. The process starts from the detector readout, and transforms this into something more understandable i.e.\ ``hits" identifying where the particle passed. The ``hits" are then connected to build segments of tracks in e.g.\ silicon tracking detectors, or clustered to reconstruct an energy deposit e.g.\ in a calorimeter or scintillating detector, after which the information from the various sub detectors is combined to reconstruct an electron, charged hadron, muon, etc.}.
Moreover, for time-dependent studies, excellent {\it time resolution}~\footnote{The time resolution depends on the measurements, however, to give a rough idea, the typical decay-time resolution at LHCb is about 40-50 fs, and the average decay-time resolutions for $D^0$ and $ D^+$ are of about 70 fs and 60 fs~\cite{Belle-II:2021cxx}, 108 fs for $D^+_s$ decays~\cite{Belle-II:2023eii} at Belle II.} is imperative, 
as well as a significant boost of the particles. 
Finally, for studies of final states with neutral particles, precise {\it reconstruction of the energy and position of photon and neutral particles}, such as $\pi^0$, is essential.  \\

\noindent
Currently, the main active charm-physics experiments are BESIII~\cite{BESIII:2018iev}, Belle/Belle II~\cite{Belle:2007qxm} and LHCb~\cite{LHCb:2015swx}. 
They show both similarities and differences, providing complementary information on the study of charmed hadrons. 
While at BESIII and Belle II the cross-sections are of the order of a few nb, at LHCb they are $10^6$ times higher, see Figs.~\ref{sec2.2:fig7}, \ref{sec2.2:fig8}. Moreover, at BESIII the charm mesons are produced at threshold, and practically at rest, so the decay-time information is not accessible. 
Even though the collision energy is relatively low at the Belle/Belle II experiments, the produced particles are highly boosted by the use of asymmetric electron-positron beams. On the other hand, at LHCb, the boost is very high, a consequence of both the hadron collider production, see Fig. \ref{sec2.2:fig1}, and because the centre-of-mass energy is in the TeV range. 
Due to the high boost that causes the charm hadrons to fly a certain distance before decaying, Belle/Belle II and LHCb are excellent for time-dependent studies such as for measurements of lifetimes, neutral-meson mixing, and CPV, see Section~\ref{sec3}. 
At LHCb, pairs of $c\bar{c}$ quarks are produced in the forward (and in the backward) direction. This is the reason for designing LHCb as a forward spectrometer. In contrast, BESIII and Belle/Belle II are detectors with cylindrical symmetry and 4$\pi$ acceptance coverage surrounding the interaction point. At BESIII and Belle/Belle II, the environment is very clean and the background is low. This is particularly true for BESIII, because of the production at threshold, and the constraints on the full kinematics. 
The clean environment of both experiments makes them particularly suitable for studies with neutral particles. 
In summary, a comparison of the main features of the LHCb, Belle/Belle II and BESIII experiments is shown in Table~\Ref{sec2.2:tab1}. 

\begin{table}[h]
    \centering
    \small
    \renewcommand*{\arraystretch}{1.3}
    \begin{tabular}{|c||c|c|c|c||c|c|c|c|}
\hline
        Exp. & Collider & Centre-of-mass & Cross- & Integrated lumi. & Time-dep. & Full kin. & Backgr. & Neutrals \\
         & type &  energy & section & [$\sim \# c\bar{c}$ pairs] &studies & constr. & & efficiency\\
         \hline
         \hline
        LHCb &  & 7, 8 TeV & 1.6 mb & 3 fb$^{-1}$&  &  &  &   \\
                 &  &  &  & [$\sim 5\cdot10^{12}$]
                 & & &  &   \\[2mm]
        forward & $pp$  & 13 TeV & 2.4 mb & 6 fb$^{-1}$& yes & no & high  & low  \\
        spectrometer                  &  &  &  & [$\sim 10^{13}$]
                 & & &  &   \\[2mm]
                  &  & 13.6 TeV & $>$ 2.4 mb& $ 9~ \rightarrow 50$ fb$^{-1}$& & &  &   \\
&  &  &  & [$\rightarrow \sim 10^{14}$]
                 & & &  &   \\
\hline
         \hline
        Belle/ &  & 10.58~GeV@$\Upsilon(4S)$& & 1/ $\sim 0.5 \to $  50 ab$^{-1}$ &  & &   & \\
         Belle II &  &  &  &  & & &  &   \\
 &  & $D^0$ & 1.1 nb & [$\sim 10^9/ 10^9 \to 10 ^{11}]$ & & &  &   \\
        & asymmetric  & & &  & yes & no & low & high  \\
hermetic & $e^+e^-$ & $D^+_{(s)}$ & 0.6 (0.3) nb & [$\sim 10^9/ 10^8 \to 10^{10}]$  & & &  &   \\
detector(s)        &  & & &  & & &  &   \\
 &  & $\Lambda_c$ & 0.2 -- 0.47 nb & [$\sim 10^8/ 10^7 \to 10^{9}]$ & & &  &   \\
         \hline
\hline
        BESIII & & 3.8~GeV$@\psi(3770)$ &  & 20 fb$^{-1}$ &  &  &  &    \\
 %        \hline        
         
         &   & $D^0\bar{D}^0$, $D^+D^-$ & 3.6, 2.8 nb & $[\sim10^8]$ &  & &  &    \\[2mm]
 hermetic & symmetric  & 4.13 -- 4.23 GeV &  & $7.3$ fb$^{-1}$ & no & yes & low  & high  \\
          detector & $e^+e^-$  & $D^*_sD_s$ & 1 nb & $[\sim 5\cdot10^6]$ &  &  &  &   \\[2mm]
 %          \hline        
         
         &  &   4.6 -- 4.7 GeV &  & 4.5 fb$^{-1}$ & & &  &    \\
                     &  & $\Lambda_c\bar{\Lambda}_c$ & $\sim$ 0.2 nb & $[\sim 0.8\cdot10^6]$ & & &  &   \\
%                    &  & & & & &  &  &  &   \\

\hline
    \end{tabular}
    \caption{A summary of the main active charm physics experiments LHCb, Belle/Belle II and BESIII, comparing features such as the type of collider, the centre-of-mass energy, the cross-section either for $c\bar{c}$ pairs, in the case of LHCb, or for specific charm particle species, in the case of Belle/Belle~II and BESIII, and the integrated luminosity of the collected or  planned (denoted by $\rightarrow$) data samples with the approximate number of $c\bar{c}$ pairs produced (the numbers in brackets accompanied by $\rightarrow$ correspond to future planned data samples). In the last four columns further properties of the detectors are compared, namely whether they are suitable for time-dependent studies, if they allow for full kinematic constraints, how large is the background level, and the efficiencies in analysing final states with neutral particles. The number of $c\bar{c}$ pairs is taken from~\cite{Li:2024bmx, MarcoNeckarzimmern}. 
    }
    \label{sec2.2:tab1}
\end{table}

\noindent
\begin{center}
$\sim \circ \sim \circ \sim \circ $ \,{\bf Overview of early charm physics experiments}\,
$\circ \sim \circ \sim \circ \sim $
\end{center}

\noindent
Early charm physics experiments were important to establish the charm particles production, behaviour and properties such as mass, lifetime, decays and branching fractions, etc.
A great number of experiments took place at  Fermilab, USA, briefly summarised in~\cite{Appel:2000rq}. Among these, a series of experiments was built~\cite{Bodnarczuk:2008eh},  where each of the later versions was an upgrade of the previous detector, starting with the E516~\cite{Estabrooks:1983hhp}, upgraded to E691, also known as the Tagged Photon Spectrometer~\cite{TaggedPhotonSpectrometer:1987poq}, the E678~\cite{E-687:1990ses}, and the E791~\cite{E791:1992tvf}. The experiments E516 and E691 used photo-production, where charm particles were produced as a result of shooting a high-energy photon beam incident on a beryllium target, while the E769 and E891 experiments used hadron beams. Other early charm experiments at Fermilab were the E400~\cite{Coteus:1986ar}, which studied charmed particle production from neutrons, the E653~\cite{Kodama:1990zy}, that studied charm particles produced by high-energy pion and proton beams with a hybrid emulsion spectrometer, the photo-production experiment E687~\cite{E-687:1990ses}, later upgraded to the high-intensity FOCUS experiment (also known as E831)~\cite{FOCUS:2000kxx}, and the E743 experiment, that used a proton beam on a fixed target and a liquid hydrogen bubble chamber (LEBC) with the Fermilab multi-particle spectrometer~\cite{LEBC-MPS:1986hmu}. The SELEX experiment E781 was built to study charm baryon physics, for which a charged hyperon beam at 600 GeV was used on copper or diamond targets.\\

\noindent
At CERN, several experiments carried out studies of charm particles production and of their properties. These were in the North Area, at the SPS accelerator, e.g.\ the photo-production NA1 experiment~\cite{NA1:1987osl}, the NA11 and the NA32~\footnote{The NA11 and NA32 experiments have been credited for having introduced for the first time the use of silicon detectors, without which modern charm physics would be unthinkable.} (and its upgrade) experiments of the ACCMOR collaboration~\cite{ACCMOR:1983pmd, ACCMOR:1987rcu}, the NA14~\cite{Filippas:1987ky}, the NA25 (HOBC)~\cite{Cobbaert:1987am}, and the NA27 (LEBC-EHS collaboration)~\cite{LEBC-EHS:1988oic}. In the West Area, there were the experiments WA04~\cite{Roudeau:1980gr}, WA58~\cite{Bologna-CERN-Florence-Genoa-Madrid-Moscow-Paris-Santander-Valencia-Rome:1983yib}, WA75~\cite{Aoki:1988pm}, WA82~\cite{WA82:1992fdk}, WA89~\cite{Simon:1994vw}, WA92 (a.k.a BEATRICE)~\cite{BEATRICE:1998hbq}, etc. 
At DESY, Hamburg, after the early experiments DASP~(Double-Arm-Spectrometer)~\cite{Braunschweig:1974qj} at
the $e^+e^-$ storage ring
DORIS, and PLUTO and TASSO~\cite{TASSO:1983hay} at the $e^+e^-$ storage ring PETRA, the ARGUS experiment at the DORIS-II $e^+e^-$ storage ring~\cite{ARGUS:1988bds} was built to study the properties of beauty and charm quarks.
The KEDR experiment at the VEPP-4M $e^+e^-$ collider at the Budker Institute of
Nuclear Physics in Novosibirsk, Russia~\cite{Anashin:2013twa} was also designed for studying the $c$- and $b$-quarks. \\

\noindent
SLAC was a home of multiple experiments contributing to our understanding of charm physics, which include
the HRS experiment at PEP (Positron-Electron Project)~\cite{Weiss:1982sj}, the Crystal Ball experiment at the SPEAR $e^+e^-$ storage ring~\cite{Partridge:1979tn} (later moved to DESY and BNL), and the 
MARK-I--III experiments at SPEAR~\cite{Bernstein:1983wk}, where several of the first observations of charm particles were made. The MARK-III experiment was optimised to study the decays of $J/\psi$, higher $\psi$ resonances and open charm decays. It pioneered the tagging technique, see Section~\ref{subsec:tagging}, with $D$ meson pairs produced at threshold, and later used at CLEO-c and BESIII.
The main goal of the CLEO experiment was to study
the properties of $B$ mesons at an $e^+e^-$ 
collider, but  with the rise of the $b$-factories 
Belle and BaBar, CLEO's focus shifted to charm
physics and the studies of other, lighter
particles. 
The experiment was upgraded several times and the
last upgrade was dubbed CLEO-c. 
In the years before CLEO-c, the experiment took 
data mainly at centre-of-mass energies 
corresponding to the mass of the $\Upsilon(4S)$ 
resonance.
Smaller data samples were taken at other thresholds
i.e.\ other centre-of-mass energies corresponding 
to the masses of the narrow resonances $\Upsilon(nS)$, $n=1,2, \ldots, 5$,  
or just below those thresholds, i.e.\ via continuum
charm-quark production~\cite{CLEOdata},
see Fig.~\ref{sec2.2:fig3}. 
During the CLEO-c operation, the experiment took 
charm data at threshold. 
Neutral charm meson pairs produced in this way are quantum correlated, and the interesting information which can be measured uniquely in this mode of production will be discussed in Section~\ref{subsec:entanglement}. 
These experiments were also excellent to study decays with missing particles/energy in the final state, e.g.\ neutrinos, $K_L$, etc., as 
the beam energies are precisely known and can be used as a kinematical constraint.
At CLEO-c, several different samples for studying
 open and hidden charm were collected. Data was
 taken at centre-of-mass energy of $\sqrt{s}=3.77$ GeV,  in order to study the properties of the $D^0/D^+$ mesons and at $\sqrt{s}=4.17$ GeV for $D_s$ physics. While $\psi(3770)$, and higher mass resonances, can decay via two charm mesons, the $J/\psi$ and the $\psi(2S)$ resonances are too light, see Fig.~\ref{sec1:fig1p}, and  decay via a $c\bar{c}$ annihilation. 
 To study them, a sample of collision data was taken at the heavier $\psi(2S)$ resonance.  
Moreover, an energy scan in the region [3.97 - 4.26] GeV was also taken.\\

\noindent
Although not primarily designed 
for charm physics, several other experiments studied charm production. Charm quark production, and forward-backward asymmetries for charm pairs production in $e^+e^-$ annihilations were also studies at the VENUS, TOPAZ and AMY experiments at TRISTAN, KEK, Japan~\cite{Takahashi:1985ji} and at LEP, CERN, (at the $Z$ resonance) by the ALEPH~\cite{ALEPH:1998quf}, DELPHI~\cite{DELPHI:1999ets}, OPAL~\cite{OPAL:2003pfe}, L3 experiments~\cite{L3:1992fsb}. At LEP, the production of charm particles was also studied in photon photon collisions~\cite{Hess:2004kk, Ngac:2003cd, L3:2002itq} (ALEPH, L3). Charm spectroscopy and (semi)leptonic branching fractions measurements were performed as well. Charm production was also studied at ZEUS and H1 experiments at the lepton (positron or electron) proton collider HERA, in deep inelastic scattering as well as in photoproduction processes~\cite{H1:2011myz, H1:2005joz, ZEUS:2009nyc, Chekanov:2003xf}.  \\

\noindent
In addition to the above experiments, other interesting cases of charm production are for example the charm studies done at the CHORUS experiment at CERN, primarily built to study $\nu_\mu\rightarrow\nu_\tau $ oscillations. At CHORUS, the charm production induced by high energy neutrino interactions was studied in nuclear emulsions with high spatial resolution~\cite{Panman:2009ica}. The hybrid emulsion detectors were also used at E531 at Fermilab~\cite{Turcotte:1986ur} which, at the time, lead to the best measurement of charmed particle lifetimes.

%%%%%%%%%%%%%%%%%%%%%%%%%%%%%%%%%%%%
%%%%%%%%%%%%%%%%%%%%%%%%%%%%%%%%%%%%
%%%%%%%%%%%%%%%%%%%%%%%%%%%%%%%%%%%%
%%%%%%%%%%%%%%%%%%%%%%%%%%%%%%%%%%%%
\subsection{Identifying the flavour of the charm hadron species}
\label{subsec:tagging}
The flavour of the hadrons produced at the experiments is identified through {\bf tagging}, a procedure which allows us to reconstruct the properties of a specific particle, as well as to reduce the background contributions. For instance, being able to correctly identify the flavour of neutral $D$ mesons is mandatory for CPV and meson mixing studies. 
\\

\begin{figure}[t]
	\centering
       \includegraphics[width=0.35\textwidth]{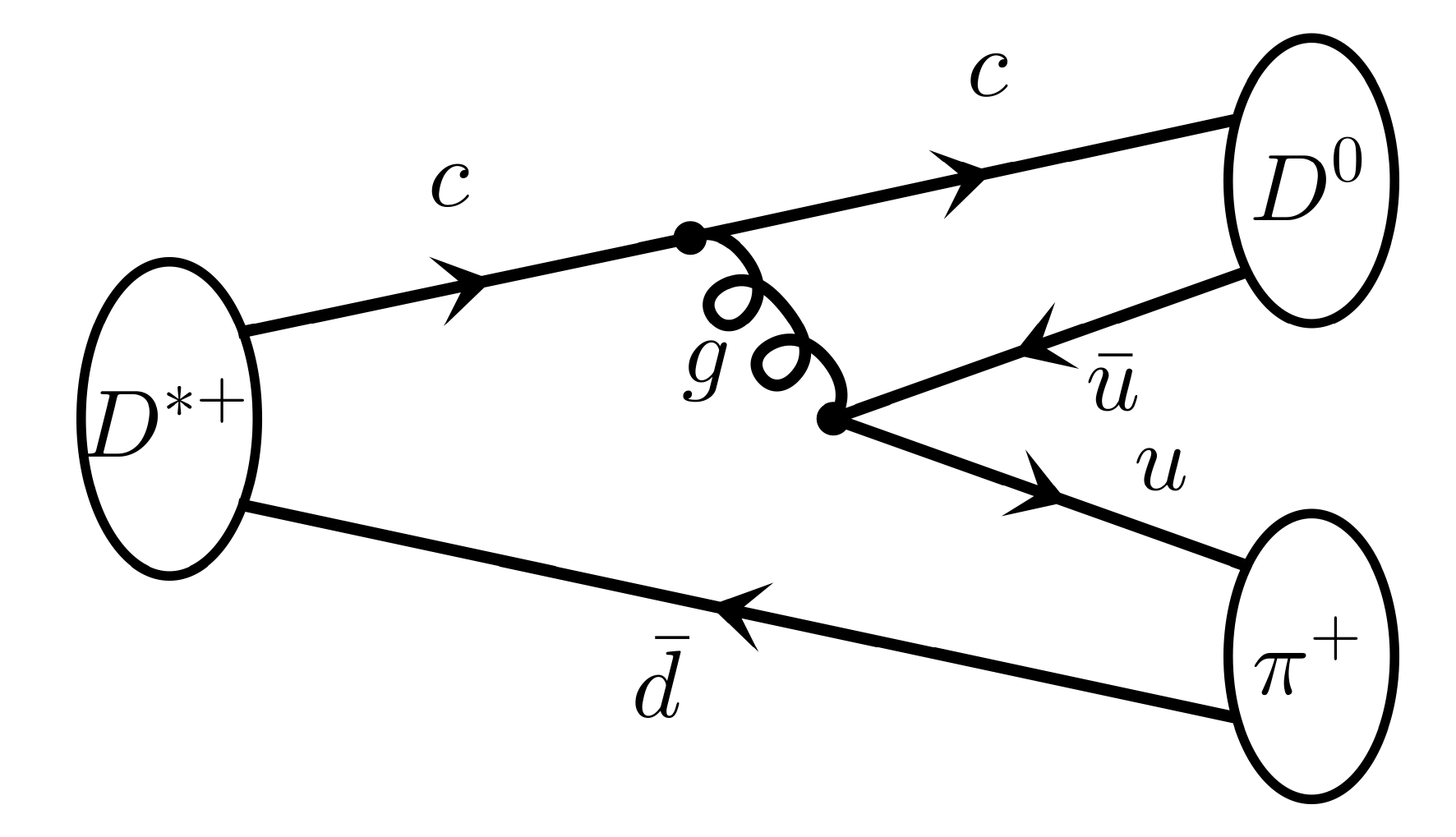}
	\caption{Example of Feynman diagram describing the strong decay of an excited $D^{*+}$ meson into ground state neutral $D^0$ and charged pion $\pi^+$ states.}
	\label{sec2.3:fig0}
\end{figure}

\begin{figure}[t]
	\centering       \includegraphics[width=0.66\textwidth]{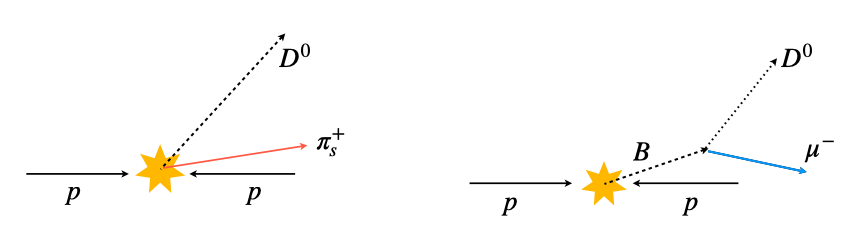}
       \hfill
    \includegraphics[width=0.32\textwidth]{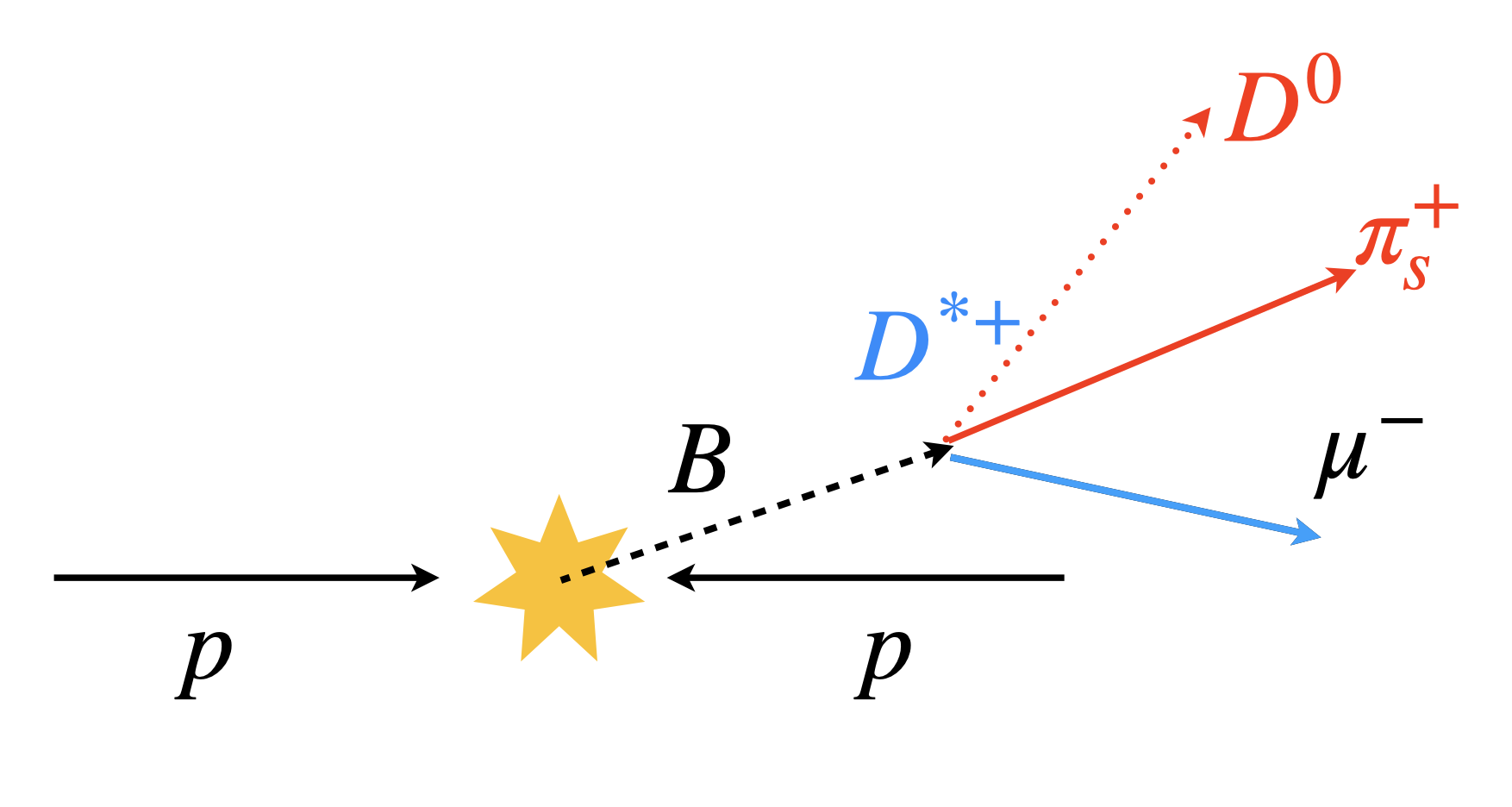}
	\caption{Tagging of prompt (left), secondary (middle) and doubly tagged (right) charm decays in proton-proton collisions. Prompt decays (i.e.\ decays of promptly produced charm mesons) are usually tagged by the charge of the soft pion in the strong decay $D^{*+}\rightarrow D^0 \pi^+$. Decays from secondary production can be efficiently tagged with the charge of the muon in semileptonic $B$-meson decays such as $B \to D^0 \mu^- \bar \nu_\mu X$. These are both examples of singly-tagged decays. Doubly tagged decays such as $ B \rightarrow D^{*+}(\to D^0\pi^+) \, \mu^- \nu_\mu X$ are tagged by both the charge of the soft pion and of the muon. }
	\label{sec2.3:fig1}
\end{figure}

\noindent
The tagging methods depend on the different production mechanisms of the charmed mesons. 
The hadronisation of $c \bar{c}$ into charmed hadrons, one of which could be an excited $D^{*}$ meson, as well as the decay of the excited $D^*$ mesons into ground state $D$ mesons and pions - see Fig.~\ref{sec2.3:fig0} - are QCD processes that happen at times scales below $10^{-20}$\,s, which are far below the experimental time-scale resolution. These interactions will thus look point-like in the detector.
Nevertheless, the flavour of promptly produced $D^0$ mesons in the collisions can be identified via the charge of the pion which is created in the strong decay $D^{*+} \rightarrow D^0 \pi^+$ - and similarly
$D^{*-}\rightarrow \bar{D}^0 \pi^-$ for the $\bar D^0$ mesons, as shown in Fig.~\ref{sec2.3:fig1} (left).
Because of the small mass difference between the masses of the $D^{*+}$ and the $D^0$ mesons (slightly above 140 MeV), the transferred energy is sufficient to produce only one soft pion, where $m_{\pi^\pm} \approx 139.57$ MeV, significantly reducing the background level. 
Note that prompt charm production  actually produces considerably more  $D^0$ mesons, compared to excited  $D^{*\pm}$ mesons. However, tagging is much more difficult for directly produced 
 $D^0$ mesons.
\\

\noindent
In the case of secondary charm production, first,  $B$ mesons are produced in the initial $e^+ e^-$  or hadron collision, then they subsequently decay into $D$ mesons, cf.\ Fig.~\ref{sec2.2:fig3p}.
These latter decays can be, for example, semileptonic $B$-meson decays with large branching fractions of the order of 10$\%$, such as $B \rightarrow D^0 \mu^- \bar \nu_\mu X$ - and $B \rightarrow \bar D^0 \mu^+ \nu_\mu X$, where $X$ stands for a arbitrary
set of unreconstructed particles. Now the charge of the muon arising in the semileptonic decay can be used 
to tag the flavour of the charm meson, see Fig.~\ref{sec2.3:fig1} (middle). 
Both at LHCb and at Belle II, the $B$ mesons are heavily boosted and their relative long lifetime (of the order of 1.5 ps) can be experimentally resolved. Thus the decay vertex of the semileptonic $B$ decay is displaced from the original interaction point.\\

\noindent
Decays from the prompt and secondary production processes described above are called {\bf singly tagged}. In addition, one can consider {\bf doubly tagged} decays, that is
secondary charm production processes where the $B$ meson decays into excited charm mesons,  e.g.\ $B \rightarrow D^{*+}(\to D^0\pi^+) \, \mu^- \nu_\mu X$, which can then be tagged by two particles, namely {\it both} by the charge of the soft pion and  
by the charge of the muon, see Fig.~\ref{sec2.3:fig1} (right). 
\\

\noindent
The $D^{*+}$ tagging technique is used at the LHCb experiment (and historically at CDF), and at the Belle/Belle II (also at BaBar). Only the LHCb experiment uses secondary production --- the relative yields and advantages of using both tagging techniques being outlined below --- although, in principle, this technique could also be used by the $b$-factories.
 To increase the yield of prompt charm samples~\footnote{It is expected that the yield of $D^{*+}\rightarrow D^0(\rightarrow K^-\pi^+)\pi^+$ decays is about 5 times smaller than that of untagged $D^0\rightarrow K^-\pi^+$ decays at Belle/ Belle II~\cite{Belle-II:2023vra}.}, the Belle II collaboration has developed a more complex tagging technique based on machine learning methods. The algorithm looks not only at the flavour at production ($D^{*+}$-tagged events) but also takes into account the electric charge of the particles reconstructed in the rest of the event ($e^+e^-\rightarrow c\bar{c}$)~\cite{Belle-II:2023vra}.
\\

\noindent
\begin{center}
$\sim \circ \sim \circ$\, {\bf{A note on the use of prompt and secondary production in CPV and mixing analyses at LHCb }}\, $\circ \sim \circ \sim$
\end{center}
Neutral charm meson identification 
 is indispensable for measurements---presented in the suceeding chapter---of CPV, measurements of important quantities such as the asymmetry, $\Delta A_{\rm CP}$, and measurements of the mixing parameters $x$ and $y$. Measurements of the mixing parameters were performed using both prompt and secondary decays, with prompt and the secondary samples being independent and complementary.
 The prompt and secondary samples are usually separated by e.g.\ the ``impact parameter"~(IP). In prompt decays, the trajectory of the reconstructed $D$ meson points to the primary vertex where the charm meson was produced, while for secondary charm mesons, this is generally not the case. The chi-squared value $\chi^2_{\rm IP}$ is a measure of how well the trajectory of the decay product aligns with the primary vertex. 
 Decays from prompt production are typically characterised by a small $\chi^2_{\rm IP}$. 
 In CPV and mixing analyses, it is necessary to distinguish between prompt and secondary production as the decays stemming from the latter production can introduce a bias in the decay-time distribution if they are mistaken for decays from prompt production, since the $B$-meson flight path could be taken as contributing to the $D$-meson flight path. In this type of analyses, one often needs to correct for the 
 production asymmetries; where in the prompt case this means correcting for $D$-meson production asymmetries, and for the secondary case it means correcting for $B$-meson production asymmetries.\\ 

\noindent
So far, results using secondary charm production have only been reported by the LHCb collaboration. Decays from secondary production are delayed by the lifetime of the $B$ meson, therefore the $D$-meson decay vertex is displaced from the busy proton proton interaction region. 
This is an advantage with respect to prompt decays which can be much more difficult to isolate. However, during analysis one requires that the $D$-meson decay vertex does not coincide with the proton proton interaction region. 
This requirement leads to the removal of the $D$ mesons that decay very close to the interaction regions, indicated by very small values of the decay time in Fig.~\ref{fig:mixing-secondary}. 
This is not a problem for secondary production because of the preceding $B$-meson decay, and these decays give access to low decay times as demonstrated in Fig.~\ref{fig:mixing-secondary}. \\

\noindent
The prompt charm production from $D^*$ gives access to large yields but it is important to carefully separate the contributions from secondary charm productions for the reasons outlined above. 
With muon-tagged charm decays, there are lower yields and higher levels of combinatorial backgrounds. Additionally, possible mistag effects need to be taken into account, where mistag refers to the association of the incorrect tagging particles to the decay, thus identifying the $D^0$ incorrectly. However, doubly-tagged charm decays, where charm mesons are produced via the decay of the $D^*$, in e.g.\ $\bar B^0 \rightarrow D^{*+}(\to D^0\pi^+) \, \mu^- \nu_\mu X$
have an extremely clean signature, see Fig.~\ref{sec2.3:fig1} (right).  
While the muon and double tagged samples are smaller, they provide additional information and are independent from the prompt samples. 
To get an idea about the relative size of these samples at LHCb, the prompt sample of $D^0\rightarrow K^+K^-$ decays contained\ $\sim$ 44 M decays, while the muon single tagged sample contained\ $\sim$ 9 M decays~\cite{LHCb:2019hro}. In both cases these numbers represent the data samples collected in Run 2 by LHCb.\\

\begin{minipage}{\textwidth}
\begin{minipage}{0.4\textwidth}
 \centering
\begin{overpic}[width=1\linewidth]   
{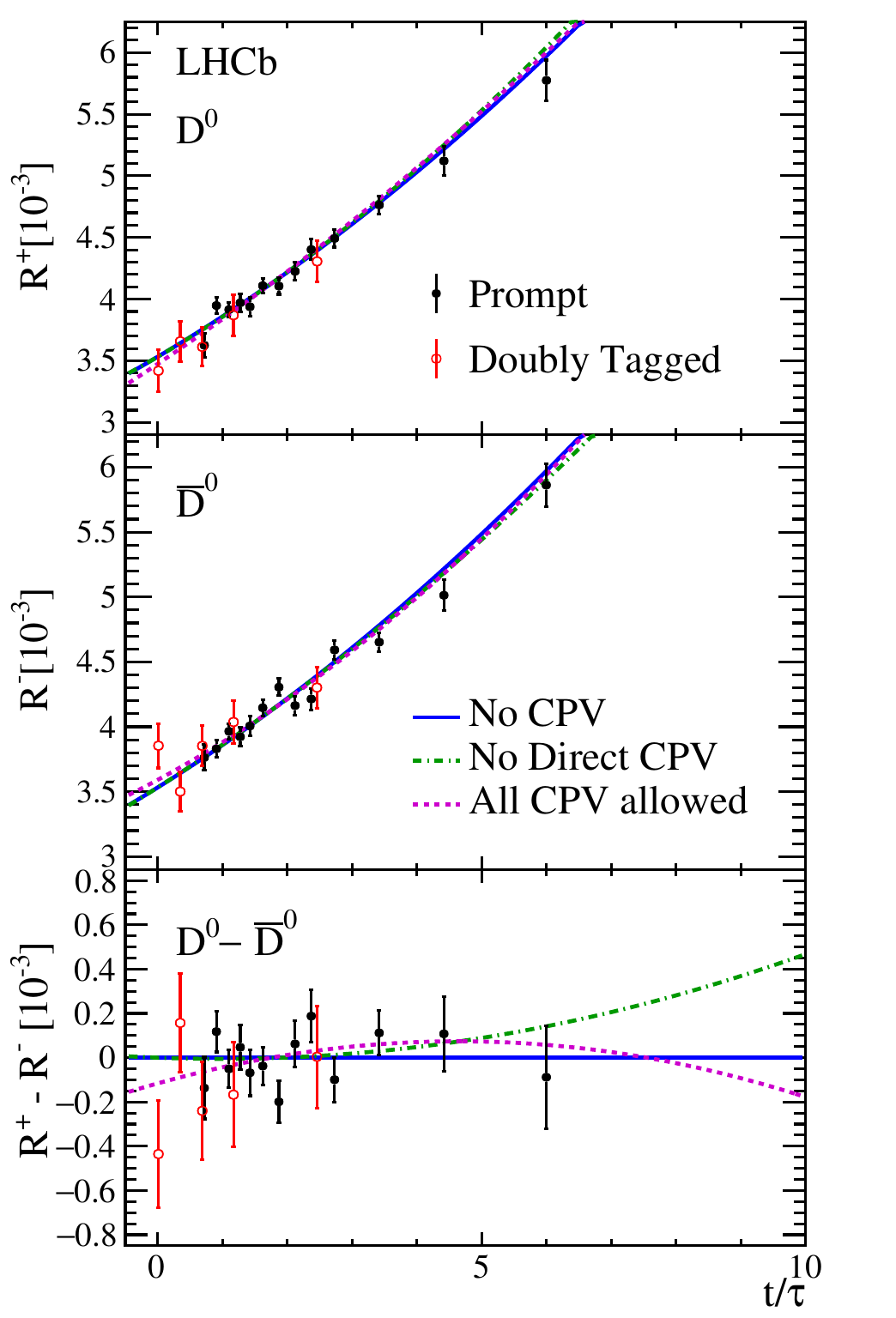}
\put(130,246){\includegraphics[width=0.15\linewidth]{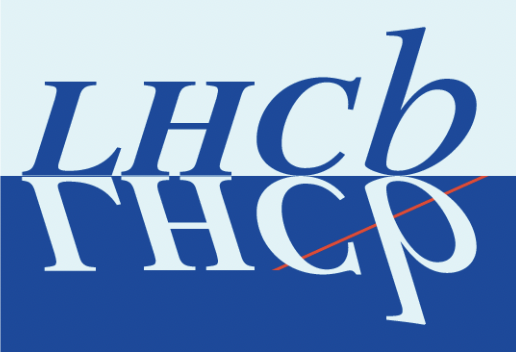}}
  \end{overpic} 
\captionof{figure}{The mixing observables $R^+$ (top), $R^-$ (middle) and their difference (bottom) (see ~\cite{LHCb:2016qsa}). The data-points in red correspond to secondary (doubly-tagged) charm, and they can reach very low decay times, while this is not possible for prompt decays (in black) because of the kinematic requirements imposed to reduce the background levels. Taken from~\cite{LHCb:2016qsa}.}
\label{fig:mixing-secondary}
\end{minipage}
\hspace*{5mm}
\begin{minipage}{0.5\textwidth}
\vspace*{-12mm}
\centering
\renewcommand*{\arraystretch}{1.7}
 \begin{tabular}[t]{|c||c|}
\hline
      Particle & Tag modes\\
      \hline 
      \hline
        & 
      $D_s^-\rightarrow K_S^0K^-$, $D_s^-\rightarrow K^-K^+\pi^-$, $D_s^-\rightarrow K_S^0K^-\pi^0$, \\ 
     $D_s^+$  & $D_s^-\rightarrow \pi^-\eta$,
      $D_s^-\rightarrow \rho^-\eta$, $D_s^-\rightarrow K^-K^+\pi^-\pi^0$, \\
      & 
      $D_s^-\rightarrow K_S^0K^-\pi^+\pi^-$, 
       $D_s^-\rightarrow K_S^0K^+\pi^-\pi^-$, $D_s^-\rightarrow \pi^-\pi^+\pi^-$, \\
       & $D_s^-\rightarrow \pi^-\eta'$, $D_s^-\rightarrow K^-\pi^+\pi^-$, \ldots \\
\hline
        & $D^-\rightarrow K^+\pi^-\pi^-$, $D^-\rightarrow K_S^0\pi^-$, $D^-\rightarrow K_S^-$, \\
      $D^+$ & $D^-\rightarrow K^+K^-\pi^-$, $D^-\rightarrow K^+\pi^-\pi^-\pi^0$, $D^-\rightarrow \pi^+\pi^-\pi^-$, \\
       & $D^-\rightarrow K^0_S\pi^-\pi^0$,  $D^-\rightarrow K^0_S\pi^-\pi^-\pi^+$,\\
       & $D^-\rightarrow K^+\pi^-\pi^-\pi^-\pi^+$, \ldots \\
        \hline
         & $\Lambda_c^-\rightarrow\bar{p}K_S^0$,
       $\Lambda_c^-\rightarrow\bar{p}K^+\pi^-$, $\Lambda_c^-\rightarrow\bar{p}K_S^0\pi^0$, \\
      $\Lambda_c^+$  & $\Lambda_c^-\rightarrow\bar{p}K^+\pi^-\pi^0$,  $\Lambda_c^-\rightarrow\bar{p}K_S^0\pi^+\pi^-$, \\
      & $\Lambda_c^-\rightarrow\bar{\Lambda}\pi^-$, 
        $\Lambda_c^-\rightarrow\bar{\Lambda}\pi^-\pi^0$,  $\Lambda_c^-\rightarrow\bar{\Lambda}\pi^-\pi^+\pi^-$, \\
        & $\Lambda_c^-\rightarrow\bar{\Sigma}^0\pi^-$, $\Lambda_c^-\rightarrow\bar{\Sigma}^-\pi^0$, $\Lambda_c^-\rightarrow\bar{\Sigma}^-\pi^+\pi^-$, \ldots \\
\hline

$D^0$ & $D^0\rightarrow K^- e^+\nu_e$, $D^0\rightarrow K^-\pi^+$, \\
& $D^0\rightarrow K^-\pi^+\pi^0$, $D^0\rightarrow K^-\pi^+\pi^-\pi^+$, ...\\
\hline
\end{tabular}
    \captionof{table}{Examples of tag modes for the $D_s^+$, $D^+$ mesons and the $\Lambda_c^+$ baryons. 
}\label{sec2.3:tab1}
\end{minipage}
\end{minipage}\\

\begin{center}
$\sim \circ \sim \circ \sim \circ \sim $
\end{center}

\begin{figure}[t]
	\centering
       \includegraphics[width=0.37\textwidth]{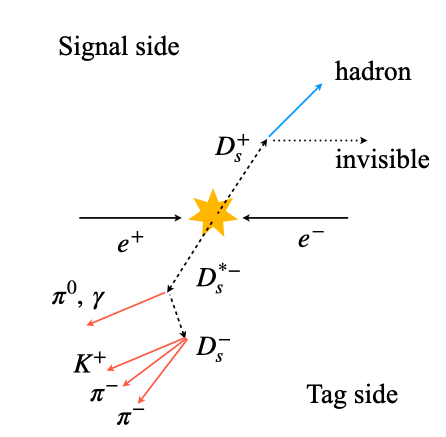}
	\caption{Example of tagging used for charm production at threshold at e.g.\ BESIII. The information from the tag side is used to pin down the properties of the $D$ meson decaying on the signal side.}
	\label{sec2.3:fig2}
\end{figure}

\noindent
For charm particles produced at threshold in electron-positron collisions, the tagging method is different. 
At threshold, two charm particles are produced back to back, see Fig.~\ref{sec2.3:fig2}, and one side defines the tagging side, while 
the other one is the signal side. In this case, two tagging procedures are possible: i) Only the signal side is considered, which is called {\bf single tag}~\footnote{In the literature, it is also common to refer to single-tagged analyses as {\it untagged}, and double-tagged analyses as {\it tagged}. Note however, that the terms single tag and double tag have a different meaning for charm physics analyses done at LHCb and at threshold: at LHCb this refers to tagging the flavour of a secondary neutral charm meson with two tagging particles (soft $\pi$ and $\mu$), and at BESIII, the double tag refers to using information from both sides of the event, and reconstructing both $D$ mesons produced simultaneously. }.
A single tag can be used for a very clean decay mode, simply meaning that the signal final state is fully reconstructed.
ii) Since at electron-positron colliders the energy of the beams is known with very high precision, one can take advantage of the full kinematic constraint and use a {\bf double tag} for $D_{(s)}$ mesons or $\Lambda_c$ baryons, meaning that both sides are reconstructed. 
This double tag technique was pioneered by the MARK III collaboration at the SLAC’s SPEAR storage ring~\cite{MARK-III:1985hbd}. The background level is typically reduced by selecting events within a certain interval of the distribution of the beam-constrained mass of the tag-side charm meson, defined as $M_{BC}=(E^2_{\mathrm{beam}}-|\vec{p}_{\mathrm{tag}}|^2)^{1/2}$, where $E_{\mathrm{beam}}$ is the beam energy, and $\vec{p}_{\mathrm{tag}}$ is the momentum of the $D$ meson on the tag side in the $e^+e^-$ centre-of-mass frame. Another useful kinematic variable is the energy difference defined as $\Delta E=E_{\mathrm{tag}} - E_{\mathrm{beam}}$, where $E_{\mathrm{tag}}$ is the energy of the tag-side $D$ meson, typically used for $D^0$ and $D^+$ decays. For $D^+_s$ decays, as most data is collected via $D_s^{*\pm}D_s^\mp$ pair production, where the energy is not evenly distributed between the two candidates, the recoil mass $M_{\mathrm{rec}} =\left( \big(E_{\mathrm{CM}}-\sqrt{|\vec{p}_{\mathrm{tag}}|^2+m^2_{D_s^-}}\big)^2-|\vec{p}_{\mathrm{tag}}|^2 \right)^{1/2}$, where $E_{\mathrm{CM}}$ is the centre-of-mass energy, and $m_{D_s^-}$ is the known mass of the tag-side $D^-_s$ meson, is used to suppress multiple candidates. 
For leptonic and semileptonic decays on the signal side, e.g.\ $D\rightarrow (X)\ell^+\nu_\ell$, where the neutrino escapes undetected---or generally for events with missing energy---the kinematic variables used are the missing mass $M_{\mathrm{miss}}$ or $U_{\mathrm{miss}}= E_{\mathrm{miss}}-p_{\mathrm{miss}}$. The missing energy, $E_{\mathrm{miss}}$, and momentum, $p_{\mathrm{miss}}$, can be calculated because the energies of the electron positron beams, and the four-momenta of the final state particles leaving signal in the detectors (both on the tag side and the signal side) are measured precisely.
A list of tag modes for the $D^0$, $D_s^+$, $D^+$ mesons and the $\Lambda_c^+$ baryons is presented in Table~\ref{sec2.3:tab1}. 
The double tag technique at threshold allows for measurements of absolute branching fractions, decay constants/form-factors, CKM parameters, lepton flavour universality tests, etc., as it will be discussed later in the succeeding chapter.\\

%%%%%%%%%%%%%%%%%%%%%%%%%%%%%%%%%%%%%%%%%%%%%%%%%%%%%%%%%%%%%%%%%%%%%%%%%%%%%%%%%%%%%%%%%%%%%%%%%%%%%%%%%%%%%%%%%%%%%%%%%%%%%%%%
\subsection{Spectroscopy - the zoo of charmed particles}
\label{subsec:spectroscopy}
Because of confinement, hadrons must be colourless i.e.\ they have to transform as singlets under the action of the QCD  symmetry group SU(3)$_c$, whereas quarks are colour charged and transform as triplets under SU(3)$_c$. 
Group theory tells us that combining a quark and an anti-quark results in single and octet representations i.e.\ 
$\bf 3_c \otimes \bar 3_c \equiv 1_c \oplus 8_c$, while combining three quarks gives
$\bf 3_c \otimes 3_c \otimes 3_c \equiv 1_c \oplus 8_c\oplus 8_c\oplus 10_c$, where again the colour singlet representation can arise. Thus, both QCD
bound states composed of a quark and an anti-quark i.e.\ {\bf mesons}, and of three quarks i.e.\ {\bf baryons}, are allowed by the colour symmetry, and the hadron classification into  
baryons and mesons 
forms the basics of the so-called {\it naive quark model}. However, more complex structures are also permitted by the QCD symmetry group and states composed of e.g.\ two quarks and two anti-quarks, so-called {\bf tetraquarks}  ($\bf 3_c \otimes \bar 3_c \otimes 3_c \otimes \bar 3_c \equiv 1_c \oplus \ldots $) 
or four quarks and an anti-quark, so-called {\bf pentaquarks} ($\bf 3_c \otimes 3_c \otimes 3_c  \otimes 3_c \otimes \bar 3_c \equiv 1_c \oplus ...$) are also in principle possible. These QCD bound states are called {\bf exotic hadrons}.\\

\begin{figure}[b]
\centering
\begin{minipage}{0.45\textwidth}
\centering
\includegraphics[width=0.8\textwidth]{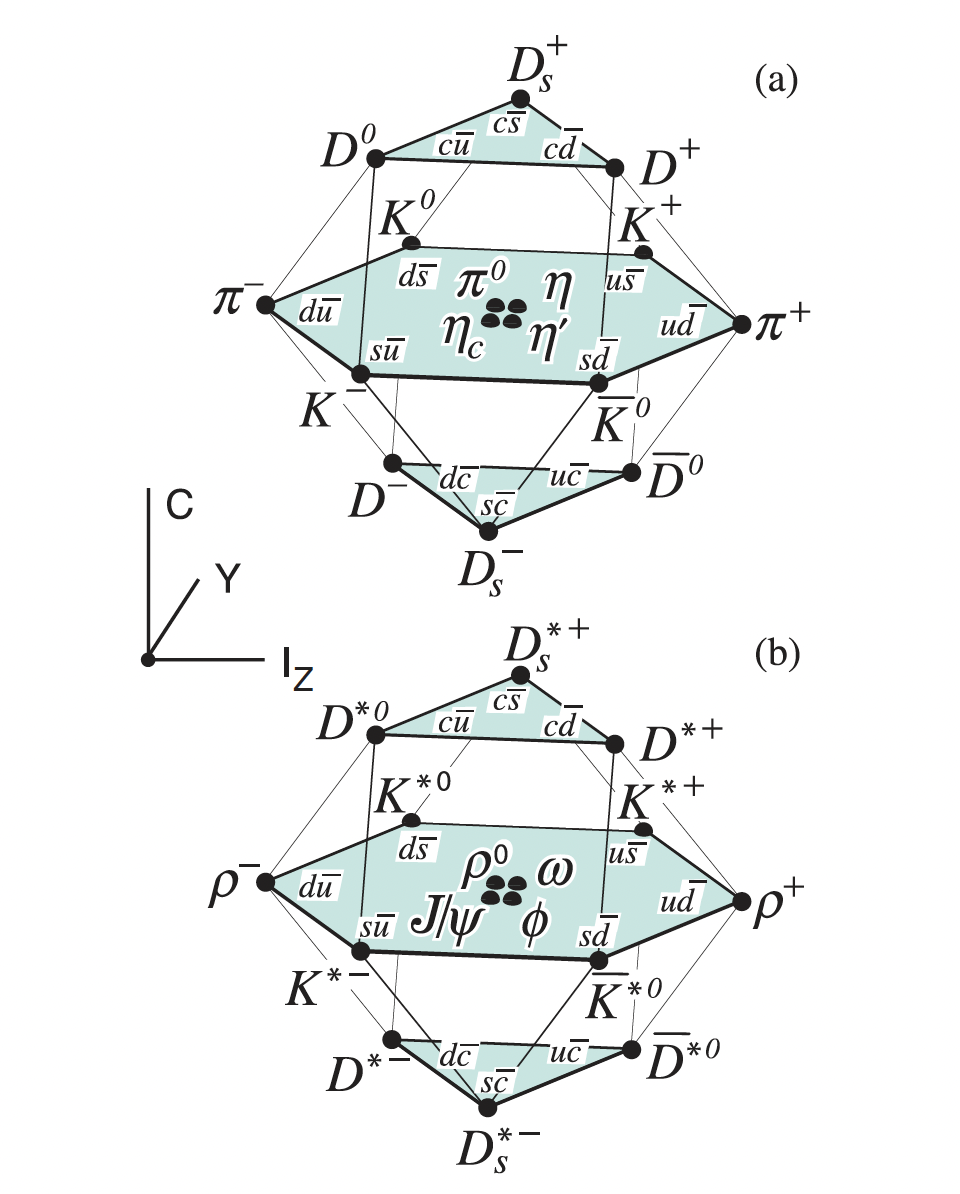}
\caption{SU(4)$_F$ multiplets $ {\bf 1}_F \oplus {\bf 15}_F$ of the pseudoscalar (a) and vector (b) mesons made of $u$, $d$, $s$ and $c$ quarks. The lightest mesons and the $c\bar{c}$ states are in the central plane. Taken from~\cite{ParticleDataGroup:2024cfk-qm} with permission.}
\label{sec2.4:fig1}
\end{minipage}
\qquad
\begin{minipage}{0.45\textwidth}
\centering
\includegraphics[width=1\textwidth]{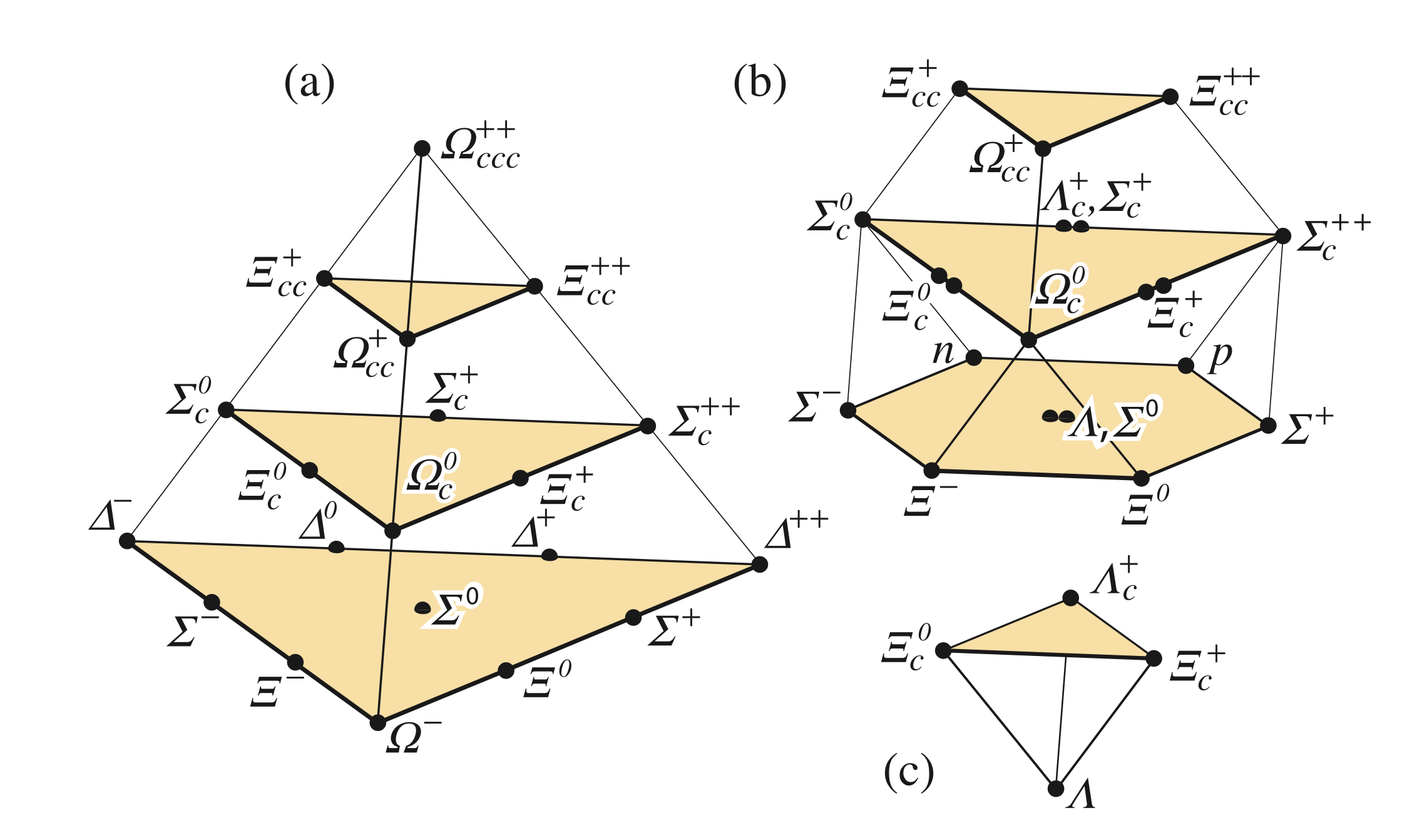}
\caption{The SU(4)$_F$ multiplets ${\bf 20}_F$ with $J^P = 3/2^+$~(a), ${\bf 20}_F$ with $J^P = 1/2^+$~(b) and ${\bf 4}_F$~(c) of baryons composed of $u$, $d$, $s$ and $c$ quarks, where $J$ and $P$ denote the particle spin and parity, respectively. Baryons with one $c$ quark are one level up from the lowest level, followed by a level of baryons with two charm quarks (only one of which has been experimentally confirmed). On the top of the pyramid in (a) there is a single baryon with three charm quarks which has not yet been observed.  
Taken from~\cite{ParticleDataGroup:2024cfk-bar} with permission.}
\label{sec2.4:fig2}
\end{minipage}
\end{figure}

 \noindent
 Hadronic bound states can be further classified according to {\bf flavour symmetries}, see e.g. the review~\cite{ParticleDataGroup:2024cfk-qm}. Assuming the masses of the up and down quarks to be the same leads to 
 the SU(2)$_F$ symmetry group, this is the 
 isospin symmetry, i.e.\ the symmetry under the exchange $u \leftrightarrow d$---assuming equal masses for the up, down and strange quarks---results in a larger flavour symmetry group, the SU(3)$_F$. This has three SU(2) subgroups. One corresponds to the isospin symmetry, one to the so-called U-spin symmetry, i.e.\ the symmetry under the exchange $d \leftrightarrow s$, and one to the V-spin symmetry, i.e.\ the symmetry under the exchange $u \leftrightarrow s$. 
 Enlarging the symmetry group to also include the charm quark, further leads to the 
 SU(4)$_F$ symmetry group. 
 Note that, while isospin symmetry is a good symmetry of QCD due to $m_d \approx m_u $, U-spin is only an approximate symmetry, broken by the larger value of the strange quark mass with respect to that of the down quark. The size of SU(3)$_F$ breaking is parametrically of the order ${\cal O} (m_s/\Lambda_{\rm QCD})$. As for the SU(4)$_F$ symmetry, this is badly broken due to the much larger value of the charm quark mass. 
 However, the SU(4)$_F$ symmetry group can be used to obtain a classification of the possible charm multiplets. 
 \\

\noindent
Following the SU(4)$_F$ flavour symmetry group, we find for the charmed mesons 16 possible configurations i.e.\ ${\bf 4}_F \otimes {\bf \bar 4}_F \equiv {\bf 1}_F \oplus {\bf 15}_F$, the corresponding ground and excited meson states are depicted in Fig.~\ref{sec2.4:fig1}.
Experimentally, the ground states of the open charm meson system, 
the $D^0$, $D^+$, $D_s^+$ mesons, are well known. 
The two lightest $D$ mesons, the $D^0$ and the $D^+$ - combinations of the charm quark with the $u$ and $d$ anti-quarks, were also first observed by MARK-I in 1976, about a year and a half after the $J/\psi$ discovery~\cite{Goldhaber:1976xn, Peruzzi:1976sv}. 
The first reported decay of an excited charm meson~\footnote{When the $D^+$ meson was discovered in electron-positron annihilations at centre-of-mass energy of 4.03 GeV~\cite{Peruzzi:1976sv}, the presence of a state with a mass of about 2 GeV/c$^2$ was also noted. It was thought that this could be the corresponding excited state but no further information was available.} 
is the strong decay $D^{*+} \rightarrow D^0 \pi^+$~\footnote{In this case, and in many other cases in this paper, we have specified only one of the charges: $D^{*+} \rightarrow D^0 \pi^+$ while we implicitly imply the usage of the charge conjugated mode $D^{*-} \rightarrow \bar{D}^0 \pi^-$ as well.} observed at SLAC in 1977~\cite{Feldman:1977ir}, which is nowadays very important for the identification of the flavour of neutral charm mesons.
First evidence for the $D_s^+$ meson, the combination of a charm quark with strange antiquark, was reported by the DASP collaboration at DESY also in 1977~\cite{DASP:1977eel}, with a mention of excited states in~\cite{DASP:1978gcx}. 
However, it was not until 1983 that the discovery of the $D_s^+$ meson was announced by the CLEO collaboration~\cite{CLEO:1983dsb}.  
\\

\begin{figure}[t]
    \centering
    \includegraphics[width=0.58\textwidth]{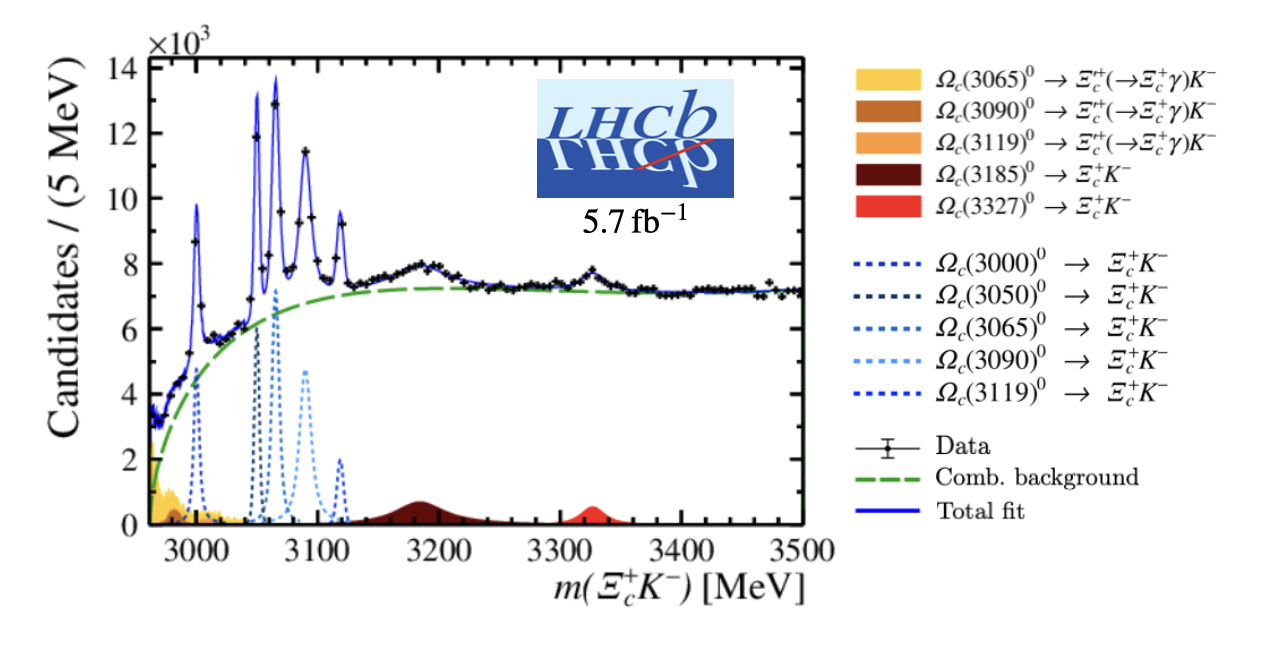}
    \includegraphics[width=0.4\textwidth]{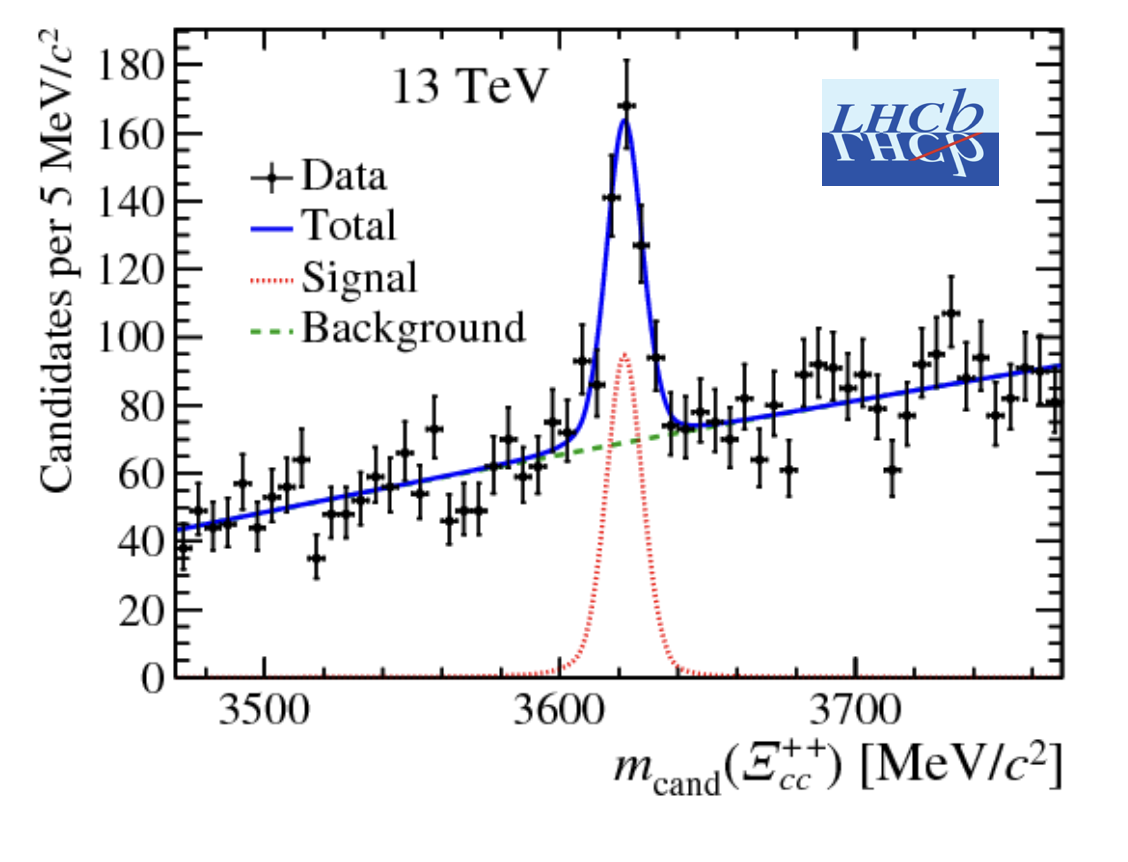}
	\caption{(Left) Several narrow excited $\Omega_c$ resonances from~\cite{LHCb:2023sxp}, the contributions are described in the legend: the five narrow peaks in blue are excited $\Omega_c$ resonances seen in an earlier version of this study, and the brown and red peaks correspond to newly discovered excited $\Omega_c$ resonances; the contributions in yellow are feed-downs from partially reconstructed decays as shown in the legend. 
    (Right) Invariant mass distribution of $\Xi_{cc}^{++}$ baryons reconstructed in the $\Lambda_c^+ K^-\pi^+\pi^+$ final state, taken from~\cite{LHCb:2017iph}.
}
 \label{sec2.4:fig3}
    \end{figure}

\noindent
According to the SU(4)$_F$ flavour symmetry,  for charmed baryons the possible representations are ${\bf 4}_F \otimes  {\bf 4}_F \otimes  {\bf 4}_F \equiv {\bf 4}_F \oplus {\bf 20}_F \oplus {\bf 20}_F\oplus {\bf 20}_F$, see Fig.~\ref{sec2.4:fig2}.
The lightest ground state charm baryon $\Lambda_c^+(udc)$ was observed for the first time in 1979 by the MARK II experiment~\cite{Abrams:1979iu}. The $\Lambda_c^+$ baryons are particularly interesting because they can be described theoretically as a system of a light di-quark ($ud$) with isospin zero coupled to a heavier quark ($c$), therefore the dynamics of the charm quark can be deduced from the properties of the $\Lambda_c^+$. 
All the nine ground state charm baryons with one charm quark have been observed. They can be classified into
$\Lambda_c$ baryons, composed of two light quarks with isopin zero 
i.e.\  
$\Lambda_c^+ (udc)$, 
$\Sigma_c$ baryons, composed of two light quarks with isopin one
i.e.\  
$\Sigma_c^0 (ddc)$, 
$\Sigma_c^+ (udc)$, 
$\Sigma_c^{++} (uuc)$,
the 
$\Xi_c$ baryons, composed of  
one light quark and one strange quark,
i.e.\ 
 $\Xi^{(')0}(dcs)$, $\Xi^{(')+} (ucs)$, where the primed baryons are needed to complete one of the flavour symmetry representations,
 and the 
 $\Omega_c$ baryons, composed of two strange quarks i.e.\ $\Omega_c^0 (ssc)$. 
 Among the baryons with two charm quarks, only the $\Xi_{cc}^{++}(ccu)$ has been observed experimentally by the LHCb collaboration~\cite{LHCb:2017iph}, by now in several decay modes, see the right plot of Fig.~\ref{sec2.4:fig3}. The SELEX experiment has reported the observation of the $\Xi_{cc}^{+}(ccd)$ baryon~\cite{SELEX:2002wqn}, however, these have not been experimentally confirmed by another experiment~\cite{LHCb:2021eaf, BaBar:2006bab, Belle:2006edu}. 
 The $\Omega_{cc}^+(ccs)$ and the $\Omega_{cc}^{++}(ccc)$ baryons have not been seen yet. 
Many of the excited charm baryon states have been seen, one particularly interesting case is the observation of very narrow $\Omega_c$ baryon states~\cite{LHCb:2017uwr} - later confirmed in~\cite{LHCb:2023sxp, Belle:2017ext}, see the left plot of Fig.~\ref{sec2.4:fig3}. For a detailed overview of the observed charm baryons, and their excited states see~\cite{ParticleDataGroup:2024cfk-qm}.\\

\noindent
As already introduced above, all particles beyond the baryons and meson states postulated in the naive quark model~\cite{Gell-Mann:1964ewy, Zweig:1964jf} are referred to as {\bf exotic hadrons}. 
Both tetraquarks  
and pentaquarks  
have by now been experimentally observed, while searches for hexaquarks are still ongoing.
These exotic states were referred to as $XYZ$ states, but in recent years new naming conventions have been proposed~\cite{Gershon:2022xnn}. 
The first exotic hadron that did not fit the naive quark model  was the charmonium-like state $\chi_{c1}(3872)$,~\footnote{Here the number between the brackets indicates the approximate mass of the particle in MeV$^2$.} also known as $X(3872)$, discovered in 2003 by Belle~\cite{Belle:2003nnu} in $B^\pm\rightarrow J/\psi K^\pm\pi^+\pi^-$ decays. 
This particle is compatible with a $c\bar{c}$ state but it is not predicted as a part of the charmonia spectrum. 
It is interesting, as it is produced at the $D^0\bar{D}^{0*}$ threshold. 
The first pentaquark state was discovered in 2015 by LHCb~\cite{LHCb:2015yax} in $\Lambda_b\rightarrow J/\psi K^- p$ decays, it was interpreted as a ($uc\bar{c}ud$) state and named $P_c(4450)^+$~\footnote{In more recent studies~\cite{LHCb:2019kea}, it was confirmed that instead of a single peak there are two narrow overlapping peaks, $P_c(4440)^+$ and $P_c(4457)^+$.}~\footnote{Note that the papers reporting these observations are the top cited Belle and LHCb physics papers.}. 
Common interpretations of hadrons with at least four quarks have not yet reached consensus whether they are composed of tightly bound quarks and antiquarks, or they can be interpreted as weakly bound molecules of e.g.\ mesons; analogously, pentaquarks can be interpreted as a bound state as well, or as a meson-baryon molecule.
Other interesting exotic hadrons with charm quarks are the $T_{cc}(3875)^+$ state with two charm quarks $(cc
\bar{u}\bar{d})$ at the $D^0D^{*
0}$ threshold, and of a $T_{cc\bar{c}
\bar{c}}(6900)$ decaying into a pair of $J/\psi$ mesons. \\

\begin{figure}[b]
    \centering
    \includegraphics[width=0.8\textwidth]{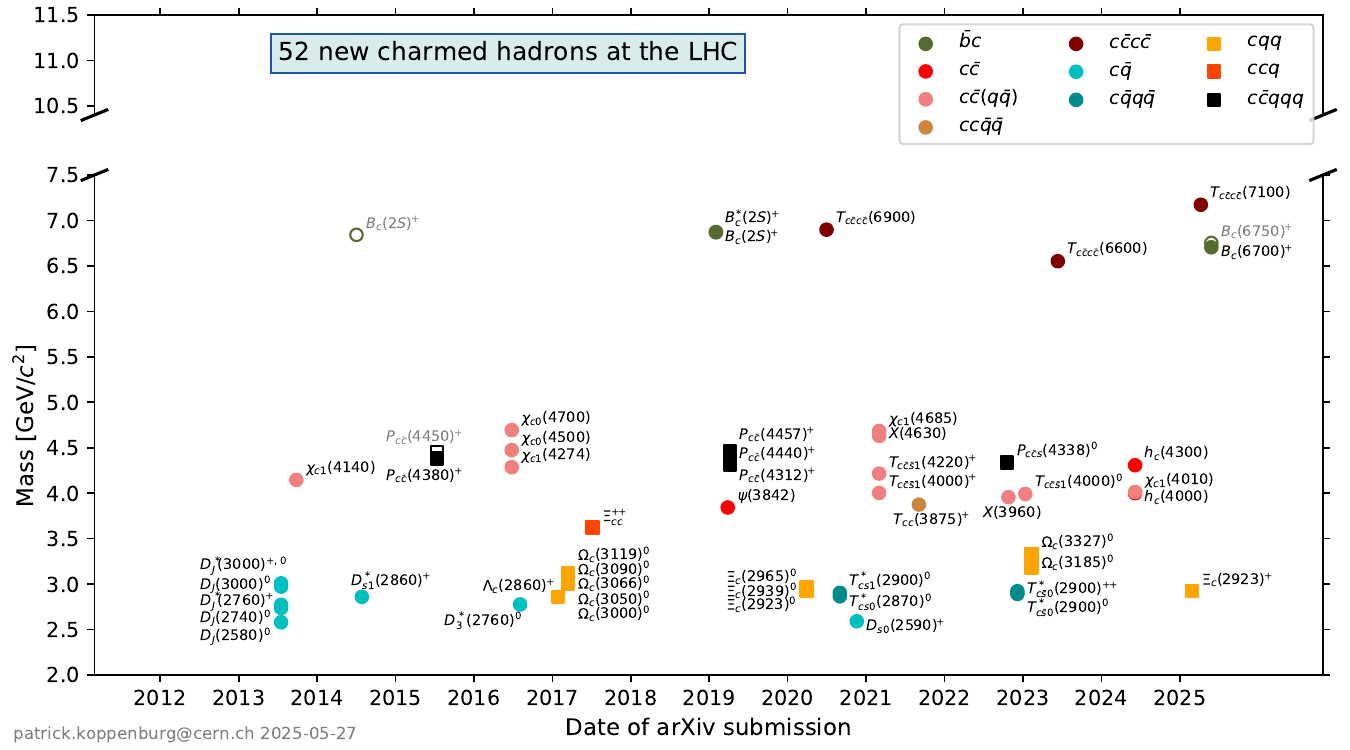}
	\caption{Besides the Higgs boson, 79 new particles have been discovered at the LHC, and 52 contain hidden or open charm. Taken from~\cite{pkoppenburg} with permission. See the Quantum Working Group Exotics hub~\cite{qwg-exotics} for similar interactive plots with results from all experiments.}
\label{sec3.4:fig4}
\end{figure}

\noindent
Charm, charmonia and exotic hadron spectroscopy is a very active field. For detailed theory reviews we refer e.g.\ to~\cite{Brambilla:2010cs, Esposito:2016noz, Chen:2016qju, Guo:2017jvc, Brambilla:2019esw, Chen:2022asf}, while recent experimental results and summary plots can be found in~\cite{ParticleDataGroup:2024cfk, pkoppenburg}, and at the Quantum Working Group Exotics hub~\cite{qwg-exotics}. 
For instance, among the recently discovered 79 particles at the LHC, 52 contain charm quarks, see Fig.~\ref{sec3.4:fig4}; 
additionally, 10 and 16 particles with charm content have been discovered by BESIII and Belle/Belle II experiments~\cite{qwg-exotics}, respectively. 
\\

\noindent
Note that in the remainder of this chapter, we will focus on open charm physics.

%%%%%%%%%%%%%%%%%%%%%%%%%%%%%%%%%%%%%%%%%%%%%%%%%%%%%%%%%%%%%%%%%%%%%%%%%%%%%%%%%%%%%%%%%%%%%%%%%%%%%%%%%%%%%%%%%%%%%%%%%%%%%%%%
\subsection{Quantum entanglement}
\label{subsec:entanglement}
\begin{figure}[t]
	\centering
\includegraphics[width=0.65\textwidth]{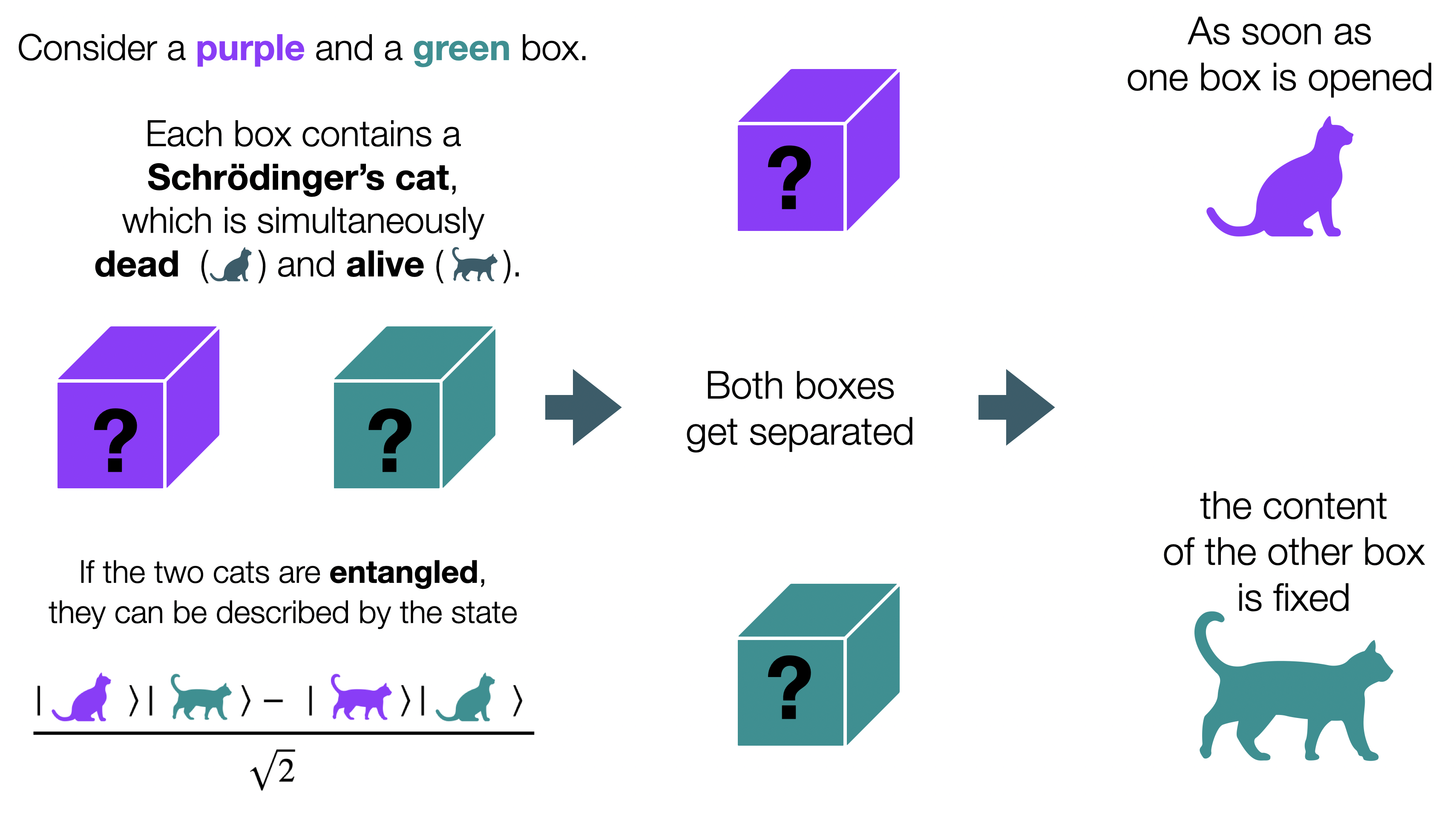}
	\caption{Using Schr\"odinger's cat to visualise the concepts of non-locality and quantum entanglement for the state $(1/\sqrt{2}) \left(|D^0\rangle|\bar{D}^0\rangle - |\bar{D}^0\rangle|D^0\rangle \right) $.
    This entangled state can be created via the decay of a $\psi(3770)$ resonance into two $D$ mesons, denoted by $D_1$ and $D_2$. A priori we do not know whether $D_1$ (and $D_2$) is a $D^0$ or $\bar D^0$ meson, but as soon as we measure e.g.\ $D_1$ to be a $D^0$ meson, then immediately $D_2$ is fixed to be a $\bar D^0$ meson.}
	\label{sec2.5:fig1}
\end{figure}

Based on earlier ideas of Einstein, Hermann, Schrödinger and  Weyl,  Einstein, Podolsky and Rosen proposed in 1935  {\bf quantum entanglement}~\cite{PhysRev.47.777} as a hypothetical concept,
in order to prove that quantum mechanics is incomplete. However, by now this concept is experimentally well established and was even awarded with the Nobel prize, e.g.\ in 2022 to Aspect, Clauser and Zeilinger. In a nutshell entanglement means that a measurement performed in one system can give access to information about another system, at a different location, instantaneously.
Quantum entanglement is a fascinating phenomenon dubbed ``spooky action at a distance'' and the basic idea behind it is sketched in Fig.~\ref{sec2.5:fig1}. 
This intriguing property is exploited by experiments such as  KLOE and KLOE-2 running at the $\phi$ resonance to create entangled $K$ mesons,
MARK III, CLEO-c and BESIII  to create entangled $D^0$ mesons,
and BaBar, Belle and  Belle II~\cite{Belle:2007ocp}
running at the $\Upsilon(4S)$ resonance to create 
entangled $B^0$ mesons.
These experiments can be used 
for tests of the fundamental discrete symmetries T (time), CP (charge-parity), and CPT symmetries.\\ 

\noindent
When pairs of neutral charm mesons are produced at threshold via the strong decay of a $c \bar{c}$ resonance $\psi(3770)$, the resulting charm mesons are in a quantum entangled state
i.e.\ $e^+e^-\rightarrow \psi(3770)\rightarrow (1/\sqrt{2})(|D^0\rangle|\bar{D}^0\rangle - |\bar{D}^0\rangle|D^0\rangle)$, see Figs.~\ref{sec2.5:fig2} and \ref{sec2.5:fig1}. 
The $\psi(3770)$ is a C-odd state, where here C stands for charge conjugation. The charm mesons are produced with a coherent anti-symmetric wavefunction with an odd charge parity C~=~-1. 
After production, the charm mesons evolve coherently.
Measurements of the properties of one charm meson give immediate information about the properties of the other meson in the pair. 
 The two neutral mesons can be detected either as a $D^0\bar{D}^0$ pair: states with well defined flavour, or alternatively they can be detected as states with well defined CP quantum number:
 $D_{CP+}D_{CP-}$, where the CP eigenstates are defined as a linear combination of the flavour eigenstates, i.e.\ $D_{CP\pm}=(1/\sqrt{2})(D^0\mp\bar{D}^0)$. Data from quantum correlated charm pairs produced at experiments like CLEO-c and BESIII provides important information for measurements of the CKM angle $\gamma$, a key parameter of the SM, and for measurements of neutral charm mixing parameters. 
%%%%%%%%%%%%%%%%%%%%%%%%%%%%%%%%%%%%%%%%%%%%%%%%%%%%%%%%%%%%%%%%%%%%%%%%%%%%%%%%%%%%%%%%%%%%%%%%%%%%%%%%%%%%%%%%%%%%%%%%%%%%%%%%
\section{Basics of charm physics}
\label{sec3}
\subsection{Peculiarities of charm 
}
\label{subsec:peculiarities}
\noindent
The physics of the charm quark shows several interesting peculiarities, with very different properties than, e.g.\ the $B$ and $K$ systems:\\

\begin{enumerate}
    \item {\bf The charm quark is neither too heavy nor too light:} the charm quark is not really heavy with respect to the typical hadronic scale $\Lambda_{\rm QCD}$, which we can take to be of the order of few hundreds MeV~\footnote{Note that $\Lambda_{\rm QCD} $ 
    does not 
     necessarily 
    coincide with 
    the pole of the QCD beta function, namely the scale at which the strong coupling becomes non-perturbative.}. This implies that the expansion of an observable $A$ in inverse powers of the charm-quark mass may not necessarily converge.
    Such an expansion would for instance look like:
\begin{equation}
        A = A_0 + \frac{\Lambda_{\rm QCD}}{m_c} A_1  
        + \left(\frac{\Lambda_{\rm QCD}}{m_c}\right)^2 A_2 + \ldots 
        \approx 
        A_0 +
        0.33 A_1  + 0.11 A_2 + \ldots 
         \, ,
    \end{equation} 
    where, for $m_c \sim 1.5$ GeV and $\Lambda_{\rm QCD} \sim 0.5$ GeV, the expansion factor naively reads $\Lambda_{\rm QCD} / m_c \approx  1/3$. Hence, if all the coefficients $A_i$ are of similar size, the series may converge but the theory prediction will be very sensitive to higher power corrections in $1/m_c$.

    \item {\bf The strong coupling is large:} the strong coupling at the scale of the charm-quark mass is quite large, i.e.\ $\alpha_s (m_c) \approx 0.3$.
    Thus the perturbative expansion for a charm observable $B$ in the strong coupling may not converge well enough. The latter, would read schematically:
     \begin{equation}
        B = B_0 + \alpha_s B_1  + \alpha_s^2 B_2 + \ldots
        \approx 
        B_0 + 0.3 B_1  + 0.09 B_2 + \ldots \,.
    \end{equation}
    Again, if all the coefficients $B_i$ are of similar size, then the series might converge, but in any case the theory prediction for $B$ will be very sensitive to higher order corrections in the strong coupling. When the coupling becomes too large, perturbation theory breaks down and one has to rely on non-perturbative methods like lattice QCD~\footnote{The method of lattice QCD~\cite{Wilson:1974sk} allows a direct numerical evaluation of the path integrals 
    in quantum field theory, obtained by discretising the space time and without performing a perturbative expansion, see e.g.\ the textbook~\cite{Gattringer:2010zz}.} or QCD sum rules\footnote{QCD sum rules~\cite{Shifman:1978bx} represents a phenomenological approach to estimate hadronic matrix elements, see e.g.\ the textbook~\cite{Khodjamirian:2020btr}.}.

    \item {\bf Cancellations:}  
    in many observables of the charm system severe cancellations arise. Because of this, it may be particularly challenging
    to verify whether an expansion in inverse powers of the charm-quark mass or in the strong coupling actually converges. To this end, it is convenient to classify charm observables according to the strength of the cancellations entering their SM predictions, namely
    
    \begin{enumerate}
        \item No cancellations --- This is the case for quantities like the total decay rate of the $D^0$ meson or the tree-level decay amplitude for the decay $D^0 \to K^+ K^-$. 
        \item Strong cancellations --- This is the case for quantities such as the total decay rate of the $D^+$ meson or the tree-level decay amplitude for the decay $D^0 \to \pi^0 \pi^0$. For the former, in the theory prediction obtained within the heavy quark expansion, 
        the leading contribution from the free charm-quark decay almost exactly cancels the one due to the dimension-six four-quark operators - the so-called ``Pauli interference". For the latter,
        the corresponding combination of Wilson coefficients entering the effective Hamiltonian, see Section~\ref{subsec:Heff}, cancels to a large extent at LO-QCD, making this observable particularly sensitive to higher-order QCD corrections.
         \item Extremely pronounced cancellations --- This is the case for quantities like $D$-mixing, see Section~\ref{subsec:basics_mixing} and the succeeding chapter. In the theory prediction, obtained within the heavy quark expansion, the sum of the relevant contributions amounts to about 0.1\textperthousand\ of any of the individual contributions, making the final result extremely suppressed.  
         \end{enumerate}
\item {\bf Tiny amount of CPV:} 
in the SM, CPV in the charm system is proportional to the small parameter ${\rm Im}[(V_{cb}V^*_{ub})/(V_{cd} V^*_{ud})] \approx 6 \cdot 10^{-4}$~\cite{ParticleDataGroup:2024cfk}. A measurement of sizeable CPV effects could thus be a strong indication for BSM physics.
\item {\bf Complementarity to the bottom- and strange-quark system:} 
loop-induced charm-quark decays and $D^0$-meson mixing proceed via virtual down-type quark contributions providing a unique sensitivity to possible BSM effects in up-type quark decays and complementing bottom- and strange-quark decays.
\item {\bf Clean experimental signature of \boldmath $D^0$ mesons:} experimentally it turns out that the difference between the mass of $D^{*+}$ and the $D^0$ mesons is just above the charged $\pi$ mass. Pions produced in the strong decay $D^{*+}\rightarrow D^0\pi^+$ are very soft but the signature of the decay is very clean and it allows for extremely efficient tagging and background reduction.\\
\end{enumerate}

\noindent
At first sight, the peculiarities 1.~- 5. make the theoretical
description of the charm system more challenging, which is one 
of the reasons why, often, theory predictions rely on the use of
symmetries like SU(3)$_F$,
the symmetry under the exchange $u \leftrightarrow d \leftrightarrow s$, or U-spin symmetry, the symmetry under the exchange $d \leftrightarrow s$. Note however, that these are only approximate symmetries of QCD and are broken by the sizable value of the strange quark mass as compared to the one of the up and down quarks. 
At the same time, the above properties can also become virtues. As some observables are very sensitive to higher orders in the expansion parameters, they constitute an ideal testing ground for the available QCD-based methods. Moreover, being many quantities very suppressed or even zero in the SM, they provide important null-tests of the SM, with a high sensitivity to NP. 
The most promising approach remains therefore to confront the current theoretical methods for the study of charm-quark decays with the corresponding measurements, to try to answer the question of whether or not an expansion in the strong coupling or in inverse powers of the charm-quark mass can be applicable.
%%%%%%%%%%%%%%%%%%%%%%%%%%%%%%%%%%%%%%%%%%%%%%%%%%%%%%%%%%%%%%%%%%%%%%%%%%%%%%%%%%%%%%%%%%%%%%%%%%%%%%%%%%%%%%%%%%%%%%%%%%%%%%%%
\subsection{The effective Hamiltonian}
\label{subsec:Heff}
\begin{figure}[t]
	\centering
    \includegraphics[width=0.35\textwidth]{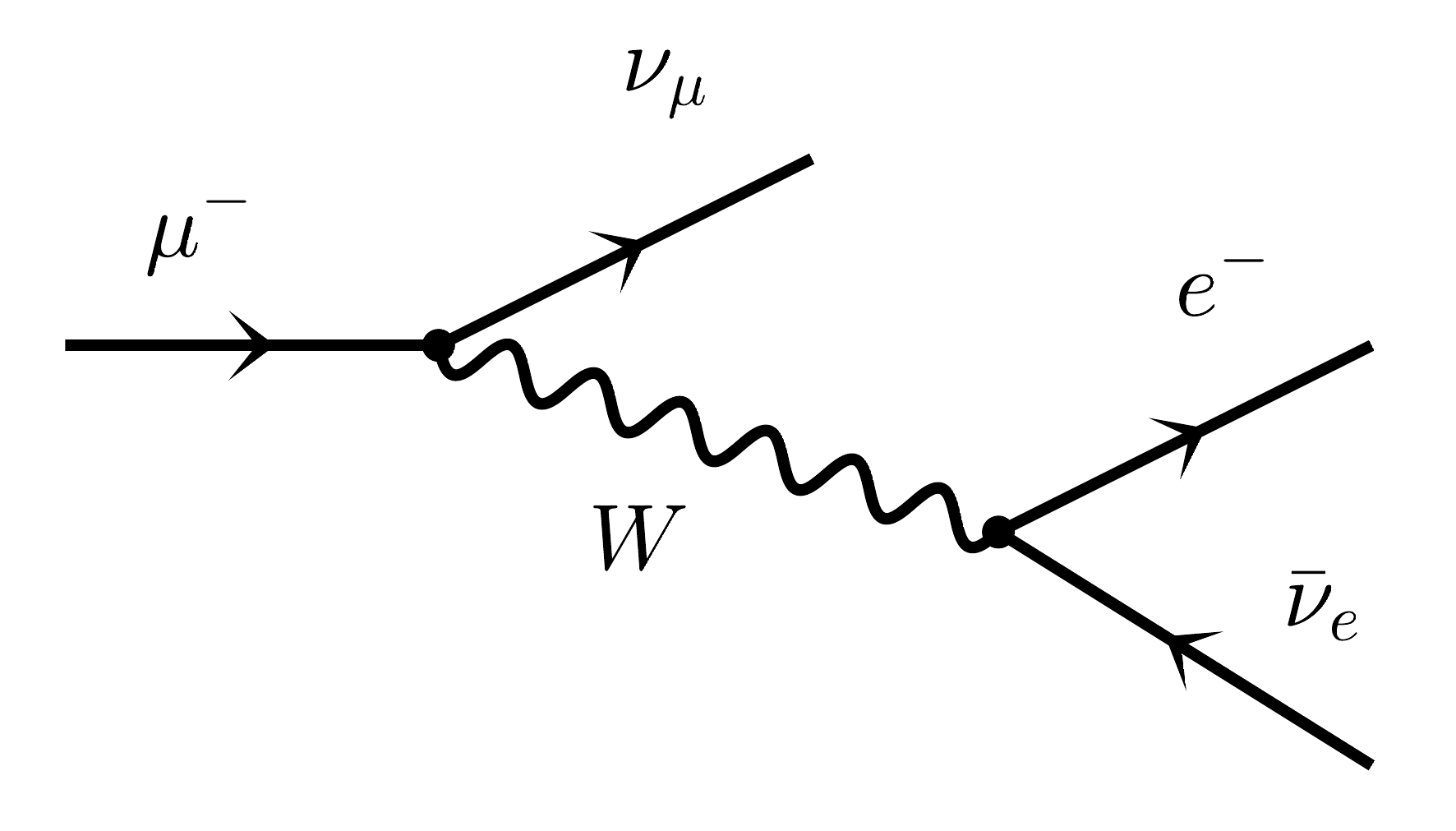}
	\caption{Feynman diagram describing the decay of a muon via the weak interactions.}
	\label{sec3.2:fig1}
\end{figure}

In this section, we give a brief introduction to the concept of the effective Hamiltonian, a fundamental tool for the study of weak charm-quark decays. It is instructive to start from the simplest example of a  weak decay, the muon decay $\mu^-  \to \nu_\mu + e^- + \bar{\nu}_e$, since, to 
good approximation,
QCD effects can be neglected~\footnote{QCD effects arise first at the two-loop order.
}.
The Feynman diagram describing this process is sketched in Fig.~\ref{sec3.2:fig1}.
Neglecting in the $W$ propagator the virtual momentum $k$ of the $W$ boson, which is of the order of the muon mass, i.e. using $m_\mu^2 \ll m_W^2$, the resulting decay rate, which equals the total width $\Gamma$ of the muon, reads - see e.g.~\cite{Michel:1949qe} 
for an early reference:
\begin{equation}
\Gamma(\mu^- \to \nu_\mu e^- \bar{\nu}_e) =
\frac{G_F^2 m_\mu^5}{192 \pi^3} f \left(\frac{m_e}{m_\mu} \right) 
\, ,
\label{sec3.2:eq1}
\end{equation}
where $G_F = g_2^2/(4 \sqrt{2} m_W^2)$ denotes the Fermi constant and 
$f(x)$ is the phase space-function for one massive particle in the final state, namely
\begin{equation} 
f(x)  =  1 - 8 x^2 + 8 x^6 - x^8 - 24 x^4 \ln (x)\,.
\label{sec3.2:eq2}
\end{equation}
The result in Eq.~\eqref{sec3.2:eq1} can be compared to $\tau(\mu)^{\rm exp}$, the experimental value of the muon lifetime~\cite{ParticleDataGroup:2024cfk}, that is
\begin{equation}
\frac{1}{\Gamma} =  \tau(\mu)^{\rm th} = 2.18776 \cdot 10^{-6} 
\, \mbox{s}
\,, \quad 
\tau(\mu)^{\rm exp} = 2.1969811(22) \cdot 10^{-6} \, \mbox{s} \,,
\label{sec3.2:eq3}
\end{equation}
which already shows a good agreement between the theoretical and the experimental determinations. Including also higher order 
electro-weak corrections to the total decay rate in Eq.~\eqref{sec3.2:eq1},
see e.g.\ the review~\cite{Sirlin:2012mh},
the experimental value is perfectly reproduced.\\

\begin{figure}[t]
	\centering
    \includegraphics[width=0.7\textwidth]{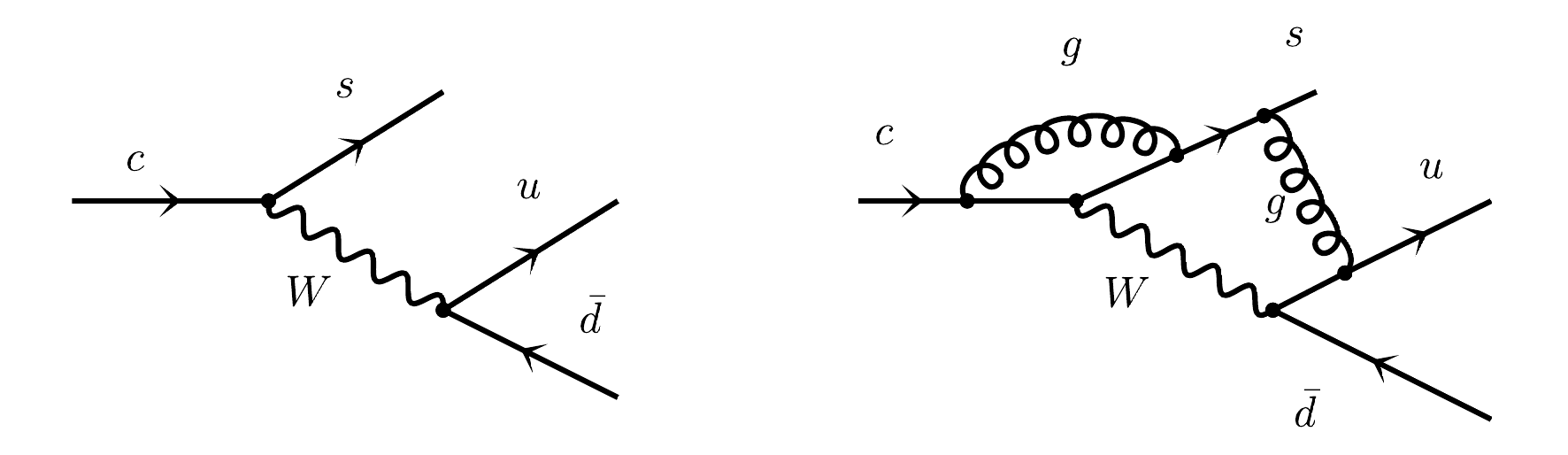}
	\caption{Examples of Feynman diagrams describing the weak decay $ c \to s \bar{d} u$ at tree-level (left) and at 2-loop in the strong coupling (right).}
	\label{sec3.2:fig2}
\end{figure}

\noindent
We now turn to weak charm-quark decays. Consider, for instance, the non-leptonic decay $c \to s + W^- \to s + \bar{u} + d$, see Fig.~\ref{sec3.2:fig2} (left), where, for simplicity, we neglect the masses of the final state quarks~\footnote{The up and down quark masses can be safely neglected, $m_{d} = m_{u} \ll m_c$; the strange quark, $m_s \approx 0.1 m_c$, gives sizeable contributions and should be retained in actual computations, however for the purpose of this section it is sufficient to omit them.}. In this decay, two very different scales arise, namely
the mass of the $W$-boson $m_W \approx$ 80 GeV, and the mass of the charm quark $m_c \approx$ 1.5 GeV, i.e.\ $m_W \gg m_c$. Now, unlike the muon decay, also QCD contributions must be taken into account, gluons can be emitted from all fermion lines, as sketched in Fig.~\ref{sec3.2:fig2} (right), and these corrections, due to the large value of the strong coupling at the charm-quark scale, $\alpha_s(m_c) \approx 0.3$, can be sizeable.
Naively, one might expect one-loop diagrams to give $\alpha_s$-corrections of $\approx 0.3$, two-loop diagrams $\alpha_s^2$-corrections of $\approx 0.09$ etc. 
However, e.g.\ the computation of one-loop corrections shows that, in addition to terms of order $\alpha_s$, also {\bf large logarithmic terms} of the form
$ \alpha_s \ln \left( m_c^2/m_W^2 \right)$ arise.
As a result, one does not obtain a Taylor expansion in $\alpha_s (m_c) $ but an expansion in
$ \alpha_s \ln \left( m_c^2/m_W^2 \right) \approx - 8 \alpha_s$, which clearly seems to spoil the perturbative approach.\\

\noindent
The solution to this problem lies in the introduction of the
{\bf effective Hamiltonian}, described in detail e.g.\ in the lecture notes~\cite{Buras:1998raa} - see also the review~\cite{Buchalla:1995vs}. The basic idea is to derive an effective theory valid at scales of the order of $m_c$, where the heavy $W$ boson, which triggers the weak decay, is {\bf integrated out} and does not appear as a dynamical degree of freedom.  
\begin{figure}[h]
	\centering
    \includegraphics[width=0.7\textwidth]{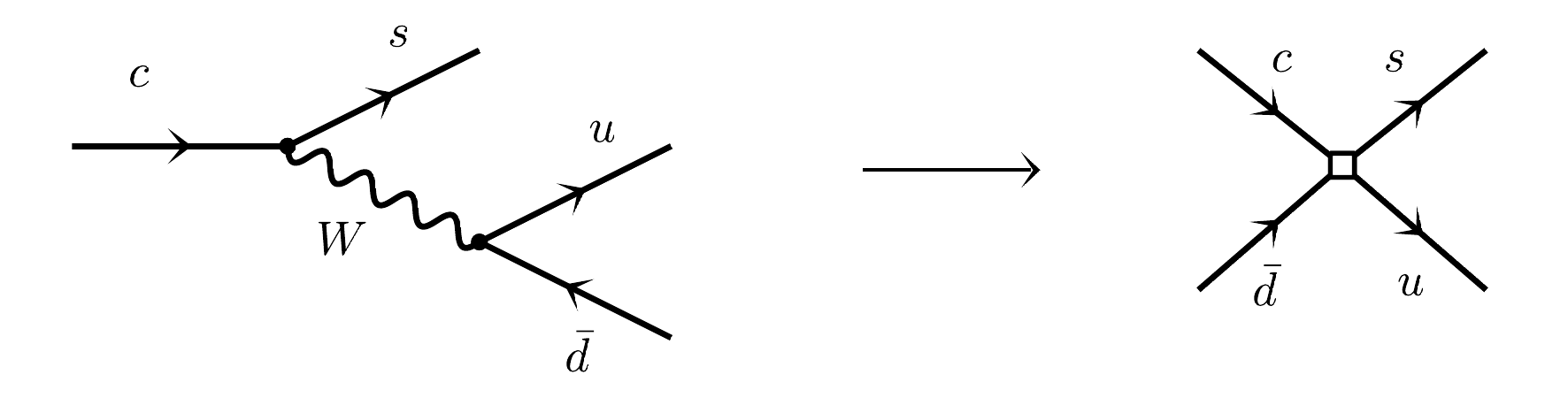}
	\caption{By expanding the $W$-boson propagator at leading order in  $1/m_W^2$, the $c \to s \bar d u$ amplitude in the SM (full
theory) (left) is matched into a local four-quark interaction in the weak effective theory (right).
    }
	\label{sec3.2:fig3}
\end{figure}
Schematically, this is obtained by expanding the decay amplitude 
in the small quantity $k^2/m_W^2$, here $k$, the transferred momentum to the $W$ boson, is of the order of the mass of the decaying quark and thus $k^2 \sim m_c^2 \ll m_W^2$. The leading term in this Taylor expansion would then correspond to a theory in which the propagator of the $W$ boson is contracted to a point and new effective four-fermion interactions are generated, see Fig.~\ref{sec3.2:fig3}. Such a theory would be described by an effective Hamiltonian ${\cal H}_{\rm eff}$ given by  
\begin{equation}
{\cal A} = \left[ \bar{s} \left(\frac{i g_2 V_{cs}^*}{2 \sqrt{2}}\right)  \gamma^\mu ( 1- \gamma_5)c\right]  
   \left[\frac{-i}{k^2-m_W^2} \right]
  \left[ \bar{u} \left( \frac{ig_2 V_{ud}}{2 \sqrt{2}}\right) \gamma_\mu ( 1- \gamma_5)d \right] 
  \quad 
   \overset{k^2 \ll m_W^2}{\longrightarrow}  
  \quad
  {\cal H}_{\rm eff}  =   
 \frac{G_F}{\sqrt{2}} V_{cs}^* V_{ud} C_2 Q_2  + {\cal O}\left(\frac{k^2}{m_W^2}\right) \,,
\end{equation}
where
the colour singlet four-quark operator $Q_2$ is defined as
\begin{equation}
Q_2  =  \left(\bar{s}_\alpha \gamma_\mu (1 - \gamma_5) c_\alpha  \right) 
          \left(\bar{u}_\beta  \gamma^\mu (1 - \gamma_5) d_\beta  \right) 
\equiv 
\left(\bar{s}_\alpha c_\alpha \right)_{V-A} 
          \left(\bar{u}_\beta  d_\beta  \right)_{V-A} 
           \, ,
\end{equation}
with $\alpha$ and $\beta$ being colour indices, and the {\bf Wilson coefficient} $C_2 = 1$. 
Including also QCD corrections, as shown in Fig.~\ref{sec3.2:fig2},  
the above description is generalised as follows:\\

\begin{itemize} 
    \item[$\diamond$] The Wilson coefficient $C_2$ deviates from one and acquires a dependence on the renormalisation
      scale $\mu$ of the strong coupling, i.e.\  $C_2(\mu) = 1 + {\cal O}(\alpha_s \ln)$, with $a_s \ln \equiv \alpha_s(\mu) \ln (m_W^2/\mu^2)$, where the matching condition at $\mu = m_W$ reproduces $C_2(m_W) = 1$.
    \item[$\diamond$] A second four-quark operator arises, the colour-rearranged operator $Q_1$, defined as
      \begin{equation}
      Q_1 \equiv \left(\bar{s}_\alpha c_\beta \right)_{V-A} 
          \left(\bar{u}_\beta  d_\alpha  \right)_{V-A} 
           \, ,
           \label{eq:Heff_Q1}
      \end{equation}
    with the corresponding Wilson coefficient being $C_1(\mu) = {\cal O}(\alpha_s \ln) $ and $C_1(m_W) = 0$.
    Including also $Q_1$, the general structure of the effective Hamiltonian describing tree-level weak decays of the charm quark then reads
    \begin{equation}
    {\cal H}_{\rm eff} =    \frac{G_F}{\sqrt{2}} V_{cs}^*V_{ud} \bigl( C_1 Q_1 + C_2  Q_2 \bigr) \, .
    \label{sec3.2:eq4}
     \end{equation}
     \item[$\diamond$]
     Due to the presence of large logarithms, the series in powers of $\alpha_s \ln \sim {\cal O}(1)$ cannot be truncated. However, by using the renormalisation group equations (RGEs) satisfied by the Wilson coefficients, these large logarithms can be {\bf resummed to all orders} in perturbation theory, yielding again a finite result of ${\cal O}(1)$.
      Including also higher order QCD corrections, 
      the structure of the $\alpha_s$-expansion at leading logarithmic~(LL) approximation, next-to-leading logarithmic~(NLL) approximation, etc., for a given loop order, is 
      shown in Table~\ref{sec3.2:tab1}.
\begin{table}
\centering
\renewcommand*{\arraystretch}{1.6}
\begin{tabular}[t]{|c||c|c|c|c|}
 \hline
   & LL & NLL & NNLL & NNNLL \\
  \hline
  \hline
   Tree & 1 & - & - & -  \\
   \hline
   1-loop & $\alpha_s \ln$ & $\alpha_s$  & - & -  \\
   \hline
   2-loop & $\alpha_s^2 \ln^2$ & $\alpha_s^2 \ln$  & $\alpha_s^2$  & -  \\
   \hline
   3-loop & $\alpha_s^3 \ln^3$ & $\alpha_s^3 \ln^2$  & $\alpha_s^3 \ln$ & $\alpha_s^3 $ \\
   \hline
   \ldots & \ldots & \ldots & \ldots & \ldots \\
   \hline
\end{tabular}
\caption{The general structure of the perturbative expansion in the strong coupling $\alpha_s$ of the weak charm decay amplitude. The progression of the loop expansion is shown row by row: at $n$-loop order, contributions of the type $\alpha_s^n \ln^n$,  $\alpha_s^n \ln^{n-1}$, \ldots , $\alpha_s^n$ arise, where the first term $\alpha_s^n \ln^n$ (LL) spoils the convergence of the series for $\alpha_s \ln \sim {\cal O}(1)$. The transition to the effective theory allows one to resum at LO-QCD all the LL contributions in the first column, at NLO-QCD all the NLL contributions in the second column etc., restoring thus the convergence of the whole series.}
\label{sec3.2:tab1}
\end{table}
      Computations performed in the full SM correspond to calculating the amplitude row by row along this table, namely computing the tree-level contribution, the 1-loop contribution etc. On the other hand, by using the effective Hamiltonian and employing the RGEs, the amplitude can be calculated column by column, that is by computing the LL contributions $(\alpha_s \ln)^n$, the NLL contributions $ \alpha_s (\alpha_s \ln)^n$, etc., to all orders, i.e.\ for $n = 0, \ldots, \infty$. 
      \item[$\diamond$] The weak effective theory (WET) constitutes an example of operator product expansion~(OPE).
      This construction leads to 
      an effective {\bf separation of scales}, with the renormalisation scale $\mu$ acting as a factorisation scale. The short-distance contributions at scales $\mu > m_c$ are encoded in the Wilson coefficients $C_i$, which can be computed perturbatively, while long-distance effects at scales $\mu < m_c$ are described by the matrix elements of the local operators $Q_i$. In charm hadrons decays, one encounters hadronic matrix elements of these effective operators, for which perturbative calculations cannot be applied. In this case
      non-perturbative methods like lattice QCD~\cite{Wilson:1974sk}, see e.g.~\cite{RBC:2015gro} for an application to the decay $K \to \pi \pi$, or QCD sum rules~\cite{Shifman:1978bx} and light-cone sum rules (LCSRs)~\cite{Balitsky:1989ry}, see e.g.~\cite{Lenz:2023rlq} for an application to the decays $D^0 \to \pi^+ \pi^-, K^+ K^-$,
      must be employed to estimate these hadronic parameters.\\ 
\end{itemize}

\begin{figure}[b]
	\centering
    \includegraphics[width=0.5\textwidth,angle=0]{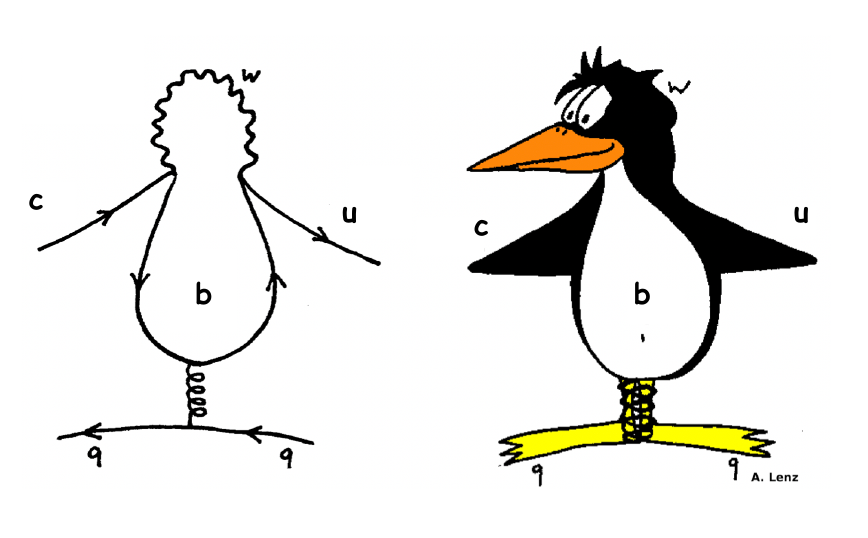}
	\caption{Penguin diagram describing the quark-level decay $c \to u q \bar  q$, contributing to e.g.\ $D^0 \to K^+ K^-$ ($q=s$) and $D^0 \to \pi^+ \pi^-$ ($q=d$). }
	\label{sec3.2:fig4}
\end{figure}

\noindent
So far, we have considered weak decays of the charm quark which are mediated by a tree-level exchange of a $W$ boson; however there are also loop-induced weak decays, like those described by the penguin diagram sketched in Fig~\ref{sec3.2:fig4}~\footnote{The origin of the name penguin diagram can be found in~\cite{Shifman:1995hc} and in the review  {\it How penguins started to fly} \cite{Vainshtein:1999ji}.}.
Analogously to the case of tree-level decays, the generalisation of the effective theory to include these new contributions is obtained by matching the amplitude computed in full SM into the WET, where all the degrees of freedom which are heavier than the charm quark are integrated out. Note that 
now, in addition to the $W$ boson, also other heavier particles like the $Z$ boson and the bottom-quark can contribute to the loop-processes.
As a result, the most general expression of the effective Hamiltonian describing weak decays of the charm quark $c \to q_1 \bar q_2 u $, with $q_i = u, d, s$, is given by~\cite{Buchalla:1995vs}
\begin{equation}
 {\cal H}_{\rm eff} = \frac{G_{F}}{ \sqrt{2}} \left[ \sum_{q_1, q_2= d,s} \lambda_{q_1 q_2}
 \left(  C_{1}  Q_{1}^{q_1 q_2} + 
   { C_{2}} Q_{2}^{q_1 q_2} \right) - \lambda_b \sum_{j \geq 3} 
   { C_{j}} { Q_{j} } \right] + {\rm h.c.} \, .
   \label{sec3.2:eq5}
\end{equation}
where $\lambda_{q_1 q_2} = V_{c q_1}^* V_{u q_2}$ and $\lambda_b = V_{cb}^* V_{ub}$ are combinations of CKM matrix elements, and the $\Delta C = 1$ operator basis includes:\\

\begin{itemize}
\item[$\diamond$] The current-current operators already introduced above:
\begin{eqnarray}
    Q_1^{q_1 q_2} = (\bar q_{1\alpha} c_\beta)_{V-A} (\bar u_{\beta} q_{2\alpha})_{V-A} \,, \quad 
    Q_2^{q_1 q_2} = (\bar q_{1\alpha} c_\alpha)_{V-A} (\bar u_{\beta} q_{2\beta})_{V-A} \,.
    \label{sec3.2:eq6}
\end{eqnarray}
\item[$\diamond$] The QCD penguin operators:
\begin{eqnarray}
    Q_3 = (\bar u_{\alpha} c_\alpha)_{V-A} \sum_{q = u,d,s,c} (\bar q_{\beta} q_\beta)_{V-A} \,, \quad 
     Q_4 = (\bar u_{\alpha} c_\beta)_{V-A} \sum_{q = u,d,s,c} (\bar q_{\beta} q_\alpha)_{V-A} \,, \\
     Q_5 = (\bar u_{\alpha} c_\alpha)_{V-A} \sum_{q = u,d,s,c} (\bar q_{\beta} q_\beta)_{V+A} \,, \quad 
     Q_6 = (\bar u_{\alpha} c_\beta)_{V-A} \sum_{q = u,d,s,c} (\bar q_{\beta} q_\alpha)_{V+A} \,. 
     \label{sec3.2:eq7}
\end{eqnarray}
\item[$\diamond$] The electromagnetic and chromomagnetic dipole operators:
\begin{equation}
    Q_7 = -\frac{e}{16 \pi^2} m_c \left(\bar u_\alpha \sigma_{\mu \nu} (1+\gamma_5) c_\alpha \right) F^{\mu \nu}\,, \quad 
    Q_8 = -\frac{g_s}{16 \pi^2} m_c \left(\bar u_\alpha \sigma_{\mu \nu} t^a_{\alpha \beta} (1+\gamma_5) c_\beta \right) G_a^{\mu \nu}\,.
    \label{sec3.2:eq8}
\end{equation}
\item[$\diamond$] The semileptonic operators:
\begin{equation}
    Q_9 = \frac{e^2}{16 \pi^2} \left(\bar u_\alpha \gamma_\mu (1- \gamma_5) c_\alpha \right) (\bar \ell \gamma^\mu \ell)\,, \qquad 
    Q_{10} =\frac{e^2}{16 \pi^2} \left(\bar u_\alpha \gamma_\mu (1- \gamma_5) c_\alpha) (\bar \ell \gamma^\mu \gamma_5 \ell \right) \,.
    \label{sec3.2:eq9}
\end{equation}
\end{itemize}
In Eq.~\eqref{sec3.2:eq8}, $e$ and $g_s$ denote respectively the electromagnetic and strong coupling, $F^{\mu \nu}$ and $G_a^{\mu \nu}$ the photon and gluon field strength tensor, where $t^a_{\alpha \beta}$ with $a = 1,\ldots, 8,$ are the SU(3)$_c$ generators, and $\sigma_{\mu \nu} = (i/2)[\gamma_\mu, \gamma_\nu]$. 
The effective Hamiltonian in Eq.~\eqref{sec3.2:eq5} was derived at LL accuracy already in 1974~\cite{Altarelli:1974exa}. A comprehensive review of the NLL results for the Wilson coefficients is given in~\cite{Buchalla:1995vs}, while values at NNLL accuracy have been obtained in~\cite{deBoer:2016dcg}. An important remark is that, differently from down-type quark decays, where the penguin operators are already generated at the scale $\mu = m_W$ by integrating out the top quark running into the loop, in the case of charm-quark decays, penguin operators can only be generated at scales below the bottom-quark mass i.e.\ for $\mu < m_b$, where the $b$-quark does not appear as a dynamical degree of freedom and its effect is a short-distance contribution encoded in the corresponding Wilson coefficients. Consequently, $C_{3-9}$ receive
a non-zero contribution only below the scale $m_b$, due to the matching from the full theory to the WET and the operator mixing with $Q_{1,2}$; the corresponding matching conditions then read $C_{3, \ldots, 9}(m_W) = 0$. 
We also note that, in the literature it is customary to use the renormalisation scheme independent effective dipole Wilson coefficients $C_7^{\rm eff}$, $C_8^{\rm eff}$, see e.g.\ in~\cite{Buchalla:1995vs} for their definition. Finally, the operator $Q_{10}$ does not mix under renormalisation
and $C_{10}$ vanishes at all scales at leading order in $G_F$, that is $ C_{10} (m_W) = C_{10} (m_c)  = 0$~\cite{deBoer:2016dcg}. A summary of the values of Wilson coefficients at NLL is shown in Table~\ref{sec3.2:tab2}. \\

\noindent
To conclude, the effective Hamiltonian is the fundamental building block for the calculation of inclusive and exclusive weak decays of hadrons containing charm quarks. The Wilson coefficients play the role of new elementary couplings in the effective theory and the four-quark operators of new elementary vertices.
\begin{table}[t]
\centering
\renewcommand*{\arraystretch}{1.7}
\begin{tabular}[t]{|c||c|c|c|c|c|c|c|c|c|c|}
 \hline
  $ \mu $ (GeV)& $C_1$ & $C_2$ & $C_3$ & $C_4$ & $C_5$ & $C_6$ & $C_7^{\rm eff}$ & $C_8^{\rm eff}$ & $C_9$ & $C_{10}$ \\
  \hline
  \hline
  1.3  & -0.36 & 1.16 & 0.01 & -0.04 & 0.01 &  -0.05 & 0.05  & -0.06 & -0.31 & 0\\
   \hline
\end{tabular}
\caption{Summary of the $\Delta C = 1$ Wilson coefficients at NLL accuracy and at the renormalisation scale $\mu = 1.3$ GeV~\cite{deBoer:2016dcg}.}
\label{sec3.2:tab2}
\end{table}
%%%%%%%%%%%%%%%%%%%%%%%%%%%%%%%%%%%%%%%%%%%%%%%%%%%%%%%%%%%%%%%%%%%%%%%%%%%%%%%%%%%%%%%%%%%%%%%%%%%%%%%%%%%%%%%%%%%%%%%%%%%%%%%%
\subsection{Classification of charm hadron decays}
\label{subsec:classification}
Charm hadron decays can be classified, according to the complexity of their hadronic structure, in leptonic, semileptonic and non-leptonic decays, where the first is the simplest type of decay and the last the most complex. In addition, 
we distinguish between {\it exclusive decays} and {\it inclusive decays}. In the first case the final state, either leptonic, semileptonic or non-leptonic, is precisely
specified, while in the second case, the final state is not specified, but it consists of a sum of exclusive modes all driven by the same elementary transition.\\
\\
\begin{minipage}{\textwidth}
\begin{minipage}{0.42\textwidth}
 \centering
   \includegraphics[width=0.8\textwidth,angle = 0]{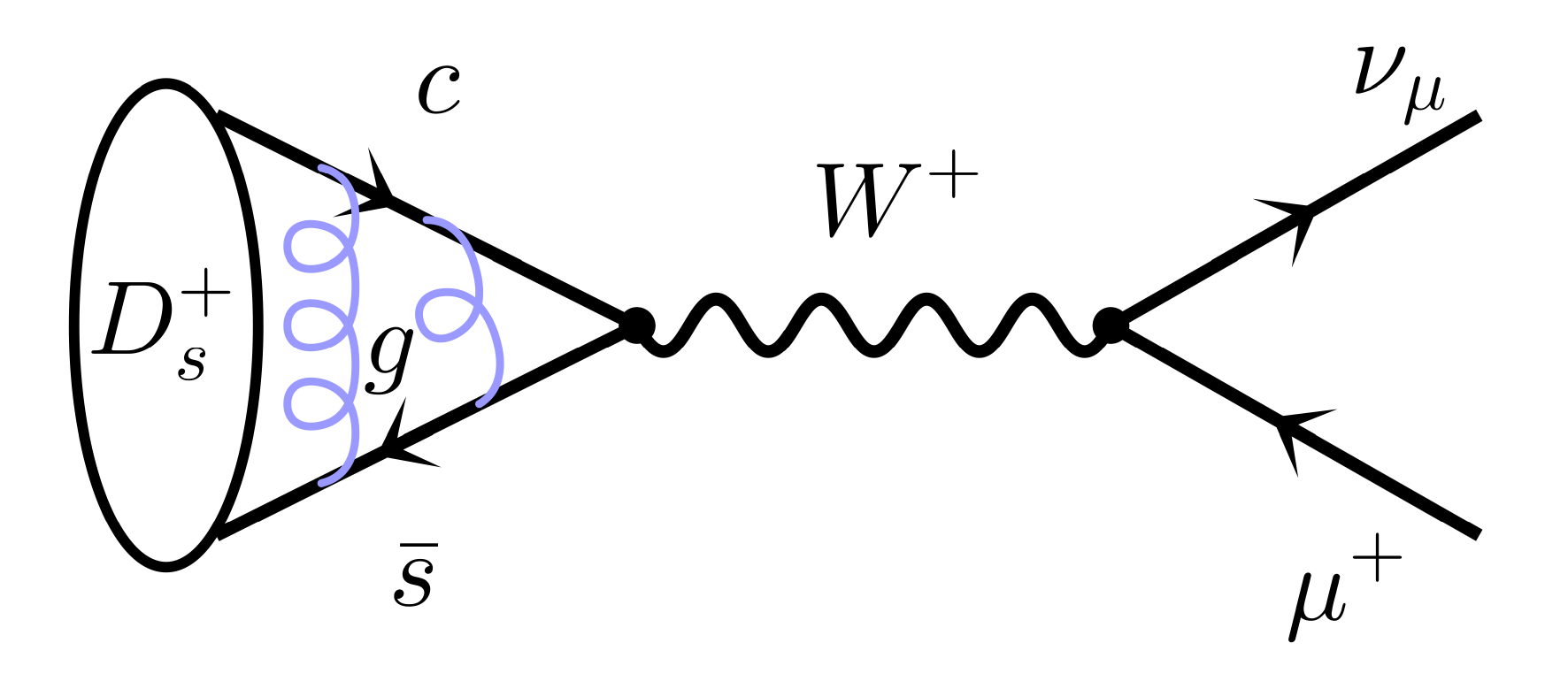}
	\captionof{figure}{Feynman diagram describing the leptonic tree-level decay $D_s^+ \to \mu^+ \nu_\mu$. Non-perturbative QCD interactions - shown in violet - are responsible for the binding of the $D_s^+$ meson.}
	\label{sec3.3:fig1}
\end{minipage}
\hspace*{5mm}
\begin{minipage}{0.5\textwidth}
\vspace*{9mm}
\centering
\renewcommand*{\arraystretch}{1.4}
\begin{tabular}[h]{|c|c|c||c|}
 \hline
  $f_D$ (MeV) & $f_{D_s}$ (MeV) & $f_{D_s}/f_{D}$ & Ref. \\
  \hline
  \hline
  212.0 (0.7) & 249.9 (0.5) & 1.1783 (0.0016) & \cite{FlavourLatticeAveragingGroupFLAG:2024oxs} \\
  \hline
\end{tabular}
\captionof{table}{Averages of Lattice QCD results for the $D$-meson decay constants based on simulations using $N_f = 2 + 1 + 1$ active flavours, as determined by the FLAG collaboration~\cite{FlavourLatticeAveragingGroupFLAG:2024oxs}. For the detailed list of references to the original computations see~\cite{FlavourLatticeAveragingGroupFLAG:2024oxs}.}
\label{sec3.3:tab1}
\end{minipage}
\end{minipage}\\
\\

\noindent
The most inclusive observable is the total decay width or equivalently the lifetime, as this includes a sum over all possible final states, leptonic, semileptonic and non-leptonic, into which a given particle can decay.\\

\noindent
{\bf Leptonic decays:}
leptonic decays have only leptons in the final state, 
an example is the tree-level decay $D_s^+ \to \mu^+ \;  {\nu}_\mu$ shown in Fig.~\ref{sec3.3:fig1}. This type of decay has the simplest hadronic structure, since from a QCD point of view the meson is decaying into nothing; leptons do not interact strongly and can thus be identified with the QCD vacuum. 
The hadronic part of the decay amplitude of $D_s^+ \to \mu^+ \;  {\nu}_\mu$ is described by the following matrix element
\begin{equation}
\langle 0 | \bar{s}  \gamma^\mu \gamma_5 c  | D_s(p_{D_s}) \rangle  =  i f_{D_s} p_{D_s}^\mu \, ,
 \end{equation}
where all non-perturbative effects that govern the binding of the quarks via gluons in the initial hadronic state 
 are described by one parameter: the {\bf decay constant}, $f_{D_s}$. In the above equation
 $p_{D_s}^\mu$ denotes the four-momentum of the $D_s^+$ meson.
  Decay constants can, nowadays, be very precisely determined by Lattice QCD simulations. Current values for the decay constants of the $D^{+,0}$ and $D_s^+$ mesons and of their ratio are shown in Table~\ref{sec3.3:tab1}, as reported by the FLAG (Flavour Lattice Averaging Group) collaboration~\cite{FlavourLatticeAveragingGroupFLAG:2024oxs}. We stress that isospin symmetry  is assumed, i.e.\  $m_u=m_d$, so that $f_{D^+} = f_{D^0} \equiv f_D$. Note that, leptonic decays can also proceed via loop-level contributions in the SM, an example is the decay $D^0 \to \mu^+ \mu^-$. \\      
%%%%%%%%%%%%%%%%%%%%%%%%%%%%%%%%%%%%%%%%%%%%%%%%%%%%%%%%%%%%%%%%%%%%%%%%%%%%%%%%%%%%%%%%%%%%%%%%%%%%%%%%%%%%%%%%%%%%%%%%%%%%%%%%
\noindent
{\bf Semileptonic decays:}
semileptonic decays have both leptons and hadrons in the final state, an example is the tree-level decay $D^+ \to \bar{K}^0 e^+  {\nu}_e$ shown in Fig.~\ref{sec3.3:fig2}.
In this case, the hadronic structure is more complicated as the 
non-perturbative QCD effects not only are responsible for
the binding of the hadrons in the initial and final states,
but also for the strong interaction between the two.
The hadronic part of the decay amplitude of $D^+ \to \bar{K}^0 e^+ \nu_e$ is now described by a matrix element between 
the $D^+$ and $\bar{K}^0$ states, triggered by a current that couples the charm to the strange quark:
\begin{equation}
\langle \bar{K}^0(p_K) | \bar{s}  \gamma^\mu  c | D^+ (p_D)\rangle = f_+^{D \to K}(q^2) 
         \left( p_D^\mu + p_K^\mu - \frac{m_D^2 - m_K^2}{q^2} q^\mu \right)
         + f_0^{D \to K}(q^2) \frac{m_D^2 - m_K^2}{q^2} q^\mu\, .
      \end{equation}
This matrix element can be generally parametrised in terms of two 
functions, the {\bf form factors} $f_+^{D \to K}(q^2)$ and $f_0^{D \to K}(q^2)$, that contain all the non-perturbative dynamics.
These are functions of $q^2$, the invariant mass squared of the leptonic pair, where $q^\mu$ is the four-momentum transferred between the initial and final state hadrons.
Form factors can be determined using lattice QCD 
 calculations or LCSRs - see e.g.\ the textbook~\cite{Khodjamirian:2020btr}. Current values for the $D \to \pi$ and $D \to K$ form factors at the fixed point $q^2 = 0$~\footnote{Note that at this point one finds that $f_+ = f_0$, as reported by the FLAG collaboration~\cite{FlavourLatticeAveragingGroupFLAG:2024oxs} as well as obtained by LCSRs studies~\cite{Khodjamirian:2009ys}.} are summarised in Table~\ref{sec3.3:tab2}. 
 As for the baryon form factors, results within LCSRs for e.g.\ the transition
 $\{\Lambda_c, \Sigma_c \}\to p$ the can be found in~\cite{Khodjamirian:2011jp}, where the latter were obtained using the non-perturbative inputs from~\cite{Braun:2001tj, Lenz:2003tq, Braun:2006hz, Lenz:2009ar}. Lattice QCD determinations for baryon form factors including $\Lambda_c \to \{ \Lambda, n, p\} $ have been obtained in~\cite{Meinel:2016dqj, Meinel:2017ggx, Zhang:2021oja, Meinel:2021mdj, Farrell:2025gis}.
 There exist again semileptonic decays that can 
 only proceed via loop-contributions in the SM, an example is the decay
 $D^0 \to \pi^0 \mu^+ \mu^-$. \\
  
 \begin{minipage}{\textwidth}
\begin{minipage}{0.43\textwidth}
 \centering
   \includegraphics[width=0.73\textwidth,angle = 0]{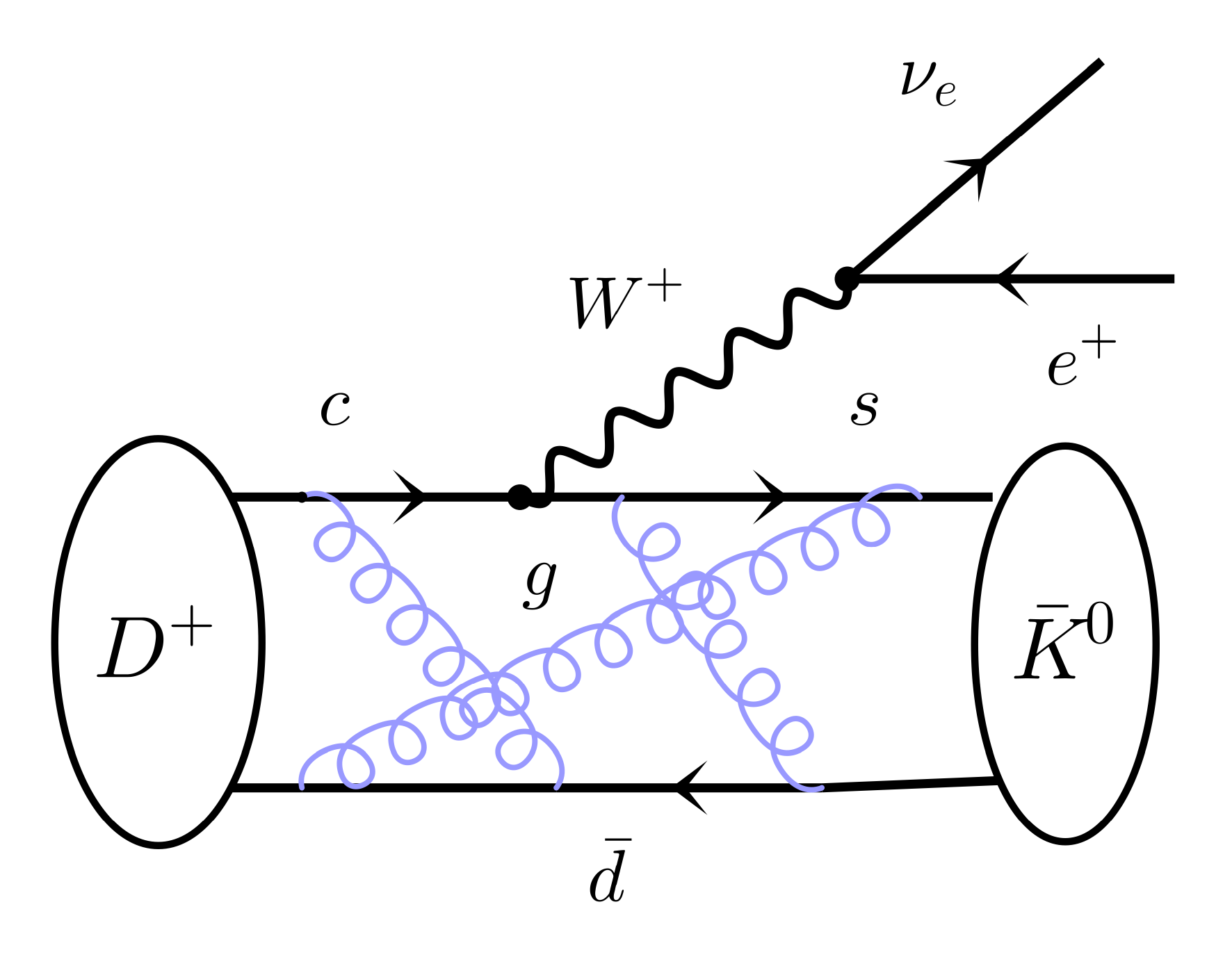}
	\captionof{figure}{Feynman diagram describing the semileptonic tree-level decay $D^+ \to \bar{K}^0 e^+ \nu_e$. Non-perturbative QCD effects - shown in violet - are responsible for
the binding of the $D_s^+$ and $\bar K^0$ mesons as well as for their interactions.}
	\label{sec3.3:fig2}
\end{minipage}
\hspace*{2mm}
\begin{minipage}{0.52\textwidth}
 \vspace*{2cm}
\centering
\renewcommand*{\arraystretch}{1.4}
 \begin{tabular}[h]{|c||c|c|} 
 \hline
  & Lattice QCD~\cite{FlavourLatticeAveragingGroupFLAG:2024oxs} & LCSRs~\cite{Khodjamirian:2009ys} \\
 \hline
 \hline
  $f_+^{D \to \pi} (0)$ & 0.6296 (50) & $0.67^{+ 0.10}_{- 0.07}$  \\
  \hline
  $f_+^{D \to K} (0)$ &  0.7430 (27) &  $0.75^{+ 0.11}_{- 0.08}$ \\
  \hline
\end{tabular}
\captionof{table}{Comparison between FLAG averages~\cite{FlavourLatticeAveragingGroupFLAG:2024oxs}, based on simulations using $N_f = 2 + 1 + 1$ active flavours, and LCSR studies~\cite{Khodjamirian:2009ys} for the $D \to \pi$ and $D \to K$ form factors at $q^2 = 0$.
For the detailed list of references to the original lattice QCD computations see~\cite{FlavourLatticeAveragingGroupFLAG:2024oxs}.}
\label{sec3.3:tab2}
\end{minipage}
\end{minipage}\\
\\

\begin{figure}[b]
	\centering
  \includegraphics[width=0.65\textwidth]{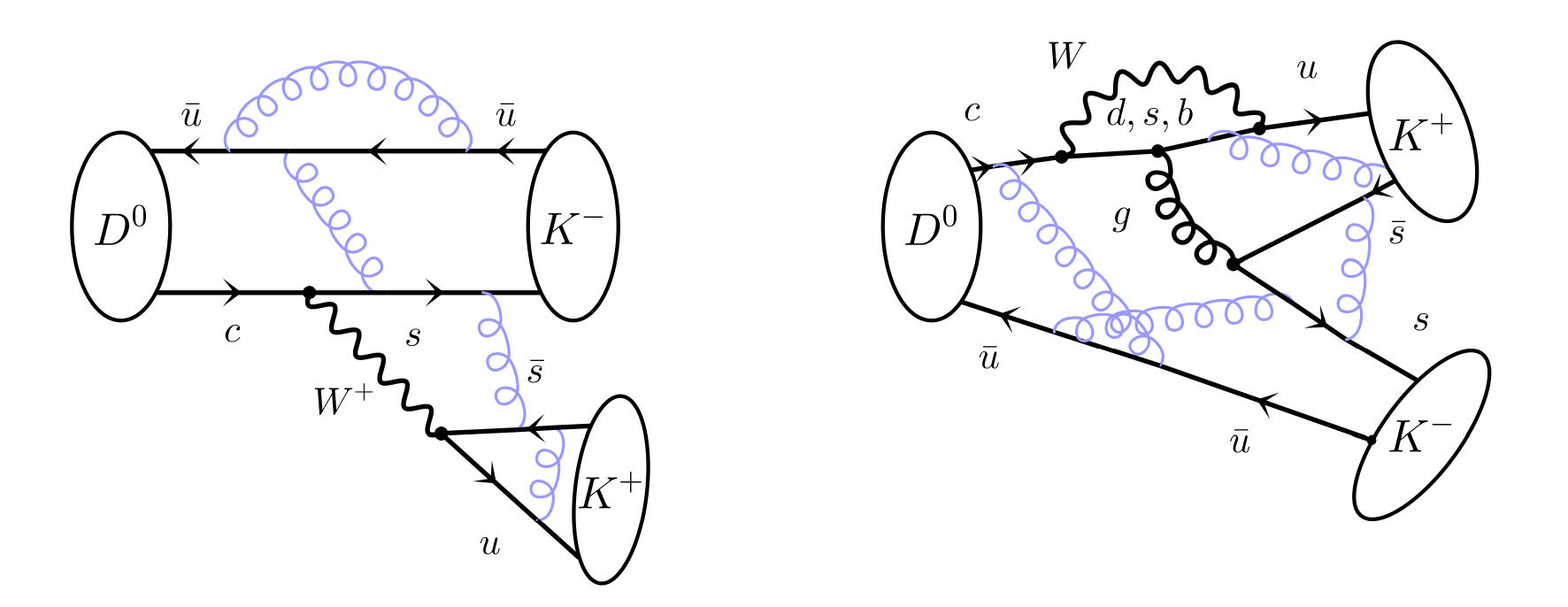}
	\caption{Example of tree-level (left) and penguin (right) Feynman diagrams contributing to the non-leptonic decay $D^0 \to K^+ K^-$. Non-perturbative QCD effects - shown in violet - are responsible for
the binding of all mesons as well as for their interactions.}
	\label{sec3.3:fig3}
  \end{figure}

%%%%%%%%%%%%%%%%%%%%%%%%%%%%%%%%%%%%%%%%%%%%%%%%%%%%%%%%%%%%%%%%%%%%%%%%%%%%%%%%%%%%%%%%%%%%%%%%%%%%%%%%%%%%%%%%%%%%%%%%%%%%%%%%
\noindent
{\bf Non-leptonic decays:}
non-leptonic decays have only hadrons in the final state, 
an example is the decay $D^0 \to  K^+  K^-$, shown in Fig.~\ref{sec3.3:fig3}.
Since in this case, QCD effects are responsible not only for the binding of all the hadronic states but also for the interactions between the initial and all final state particles, as well as among the latter, quantitative SM predictions for this class of decays are extremely challenging, and currently, no systematic description has been achieved. 
The decay amplitude for $D^0 \to K^+ K^-$ contains the matrix element of a four-quark operator between the initial state $D^0$ and the final state of the two-kaon system, which cannot be further simplified without additional assumptions. 
For certain decay topologies, a first estimate of this hadronic matrix element can be obtained using the naive factorisation~(NF) approximation, in which case the decay amplitude for a two-body decay such as 
$D^0 \to K^+ K^-$ is factorised in the product of the kaon decay constant $f_K$ and of the $D \to K$ scalar form factor $f_0^{D \to K}$, namely
      \begin{align}
      \langle K^+ K^- | (\bar{s}  \gamma_\mu  (1-\gamma_5) c ) \left(\bar{u}  \gamma^\mu  (1-\gamma_5) s \right) | D^0 \rangle|_{\rm NF} 
      &= 
     \langle K^-       | \bar{s}  \gamma_\mu  (1-\gamma_5) c  | D^0 \rangle \,\, 
      \langle   K^+ | \bar{u}  \gamma^\mu  (1-\gamma_5) s  | 0   \rangle 
      \nonumber \\[2mm]  
      &= i f_K (m_D^2 - m_K^2) f_0^{D\to K}(m_K^2)\,.
      \end{align}

\noindent
Theoretical investigations into whether or not the naive factorisation assumption is justified constitute an important topic of the current research, see e.g.~\cite{Lenz:2023rlq}. 
Note that in early 2000, the framework of QCD factorisation~(QCDf) was developed in~\cite{Beneke:1999br, Beneke:2000ry, Beneke:2001ev} for the study of exclusive $B$-meson decays in the heavy-quark limit $m_b \gg \Lambda_{\rm QCD}$. 
However, given the smaller value of the charm quark mass, as compared to the bottom quark mass, an analogous approach valid for exclusive charm hadron decays has not been established yet, see e.g.~\cite{Feldmann:2017izn, Wirbel:1985ji, Fu-Sheng:2011fji} for studies of the factorisation hypothesis in $D$-meson decays. \\

\noindent
Beyond the factorisation hypothesis, hadronic decays can also be studied using the framework of QCD sum rules, although, applications of this method to non-leptonic decays have mainly focused on the $b$-sector, see e.g.~\cite{Khodjamirian:2023wol}. Studies within LCSRs of the two-body decays $D^0 \to \pi^+ \pi^-$, $D^0 \to K^+ K^-$ can be found in~\cite{Khodjamirian:2017zdu, Lenz:2023rlq}.
Other approaches widely used in the literature for the analysis of charm-hadron decays make use of topological amplitude decompositions together with the SU(3)$_F$ symmetry, i.e.\ the limit $m_u = m_d = m_s$, see e.g.~\cite{Muller:2015lua, Hiller:2012xm, Grossman:2012ry, Cheng:2019ggx}, or of dispersive methods, see e.g.~\cite{Pich:2023kim, Omnes:1958hv, Hanhart:2012wi}.\\
%%%%%%%%%%%%%%%%%%%%%%%%%%%%%%%%%%%%%%%%%%%
%%%%%%%%%%%%%%%%%%%%%%%%%%%%%%%%%%%%%%%%%%%
%%%%%%%%%%%%%%%%%%%%%%%%%%%%%%%%%%%%%%%%%%%
\subsection{Mixing and CPV in mixing}
\label{subsec:basics_mixing}
\begin{figure}[t]
    \centering
    \includegraphics[scale=0.3]{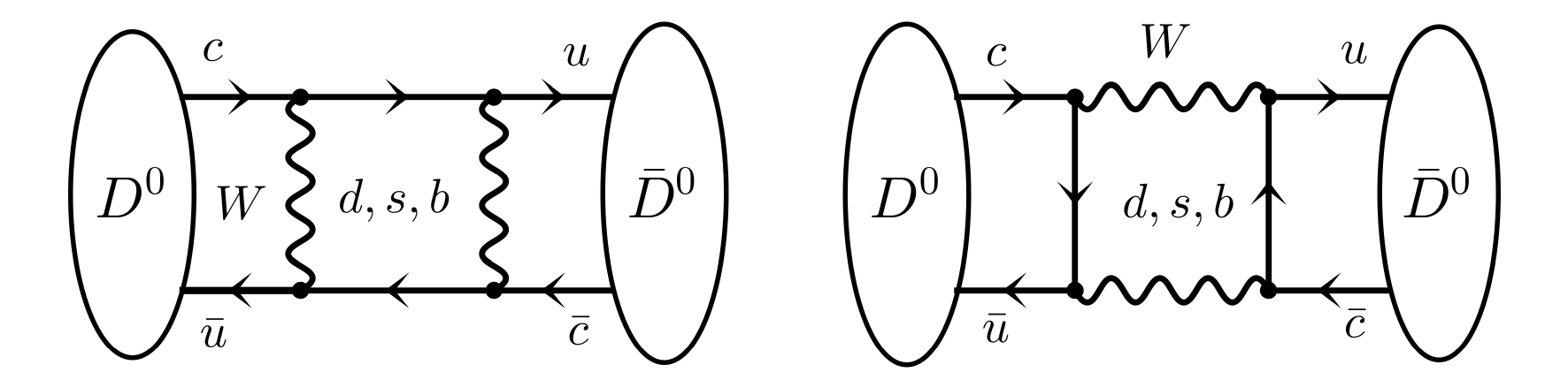}
    \caption{Box diagrams responsible for the mixing between the $D^0$ meson and its anti-particle $\bar D^0$. Analogous diagrams with up-type quarks inside the loop describe  mixing of down-type mesons i.e.\ $K^0, B_d^0,$ and $B_s^0$.}
    \label{sec3.4:fig1}
\end{figure}

The neutral ground state mesons $K^0$, $D^0$, $B_d^0$, and $B_s^0$ can transform, via the weak interaction, into their anti-particles, leading to the phenomenon of {\bf meson mixing}~\footnote{For recent reviews of charm mixing and CPV see also~\cite{Pajero:2022vev, Lenz:2020awd}.}, see Fig.~\ref{sec3.4:fig1}. 
The time evolution of e.g.\ the two-particle system $\{D^0, \bar D^0\}$ is governed by a $2\times 2$ Hamiltonian matrix $\hat {\cal H}$ and can be described by a Schr\"odinger like equation:
\begin{equation}
    i \frac{d}{dt} 
    \begin{pmatrix}
     |D^0(t) \rangle \\
     |\bar D^0(t) \rangle
     \end{pmatrix}
     = 
     \underbrace{\begin{pmatrix}
M_{11} - \frac{i}{2} \Gamma_{11} & M_{12} - \frac{i}{2} \Gamma_{12}\\
M_{21} - \frac{i}{2} \Gamma_{21} & M_{22} - \frac{i}{2} \Gamma_{22}
     \end{pmatrix}}_{\hat {\cal H} =\hat M - \frac{i}{2}\hat \Gamma}
     \begin{pmatrix}
     |D^0(t) \rangle \\
     |\bar D^0(t) \rangle
     \end{pmatrix}\,,
     \label{sec3.4:eq0}
\end{equation}
where $\hat M$ and $\hat \Gamma$ denote the mass and decay matrices, respectively~\footnote{
Note that $M_{11} = M_{22}$ and $\Gamma_{11} = \Gamma_{22}$, which follow from the requirement of invariance of the Hamiltonian under the action of CPT i.e.\ the combined discrete symmetries charge-conjugation, parity and time; moreover, $M_{21} = M_{12}^*$ and $\Gamma_{21} = \Gamma_{12}^*$, as consequence of the hermiticity of $\hat M$ and $\hat \Gamma$.}.
From Eq.~\eqref{sec3.4:eq0} we see that the {\it weak eigenstates} $|D^0 \rangle = |c \bar{u}\rangle$ and $|\bar{D}^0\rangle = |\bar{c} u\rangle$, 
which have a definite quark-flavour content, are not eigenstates of $\hat {\cal H}$ and are thus not characterised by a definite mass and decay width. The {\it mass eigenstates} $|D_{1,2}\rangle$ are obtained by diagonalising the Hamiltonian matrix and correspond to a linear combination of the flavour eigenstates. At $t=0$, their expression is given by
\begin{align}
|D_1\rangle =  p |D^0\rangle - q |\bar D^0 \rangle\,, 
\label{sec3.4:eq1}\\
|D_2\rangle = p |D^0\rangle + q |\bar D^0\rangle \,,
\label{sec3.4:eq2}
\end{align}
with $p,q$ being, in general, complex coefficients satisfying $|p|^2 + |q|^2 = 1$. At $t \neq 0$ the flavour eigenstates will start to mix and 
the characteristic properties of this phenomenon depend on the value of the mass difference $\Delta M_D \equiv M_1 - M_2$ and of the decay rate difference $\Delta \Gamma_D \equiv \Gamma_1 - \Gamma_2$ between the mass and decay width of the mass eigenstates $M_{1,2}$ and $\Gamma_{1,2}$.
The general expression for the time evolution of the weak eigenstates is
   \begin{eqnarray}
   |D^0 (t)\rangle       
   & = &             
   g_+(t) |D^0 \rangle  
   + 
   \frac{q}{p} g_-(t) |\bar{D}^0 \rangle \; ,
   \label{sec3.4:eq3} 
   \\
   |\bar{D}^0 (t)\rangle 
   & = & 
   \frac{p}{q} g_-(t) |D^0 \rangle  
   +             
   g_+(t) |\bar{D}^0 \rangle  \; ,
   \label{sec3.4:eq4} 
   \end{eqnarray}
where the functions $g_{\pm}(t)$ read
   \begin{eqnarray}
   g_+ (t) & = &   e^{- i M_{D} t} e^{- \Gamma_{D} /2 t}
    \left[ \cosh \frac{\Delta \Gamma_D t }{4} \cos \frac{\Delta M_D t}{2} -
         i \sinh \frac{\Delta \Gamma_D t }{4} \sin \frac{\Delta M_D t}{2} 
   \right] \, ,
   \label{sec3.4:eq5} 
   \\
   g_- (t) & = &   e^{- i M_{D} t} e^{- \Gamma_{D}/2 t}
    \left[ - \sinh \frac{\Delta \Gamma_D t }{4} \cos \frac{\Delta M_D t}{2} +
           i \cosh \frac{\Delta \Gamma_D t }{4} \sin \frac{\Delta M_D t}{2} 
   \right] \, ,
   \label{sec3.4:eq6} 
   \end{eqnarray}
and the averaged mass $M_{D} $ and decay rate $\Gamma_D$ are defined as
   \begin{equation}
    M_{D} = \frac{M_1 + M_2}{2} \,   ,
   \hspace{1cm}
   \Gamma_{D} = \frac{\Gamma_{1} + \Gamma_{2}}{2} \, . 
   \end{equation} 
The mixing observables $\Delta M_D$ and $\Delta \Gamma_D$ are computed from the off-diagonal entries of $\hat {\cal H}$, that is $M_{12}$ and $\Gamma_{12}$. The former describes the dispersive part of the $D^0 \to \bar D^0$ amplitude, namely the off-shell contributions to the box diagrams in Fig.~\ref{sec3.4:fig1} due to internal down, strange and bottom quarks; the latter describes the absorptive part of the $D^0 \to \bar D^0$ amplitude, i.e.\ the on-shell contributions to the box diagrams in Fig.~\ref{sec3.4:fig1} due to internal down and strange quarks.\\

\noindent
Turning to CP violation, we adopt the convention for the flavour eigenstates CP$|D^0\rangle = - |\bar D^0\rangle$. 
Acting with the CP operator on the mass eigenstates in Eqs.~\eqref{sec3.4:eq1}, \eqref{sec3.4:eq2}, we see that these are invariant under CP only if 
$|p/q| = 1$~\footnote{With the adopted phase convention, it follows that in the CP symmetric case, with $q = p$,  $D_1$ is the CP even final state and $D_2$ is CP odd.}. Thus, a non-zero value of $|p/q| - 1$ is a measure of {\bf CPV in mixing}, which is quantified by the relative phase between $M_{12}$ and $\Gamma_{12}$, that is, $\phi_{12} = \arg(- M_{12}/\Gamma_{12})$. 
Taking into account that experimental determinations indicate a small value for $\phi_{12}$, it is possible to derive a simple expression for the mass and decay rate differences $\Delta M_D$ and $\Delta \Gamma_D$, in terms of $M_{12}$ and $\Gamma_{12}$, as a Taylor expansion in the small parameter $\phi_{12}$ i.e.
\begin{eqnarray}
    x \equiv \frac{\Delta M_D}{\Gamma_D}
      \approx  \frac{2 | M_{12} |}{\Gamma_D}
      \, ,
      &&
    y \equiv \frac{\Delta \Gamma_D}{2 \Gamma_D}
      \approx  \frac{ | \Gamma_{12} |}{\Gamma_D}
      \, ,
      \label{sec3.4:eq7}
\end{eqnarray}
where the notation in terms of $x$ and $y$ is usually adopted in the literature. In this case, CPV in mixing takes the approximate expression
\begin{equation}
\left| \frac{p}{q}\right| \approx 1 - \frac{1}{2} \left|\frac{\Gamma_{12}}{M_{12}}\right|\sin{\phi_{12}} \,.
\label{sec3.4:eq8}
\end{equation}
The theoretical determination of the mixing parameters $x$, $y$, and $\phi_{12}$ is a very difficult task. Box diagrams, such as those in Fig.~\ref{sec3.4:fig1}, are severely affected by the GIM mechanism, i.e.\ the amplitude vanishes in the limit of equally massive internal quarks as consequence of the unitarity of the CKM matrix.
Due to the fact that the masses of all down-type quarks are much smaller than that of the $W$ boson, and that for charm decays the unitarity of the CKM matrix is already satisfied to a good approximation by the Cabibbo matrix, the GIM mechanism is extremely effective in the charm sector, causing a strong suppression of the amplitude for $D^0$ mixing. This makes the theoretical determination of $M_{12}$ and $\Gamma_{12}$ particularly challenging. This is in strong contrast with mixing of down-type mesons, i.e.\ $K^0$, $B^0_d$ and $B^0_s$ mixing, where the corresponding box diagrams contain internal up, charm and top quarks. In these cases, the large top-quark mass breaks the GIM mechanism leading to a sizeable mixing amplitude. \\

\noindent
Finally, starting from the time evolution of the weak eigenstates given Eqs.~(\ref{sec3.4:eq5}), (\ref{sec3.4:eq6}), 
we can derive the expression for the time-dependent decay rate of a
$D^0$ meson that was at $t=0$ tagged as a $D^0$, into an arbitrary final state $f$. That is:
\begin{equation}
\Gamma (D^0 (t) \to f )  = N_f \left|{\cal A}_f\right|^2
\left(1+|\lambda_f|^2 \right)
e^{-\Gamma_D t}
\left\{
\frac{\cosh \left( \frac{\Delta \Gamma_D}{2} t\right)}{2}
+ \frac{1-|\lambda_f|^2 }{1+|\lambda_f|^2 }\frac{\cos  \left( \Delta M_D t\right)}{2}
+ \frac{2 \Re (\lambda_f )}{1+|\lambda_f|^2 }
  \frac{\sinh \left( \frac{\Delta \Gamma_D}{2} t\right)}{2}
- \frac{2 \Im (\lambda_f )}{1+|\lambda_f|^2 }
  \frac{\sin  \left( \Delta M_D t\right)}{2}
\right\}
\label{sec3.4:eq9} \,.
\end{equation}
Here $N_f$ denotes a time-independent normalisation factor, which 
includes e.g.\ phase-space effects, and ${\cal A}_f$ is the decay amplitude describing the transition of the flavour 
eigenstate $D^0$ into the final state $f$; using, for the decay of a $\bar{D}^0$ state into 
$f$, the notation $\bar{\cal A}_f$, we have:
\begin{equation}
{\cal A}_f = \langle f | {\cal H}_{\rm eff} |D^0 \rangle 
\, ,
\qquad 
\bar{\cal A}_f = \langle f | {\cal H}_{\rm eff} | \bar{D}^0 \rangle \,,
\qquad 
\mbox{and}
\qquad
\lambda_f = \frac{q}{p} \frac{\bar{\cal A}_f}{{\cal A}_f}
\, ,
\label{sec3.4:eq10}
\end{equation}
with the effective Hamiltonian described in 
Section~\ref{subsec:Heff}.
Analogous expressions to Eq.~\eqref{sec3.4:eq9} for the time-dependent
decay rates $\Gamma(\bar{D}^0 (t) \to f) $,
 $\Gamma (D^0 (t) \to \bar{f}) $ and
$\Gamma ( \bar{D}^0 (t) \to \bar{f} )$
can be found e.g.\ in~\cite{Artuso:2015swg, Proceedings:2001rdi, AodhanBurke}.
%%%%%%%%%%%%%%%%%%%%%%%%%%%%%%%%%%%%%%%%%%%
%%%%%%%%%%%%%%%%%%%%%%%%%%%%%%%%%%%%%%%%%%%
%%%%%%%%%%%%%%%%%%%%%%%%%%%%%%%%%%%%%%%%%%%
\subsection{CPV in the decay}
\label{subsec:basics_CPV}
CP violation can also affect the decay amplitude. In fact, {\bf direct CPV} occurs if the decay of a charm hadron into a certain final state has a different probability than the corresponding CP conjugated mode. This phenomenon can be studied via the following asymmetry
    \begin{equation}
        a_{\rm CP}^{\rm dir}(f) \equiv  
\frac{ |{\cal A}(D \to f)|^2 - | {\cal A}(\bar D \to \bar f)|^2}
     {|{\cal A}(D \to f)|^2 + |{\cal A}(\bar D \to \bar f)|^2}
     \, ,
     \label{sec3.5:eq1}
    \end{equation}
    where, without loss of generality, we have considered the decay of a $D$ meson into a final state $f$. Starting from the general decomposition of the decay amplitude in terms of the weak and strong phases $\phi^w$ and $\phi^s$, i.e.~\footnote{Due to the unitarity of the CKM matrix, the general expression of the total decay amplitude, also in the presence of more than two contributing amplitudes, can be recast as in Eq.~\eqref{sec3.5:eq2}.}
    \begin{equation}
        {\cal A} (D \to f)  = |A_1| e^{i \phi_1^w} e^{i \phi_1^s} +  |A_2| e^{i \phi_2^w} e^{i \phi_2^s}
        \, ,
     \label{sec3.5:eq2}        
    \end{equation}
and taking into account that the weak interaction violates the CP symmetry and that the strong interaction preserves it, meaning that under a CP transformation the weak phase changes sign whereas the strong phase does not, the direct CP asymmetry in Eq.~\eqref{sec3.5:eq1} takes the form
    \begin{equation}
        a_{\rm CP}^{\rm dir}(f) = - \frac{2  \left|\frac{A_2}{A_1}\right| \sin (\phi_1^s - \phi_2^s) \sin (\phi_1^w - \phi_2^w)}{1 + 2 \left|\frac{A_2}{A_1}\right| \cos (\phi_1^s - \phi_2^s) \cos (\phi_1^w - \phi_2^w) + \left|\frac{A_2}{A_1}\right|^2}
     \, .
          \label{sec3.5:eq3}
    \end{equation}
    From Eq.~\eqref{sec3.5:eq3}, we find that a non-zero value of $a_{\rm CP}^{\rm dir}(f)$ can only be obtained if there are at least two different amplitudes with different strong and weak phases that contribute to the decay i.e. $|A_1| \neq |A_2|$, and $\phi_1^w \neq \phi_2^w$, $\phi_1^s \neq \phi_2^s$ (mod $\pi$). Thus,
    in order to obtain robust theoretical predictions of direct CPV, it is necessary to have control over multiple hadronic amplitudes and in particular over their strong phases, a task that is currently extremely difficult to achieve.\\ Note that if one of the two amplitudes in Eq.~\eqref{sec3.5:eq3} dominates, say $|A_1| \gg |A_2|$, the expression for $a_{\rm CP}^{\rm dir}(f)$ simplifies as follows
    \begin{equation}
        a_{\rm CP}^{\rm dir}(f)  \approx - 2  \left|\frac{A_2}{A_1}\right| \sin (\phi_1^s - \phi_2^s) \sin (\phi_1^w - \phi_2^w)
     \, .
          \label{sec3.5:eq4}
    \end{equation}
%%%%%%%%%%%%%%%%%%%%%%%%%%%%%%%%%%%%%%%%%%%%%%%%%%%%%%%%%%
%%%%%%%%%%%%%%%%%%%%%%%%%%%%%%%%%%%%%%%%%%%%%%%%%%%%%%%%%%%%%%%%%%%%%%
\section{Inclusive charm hadron decays}
\label{sec4}
\subsection{The heavy quark expansion}
\label{subsec:HQE}
Hadronic weak decays proceed through the interplay between the weak and the strong interactions that are responsible for the decay and hadronisation dynamics, and this occurs over very different scales, spanning from $m_W$ to $\Lambda_{\rm QCD}$. Furthermore, in the case of weak
decays of heavy hadrons, i.e.\ of hadrons containing a heavy quark $Q$, with $m_Q \gg \Lambda_{\rm QCD}$, an additional intermediate scale emerges, $m_Q$. 
As a result, the underlying dynamics is characterised by the large hierarchy $m_W \gg m_Q \gg \Lambda_{\rm QCD}$. Such a multi-scale problem can be addressed by resorting to effective field theories, in which only the degrees of freedom that are dynamical at a given scale appear as active fields, while heavy particles are integrated out and leave a trace only in the effective coefficients of the theory. For instance, integrating out the $W$ boson as a dynamical degree of freedom leads to the weak effective theory (WET) already discussed in the preceding chapter. \\

\begin{figure}[t]
\centering
\includegraphics[scale=0.47]{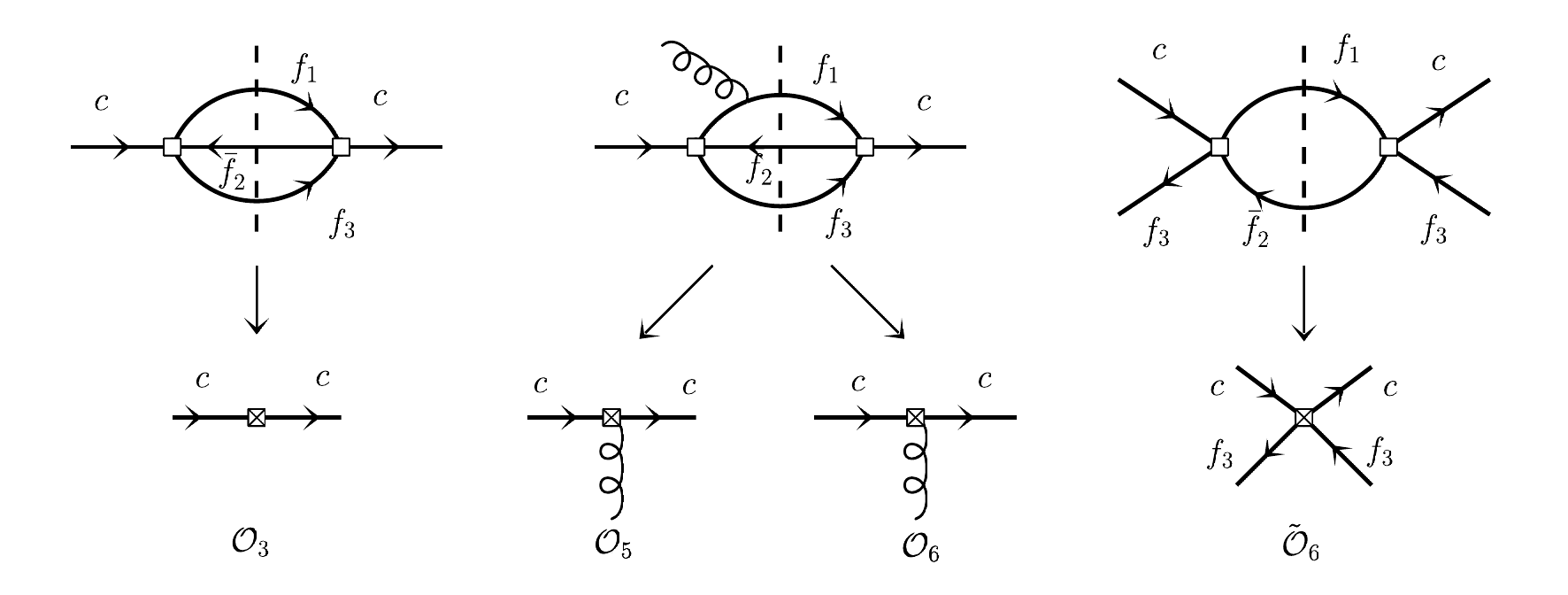}
\caption{Schematic representation of the HQE in Eq.~\eqref{sec4.1:eq6}. By assuming that the charm quark is heavy, the discontinuity of two- and one-loop diagrams obtained from the double insertion of the effective Hamiltonian in Eq.~\eqref{sec4.1:eq5}, is matched into local two- and four-quark operators of increasing dimension. Specifically, the diagrams show the LO-QCD contribution to two-quark operators due to the free charm-quark decay at dimension-three (left), due to soft-gluon emission at dimension-five and -six (middle) and the LO-QCD contribution to four-quark operators at dimension-six due to spectator-quark effects (right). The labels $f_{1}$, $f_2$ and $f_3$ denote all possible fermions - quarks and leptons, into which the charm quark can decay, i.e.\ $c \to f_1 \bar f_2 f_3$ with $f_1 = d,s$, $f_2 = d, s, e, \mu$, and $f_3 = u, \nu_e, \nu_\mu$.
}
\label{sec4.1:fig1}
\end{figure}

\noindent
An effective description for inclusive weak decays of heavy hadrons can be obtained within the framework of the {\bf heavy quark expansion~(HQE)}, which was established by Shifman and Voloshin in 1984 precisely to study inclusive weak decays of charmed particles~\cite{Shifman:1984wx} - see the review~\cite{Lenz:2014jha} for an overview of the historical development. The basic assumption of the HQE is that a heavy quark inside the hadronic state carries most of the hadron momentum and is almost on-shell. Its momentum can be then parametrised in terms of a large contribution proportional to the heavy-quark mass and a residual component $k^\mu$ originating from the non-perturbative interactions with soft quarks and gluons due to the strong interaction. That is 
\begin{equation}
p_Q^\mu = m_Q v^\mu + k^\mu\,,
\qquad 
k^2 \ll m_Q^2  \,,
\label{secL4.1:eq1}
\end{equation}
where $v^\mu = p_{H_Q}^\mu/m_{H_Q}$ is the hadron velocity and $k \sim \Lambda_{\rm QCD}$. The heavy-quark field is also redefined 
as
\begin{equation}
Q(x) = e^{- i m_Q v \cdot x} Q_v(x)\,,
\label{secL4.1:eq2}
\end{equation}
in order to factorise the dependence on the heavy-quark mass. The rescaled field $Q_v$ is related to the heavy-quark effective theory (HQET) field $h_v$, describing an infinitely heavy quark, see e.g.~\cite{Neubert:1993mb}, by
\begin{equation}
     Q_v =  h_v (x) + \frac{i \slashed D_\perp}{2 m_Q} h_v (x) + {\cal O}\left( \frac{1}{m_Q^2} \right) \,,
     \label{secL4.1:eq3}
\end{equation}
here $D_\perp^\mu = D^\mu - (v \cdot D) v^\mu$ and the covariant derivative $D^\mu = \partial^\mu -  i g_s A^\mu$, with $g_s$ being the strong coupling and $A^\mu$ the gluon field.
Eqs.~\eqref{secL4.1:eq1} - \eqref{secL4.1:eq3} represent the main ingredients that allow the total width $\Gamma$ to be expressed in terms of a series expansion in powers of $\Lambda_{\rm QCD}/m_Q$ or, in operator form, in terms of $D^\mu/m_Q$. To see how this works in practice,
let us consider the case of inclusive charm-hadron decays, with the important caveat that, to justify this formulation, we have to assume that the charm quark is heavy, i.e.\ $m_c \gg \Lambda_{\rm QCD}$. In the following, we consider this to be the case, but we will comment on its validity later in the text; see also the discussion in the preceding chapter.\\  
The starting point for the construction of the HQE is the following expression, which gives the definition of the total decay width of a particle:
\begin{equation}
\Gamma(D) = \frac{1}{2 m_{D}} \sum_n \int_{\rm PS} (2 \pi)^4 \delta^{(4)} (p_n - p_{D}) |\langle n| {\cal H}_{\rm eff}| D \rangle |^2\,,
\label{sec4.1:eq4}
\end{equation}
here we assume $D$ to be a charmed meson, although the description generalises to an arbitrary hadron. In Eq.~\eqref{sec4.1:eq4} a summation over all kinematically allowed final states $|n\rangle$ into which the $D$ meson can decay is explicitly shown, PS indicates the corresponding phase-space integration, ${\cal H}_{\rm eff}$ is the effective Hamiltonian describing the non-leptonic and semileptonic weak decays of the charm quark, and $p_X$ denotes the four-momentum of the particle $X$. Using the optical theorem, Eq.~\eqref{sec4.1:eq4} can be recast in the form:
\begin{equation}
{\Gamma(D) = \frac{1}{2 m_{D}} {\rm Im} \, \langle D|\,  i \!\! \int \!\! d^4x 
\,  {\rm T} \Big \{ {\cal H} _{\rm eff} (x) \, ,
 {\cal H} _{\rm eff} (0)  \Big \} | D \rangle}\,,
 \label{sec4.1:eq5}
\end{equation}
where the sum over the modulus squared of each decay amplitude has been replaced by the imaginary part of the forward scattering amplitude of the $D$ meson via the effective Hamiltonian. 
As a consequence, differently to Eq.~\eqref{sec4.1:eq4}, now in Eq.~\eqref{sec4.1:eq5} a non-local operator appears, the time-ordered product of the effective Hamiltonian at two different space-time points. 
Taking into account Eqs.~\eqref{secL4.1:eq1} - \eqref{secL4.1:eq3}, the non-local amplitude in Eq.~\eqref{sec4.1:eq5} can be computed as a series expansion in inverse powers of the charm-quark mass. The result, which defines the HQE, reads schematically: 
\begin{equation}
\Gamma(D) = 
\Gamma_3  +
\Gamma_5 \frac{\langle {\cal O}_5 \rangle}{m_c^2} + 
\Gamma_6 \frac{\langle {\cal O}_6 \rangle}{m_c^3} + \ldots
 + 16 \pi^2 
\left[ 
  \tilde{\Gamma}_6 \frac{\langle \tilde{\mathcal{O}}_6 \rangle}{m_c^3} 
+ \tilde{\Gamma}_7 \frac{\langle \tilde{\mathcal{O}}_7 \rangle}{m_c^4} + \dots 
\right]\,,
\label{sec4.1:eq6}
\end{equation}
where
${\cal O}_d$, $\tilde {\cal O}_d$, are local operators of increasing dimension $d$, and we have introduced the notation $\langle {\cal X} \rangle \equiv \langle D| {\cal X} | D\rangle / (2 m_D)$. A graphical representation of Eqs.~\eqref{sec4.1:eq5}, \eqref{sec4.1:eq6} is shown in Fig.~\ref{sec4.1:fig1}. 
Note that two possible ways of contracting the internal fermion lines are possible, and at LO-QCD the imaginary part of both two- and one-loop diagrams must be computed. In the first case, the resulting operators will be composed of two charm-quark fields and a string of $d-3$ covariant derivatives,  whereas in the second case the resulting operators will be composed of four-quarks, respectively of two charm-quark and two light-quark fields, with $q = u,d,s$, plus a string of $d-6$ covariant derivatives,~i.e. 
\begin{equation}
{\cal O}_d \propto [\bar c_v  \underbrace{(i D_{\mu} )\ldots (i D_{\nu})}_{d-3} c_v]\,,
\qquad
\tilde {\cal O}_d \propto [\bar c_v q] [\bar q \underbrace{(i D_{\mu}) \ldots (i D_{\nu} )}_{d-6} c_v ]\,.
\end{equation}
The fact that four-quark operators are generated already at one loop, starting at dimension-six, is reflected in the factor 16$\pi^2$ in front of the square brackets in Eq.~\eqref{sec4.1:eq6}. 
Note that
Eq.~\eqref{sec4.1:eq6} is another example of an OPE and, as such, it allows for a systematic separation of the scales relevant for the decay process. 
On the one hand, the matrix elements of the local operators parametrise the low-energy dynamics at the scale $\mu < m_c$, 
on the other hand, the short-distance contributions at the scale $\mu > m_c$ are encoded in the functions $\Gamma_d$, $\tilde \Gamma_d$. 
The latter can be computed perturbatively in QCD, namely
\begin{equation}
\Gamma_d = \Gamma_d^{(0)} + \left( \frac{\alpha_s(m_c)}{4 \pi}\right) \Gamma_d^{(1)} +  \left( \frac{\alpha_s (m_c)}{4 \pi}\right)^2 \Gamma_d^{(2)} +  \left( \frac{\alpha_s (m_c)}{4 \pi}\right)^3 \Gamma_d^{(3)} + \ldots \,,
\label{sec4.1:eq7}
\end{equation}
and similarly for $\tilde \Gamma_d$. 
The current status of the perturbative contributions is summarised in Table~\ref{Sec4.1:tab1} for semileptonic charm quark decays, i.e.\ $c \to q \ell^+ \nu_\ell$, with $q = d,s$, and in Table~\ref{Sec4.1:tab2} for non-leptonic decays i.e.\ $c \to q_1 \bar q_2 u$ with $q_{1,2} = d,s$. For a complete list of references see e.g.~\cite{Egner:2024lay}. We stress that, although most of the results have been obtained in the case of bottom-quark decays, they can be easily generalised also to the case of charm-quark decays by means of straightforward replacements i.e.\ $m_b \to m_c$, $m_c \to m_s$, etc. This is however, not possible for the coefficient of the Darwin operator $\tilde \Gamma_6^{(0)}$, since, compared to the case of bottom-quark decays, further contributions due to mixing with four-quark operators with external strange quarks must be taken into account; the corresponding expressions have been derived in~\cite{King:2021xqp}. \\

\noindent
Moving to the hadronic matrix elements, the non-perturbative parameters for charm-quark decays are currently poorly constrained and first principle determinations are for the most part largely missing. 
In many cases symmetry relations must be used to relate the parameters in the charm system to the corresponding ones for bottom-quark decays. For the latter, in fact, the status is more advanced; the parameters of the two-quark operator matrix elements have been determined performing fits to semileptonic $B$-meson decays~\cite{Finauri:2023kte}, using non-perturbative methods like Lattice QCD~\cite{FermilabLattice:2018est, Gambino:2017vkx}, and HQET sum rules~\cite{Falk:1992wt}, or from spectroscopy relations~\cite{Bigi:2011gf}, while matrix-elements of dimension-six four-quark operators have been obtained using HQET sum rules~\cite{Kirk:2017juj, Black:2024bus, King:2021jsq} and first results are also available from Lattice QCD simulations~\cite{Black:2024iwb} - see~\cite{Egner:2024lay} for further references. Current prospects for measuring the inputs of two-quark operators matrix elements from inclusive semileptonic $D$-meson decays at BESIII have been discussed in~\cite{Bernlochner:2024vhg}~\footnote{Note that recently a first attempt to determine the HQE two-quark operators non-perturbative parameters using data on inclusive semileptonic $D$-meson decays by the CLEO~\cite{CLEO:2009uah} and BESIII~\cite{BESIII:2021duu} collaborations has been performed in~\cite{Shao:2025vhe}.}. In the case of charmed baryon decays, the dimension-six matrix elements of four-quark operators can be estimated recurring to constituent quark models, namely simplified models of QCD which allow to parametrise the hadronic matrix elements in terms of the baryon wave function. In the literature, both relativistic and non-relativistic constituent quark models are employed, see e.g.~\cite{PhysRevD.9.3471, PhysRevD.12.147}. A recent study of the two-quark operator hadronic matrix elements for the $\Lambda_b$ baryon has been performed in~\cite{Melic:2025tsr}. \\

\noindent
As stated above, the validity of the HQE for charmed hadrons decays is based on the assumption that $m_Q \gg \Lambda_{\rm QCD}$, in which case we can expect both the expansions in $\Lambda_{\rm QCD}/m_c$ and $\alpha_s(m_c)$
to converge
fast enough and the first few terms in Eq.~\eqref{sec4.1:eq6} to be sufficient to describe $\Gamma$. This picture, however, may not hold in practice, as the charm-quark mass $m_c \sim 1$ GeV is dangerously close to the hadronic scale $\Lambda_{\rm QCD} \sim 400$~MeV. Ultimately, to understand if such an expansion is well justified for the charm system or not, we 
must compare the corresponding HQE predictions with the available data. 
\begin{table}[t]
\centering
\renewcommand{\arraystretch}{1.9}
\begin{minipage}{0.4\textwidth}
\centering
\begin{tabular}{|c||c|}
\hline
\multicolumn{2}{|c|}{Semileptonic modes} \\
\hline 
\multirow{2}{*}{$\Gamma_3^{(3)}$} & 
{\scriptsize \it Fael, Sch\"onwald, Steinhauser \cite{Fael:2020tow}} \\[-2mm]
& 
{\scriptsize \it Czakon, Czarnecki, Dowling \cite{Czakon:2021ybq}} \\
\hline
\multirow{2}{*}{$\Gamma_5^{(1)}$} & 
 {\scriptsize \it  Alberti, Gambino, Nandi \cite{Alberti:2013kxa}} \\[-2mm]
& {\scriptsize \it  Mannel, Pivovarov, Rosenthal \cite{Mannel:2015jka}}\\
\hline
\multirow{1}{*}{$\Gamma_6^{(1)}$} &
 {\scriptsize \it  Mannel, Moreno, Pivovarov \cite{Mannel:2019qel, Mannel:2021zzr, Moreno:2022goo, Moreno:2024bgq}}
 \\
\hline
\multirow{1}{*}{$\Gamma_7^{(0)}$} & 
{\scriptsize \it  Dassinger, Mannel, Turczyk \cite{Dassinger:2006md}} \\
\hline
\multirow{1}{*}{$\Gamma_8^{(0)}$} & 
 {\scriptsize \it  Mannel, Turczyk, Uraltsev \cite{Mannel:2010wj}} \\
\hline
\multirow{1}{*}{$\tilde \Gamma_6^{(1)}$} & 
{\scriptsize \it  Lenz, Rauh \cite{Lenz:2013aua}} \\
\hline
\end{tabular}
\caption{Status of the perturbative corrections in the HQE for semileptonic charm-quark decays.}
\label{Sec4.1:tab1}
\end{minipage}
\quad 
\begin{minipage}{0.4\textwidth}
\centering
\vspace*{-1mm}
\begin{tabular}{|c||c|}
\hline
\multicolumn{2}{|c|}{Non-leptonic modes} \\
\hline 
\multirow{1}{*}{$\Gamma_3^{(2)}$} &  
{\scriptsize \it  Egner, Fael, Sch\"onwald, Steinhauser \cite{Egner:2024azu}} \\
\hline
\multirow{1}{*}{$\Gamma_5^{(1)}$} &  
{\scriptsize \it  Mannel, Moreno, Pivovarov \cite{Mannel:2023zei, Mannel:2024uar, Mannel:2025fvj}}
\\
\hline
\multirow{2}{*}{$\Gamma_6^{(0)}$} &  
{\scriptsize \it  Lenz, Piscopo, Rusov \cite{Lenz:2020oce, King:2021xqp}} \\[-2mm]
&
\scriptsize{\it Mannel, Moreno, Pivovarov \cite{Mannel:2020fts}}
 \\
\hline
\multirow{2}{*}{$\tilde \Gamma_6^{(1)}$} & 
{\scriptsize \it Beneke, Buchalla, Greub, Lenz, Nierste \cite{Beneke:2002rj}} \\[-2mm]
& {\scriptsize \it  Franco, Lubicz, Mescia, Tarantino  \cite{Franco:2002fc}} \\
\hline
\multirow{1}{*}{$\tilde \Gamma_7^{(0)}$} & 
 {\scriptsize \it Gabbiani, Onishchenko, Petrov \cite{Gabbiani:2003pq}} \\
\hline
\end{tabular}
\caption{Status of the perturbative corrections in the HQE for non-leptonic charm-quark decays. Note that for $\Gamma_5^{(1)}$, the corresponding $\alpha_s$-corrections to the $b \to c \bar c s$ mode have been very recently computed in~\cite{Mannel:2025fvj}.}
\label{Sec4.1:tab2}
\end{minipage}
\end{table} 
%%%%%%%%%%%%%%%%%%%%%%%%%%%%%%%%%%%%%%%%%%%%%%%%%%%%%%%%%%%%%%%%%%%%%%%%%%%%%%%%%%%%%%%%%%%%%%%%%%%%%%%%%%%%%%%%%%%%%%%%%%%%%%%%
\subsection{Lifetimes and inclusive semileptonic decays}
\label{subsec:Lifetimes}

\noindent
The validity of the HQE for charm-quark decays can be tested in inclusive observables such as semileptonic decays of charm hadrons e.g.\ $\Gamma (D \to X_{s,d} \ell^+ \nu_\ell)$, and charm hadron lifetimes e.g.\ $\tau(D) \equiv \Gamma(D)^{-1}$. In particular, the large number of data collected by e.g.\ the Belle~II,  BESIII 
and LHCb collaborations, as well as the plethora of data expected to be collected in the near future, can be used to study the structure and the convergence of the HQE and, also, in the case of semileptonic decays, open the road to an inclusive extraction of the CKM matrix elements $|V_{cs}|$ and $|V_{cd}|$~\cite{Fael:2019umf}.\\

\begin{figure}[t]
    \centering
    \includegraphics[scale=0.85]{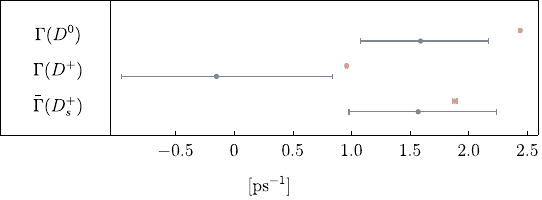} \,
    \includegraphics[scale=0.85]{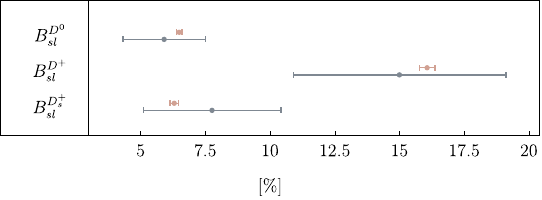} \\
    \includegraphics[scale=0.85]{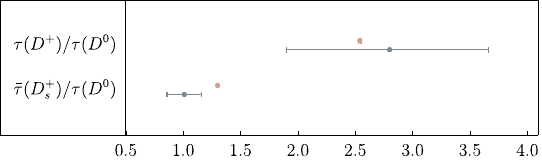} \,
    \includegraphics[scale=0.85]{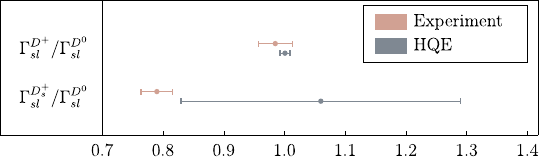}
    \caption{Comparison of HQE predictions and experimental data for charmed mesons total decay widths (top left), semileptonic branching ratios (top right), lifetime ratios (bottom left) and ratios of semileptonic widths (bottom right), based on~\cite{King:2021xqp}.}
    \label{sec4.2:fig1}
\end{figure}

\noindent
Theoretical predictions based on the HQE for the total widths of the $D^0, D^+,$ and $D_s^+$ mesons, their lifetime ratios, as well as the corresponding semileptonic branching ratios $B_{\rm sl}^D \equiv \Gamma(D \to X_s e^+ \nu_e)/\Gamma(D)$, and the ratios of semileptonic widths, taken from~\cite{King:2021xqp}, are shown in Fig.~\ref{sec4.2:fig1}~\footnote{Because the tau lepton is heavier than the charm quark, the decay $D_s^+ \to \tau^+ \nu_\tau$ cannot be reproduced within the HQE. Hence, we define $\bar \Gamma(D_s^+) \equiv \Gamma(D_s^+) - \Gamma(D_s^+ \to \tau^+ \nu_\tau)$.} together with the corresponding experimental values~\cite{ParticleDataGroup:2024cfk-qm}. Despite the large uncertainties, the HQE succeeds in reproducing the observed pattern and there is no evident signal of a possible breakdown of the theoretical framework. However, note that 
currently the HQE predicts also negative values for the lifetime of the $D^+$ meson, which are clearly not physical. Further investigations are thus still needed to resolve this puzzle, which originates from an effective cancellation of contributions of the free charm quark decay, $\Gamma_3$, with contributions from spectator effects, in particular the so-called Pauli-interference, denoted by $\tilde{\Gamma}_6$.
New insights could potentially be gained by including the recently determined $\alpha_s^{2}$-corrections to the leading term $\Gamma_3$~\cite{Egner:2024azu}, following the corresponding analysis of $B$-meson lifetimes performed in~\cite{Egner:2024lay}.\\

\noindent
A recent experimental study of inclusive $D_s^+\rightarrow Xe^+\nu_e$ decays~\cite{BESIII:2021duu}, combined with an earlier analysis of the inclusive $D^0\rightarrow Xe^+\nu_e$ decays by the CLEO collaboration~\cite{CLEO:2009uah}, results in $\Gamma(D_s^+\rightarrow Xe^+\nu_e)/\Gamma(D^0\rightarrow Xe^+\nu_e) =0.790 \pm 0.016~(\mathrm{stat}) \pm 0.020~(\mathrm{syst})$, 
quoted with statistical and systematic uncertainties, 
which is consistent with the HQE result shown in Fig.~\ref{sec4.2:fig1}. The CLEO experiment~\cite{CLEO:2009uah} also reported $\Gamma(D^+\rightarrow Xe^+\nu_e)/\Gamma(D^0\rightarrow Xe^+\nu_e) =0.985 \pm 0.015 ~(\mathrm{stat}) \pm 0.024 ~(\mathrm{syst})$, in agreement with theoretical predictions. Further potential tests of the HQE with inclusive semileptonic decays are outlined in~\cite{Bernlochner:2024vhg}. Currently, the most precise measurements of lifetimes of $D$ mesons come from Belle II. Using a sample of about 116k $D_s^+\rightarrow\phi\pi^+$ decays, they measure $\tau_{D^+_s}= (499.5 \pm 1.7 \pm 0.9)~\mathrm{fs}$~\cite{Belle-II:2023eii}, in agreement with predictions in~\cite{King:2021xqp, Gratrex:2022xpm}. Using about 171k $D^{*+} \rightarrow D^0(\rightarrow
K^-
\pi^+)\pi^+$ and about 59k $D^{*+} \rightarrow D^+(\rightarrow K^-
\pi^+\pi^+)\pi^0$ decays, the results for the lifetimes are $\tau(D^0) = (410.5 \pm
1.1 \pm 0.8)  ~ \mathrm{fs}$ and $\tau (D^+) = (1030.4 \pm 4.7  \pm 3.1)  ~ \mathrm{fs}$, and their ratio is $\tau(D^+)/\tau(D^0) = 2.510\pm0.013 \pm0.007$~\cite{Belle-II:2021cxx}, in agreement with HQE predictions. Other recent lifetime measurements come from the LHCb experiment~\cite{LHCb:2017knt}, and earlier lifetime studies have been performed by the FOCUS~\cite{FOCUS:2005gui, FOCUS:2002sso}, SELEX~\cite{SELEX:2001miq, SELEX:2000ijz}, E791~\cite{E791:1998zjs, E791:1999bzz}, CLEO~\cite{CLEO:1999xvl}, E687~\cite{E687:1993lxk, E687:1993xsd}, E691~\cite{TaggedPhotonSpectrometer:1987poq} experiments.\\

\begin{figure}[h]
\centering
\includegraphics[scale=0.85]{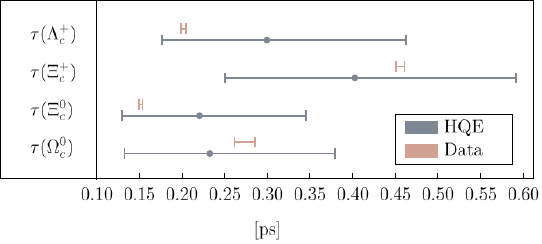}
\caption{Comparison of HQE predictions and experimental data for charmed baryons lifetimes, based on~\cite{Gratrex:2022xpm}.}
\label{sec4.2:fig2}
\end{figure}

\noindent
The results in Fig.~\ref{sec4.2:fig1} have also been confirmed by another independent study~\cite{Gratrex:2022xpm}, where also singly charmed baryons have been investigated within the HQE. The results of~\cite{Gratrex:2022xpm} for the lifetimes of the $\Xi_c^0, \Lambda_c^+, \Omega_c^0$ and $\Xi_c^+$, and of their ratios are shown in Fig.~\ref{sec4.2:fig2}. Also in this case the agreement with the experimental data appears to be good and in particular the HQE reproduces the experimentally observed hierarchy $\tau(\Xi_c^0) < \tau(\Lambda_c^+)<\tau(\Omega_c^0) < \tau(\Xi_c^+)$~\footnote{It is curious to note that from 2004 until 2018 the observed pattern was $ \tau(\Omega_c^0) <\tau(\Xi_c^0) < \tau(\Lambda_c^+) < \tau(\Xi_c^+)$~\cite{PhysRevD.98.030001}. However, the measurement by the LHCb collaboration with semileptonic $b$-hadron
decays~\cite{LHCb:2018nfa} dramatically changed the situation and revealed a new pattern $\tau(\Xi_c^0) < \tau(\Lambda_c^+)<\tau(\Omega_c^0) < \tau(\Xi_c^+)$, in tension with the previous measurements.
% theoretical predictions. 
In fact, the lifetime of the $\Omega_c^0$ baryon was about 4 times larger than the previous world average, and there was a tension of about 3 standard deviations for the lifetime of the $\Xi_c^0$ baryon.
Later measurements by LHCb~\cite{LHCb:2021vll} and by Belle II~\cite{Belle-II:2022ggx, Belle-II:2022plj} showed agreement with~\cite{LHCb:2018nfa} and confirmed the newly established hierarchy.}. Similar results are also found in~\cite{Cheng:2023jpz}.
Recently, an updated study of doubly-charmed baryons within the HQE has been performed in~\cite{Dulibic:2023jeu}. The authors confirm the expected hierarchy  $\tau(\Xi_{cc}^+) < \tau(\Omega_{cc}^{+}) < \tau(\Xi_{cc}^{++})$ and HQE prediction for $\tau(\Xi_{cc}^{++})$ is found to be consistent with the LHCb determination of $\tau (\Xi_{cc}^{++})= (0.256^{+0.024}
_{-0.022} \pm 0.014) ~\mathrm{ps}$~\cite{LHCb:2018zpl}. 

%%%%%%%%%%%%%%%%%%%%%%%%%%%%%%%%%%%%%%%%%%%%%%%%%%%%%%%%%%%%%%%%%%%%%%%%%%%%%%%%%%%%%%%%%%%%%%%%%%%%%%%%%%%%%%%%%%%%%%%%%%%%%%%%
\section{Exclusive charm hadron decays}
\subsection{Leptonic and semileptonic decays}
\label{subsec:leptonic}
Studies of leptonic and semileptonic charm hadron decays provide invaluable information on the properties of the strong and weak interactions. 
As already discussed in the preceding chapter, in the case of leptonic decays, the strong interaction is described by a single constant non-perturbative input, the decay constant, whereas, for semileptonic decays, the non-perturbative dynamics is encoded in a set of $q^2$-dependent functions, the form factors. In the SM, (semi)leptonic decays can proceed through the tree-level exchange of a $W$ boson or through loop transitions, that is via penguin and box diagrams. The latter, i.e.\ loop induced (semi)leptonic charm decays are strongly suppressed in the SM and commonly referred to as {\it rare charm decays}, with a typical branching ratio $B < 10^{-6}$~\cite{Gisbert:2020vjx}. In general, experimental measurements have been reported by the MARK-III, CLEO-c, BaBar, Belle/Belle II and BESIII collaborations. One fundamental experimental complication is the presence of a neutrino which escapes undetected. The double tag technique used at experiments like BESIII allows for a clean sample of (semi)leptonic decays to be reconstructed as will be discussed in this chapter. Belle has used a similar "double-tag" technique albeit for continuum charm production from the $\Upsilon(4S)$ region where the flavour and momentum of the neutral charm meson are tagged
through a full reconstruction of the recoiling charm meson (tag side) and additional mesons from fragmentation~\cite{Belle:2006idb}. This method improves the momentum resolution of the missing neutrino. We note that at the LHCb experiment measurements of tree-level semileptonic charm decays 
are possible but complicated due to the higher background level. 
\\

\noindent
{\bf Tree-level leptonic and semileptonic decays:}
The cleanest leptonic decays in the SM are those that can proceed via the tree-level exchange of a $W$ boson, namely the leptonic decays of a  charged meson such as the mode $D_s^+ \to \ell^+ \nu_\ell$ (Feynman diagram can be seen in the preceding chapter). 
In this case, the theory prediction has a very simple structure, i.e.
\begin{equation}
 B(D_{s}^+ \to \ell^+ \nu_\ell)|_{\rm SM}   =  
 \frac{G_F^2 m_{D_{s}^+} m_{\ell}^2}{8 \pi}
 \left(
 1- \frac{m_\ell^2}{m_{D_{s}^+}^2}
 \right)f_{D_{s}}^2 |V_{cs}|^2
 \tau (D_{s}^+) 
 \left(1 + \delta_{\rm EM} 
 \right)
 \, ,
\label{sec5.1:eq1}
\end{equation}
and is directly proportional to the modulus of the CKM matrix element $V_{cs}$ and to the decay constant $f_{D_{s}}$, where the value of the latter is by now determined with a precision of less than $1\%$ from Lattice QCD studies, see the FLAG report~\cite{FlavourLatticeAveragingGroupFLAG:2024oxs} and in the preceding chapter. In Eq.~\eqref{sec5.1:eq1},
$\delta_{\rm EM}$ denotes the effect of channel-dependent electromagnetic radiative corrections with an approximate size of $1\%$, see e.g.~\cite{Burdman:1994ip}.
The remaining parameters in Eq.~\eqref{sec5.1:eq1} i.e.\ $G_F$, $m_{\ell}$, $m_{D_{s}}$ and the lifetime of the $D^+_{s}$ meson are all extremely precisely known~\cite{ParticleDataGroup:2024cfk}. The generalisation of Eq.~\eqref{sec5.1:eq1} to an arbitrary charm meson decay can be obtained with the proper replacement of masses and CKM matrix elements~\footnote{Note that since baryon number is conserved in the SM, leptonic decays of charm baryons are forbidden.}.\\

\noindent
Experimentally, charged leptonic decays have been measured with muons and tauons in the final state. Due to the helicity suppression, namely the proportionality of the branching ratio to the mass of the final state lepton, the decays with electrons in the final state are still too suppressed to be observed
and, so far, only upper limits have been obtained. Current averages read~\cite{ParticleDataGroup:2024cfk-qm}
\begin{eqnarray}
 B (D_s^+ \to \tau^+ \nu_\tau)|_{\rm exp}   =  (5.36 \pm 0.10) \cdot 10^{-2}
 \, ,
 &&
B(D^+ \to \tau^+ \nu_\tau)|_{\rm exp}   = (1.20 \pm 0.27) \cdot 10^{-3} 
\, ,
 \label{sec5.1:eq2}
\\
 B (D_s^+ \to \mu^+ \nu_\mu)|_{\rm exp}   =  (5.35 \pm 0.12) \cdot 10^{-3} 
 \, ,
 &&
B (D^+ \to \mu^+ \nu_\mu)|_{\rm exp}   =  (3.74 \pm 0.17) \cdot 10^{-4} \, ,
\\
 B (D_s^+ \to e^+ \nu_e)|_{\rm exp}    < 8.3           \cdot 10^{-5}  \, ,
 &&
B (D^+ \to e^+ \nu_e)|_{\rm exp}     < 8.8           \cdot 10^{-6}  \, .
 \label{sec5.1:eq3}
\end{eqnarray}
We note however that the previous results do not yet include the recent and more precise measurements by the BESIII collaboration~\cite{BESIII:2024kvt, BESIII:2024vlt, BESIII:2024dvk,BESIII:2025ows}, namely
\begin{eqnarray}
 B (D_s^+ \to \tau^+ \nu_\tau)|_{\rm exp}   =  (5.38 \pm 0.09) \cdot 10^{-2}
 \, ,
 &&
B(D^+ \to \tau^+ \nu_\tau)|_{\rm exp}   = (0.99 \pm 0.12) \cdot 10^{-3} 
\, ,
 \label{sec5.1:eq2b}
\\
 B (D_s^+ \to \mu^+ \nu_\mu)|_{\rm exp}   =  (5.39 \pm 0.09) \cdot 10^{-3} 
 \, ,
 &&
B (D^+ \to \mu^+ \nu_\mu)|_{\rm exp}   =  (3.981 \pm 0.089) \cdot 10^{-4} \,  ,
 \label{sec5.1:eq3b}
\\
 &&
B (D^+ \to e^+ \nu_e)|_{\rm exp}   <  9.7 \cdot 10^{-7} \,  .
 \label{sec5.1:eq3c}
\end{eqnarray}
To reconstruct a (semi-)leptonic decay with a missing neutrino, experiments like BESIII (and previously CLEO-c) make use of the ``double-tag technique'' 
introduced in the preceding chapter: 
one side (tag) is fully reconstructed, while the signal decay is searched for in the recoil system. The presence of the neutrino is inferred from energy and momentum conservation as the energies of the $e^+e^-$ beams are well known. Typical kinematic variables used for the signal decay reconstruction are the missing mass, $M_{\mathrm{miss}}^2 = E^2_{\mathrm{miss}}-p^2_{\mathrm{miss}}$, or  $U_{\mathrm{miss}}=E_{\mathrm{miss}}-p_{\mathrm{miss}}$, where $E_{\mathrm{miss}}$ and $p_{\mathrm{miss}}$ are the missing energy and momentum, respectively. \\

\noindent
The double-tag method at BESIII allows to measure absolute branching fractions. For example, for a generic (semi-)leptonic decay, the branching fraction is defined as 
\begin{equation}
B (D \to (X)\ell \nu_\ell) = \frac{\Sigma_i N^i_{\mathrm{signal}}}{\Sigma_i N^i_{\mathrm{tag}}.(\epsilon^i_{\mathrm{signal}}/\epsilon^i_{\mathrm{tag}})} 
\, ,
\label{eq:doubletag}
\end{equation} 
where $N^i_{\mathrm{tag}}$ and $\epsilon^i_{\mathrm{tag}}$ are the yield and selection efficiency for a given tag mode $i$, respectively; $N^i_{\mathrm{signal}}$ and $\epsilon^i_{\mathrm{signal}}$ are the yield and selection efficiency, respectively, for the signal mode provided that the tag mode $i$ has been reconstructed simultaneously with the signal side. The total signal yield in a presence of a tag is then a sum of all contributions, $N_{\mathrm{signal}}=\Sigma_i N^i_{\mathrm{signal}}$.
The summation index $i$ runs over the number of tag modes used in a particular analysis.
The efficiencies are typically determined with a Monte Carlo (MC) simulation sample selected with the same criteria as the data sample, and fitted with the same fitting procedure, and the efficiencies are given by the ratio of the MC yields with respect to the total number of simulated events. The derivation of Eq.~\ref{eq:doubletag} can be found in the original paper by the MARK-III experiment where the double tag technique was first used~\cite{MARK-III:1985hbd}.\\

\begin{figure} 
\centering 
\begin{minipage}[b]{0.44\textwidth}
\centering
\vspace*{-1mm}
 \begin{overpic}[width=1\linewidth]{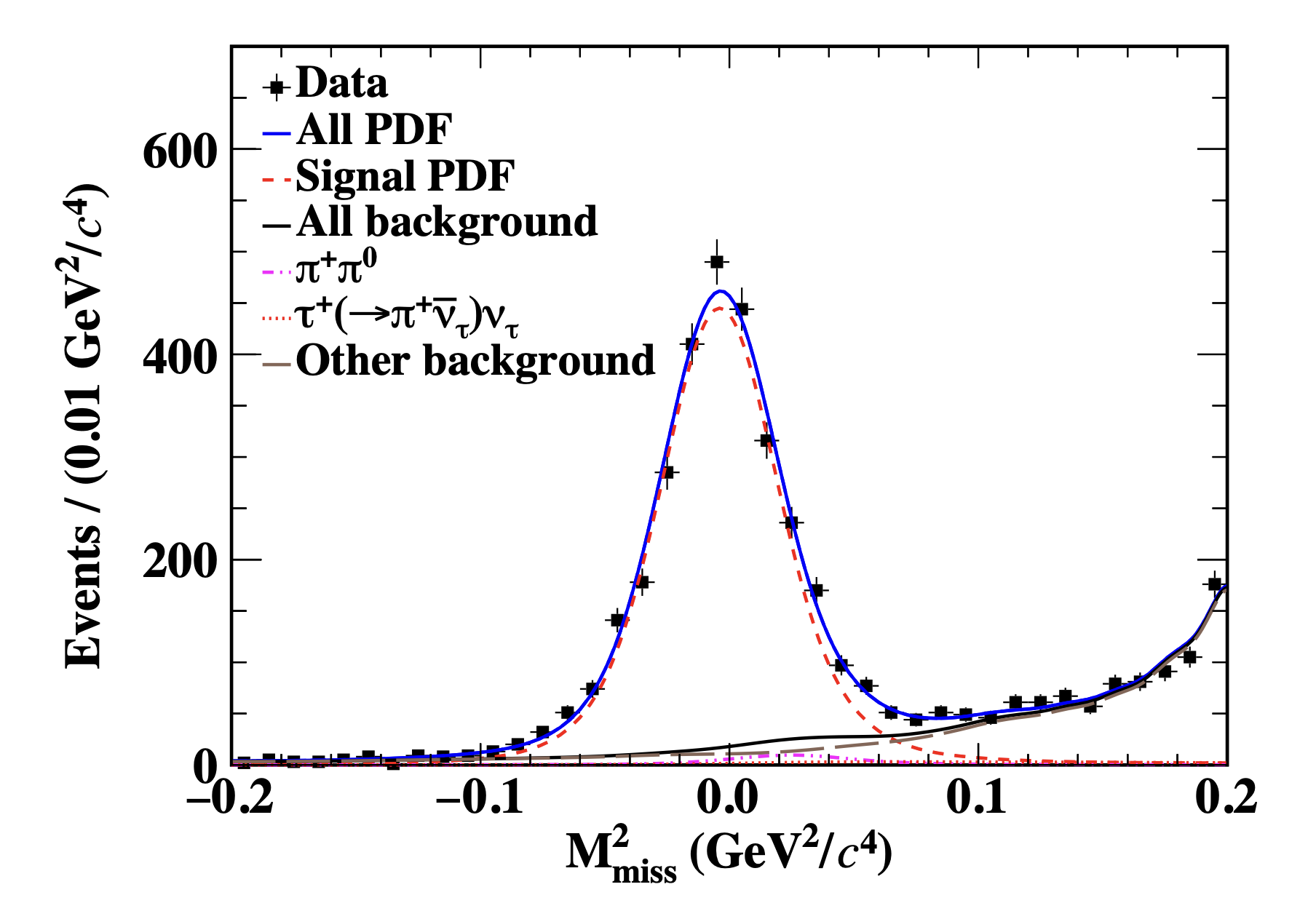} \put(160,120){\includegraphics[width=0.14\linewidth]{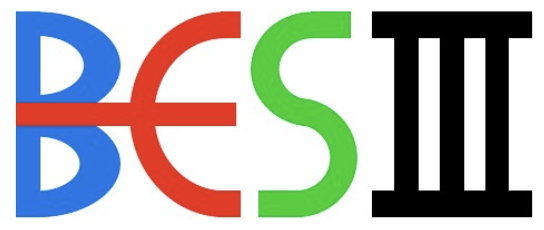}}
  \end{overpic}
 \caption{Missing mass squared distribution for the decays $D^+\rightarrow\mu^+\nu_\mu$, taken from~\cite{BESIII:2024kvt}. The data are shown as black points with uncertainty bars. The signal distribution is indicated in blue. All background contributions are described in the legends of the figures.}
 \label{sec5.1:fig1.1}
  \end{minipage} 
\hfill
\begin{minipage}[b]{0.50\textwidth}
\centering
\vspace*{-5mm}
    \begin{overpic}[width=0.93\linewidth]{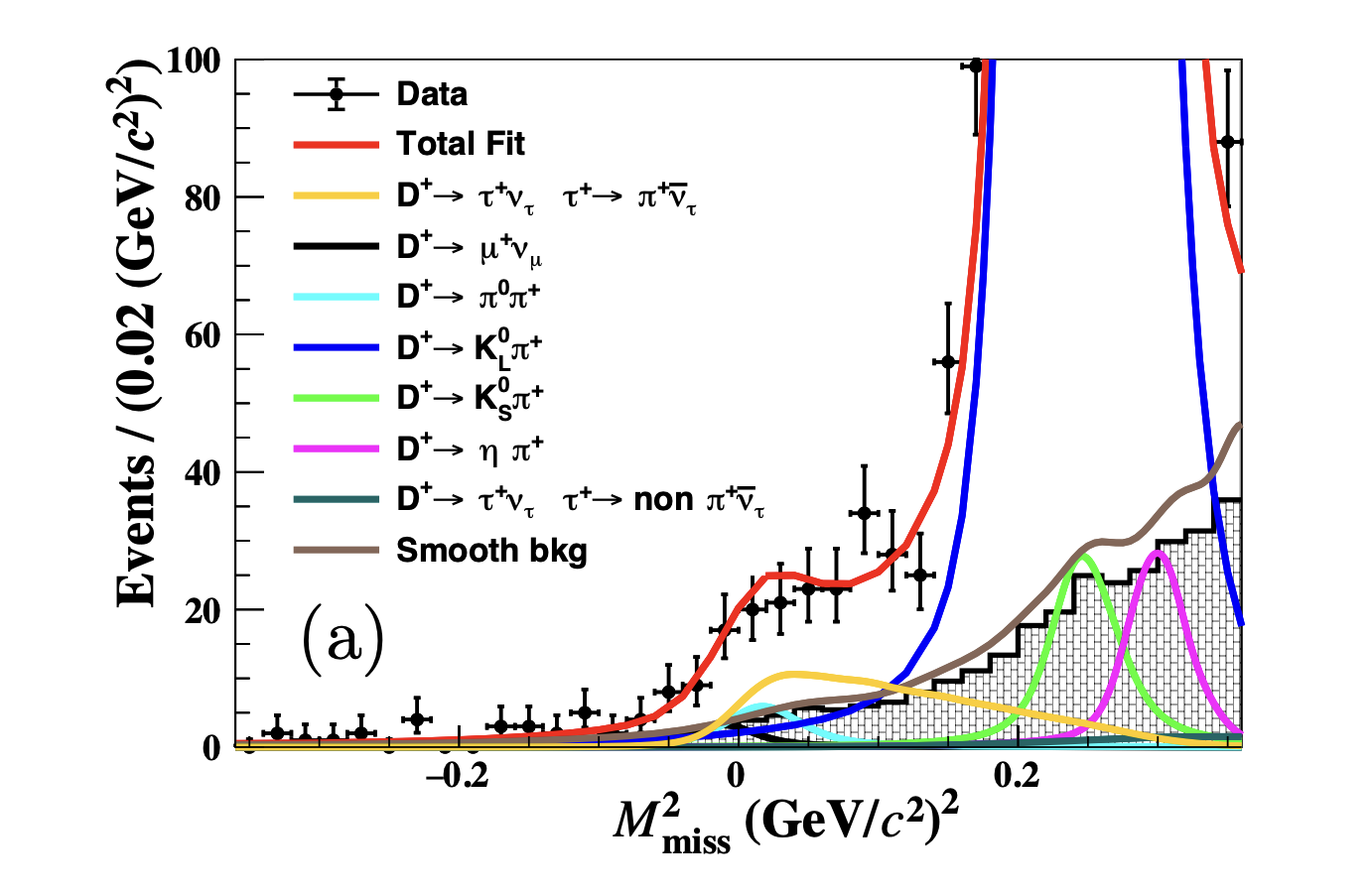} \put(167,128){\includegraphics[width=0.12\linewidth]{bes3-logo.png}}
  \end{overpic}
    
   \caption{Missing mass squared distributions for the decays $D^+\rightarrow\tau^+\nu_\tau$, taken from~\cite{BESIII:2024vlt}. The data are shown as black points with uncertainty bars. The signal distribution is described by the golden line. All background contributions are described in the legends of the figures.
    }
    \label{sec5.1:fig1.2}
\end{minipage}
\end{figure}

\noindent
To illustrate how challenging the environment in which decays with a tau lepton are reconstructed can be, Figs.~\ref{sec5.1:fig1.1}, and~\ref{sec5.1:fig1.2} show a comparison between the distribution of the missing mass for the muonic decay $D^+\rightarrow \mu^+\nu_\mu$~\cite{BESIII:2013iro} and for the tauonic decay $D^+\rightarrow \tau^+\nu_\tau$~\cite{BESIII:2019vhn}, respectively. The tau lepton is reconstructed via the $\tau^+\rightarrow\pi^+\bar{\nu}_\tau$ decay, with an additional missing particle. 
Note that in Fig.~\ref{sec5.1:fig1.2}, the signal distribution in gold is asymmetric due to the missing tau neutrino from the $D^+$ decay, and the tau anti-neutrino from the tau decay.\\

\begin{figure}
\centering
 \includegraphics[scale=0.22]{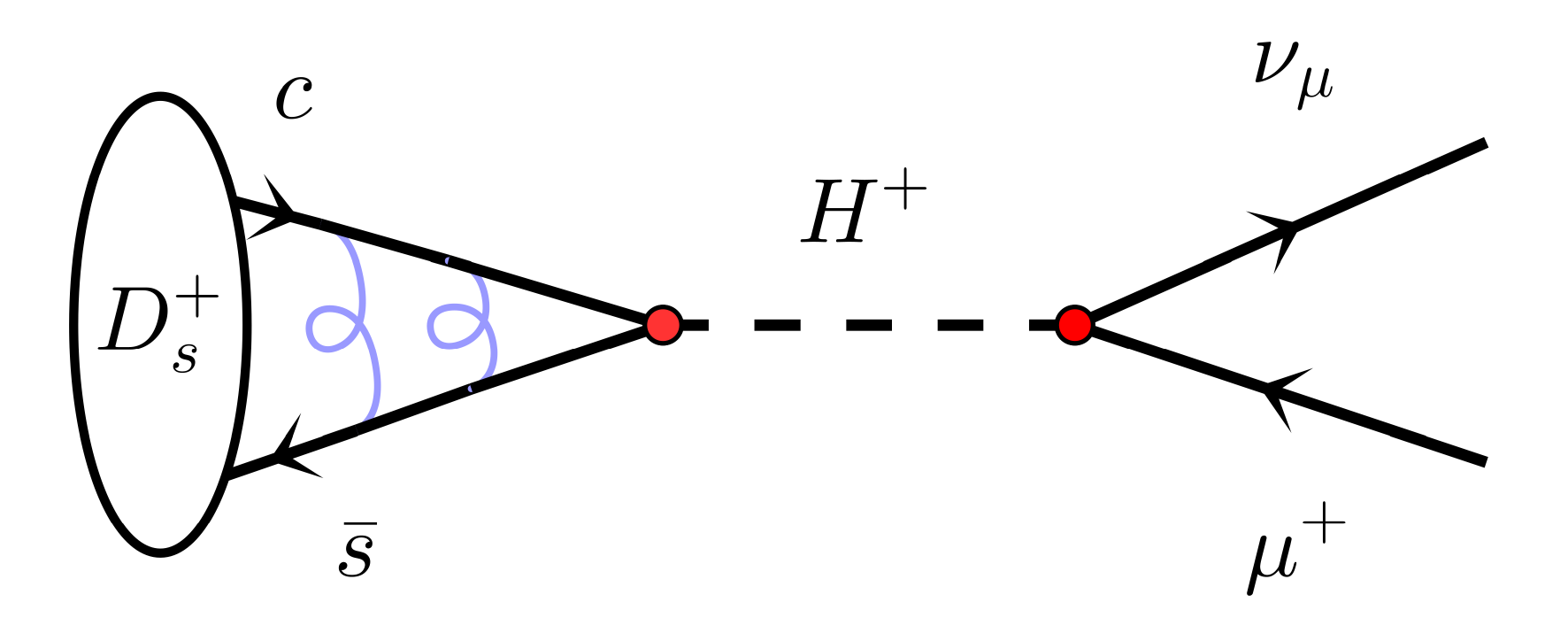}
\caption{In BSM scenarios like the 2HDM, the leptonic decay $D_s^+ \to \mu^+ \nu_\mu$ would receive additional contributions compared to those allowed in the SM, such as  due to a new charged Higgs boson.}
\label{sec5.1:fig2}
\end{figure}

\noindent
Besides offering a clean way to test the SM and determining its parameters, such as the CKM matrix elements $|V_{cs}|$ and $|V_{cd}|$~\footnote{The most precise determinations of $|V_{cs}|$ and $|V_{cd}|$ are obtained from these leptonic tree-level decays, see~\cite{HeavyFlavorAveragingGroupHFLAV:2024ctg}.}, leptonic decays can also be used to constrain BSM models. For example, in so-called 2 Higgs-doublet models (2HDM), instead of a single SM Higgs field, five physical Higgs particles appear, i.e.\ $h^0$, $H^0$, $A^0$, $H^+$ and $H^-$. The latter two charged Higgs bosons have similar couplings as the charged weak gauge bosons $W^{\pm}$, and thus, in this scenario, additional diagrams would contribute to the tree-level decays of the $D^+_{(s)}$ mesons, as showed in Fig.~\ref{sec5.1:fig2}. This would 
lead to a modification of the SM result for the branching fractions, namely~\cite{Deschamps:2009rh}
\begin{equation}
 B (D_{(s)}^+ \to \ell^+ \nu_\ell)|_{\rm 2HDM}   =  
 B (D_{(s)}^+ \to \ell^+ \nu_\ell)|_{\rm SM} \, (1 + r_H)^2
  \, ,
  \label{sec5.1:eq4}
\end{equation}
where $r_H$ depends on the parameters of the 2HDM models, such as the mass of the charged Higgs boson $m_{H^+}$.
Comparing the expression in Eq.~\eqref{sec5.1:eq4} with the experimental results in Eqs.~\eqref{sec5.1:eq2} - \eqref{sec5.1:eq3}, it is thus possible to derive constrains on the parameters of the 2HDM models, see e.g.~\cite{Atkinson:2021eox,Atkinson:2022pcn,Atkinson:2022qnl} for recent,  comprehensive studies.\\

\noindent
For semileptonic charm decays driven by a tree-level exchange of a $W$ boson, the corresponding decay width  
depends, in addition to the modulus of the CKM matrix elements $V_{cd}$ or $V_{cs}$, on the value of the hadronic form factors. 
For a pseudoscalar meson $P$ in the final state, the differential decay width with respect to the invariant mass of the lepton pair, 
$q^2 = (p_{\ell^+} + p_{\nu_\ell})^2$,
e.g.\ in the case of $c\to s \ell^+ \nu_\ell$ transition reads, 
\begin{equation}
\frac{d\Gamma}{dq^2} (D \to P \ell^+ \nu_\ell) \propto p^3 \frac{G_F^2}{24\pi^3}\big|V_{cs}\big|^2 \big|f^{D \to P}_+(q^2)\big|^2 \,.
\label{sec5.1:eq5}
\end{equation}
In Eq.~\eqref{sec5.1:eq5},
$p$ denotes the momentum of the pseudoscalar meson in the final state and $f^{D \to P}_+$ the form factor describing the non-perturbative transition $D \to P$.
A simple parametrisation of the $q^2$-dependence of the form factor can be obtained using properties originating in its analytic structure, see e.g. \cite{Khodjamirian:2020btr}\footnote{Form factors are analytic functions in $q^2$ and present isolated singularities---poles---in correspondence with values of $q^2$ equal to the mass of physical hadronic states.}.
Taking only the first pole in the variable $q^2$ into account, one finds \ $|f^{D \to P}_+(q^2)| = |f^{D \to P}_+(0)|/(1-q^2/m_{\rm pole}^2)$, 
where  
$m_{\rm pole}$ is the mass of the nearest pole~\footnote{This is typically the lowest lying vector resonance $c\bar{q}$
for a $c\rightarrow q$ transition. For example, for $D\rightarrow \pi$ transitions, $m_{\mathrm{pole}}$ will be the mass of 
the $D^{*+}$ meson.}.  
An alternative and more comprehensive parametrisation is based on the so-called {\it $z$-expansion} formalism, see~\cite{Khodjamirian:2020btr} or also~\cite{HeavyFlavorAveragingGroupHFLAV:2024ctg}.\\

\noindent
Typically, in experiments it is the product of the CKM matrix element and of the form factor to be measured, e.g.\ $|V_{cs}|^2 |f^{D \to P}_+(q^2)|^2$.
Recent measurements of the latter
for {\bf pseudoscalar mesons} have been reported for the $D \rightarrow \pi \ell^+\nu_\ell$ decay in~\cite{ BESIII:2017ylw, BESIII:2015tql, CLEO:2007ntr, CLEO:2004arv, BaBar:2014xzf, Belle:2006idb, CLEO:2009svp} and for $D \rightarrow \bar{K} \ell^+\nu_\ell$ in~\cite{BESIII:2024slx, BESIII:2024zft, BESIII:2015kin, BESIII:2018ccy, BESIII:2017ylw, BESIII:2016gbw, BESIII:2015jmz, BESIII:2015tql, CLEO:2007ntr, CLEO:2004arv, CLEO:2009svp}.
Assuming the unitarity of the CKM matrix and taking the average value of the experimental measurements of $f^{D \to K}_+(0)|V_{cs}|$ and $f^{D \to \pi}_+(0)|V_{cd}|$, the Heavy Flavour Averaging Group (HFLAV) quotes the following results for the $D\rightarrow K$ and $D\rightarrow\pi$ form factors~\cite{HeavyFlavorAveragingGroupHFLAV:2024ctg}:
\begin{equation}
f^{D\rightarrow K}_+(0) = 0.7376 \pm0.0034\,, 
\qquad
f^{D\rightarrow \pi}_+(0) = 0.6342 \pm 0.0082\,,
\label{eq:ff_from_exp}
\end{equation}
which agree with the non-perturbative predictions stated in the preceding chapter. 
Alternatively, by combining the value of the form factors obtained from theoretical Lattice QCD calculations~\cite{Lubicz:2017syv, Parrott:2022rgu, FermilabLattice:2022gku}, and the experimental determination of the product of CKM matrix element and form factor, yields \begin{equation}
|V_{cs}|= 0.9639 \pm 0.0044 \pm 0.0032\,,
\qquad
|V_{cd}|= 0.2265 \pm 0.0029 \pm 0.0018\,,
\label{sec5.1:eq6}
\end{equation}
where the first uncertainty is due to the experimental measurement and the second is due to the Lattice QCD determination of the form factors.
The extracted value of $|V_{cd}|$ agrees perfectly with the value obtained from a global CKM fit, see e.g.\ CKMfitter \cite{Charles:2004jd},
while the value of $|V_{cs}|$ differs by a little more than one standard deviation; see also the results from the global fit by the UTfit collaboration~\cite{UTfit:2022hsi}.
 Recent studies of the semileptonic decays $D\rightarrow \eta^{(')}\ell^+\nu_\ell$ \footnote{The  $\eta(')$ mesons are linear combinations of $ u \bar{u}$, $ d \bar{d}$  and  $ s \bar{s}$ states.} performed in~\cite{BESIII:2024njj, BESIII:2023ajr, Ablikim:2020hsc, BESIII:2018eom, CLEO:2010pjh} complement the suite of measurements done with pseudoscalar mesons in the final state. The corresponding results obtained for the value of $f^{D\rightarrow\eta(')}_+(0)$ and of the CKM matrix elements are also reported in~\cite{HeavyFlavorAveragingGroupHFLAV:2024ctg}~\footnote{We note that by using the decays with $\eta^{(\prime)}$ as intermediate state e.g.\ $D^+_{(s)}
 \rightarrow \eta^{(\prime)}e^+\nu_e$, the mixing angle between $\eta - \eta^\prime$ can be accessed via the ratio of the partial decay widths, $\cot \phi = \bigr(\Gamma(D^+_s \rightarrow \eta^\prime e^+\nu_e)/\Gamma(D^+_s \rightarrow \eta e^+\nu_e)\bigl)/\bigl( \Gamma(D^+ \rightarrow \eta^\prime e^+\nu_e)/\Gamma(D^+ \rightarrow \eta e^+\nu_e)\bigr)$~\cite{BESIII:2023ajr, BESIII:2018eom}.}.\\

\begin{figure}[t]
    \centering
    \includegraphics[width=0.4\linewidth]{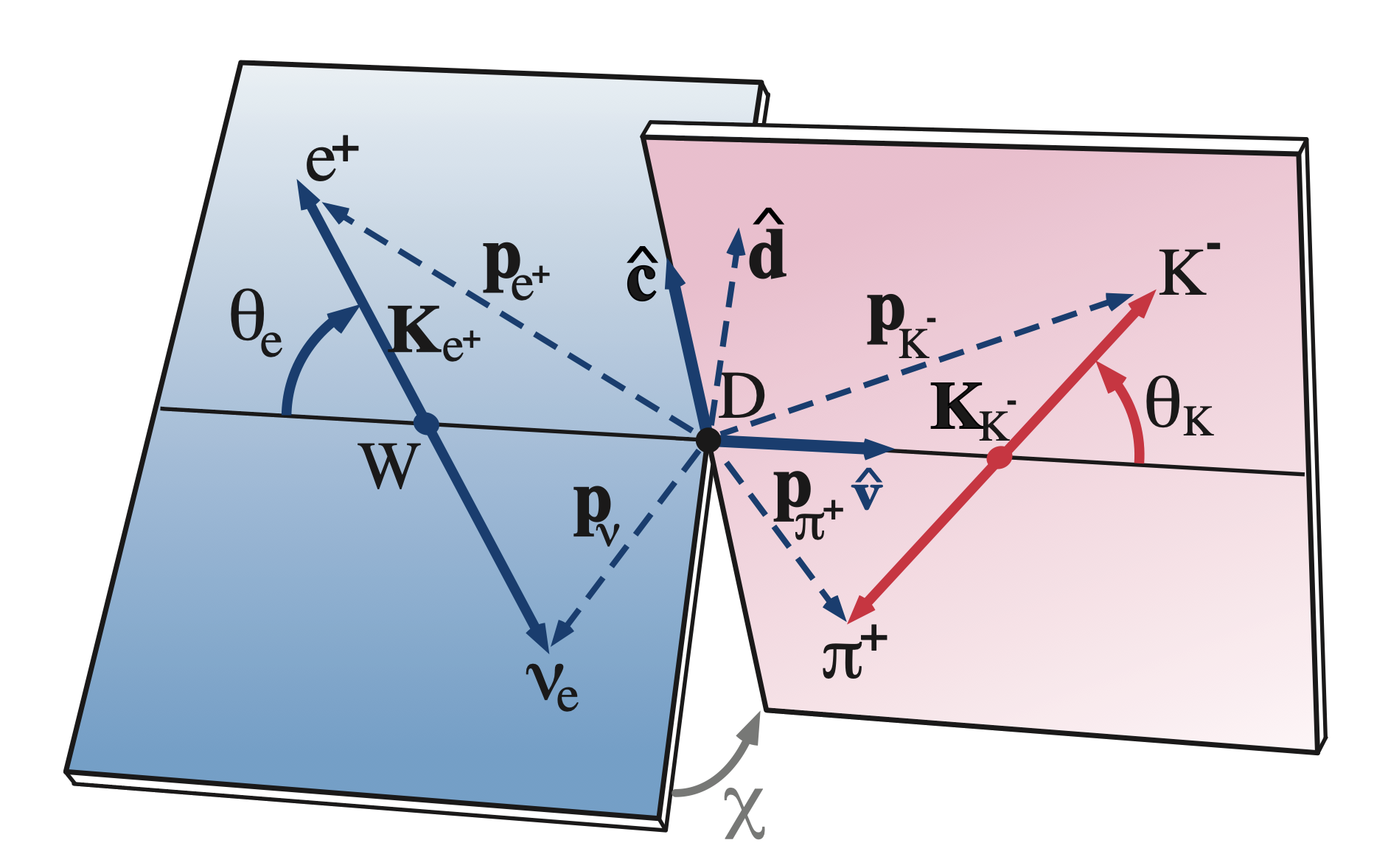}
    \caption{Definition of the variables used in four-body semileptonic decays as defined in~\cite{BaBar:2010vmf} for the $D^+\rightarrow K^-
\pi^+ e^+\nu_e$ mode. There are two rest frames, one for the lepton pair and one for the final state hadrons pair. $\theta_K$ is the angle between the pion and the $D$-meson direction in the $K^-\pi^+$ rest frame, $\theta_e$ is the angle between the neutrino and the $D$-meson direction in the lepton $e^+\nu_e$ rest frame, and $\chi$ is the acoplanarity angle between the two decay planes. Reprinted with permission from~\cite{BaBar:2010vmf}. }
    \label{sec5.1:fig3}
\end{figure}

\noindent 
Semileptonic charm-meson decays with two hadrons in the final state and a leptonic pair, $\ell\nu_\ell$, allow to study the dynamics of the meson-meson final-state interaction, since  the leptonic pair interacts only weakly. Four-body decays are in general more complex, and both $S$-wave~(spin 0) and $P$-wave (spin 1) components, cf.\ Eq.~\eqref{eq:ampl-spin},
may contribute. The different resonant and non-resonant 
contributions~\footnote{When a decay proceeds via an intermediate short-lived resonance e.g.\ $D^+ \to \bar{K}^*(892)(\rightarrow K^- \pi^+) \ell^+ \nu_\ell$, we call this a resonant contribution, whereas the direct decay $D^+ \to K^- \pi^+ \ell^+ \nu_\ell$ is referred to as non-resonant contribution.} 
can be disentangled via an amplitude analysis, see section~\ref{subsec:nonlep} for a definition of an amplitude analysis. Five-body semileptonic decays such as $D^0\rightarrow K^0_S\pi^-
\pi^0e^+\nu_e$ and $D^+ \rightarrow K^0_S\pi^+\pi^-e^+\nu_e$ have been seen~\cite{BESIII:2024ieo} but no information on the form factors is available yet.\\

\noindent
The analysis of four-body semileptonic decays allows the study of the hadronic form factors for $D$ meson transitions to vector mesons. One example of vector mesons is the $K^*$ meson which decays into a kaon pion pair. For instance, the decay $D^+\rightarrow K^-
\pi^+\ell^+\nu_\ell$ is dominated by the $D^+\rightarrow \bar{K}^{*0}(892)\ell^+\nu_\ell$
contribution, allowing an extraction of the $D^+ \to \bar K^{*0}(892)$ form factors. 
The hadronic form factors of $D^0\rightarrow \bar{K}^{*-}\ell^+\nu_\ell$ can be extracted from $D^0\rightarrow K^0_S\pi^-\ell^+\nu_\ell$ decays, while the $D^{0(+)} \rightarrow \rho^{-(0)}\ell^+\nu_\ell$ form factors are studied from a $D^{0(+)} \rightarrow \pi\pi\ell^+\nu_\ell$ sample. 
Note that the same sample allows also to study the much smaller $\omega$ contribution\footnote{The $\rho$ and $\omega$ are both vector mesons composed of the light quarks up and down but in two different isospin configurations. The $\rho$ forms an isospin triplet $\rho = (\rho^+, \rho^0, \rho^-)$, with $\rho^+ = u\bar d$, $\rho^- = d \bar u$, $\rho^0 = (u \bar u - d \bar d)/\sqrt{2}$,  while the $\omega^0$ is an isospin singlet with $\omega^0 = (u \bar u + d \bar d)/\sqrt{2}$.}. 
Final states with a $KK$ pair can be used to study decays with the vector meson $\phi$ (vector $s\bar{s}$ state), etc.
The semileptonic decay rate for a vector meson in the final state is described by five kinematic variables. 
For instance, in the case of $D^+ \to \bar K^{*0}(892) \ell^+ \nu_\ell$ decay with the  $K^{*0}(892)$ decaying into a $  K^- \pi^+$ pair, 
it reads
\begin{equation}
\frac{d\Gamma (D^+ \to K^{*0}(892)  \ell^+ \nu_\ell)}{dm^2_{hh}dq^2 d\cos \theta_K d\cos\theta_e d\chi}= X\, \beta \, \frac{G_F^2 |V_{cs}|^2}{(4\pi)^6 m_D^2} I(m^2_{hh}, q^2, \theta_K, \cos\theta_e, \chi) \,,
\label{sec5.1:eq7}
\end{equation}
where $q^2$ is the momentum transfer, $m_{hh}$ is the invariant mass of the final state hadrons, $X = p_{K\pi} m_D$, with $p_{K\pi}$ being the momentum of the $K\pi$ system in the $D$-meson rest frame,
$\beta = 2p^*/m_{hh}$, where $p^*$ is the 
$K$ momentum in the $K\pi$ rest frame 
and the function $I$ contains the information about the form factors. Note that there are three different form factors contributing to the differential decay rate in this case, because of the presence of vector and axial-vector components in the hadronic current of these decays. The remaining angular variables $\theta_K$, $\theta_e$ and $\chi$ are defined in Fig.~\ref{sec5.1:fig3}. For further details on this formalism, see e.g.~\cite{BaBar:2010vmf, HeavyFlavorAveragingGroupHFLAV:2024ctg}.  
Recent measurements for {\bf vector mesons} have been reported for the decay $D \rightarrow \rho(770)\ell^+\nu_\ell$ in~\cite{BESIII:2024lxg, CLEO:2011ab, BESIII:2021pvy, BESIII:2018qmf}, 
for $D^+_s \rightarrow \phi\mu^+\nu_\mu$ in~\cite{BESIII:2023opt}, for $D \rightarrow \bar{K}^{*}(892) \ell^+ \nu_\ell$ in~\cite{BESIII:2018jjm, CLEO:2006uty, CLEO:2010enr, BaBar:2010vmf, BESIII:2015hty, BESIII:2015kin}, and for $D\rightarrow \omega\ell^+\nu_\ell$ in~\cite{BESIII:2015kin}.
Moreover, studies of the four-body semileptonic decays, and of their $S$- and $P$-wave contributions, were recently published for the decay $D^0\rightarrow \bar{K}^0\pi^-e^+\nu_e$ in~\cite{BESIII:2024xjf} and for $D^+\rightarrow K^0_S\pi^0e^+\nu_e$ in~\cite{BESIII:2024awg}. In addition to measurements of $D$-meson form factors for pseudoscalar and vector final states, also measurements for decays with {\bf scalar} and {\bf axial-vector mesons} in the final states are actively performed, in particular at BESIII. For the former case, recent results can be found in~\cite{BESIII:2024zvp, BESIII:2024lnh, BESIII:2023wgr, BESIII:2021drk} while for the latter in~\cite{BESIII:2021uqr, BESIII:2019eao, BESIII:2024pwp}.\\

\begin{figure}[t]
    \centering
    \includegraphics[width=0.5\linewidth]{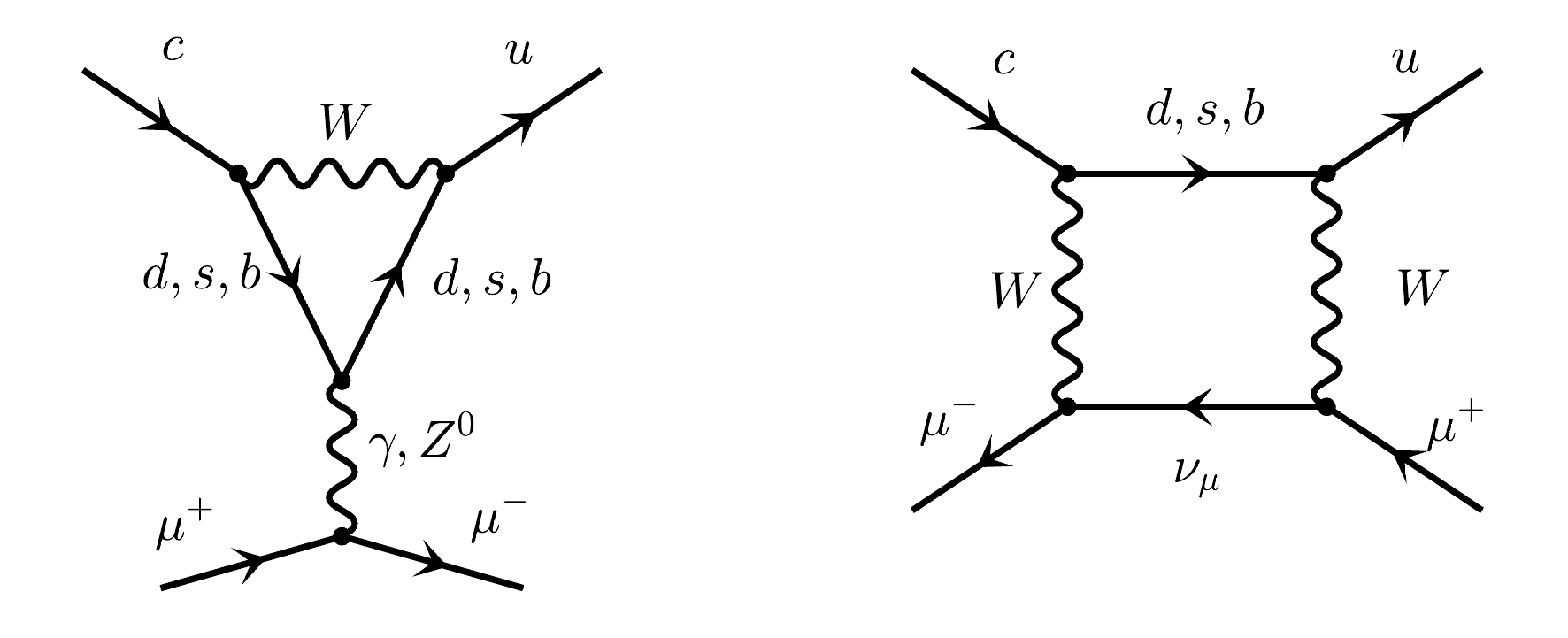}
    \caption{Example of penguin (left) and box (right) diagrams contributing in the SM to the rare decay $D^0 \rightarrow\mu^+\mu^-$.}
    \label{sec5.1:fig4}
\end{figure}

\noindent
Isospin symmetry is an approximate SU(2) symmetry of the strong interaction introduced to explain 
the nearly identical masses of the up and down quarks. A simple example of isospin partners is given by the proton ($uud$) and neutron ($udd$). Their quark content is very similar, and differs only by  
the exchange of the $u\leftrightarrow d$ quarks. Moreover, they have similar properties such as their mass.
The use of isospin symmetry means that two particles that differ only by the exchange of the $u\leftrightarrow d$ quarks can be treated as two states of one particle. {\bf Isospin symmetry tests}
can be performed in semileptonic decays by measuring the ratio of partial decay widths such as 
$\Gamma(D^0\rightarrow K^-e^+\nu_e)/\Gamma(D^+\rightarrow \bar{K}^0e^+\nu_e)$ 
and $\Gamma(D^0\rightarrow \pi^-e^+\nu_e)/2\Gamma(D^+\rightarrow \pi^0e^+\nu_e)$.
Measurements of these ratios compatible with 1 would imply results consistent with isospin symmetry. Recent studies performed with the decays   
$D\rightarrow K_1(1270)\ell\nu$~\cite{BESIII:2025yot, BESIII:2019eao}, $D^0 \rightarrow b_1(1235)e^+\nu_e$~\cite{BESIII:2024pwp}, $D \rightarrow \rho\ell^+\nu_\ell$~\cite{BESIII:2021pvy,BESIII:2018qmf} and $D\rightarrow K\ell^+\nu_\ell$~\cite{BESIII:2016hko, BESIII:2016gbw}
show results that are consistent with isospin symmetry within current uncertainties.\\

\begin{figure}[t]
    \centering
\includegraphics[scale = 0.35]{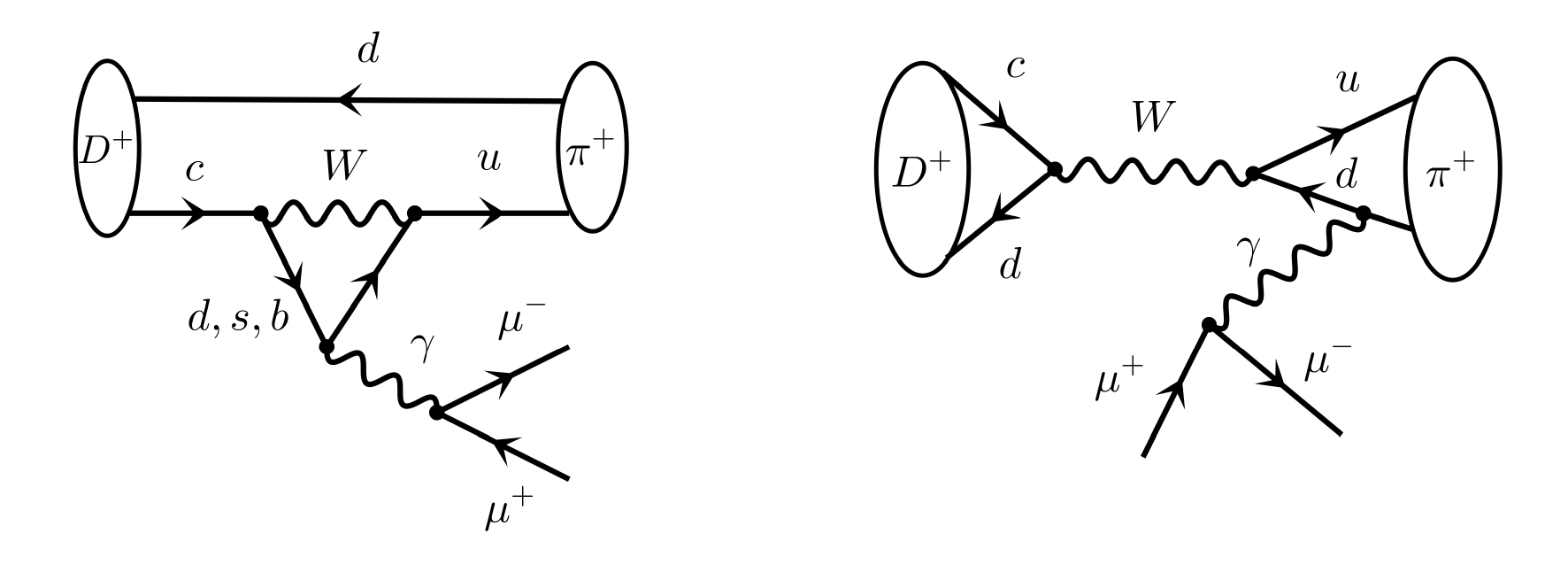}
    \caption{Examples of diagrams contributing to the rare decay $D^+\rightarrow\pi^+\mu^+\mu^-$: FCNC transitions (left) and weak annihilation (right).}
    \label{sec5.1:fig5}
\end{figure}

\noindent
{\bf Rare and forbidden leptonic and semileptonic decays:} Rare decays are {\it flavour changing neutral current} (FCNC) decays which proceed via penguin and box diagrams in the SM, while forbidden processes in the SM are the {\it lepton number violating} (LV) and the {\it lepton flavour violating} (LFV) decays. Leptonic modes such as $D^0 \to \mu^+ \mu^-$, sketched in Fig.~\ref{sec5.1:fig4}, and semileptonic modes such as $D^+ \to \pi^+ \mu^+ \mu^-$, shown in Fig.~\ref{sec5.1:fig5}, are examples of FCNC charm decays, whereas examples of LFV and LV decays are the leptonic decay $D^0 \to e^+ \mu^-$ and the semileptonic decay $D^+ \to \pi^- \mu^+ \mu^+$, respectively.\\ 

\begin{figure}
    \centering
\includegraphics[scale = 0.4]{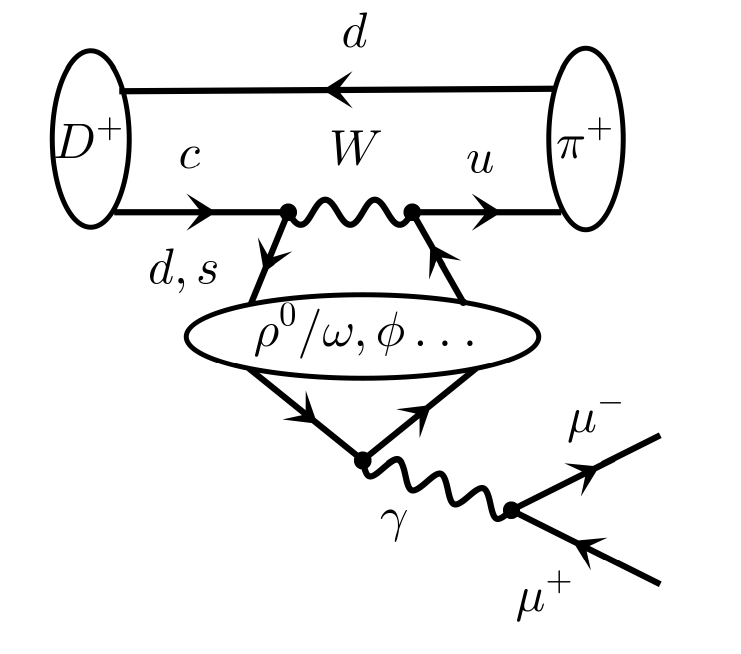}
    \caption{Example of diagram describing the contribution of intermediate vector meson resonances to the FCNC decay $D^+\rightarrow\pi^+\mu^+\mu^-$.}
    \label{sec5.1:fig6}
\end{figure}
\begin{figure}[b]
    \centering
\includegraphics[scale = 0.35]{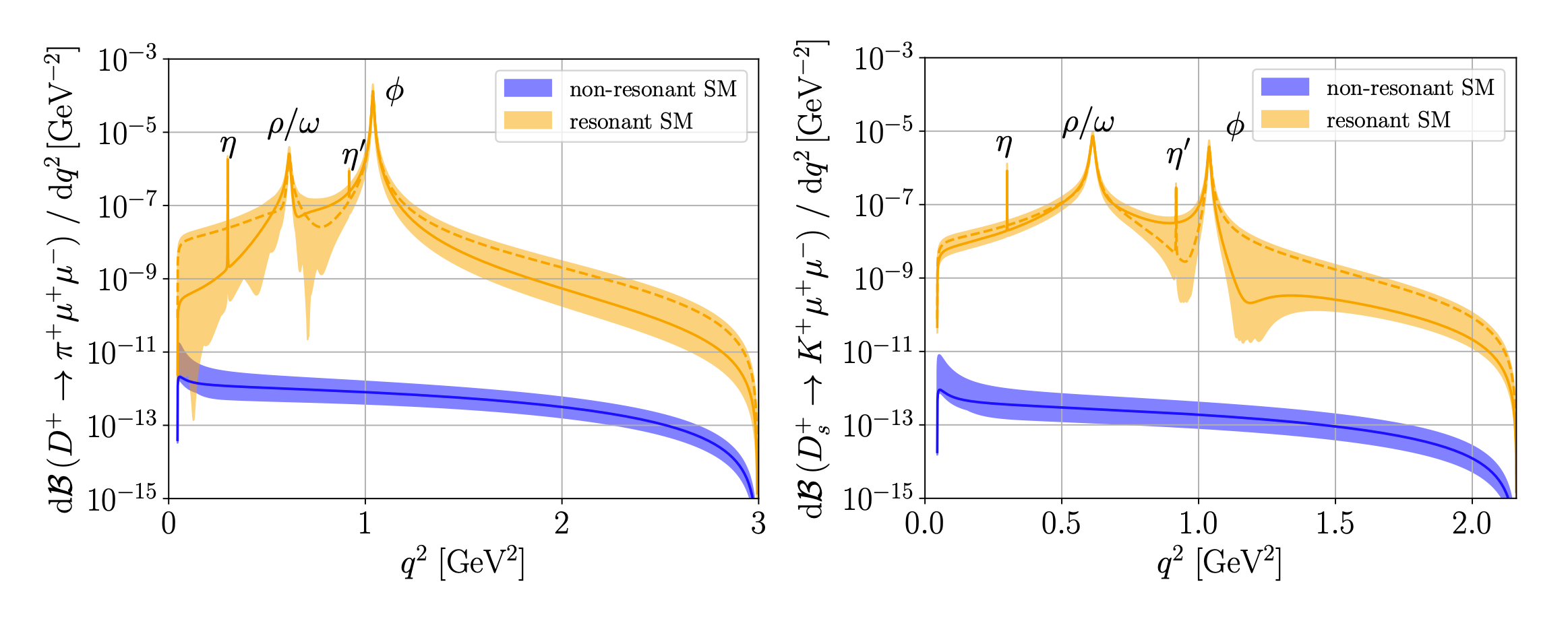}
    \caption{SM differential branching fractions of the rare semileptonic decays $D^+ \to \pi^+ \mu^+ \mu^-$ (left) and $D^+_s \to K^+ \mu^+ \mu^-$
(right). In each plot, the blue band represents the short-distance (non-resonant) contribution, whereas the orange band the long-distance (resonant) one. The solid and dashed orange curves correspond to different values of the strong phases entering the Breit-Wigner parametrisation of the resonances, namely $\delta_\rho = 0$, $ \delta_\phi = \pi$, and $\delta_\rho =  \delta_\phi = 0$, respectively. Taken from~\cite{Bause:2019vpr} with permission.}
    \label{sec5.1:fig6p}
\end{figure}
\noindent
Leptonic FCNC charm-meson decays are both helicity and GIM suppressed in the SM and are expected to have tiny branching fractions. For example, the SM prediction for the mode $D^0 \to \mu^+ \mu^-$ is of the order of $10^{-12}$~\cite{Gisbert:2024kob} while current experimental searches yield upper bounds of the order of $10^{-9}$~\cite{LHCb:2022jaa, CMS:2024smm}. This experimental measurement is complicated by the presence of hadronic background ($D^0\rightarrow \pi^+\pi^-$).
Semileptonic FCNC decays include modes such as $D \to X \ell^+ \ell^-$, where $X$ can be any light vector or pseudoscalar meson i.e.\ $K$, $\pi$, etc. These decays proceed via $c \rightarrow u \ell^+ \ell^-$ FCNC transitions and are suppressed by the GIM mechanism in the SM.
Their short-distance contribution is of the order of $\mathcal{O}(10^{-9})$, however, long-distance effects, where the two leptons originate from an intermediate vector resonance, i.e.\ $D\rightarrow XV(\rightarrow \ell^+ \ell^-)$, as shown in Fig.~\ref{sec5.1:fig6}, can lead to an enhancement, see e.g.\ \cite{Bansal:2025hcf, deBoer:2015boa, Bause:2019vpr} and Fig.~\ref{sec5.1:fig6p}.
Modelling the corresponding resonance contributions plays a crucial role in the theoretical predictions and, particularly, in the estimate of their uncertainties, and in most cases, parameterisations based on fitting Breit-Wigner distributions to experimental data are used, see e.g.\ \cite{Fajfer:2023tkp, Gisbert:2024kob}.
Alternative, more sophisticated approaches can be found in e.g.\ \cite{Bharucha:2020eup, Feldmann:2017izn}. In general, the SM predictions for these channels range from $10^{-5}$ to $10^{-9}$~\cite{Gisbert:2024kob, Fajfer:2007dy, Fajfer:2002bu}, where any signal above these values could be a sign of BSM physics. 
Experimental searches are performed in different regions of the di-lepton invariant mass, and the intervals are usually separated in low- and high- di-lepton invariant mass for the signal searches, and in regions around the dominant resonances i.e.\ $\eta$,  $\omega$, $\rho$, $\phi$,~\ldots, for the non-signal candidates, which are often used as normalisation. 
The rarest charm-hadron decays ever observed are the modes $D^0\rightarrow \pi^+\pi^- \mu^+\mu^-$ and $D^0\rightarrow K^+K^- \mu^+\mu^-$ with branching fractions of $(9.64\pm 1.20)\times 10^{-7}$ and $(1.54\pm 0.33)\times10^{-7}$~\cite{LHCb:2017uns}, respectively, in agreement with the SM expectations~\cite{Cappiello:2012vg}.
A special role is played by the FCNC decays involving di-neutrinos such as $D^0 \to \pi^0\nu\bar{\nu}$, as in this case the long-distance effects are negligible and the short-distance contributions from
penguin and box diagrams dominate, resulting in a
branching fraction of the order of $\mathcal{O}(10^{-15})$~\cite{Burdman:2001tf}. Searches for such rare decays are ongoing~\cite{BESIII:2021slf}, but they are highly complicated by the presence of two missing neutrinos in the final state.  \\

\begin{figure}
    \centering
    \includegraphics[scale=0.55]{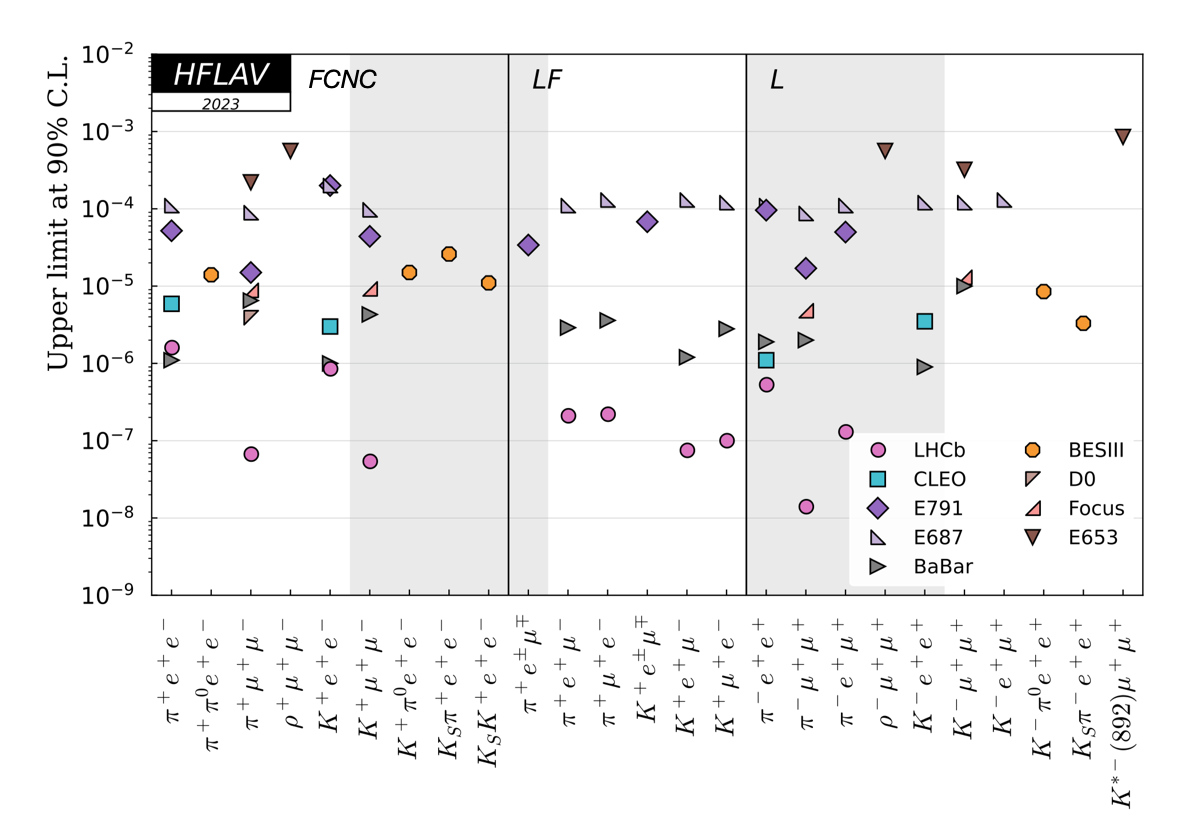}
    \caption{Branching fractions limits for rare and forbidden $D^+$-meson decays. These are classified into decays which are FCNC (left panel), lepton flavour (LF) violating (middle panel) and lepton number (L) violating (right panel) (the white and grey areas are there to simply guide the eye: there are results for five decay modes within each white or grey area). Taken from~\cite{HeavyFlavorAveragingGroupHFLAV:2024ctg}.
    }
    \label{sec5.1:fig7}
\end{figure}

\noindent
Forbidden semileptonic charm decays include both LV modes with same-sign lepton pair, namely $D \to X \ell_1^+ \ell_2^+$, where the leptons could belong to different flavour families, i.e.\ $l_1 \neq l_2$ -  lepton family number violating, or to the same family, i.e.\ $l_1 = l_2$, and LFV modes with opposite-sign lepton pair, namely $D \to X \ell_1^+ \ell_2^-$ where $l_1 \neq l_2$. The first class includes channels such as $D^+ \rightarrow \rho^-\mu^+\mu^+$, $D^+\rightarrow K^-e^+e^+$ and $D^+ \rightarrow \pi^-\mu^+e^+$. The second class contains modes such as $D^+\rightarrow \pi^+\mu^+e^-$ and $D^+\rightarrow K^+\mu^+e^-$. 
A summary of the current experimental limits for rare and forbidden $D^+$-meson decays is shown in Fig.~\ref{sec5.1:fig7}, while a comprehensive list with all available limits for the rare and forbidden decays of the $D^+$, $D^0$, $D^+_s$ and $\Lambda_c$ hadrons can be found in~\cite{HeavyFlavorAveragingGroupHFLAV:2024ctg, ParticleDataGroup:2024cfk}.\\

\noindent
Being very suppressed or forbidden processes in the SM, these charm decays provide a very sensitive candle for BSM physics and their study constitutes a growing and exciting field of research, see e.g.\ the reviews~\cite{Gisbert:2020vjx, Artuso:2008vf, Burdman:2003rs}. In fact, for the rare channels, any measurement above the SM predictions would be a strong indication of NP, whereas for the forbidden modes, any experimental signal would unambiguously require a BSM interpretation. Currently, the main strategies for NP searches include the study of (i) branching ratios of forbidden decays~\cite{LHCb:2015pce, LHCb:2020car} or of decays whose SM predictions are far below the current experimental reach, such as $D \to \mu^+ \mu^-$~\cite{Gisbert:2024kob, Fajfer:2015mia, LHCb:2022jaa,  LHCb:2023fuw}, and di-neutrino modes~\cite{Bause:2020auq, Bause:2020xzj, Burdman:2001tf,  BESIII:2021slf}, see also \cite{Fajfer:2021woc}, (ii) multi-body decays in regions of the phase-space away from resonances, in order to minimise the sensitivity to poorly known hadronic effects~\cite{Gisbert:2024kob, deBoer:2015boa, LHCb:2024ely, LHCb:2024hju, LHCb:2021yxk, LHCb:2017yqf}, (iii) null tests of the SM, which often are sensitive to the interference of NP and SM amplitudes, such as angular and CP asymmetries~\cite{Gisbert:2024kob, Fajfer:2023tkp, Golz:2022alh, Golz:2021imq, Bause:2020obd, Bause:2019vpr, Fajfer:2015ixa, LHCb:2025bfy, LHCb:2018qsd}, and (iv) radiative decays~\cite{Adolph:2022ujd, Adolph:2020ema, deBoer:2017que}.

%%%%%%%%%%%%%%%%%%%%%%%%%%%%%%%%%%%%%%%%%%%%%%%%%%
%%%%%%%%%%%%%%%%%%%%%%%%%%%%%%%%%%%%%%%%%%%%%%%%%%
%%%%%%%%%%%%%%%%%%%%%%%%%%%%%%%%%%%%%%%%%%%%%%%%%%
%%%%%%%%%%%%%%%%%%%%%%%%%%%%%%%%%%%%%%%%%%%%%%%%%%
%%%%%%%%%%%%%%%%%%%%%%%%%%%%%%%%%%%%%%%%%%%%%%%%%%
\subsection{LFU tests with leptonic and semileptonic decays}
\label{subsec:LFU}
\begin{figure}
    \centering
    \includegraphics[scale=0.45]{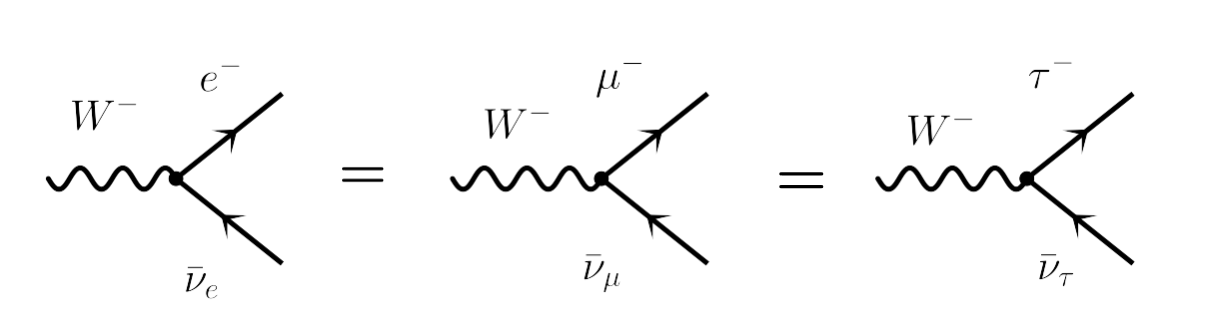}
    \caption{Lepton flavour universality principle.}
    \label{sec5.2:fig1}
\end{figure}

Lepton flavour universality (LFU) is an intrinsic feature of the SM but also accidental. The principle behind it is that the leptons of all three families, i.e. $e,\mu, \tau$, interact equally with the weak force carriers, as illustrated in Fig.~\ref{sec5.2:fig1}. 
The only difference between these interactions is given by the mass of the leptons. Experimentally, LFU is probed by measuring the ratio of (semi)leptonic decay rates with leptons of different flavour. Due to negligible mass differences in the corresponding phase-space functions, the ratio of decays to muons and electrons is expected to be close to 1,
however, mass-dependent terms become significant when one of the decay rates involves the tau lepton, in which case the ratios are not equal to 1.\\

\noindent
For interactions with the $W$ bosons, LFU has been tested to great precision by the LEP, ATLAS, CMS and LHCb experiments. 
In particular, in recent years, LFU has been extensively studied in $B$ meson decays (by the Belle, BaBar, LHCb and CMS experiments), 
but a significant number of LFU searches have also been performed
in open and hidden charm decays, which are fundamental and complementary to the $b$-sector searches.
By measuring the branching fractions for at least two lepton families, the following ratio can be constructed
\begin{equation}
R\left(D \to (X) \frac{\ell^\prime}{\ell}\nu\right) \equiv \frac{D\rightarrow (X)\ell'\nu_\ell'}{D\rightarrow (X)\ell\nu_\ell} \,,
\label{sec5.2:eq1}
\end{equation}
where most of the experimental systematic uncertainties cancel.
In Eq.~\eqref{sec5.2:eq1}, $\ell'$ and $\ell$ are two charged leptons belonging to different flavour families. In the case of leptonic decays, the SM prediction does not depend on hadronic inputs and the ratio in Eq.~\eqref{sec5.2:eq1} reads
\begin{equation}
R\left(D \to \frac{\ell^\prime}{\ell}\nu\right) = \frac{\Big( 1-\frac{m^2_{\ell'}}{m^2_D}\Big)^2 m^2_{\ell'}}{\Big( 1-\frac{m^2_{\ell}}{m^2_D}\Big)^2 m^2_{\ell}} + \mathcal{O}(10^{-2})\,,
\label{sec5.2:eq2}
\end{equation}
up to higher order electro-weak corrections. 
In the case of semileptonic decays, the corresponding ratios are built in a similar way, however, the SM predictions depend on hadronic inputs, namely the form factors and on different phase-space functions, since the phase-space for multi-body decays is at least two-dimensional. Experimentally, among all three charged leptons, electrons and muons can be detected directly. Tau leptons are detected via their hadronic or leptonic decays, e.g.\ via the
$\tau\rightarrow\pi(\pi\pi)\bar{\nu}_\tau$, or $\tau\rightarrow\ell\nu_\ell\bar{\nu}_\tau$ decays.
A summary of selected measurements is shown in Fig.~\ref{sec5.2:fig2} and in Table~\ref{sec5.2:tab1}. As of today, no tensions with the SM predictions have been observed.\\
\\

\begin{minipage}{\textwidth}
\begin{minipage}{0.43\textwidth}
\hspace*{-5mm}
        \includegraphics[width=1.\linewidth]{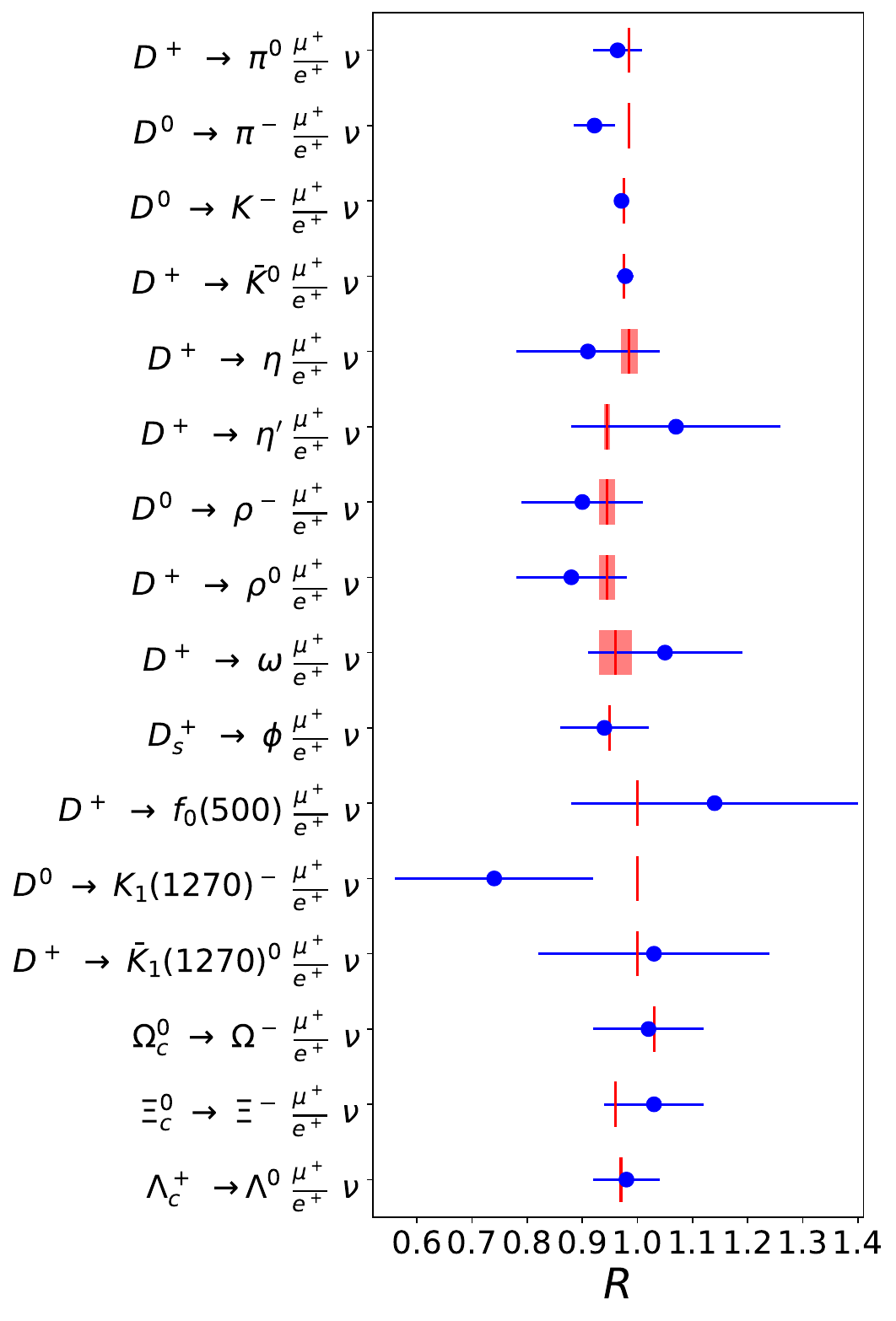}

    \captionof{figure}{LFU tests with semileptonic charm decays: the results for the ratio $R$ are shown for different decay modes (here e.g.\ $D^+\rightarrow\pi^0\frac{\tau^+}{\mu^+} \nu$ indicates that the ratio $R$ corresponds to $\frac{B(D^+\rightarrow\pi^0{\tau^+} \nu)}{B(D^+\rightarrow\pi^0{\mu^+} \nu)}$, and similarly for the other entries); the experimental results are shown in blue, with error bars, while the red boxes correspond to the range of central values predicted by theory, given in Table~\ref{sec5.2:tab1}.}
    \label{sec5.2:fig2}
\end{minipage}
\qquad
\begin{minipage}{0.48\textwidth}
 \renewcommand{\arraystretch}{1.5}
    \begin{tabular}{|c||c|c|}
       \hline
        & {\it Exp.\ value} & {\it SM value} \\
        \hline
        \hline
        $R(D^+\rightarrow\frac{\tau^+}{\mu^+} \nu )$& $2.49 \pm 0.31$~\cite{BESIII:2024vlt} 
        & 
        2.67\\
        \hline
        $R (D_s^+\rightarrow\frac{\tau^+}{\mu^+} \nu)$ & $9.72\pm 0.37$~\cite{BESIII:2021bdp} 
        & 9.74\\
        \hline
        \hline
        $R(D^+\rightarrow\pi^0\frac{\mu^+}{e^+}\nu)$ & $0.964\pm0.045$~\cite{BESIII:2018nzb} & $\sim$ 0.985\\
        \hline
        $R(D^0\rightarrow\pi^-\frac{\mu^+}{e^+}\nu)$ & $0.922\pm0.037$~\cite{BESIII:2018nzb} & $\sim$ 0.985\\
        \hline
        $R(D^0\rightarrow K^-\frac{\mu^+}{e^+}\nu)$ & $0.971 \pm 0.007$~\cite{BESIII:2024slx}

& $\sim$ 0.975\\
\hline
$R(D^+\rightarrow \bar{K}^0\frac{\mu^+}{e^+}\nu)$ & $0.978 \pm 0.015$~\cite{BESIII:2024slx}
 & $\sim$ 0.975\\
\hline
        $R(D^+\rightarrow\eta\frac{\mu^+}{e^+}\nu)$ & $0.91\pm0.13$~\cite{Ablikim:2020hsc} & 0.97 -- 1.00 \\
        \hline
     $R(D^+\rightarrow \eta'\frac{\mu^+}{e^+}\nu)$ & $1.07 \pm 0.19$~\cite{BESIII:2024njj} 
     %\pm 0.03$ 
      & $0.94 - 0.95$\\
        \hline
        $R(D^0\rightarrow\rho^-\frac{\mu^+}{e^+}\nu)$ & $0.9\pm 0.11$ ~\cite{BESIII:2021pvy}& 0.93 -- 0.96\\
        \hline
        $R(D^+\rightarrow\rho^0\frac{\mu^+}{e^+}\nu)$ & $0.88\pm 0.10$ ~\cite{BESIII:2024lnh}& 0.93 -- 0.96\\
        \hline
        $R(D^+\rightarrow\omega\frac{\mu^+}{e^+}\nu)$ & $1.05\pm0.14$~\cite{Ablikim:2020tmg} & 0.93 -- 0.99\\
        \hline
     $R(D^+_s \rightarrow\phi \frac{\mu^+}{e^+}\nu)$& $0.94 \pm 0.08$~\cite{BESIII:2023opt} & 0.95\\
     \hline
        $R(D^+\rightarrow f_0(500)\frac{\mu^+}{e^+}\nu)$ & $1.14\pm 0.26$ ~\cite{BESIII:2024lnh}&  $\sim1$ \\
        \hline
     $R(D^0\rightarrow K_1(1270)^-\frac{\mu^+}{e^+}\nu)$ & $0.74^{+0.15}_{-0.18}$~\cite{BESIII:2025yot} 
     %\pm 0.13^{+0.08}_{-0.13}$ 
     & $\sim1$\\
     \hline
     $R(D^+ \rightarrow{\bar{K}}_1(1270)^0\frac{\mu^+}{e^+}\nu)$ & $1.03 ^{+0.18}_
{-0.21}$~\cite{BESIII:2025yot}
& $\sim1$ \\
\hline
\hline
%         \hline
        $R(\Omega^0_c \rightarrow\Omega^- \frac{\mu^+}{e^+}\nu)$&   $1.02  \pm  0.10$~\cite{Belle:2021dgc}& 1.03\\
        \hline
        $R(\Xi^0_c \rightarrow\Xi^- \frac{\mu^+}{e^+}\nu) $& $1.03\pm0.09$~\cite{Belle:2021crz}&   0.96\\
        \hline
        $R(\Lambda^+_c \rightarrow\Lambda^0 \frac{\mu^+}{e^+}\nu)$& $0.98 \pm 0.06$~\cite{BESIII:2023jxv}&  $0.97$ \\
        \hline
    \end{tabular}
    \captionof{table}{Overview of LFU results with leptonic and semileptonic charm meson decays. Here $R(D^+\rightarrow\pi^0\frac{\tau^+}{\mu^+} \nu) = \frac{B(D^+\rightarrow\pi^0{\tau^+} \nu)}{B(D^+\rightarrow\pi^0{\mu^+} \nu)}$, and similarly for the other entries, see Eqs.~\eqref{sec5.2:eq1}, \eqref{sec5.2:eq1}. For decays with tau leptons, values deviate from unity because of the different available phase-space due to the sizeable difference in the lepton masses. Theoretical predictions can be found in~\cite{Wu:2006rd,Riggio:2017zwh,Soni:2018adu, Cheng:2017pcq, Ivanov:2019nqd, Sekihara:2015iha, Faustov:2019mqr, Huang:2021ots, Zhang:2021oja}. 
    Note that the SM values for the ratios of leptonic decays in the first two rows do not include electro-weak corrections, cf. Eq.~\eqref{sec5.2:eq2}. No significant deviations form the SM predictions are seen.
    }
    \label{sec5.2:tab1}
\end{minipage}
\end{minipage}
\\
\\

%%%%%%%%%%%%%%%%%%%%%%%%%%%%%%%%%%%%%%%%%%%%%%%%%%%%%%%%%%%%%%%%%%%%%%%%%%%%%%%%%%%%%%%%%%%%%%%%%%%%%%%%%%%%%%%%%%%%%%%%%%%%%%%%
\subsection{Non-leptonic decays}
\label{subsec:nonlep}
SM predictions for non-leptonic charm-hadron decays are particularly challenging because of the difficulty to precisely quantify the size of the corresponding non-perturbative effects. For this reason, different approaches can be found in the literature and often additional assumptions are needed to obtain quantitative predictions.\\ 

\noindent
{\bf Two-body non-leptonic decays:} A very common method for the study of two-body non-leptonic charm-hadron decays makes use of exact or approximate SU(3)$_F$ symmetry relations, see e.g.~\cite{PhysRevD.11.1919, Voloshin:1975yx, PhysRevLett.43.812, PhysRevD.42.1527, Bhattacharya:2012ah, Hiller:2012xm, Grossman:2012eb, Cheng:2012wr, Nierste:2015zra, Muller:2015lua, Muller:2015rna, Dery:2021mll, Gavrilova:2022hbx, Gavrilova:2024npn, Bolognani:2024zno} and the references therein. In this type of studies, the approximate SU(3)$_F$ symmetry of QCD is used to relate several hadronic structures for a vast class of decay processes, thus reducing the number of independent parameters needed to describe the corresponding decay amplitudes. The resulting set of common hadronic inputs is then extracted by performing a global fit to the available experimental data on, for example, branching ratios and CP asymmetries. Note that, in some cases, by leveraging large datasets, SU(3)$_F$ breaking effects can also be taken into account, see e.g.~\cite{Muller:2015lua, Muller:2015rna, Bolognani:2024zno}. 
Even though such analyses can provide some insight on the hadronic structure of these theoretically challenging decays, they do not constitute a calculation of the size of their branching fractions or CP asymmetries in the SM.\\

\noindent
Further theoretical approaches
are based on the use of dispersion relations~(DRs), which follow from the analyticity and unitarity of the scattering matrix. For example, the use of DRs in the large $N_c$ limit~\cite{Pich:2002xy}, where $N_c$ denotes the number of colours, was employed in~\cite{Pich:2023kim} to study the non-leptonic decays $D^0 \to \pi^+ \pi^-$ and $D^0 \to K^+ K^-$ and, specifically, to determine the size of the strong phases due to the $\pi \pi$, $KK$ rescattering, a key ingredient for the prediction of CPV in these decays. 
Other studies of two-body non-leptonic $D$-meson decays including specific modelling for rescattering effects and final state interactions can be found in~\cite{Bauer:1986bm, Buccella:1992ym, Buccella:1994nf, Buccella:2013tya, Buccella:2019kpn, Buccella:1992sg, Franco:2012ck, Bediaga:2022sxw}. We also note how first steps in determining rescattering effects in charmed meson decays from Lattice QCD have been made in~\cite{DiCarlo:2025mvt, Hansen:2012tf}, however, an exhaustive calculation of these contributions is at the present still out of reach.\\

\begin{figure}[t]
\centering
\includegraphics[scale =0.65]{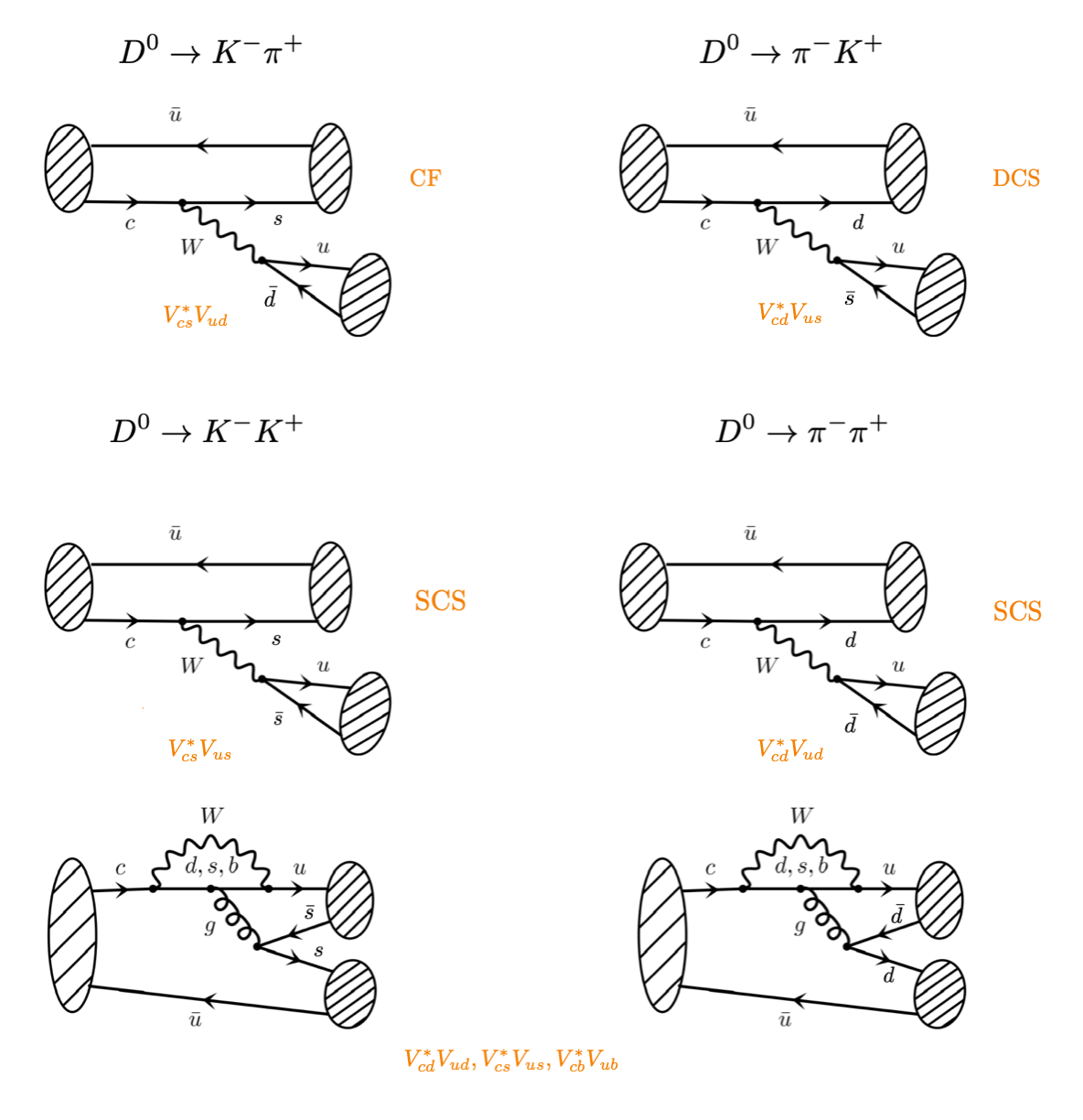}
\caption{Some of the topologies contributing to the CF, SCS, and DCS two-body $D^0$ decays. The CF and DCS proceed only through tree-level amplitudes. The SCS decays receive contributions from tree-level and penguin amplitudes, which have different strong and weak phases. To keep the diagrams more transparent additional gluon contributions are not drawn.}
\label{sec5.3:fig1}
\end{figure}

\noindent
The framework of LCSRs has proved to be a powerful method for describing hadronic matrix elements, although, in the context of non-leptonic decays, its use has mainly focused on the $B$-meson system, see e.g.~\cite{Khodjamirian:2000mi, Khodjamirian:2003eq, Piscopo:2023opf}. In the case of hadronic charmed-meson decays, LCSRs has been employed in \cite{Khodjamirian:2017zdu} to determine the size of penguin contributions in $D^0 \to \pi^+ \pi^-$ and $D^0 \to K^+ K^-$ and recently in~\cite{Lenz:2023rlq} to also estimate the  tree-level amplitude and the corresponding branching fractions for the Cabibbo favourite~(CF), singly Cabibbo suppressed~(SCS), and doubly Cabibbo suppressed~(DCS) decays $D^0 \to \pi^+ K^-, K^+ K^-, \pi^+ \pi^-, K^+ \pi^- $, sketched in Fig.~\ref{sec5.3:fig1}. The comparison between the LCSRs predictions and the experimental average determinations is given in Table~\ref{sec5.3:tab1}.  The overall agreement appears very good, particularly given the approximations made and the challenges of dealing with charm systems. The central values agree well with the data, while the theory uncertainties, conservatively estimated, are still much larger than the experimental ones. Experimental measurements allow to validate theoretical calculations and test the SU(3)$_F$ symmetry breaking in two-body hadronic charm decays. Currently, the most precise absolute branching fraction measurements for charm decays into two
pseudoscalar mesons come from the BESIII collaboration~\cite{BESIII:2018apz} where the single-tag technique has been used as the final states can be fully reconstructed. These results are then averaged with other experimental measurements in~\cite{HeavyFlavorAveragingGroupHFLAV:2024ctg} - or alternatively in~\cite{ParticleDataGroup:2024cfk}.\\

\noindent
In the ratio of branching ratios, however, due to the cancellation of common theoretical uncertainties, the theory precision becomes higher and one finds~\cite{Lenz:2024rqy}:
\begin{equation}
\frac{B (D^0 \to K^- K^+ )}
     {B (D^0 \to \pi^- \pi^+ )}\Bigg|_{\rm LCSRs}
      = 
2.63 \pm 0.86 \,, 
\qquad \quad
 \frac{B  (D^0 \to K^- K^+ )}
     {B(D^0 \to \pi^- \pi^+ )}\Bigg|_{\rm exp} 
      =  
2.760 \pm { 0.060}
 \, ,
 \label{sec5.3:eq1}
\end{equation}
where, for simplicity, the experimental uncertainty has been obtained adding the individual uncertainties in quadrature.
It is important to stress that the experimental value in Eq.~\eqref{sec5.3:eq1} indicates a severe violation of the SU(3)$_F$ symmetry in the branching fractions of the SCS decays, which overcompensates the phase space suppression of the $D^0 \to K^+ K^-$ mode. This result appears to be well captured by the LCSRs prediction. We also note that already within the naive factorisation approximation the agreement between the above-mentioned decays and the corresponding experimental data is found to be, taking into account the complicated hadronic structure, excellent~\cite{Lenz:2023rlq}, and the large size of the SU(3)$_F$ breaking in the SCS decays is reproduced by the current Lattice QCD results for the $\pi, K$ decay constants and the $D \to \{\pi, K\}$ form factors~\cite{FlavourLatticeAveragingGroupFLAG:2024oxs}. \\

\begin{table}[t]
\centering
\renewcommand*{\arraystretch}{1.7}
\begin{tabular}[t]{|c||c|c|c|c|}
 \hline
  &
  $ 10^2 \cdot  B(D^0 \to  K^- \pi^+)$  & 
  $ 10^3 \cdot B(D^0 \to  K^- K^+)$  & 
  $ 10^3 \cdot B(D^0 \to  \pi^- \pi^+)$  &
  $ 10^4 \cdot B(D^0 \to  \pi^- K^+)$ \\
  \hline
  \hline
  LCSRs~\cite{Lenz:2023rlq} &
  $2.99^{+3.26}_{-2.26}$  &
  $3.67^{+3.90}_{-2.69}$ &
  $1.40^{+1.53}_{-1.06}$ &
  $1.80^{+1.93}_{-1.33}$\\
  \hline
HFLAV~\cite{HeavyFlavorAveragingGroupHFLAV:2024ctg} &
 $3.999 \pm 0.045$ &
 $4.113 \pm 0.051$ &
 $1.490 \pm 0.027$ &
 $1.376 \pm 0.017$\\
 \hline
\end{tabular}
\caption{Comparison between the LCSRs predictions and the HFLAV values for the two-body CF, SCS and DCS $D^0$ decays.
For simplicity we have added the different experimental uncertainties in quadrature, but the individual contributions can be found in the HFLAV report. Note that the HFLAV averages have been very recently updated by taking the final state radiation consistently into account.}
\label{sec5.3:tab1}
\end{table}

\noindent
{\bf Multi-body non-leptonic decays:} Multi-body charm-hadron decays provide a very interesting system with a larger set of observables, as consequence of the richer kinematical structure. Multi-body final states can be reached through many different intermediate short-lived resonances. These decays are particularly sensitive to measurements of $D^0-\bar{D}^0$ mixing and searches for CPV as they offer a particularly interesting playground to study the interplay between weak and strong interactions. They are also important for QCD tests and spectroscopy. In fact, multi-body hadron decays provide important information about light meson-meson scattering amplitudes for example in decays like $D^+ \rightarrow K^-K^+K^+
$~\cite{LHCb:2019tdw}, $D^+ \rightarrow K^-
\pi^+\pi^+$ ~\cite{E791:2005gev, CLEO:2008jus}, and $D^+_{(s)} \rightarrow \pi^-
\pi^+\pi^+$ ~\cite{LHCb:2022pjv}. Precise theoretical predictions, however, are clearly difficult and often based on specific model assumptions.  One of the most commonly used theoretical method is the isobar approach, in which the hadronic matrix element is parametrised as a sum of interfering
decay amplitudes, each proceeding through resonant two-body decays, see e.g.~\cite{Jackson:1964zd}. Other approaches use dispersive methods~\cite{Omnes:1958hv, Hanhart:2012wi, Niecknig:2015ija}, multi-meson models with chiral Lagrangians, see e.g.~\cite{Magalhaes:2011sh, Aoude:2018zty}, and methods based on chiral unitarity, see e.g.~\cite{Oller:2004xm, Roca:2020lyi}.
\\

\begin{figure}
    \centering
\begin{overpic}[width=0.9\linewidth]{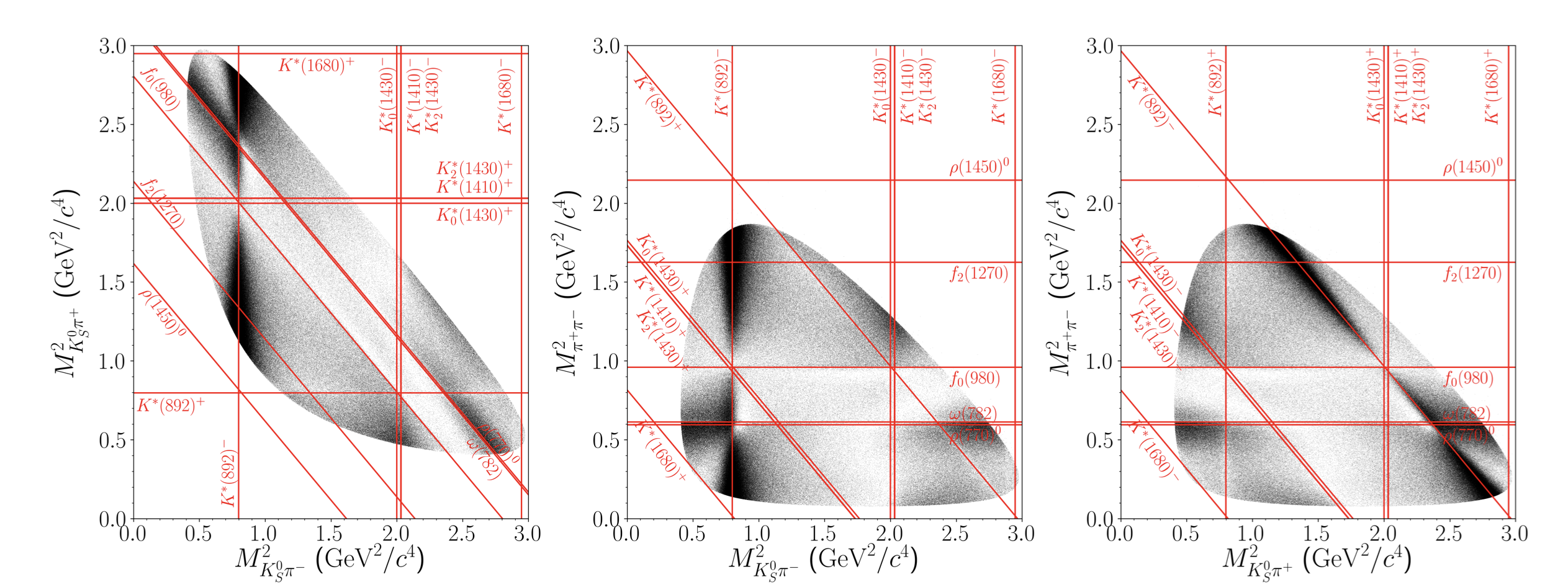} \put(340,120){\includegraphics[width=0.05\linewidth]{belle-logo.jpeg}}
  \end{overpic}
    \caption{Three different Dalitz plots of the $D^0\rightarrow K_S\pi^+\pi^-$ decay from $D^{*+}\rightarrow D^0\pi^+$ decays formed with the two-body invariant masses $M^2_{K_S\pi-}$, $M^2_{K_S\pi+}$ (left),   $M^2_{K_S\pi-}$, $M^2_{\pi+\pi^-}$ (middle),  $M^2_{\pi^+\pi-}$, $M^2_{K_S\pi+}$ (right). To guide the eye, the vertical, horizontal and diagonal lines indicate the approximate location of the two-body intermediate resonances. Note that here $M^2$ is equivalent to the $m^2$ defined in the text. Taken from~\cite{BaBar:2018cka}.}
    \label{sec5.3:fig4_dalitz}
\end{figure}

\noindent
Experimentally, multi-body analysis in charm is a challenging area of research. However large datasets available from LHCb, Belle/Belle~II and BESIII mean it is possible to explore the phase-space of $D$ meson decays in extreme detail.
For the analysis of multi-body decays constraining the kinematics of the decay is extremely important. 
The {\bf Dalitz plot}, named after the late Richard Dalitz, displays invariant quantities of a decay in a multi-dimensional 
plot~\cite{Dalitz:1953cp}. 
Originally a tool used to explore the $\tau - \theta$ problem where two nearly identical particles decayed into two and three $\pi$ respectively, this plot---more humbly named a “phase space plot” ---allowed physicists to investigate the spin structure of the {\it $\tau$ meson}~\footnote{At the time when the Dalitz plot technique was invented, the $K$ mesons were referred to as ``tau mesons''.}. 
For a three-body decay to three scalar/ pseudoscalar particles the degrees of freedom can be constrained as in Table~\ref{sec5.3:tab2}. 
A common parametrisation of the remaining degrees of freedom is $m_{12}^2$, $m_{23}^2$,
namely the two-body invariant masses, where the subscript indicates the combination of the final state four-momenta used, i.e.\ $m_{ij}^2 = (p_i + p_j)^2$. 
 The decay rate for a given three body decay is given by
\begin{equation}
d\Gamma = \frac{1}{(2\pi)^3} \frac{1}{32 m_h^3} |\mathcal{M}|^2 \, dm_{12}^2 \, dm_{23}^2 \,,
\end{equation}
\noindent
\begin{table}[t]
    \centering
    \begin{tabular}{l r}
        \toprule
        \textbf{Constraints} & \textbf{Degree of freedom} \\
        \midrule
        Three four-vectors & 12 \\
        Momentum conservation & -4 \\
        Three masses & -3 \\
        Three Euler angles & -3 \\
        \midrule
        \textbf{TOT} & \textbf{2} \\
        \bottomrule
    \end{tabular}
    \caption{Degrees of freedom for a three-body decay. These constraints are due to momentum conservation in the frame of the decaying particle, the masses of the final particles being known, and finally the rotational invariance of the system, i.e.\ conservation of angular momentum, the Euler angles.}
    \label{sec5.3:tab2}
\end{table}

\noindent
where $m_h$ is the mass of the decaying hadron, and the density of different parts of the Dalitz plot are described by the modulus squared of the invariant amplitude, $\mathcal{M}$. This tells us that if the invariant amplitude is constant the plot will have no features and would be consistent with a direct decay to the final state with no intermediate processes or final state interactions, a {\it non-resonant} decay ~\footnote{This makes the assumption that non-resonant contributions have a constant propagator. There can be small variation in $\mathcal{M}$ due to final state interactions, although these do not manifest as strongly as resonances in the Dalitz space. A full treatment can be found in~\cite{Aoude:2018zty}.}. However, nature is rarely so tidy, so we can consider intermediate states known as {\it resonances}. In general these resonant states are described through a series of two-body decays, this is the isobar formalism---first used to describe pion-nucleon, nucleon-nucleon and antinucleon-nucleon interactions. If resonant structures exist, the invariant amplitude $\mathcal{M}$ is no longer constant and structures will form in the Dalitz plot. In general two-body intermediate states are described by the relativistic Breit-Wigner function, i.e.
\begin{equation}
    \mathcal{A}_{BW} = \frac{1}{m_{12}^2 - s_{12} - i\Gamma m_{12}} \,,
    \label{eq:ampl-spin}
\end{equation}
which models the resonant structure in terms of an amplitude governed by its pole mass, $s_{12}$, decay width, $\Gamma $, and the mass of the combined particles, $m_{12}$. In the case where the decay width of the resonance tends to zero this generates a delta function passing through the Dalitz plot at the given pole mass, whereas a large decay width will generate wide structures. Similarly to how mass and the decay width of an intermediate resonance can be manifest in the Dalitz plot, so does spin. In the case of spinless final state particles, a resonant state of spin $S$ will have nodes and zeros in its propagator corresponding to Legendre polynomials that are used to describe this phenomenon. This parameterisation is known as ``helicity'' and in essence describes the correction to the propagator as,
\begin{equation}
\mathcal{A} \sim \mathcal{A}_{\rm BW} \, P_S(\cos\theta) \, ,
\end{equation}
where
\begin{equation}
P_0(\cos\theta) = 1\,, \quad 
P_1(\cos\theta) = \cos\theta\,, \quad 
P_2(\cos\theta) = \frac{1}{2}(3\cos^2\theta - 1)\,, \quad \ldots \,.
\end{equation}
with $\theta$ being the angle between the decaying particles momentum vector and the spin quantization axis —interested readers are directed to~\cite{Jacob:1959at} for a deeper discussion of the formalism. This parametrisation leads to the rich array of structures in Fig.~\ref{sec5.3:fig4_dalitz} where resonances such as the spin 2 $K^*(892)^-$ have 2 nodes in the kinematic space. As a final point, depending on the decay path and final state interactions, the resonances can be \textit{out of phase}, leading to constructive and deconstructive interference across the Dalitz plane. This is extremely important to methods sensitive to strong phase variation across the Dalitz plane.\\

\noindent
There are various approaches to using multi-body analyses in charm physics that build off the use of Dalitz space. One of these is the \textbf{amplitude analysis}, that seeks to model all the resonant and non-resonant contributions. These are summed in the form
\begin{equation}
\mathcal{A}(m) = \sum_{r} a_r e^{i\phi_r} \mathcal{A}_r(m) + a_{\mathrm{NR}} e^{i\phi_{\mathrm{NR}}} \mathcal{A}_{\mathrm{NR}}(m) ,
\label{sec5.3:eqAA}
\end{equation}
where each resonant amplitude consists of a magnitude, $a_r$, a phase, $\phi_r$ and propagator - usually a relativistic Breit-Wigner. The components with a subscript NR correspond to the non-resonant contribution, typically an $S$-wave. The K-matrix formalism~\cite{Lohse:1990ew} can be applied to the analysis of the $S$-wave, alternatively quasi-model independent methods exist which parametrise the $S$-wave component with generic complex functions and determine it from data, as done e.g.\ in~\cite{LHCb:2022pjv, LHCb:2022lja}.
\\

\begin{figure}
    \centering
\begin{overpic}[width=0.6\linewidth]{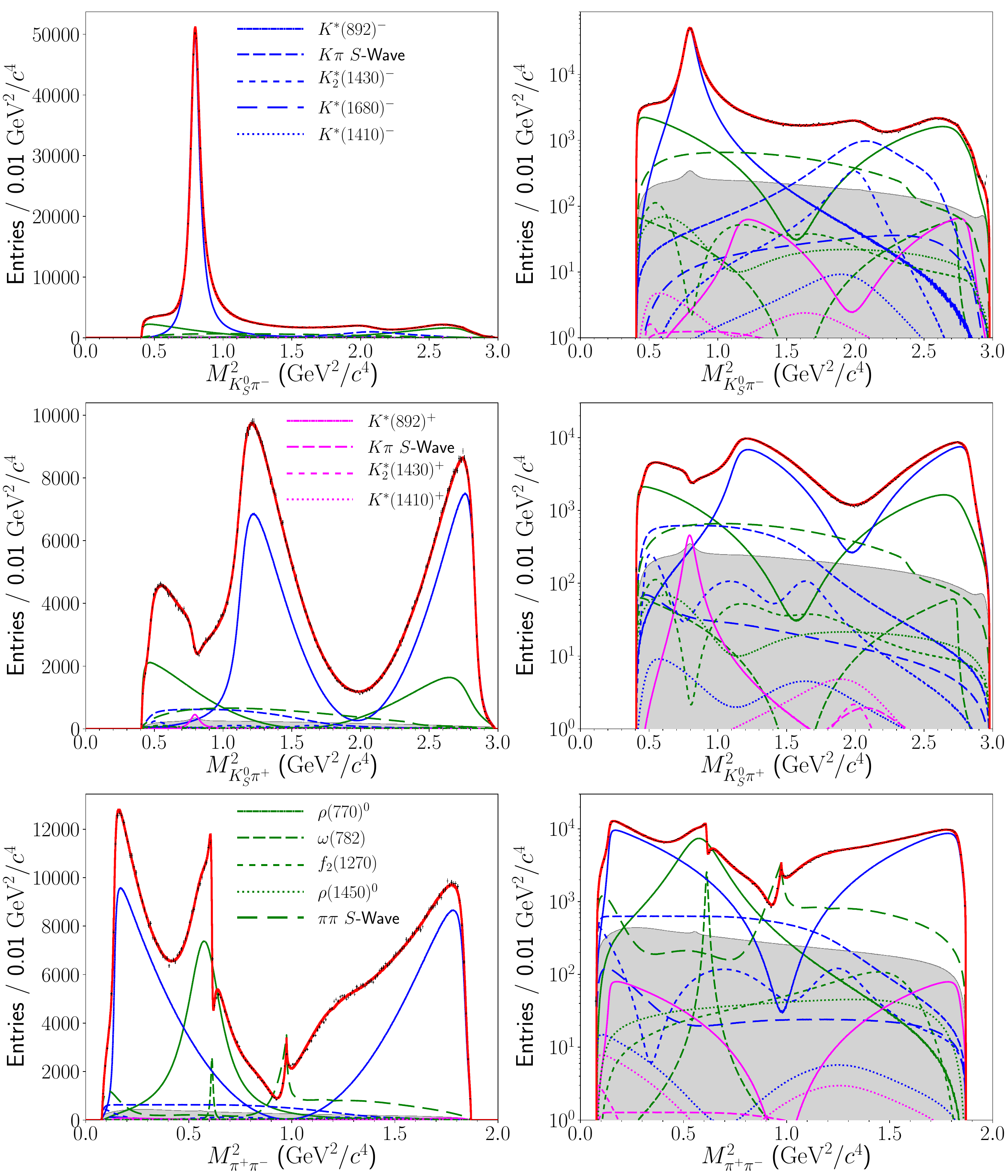} \put(250,300){\includegraphics[width=0.05\linewidth]{belle-logo.jpeg}}
  \end{overpic}    
    \caption{Three different Dalitz plot projections of the $D^0\rightarrow K_S\pi^+\pi^-$ decay from $D^{*+}\rightarrow D^0\pi^+$, formed with the two-body invariant masses $M^2_{K_S\pi-}$~(top), $M^2_{K_S\pi+}$ (middle), $M^2_{\pi+\pi^-}$ (bottom). The plots on the right use a log scale to highlight sensitivity to low fit fraction resonances. Note that here $M^2$ is equivalent to the $m^2$ defined in the text. Taken
    from~\cite{BaBar:2018cka}.}
    \label{sec5.3:fig4_dalitz_comp}
\end{figure}

\noindent
An example of modelling the phase-space in the charm sector is best shown through the analysis of $D^0\to K^0_S\pi^+\pi^-$ completed by Belle and BaBar~\cite{BaBar:2018cka}, a channel of particular importance to time-dependent CPV measurements. 
This modelling as seen in Fig.~\ref{sec5.3:fig4_dalitz_comp} takes into account 12 different resonant structures and a further 2 non resonant contributions. From an experimental perspective having datasets of order 1 million events as presented in Fig.~\ref{sec5.3:fig4_dalitz_comp} gives a unique sensitivity to resonances with sub \% level contributions to the fit. 
This level of sensitivity ensures that interferences are not mistaken as signs of CPV. This is only a \textit{single} example of amplitude analysis with a rich multi-body program presented across Belle, BaBar, BESIII, CLEO and LHCb. Other examples of model building through amplitude analysis include $D_s^+ \to \pi^+\pi^-\pi^+$ ~\cite{LHCb:2022pjv}, $D^+ \to \pi^+ \pi^- \pi^+$ ~\cite{LHCb:2022lja}, $\Lambda_c^+ \to pK^-\pi^+$ ~\cite{LHCb:2022sck}, $D^0 \to K^- K^+ \pi^0$ ~\cite{BaBar:2007soq}, $D^0 \to K^\mp\pi^\pm\pi^\mp\pi^\pm$ ~\cite{LHCb:2017swu}, $D^0 \to K^+ K^- \pi^+ \pi^-$ ~\cite{LHCb:2018mzv}, $D^+ \to K^-\pi^+\pi^+\pi^0$ ~\cite{BESIII:2024waz}, $D_s^+ \to \pi^+ \pi^- \pi^+ \pi^0$~\cite{BESIII:2024muy}, $D^0 \to \pi^+\pi^-\pi^{+(0)}\pi^{-(0)}$ ~\cite{BESIII:2023exz},and $D^+\to K^0_S \pi^+\eta$~\cite{BESIII:2023htx}. 
Searches for CPV using amplitude analysis will be discussed in Section~\ref{subsec:CPV}. Furthermore, multi-body decays can be analysed with statistical, model-independent methods to measure $D^0$-$\bar{D}^0$ mixing and CPV parameters. These will be discussed in more detail in Sections~\ref{subsec:mixing},~\ref{subsec:CPV}. 

%%%%%%%%%%%%%%%%%%%%%%%%%%%%%%%%%%%%%%%%%%%%%%%%%%%%%%%%%%%%%%%%%%%%%%%%%%%%%%%%%%%%%%%%%%%%%%%%%%%%%%%%%%%%%%%%%%%%%%%%%%%%%%%%
\section{Mixing and CPV in charm}
\subsection{Charm mixing and CPV in mixing}
\label{subsec:mixing}
Neutral meson-antimeson mixing, although established in the kaon system in the late 50's~\cite{Gell-Mann:1955ipe, Pais:1955sm, Muller:1960ph, Lande:1956pf, Mulled:1960cka, Okun:1957vy},
in the $B^0$-meson system in 1987~\cite{ARGUS:1987xtv}, and in the $B^0_{s}$-system in 2006~\cite{CDF:2006imy},
could only be observed later in the charm sector. In fact, as discussed in Section~\ref{sec:intro}, while the existence of $D^0$-$\bar D^0$ mixing could be inferred already in 2007 from a combination of measurements from different experiments, see~\cite{HFLAV:2012imy}, it was not until 2012 that the observation was made with a single measurement~\cite{LHCb:2012zll}.
The reason being that charm mixing, unlike the mixing of down-type neutral mesons, is severely suppressed by the GIM mechanism. \\

\noindent
This section presents the theory status of charm mixing and experimental searches for mixing and CPV in time-dependent measurements. Note, however, that searches for $\Delta Y$, also a time-dependent observable related to $D^0$-$\bar D^0$ mixing, are discussed in the context of CPV in charm hadron decays in Section~\ref{subsec:CPV}. For a comprehensive review of the experimental status of mixing and CPV see also~\cite{Pajero:2022vev}.
\\

\begin{figure}[b]
\centering
\includegraphics[scale = 0.3]{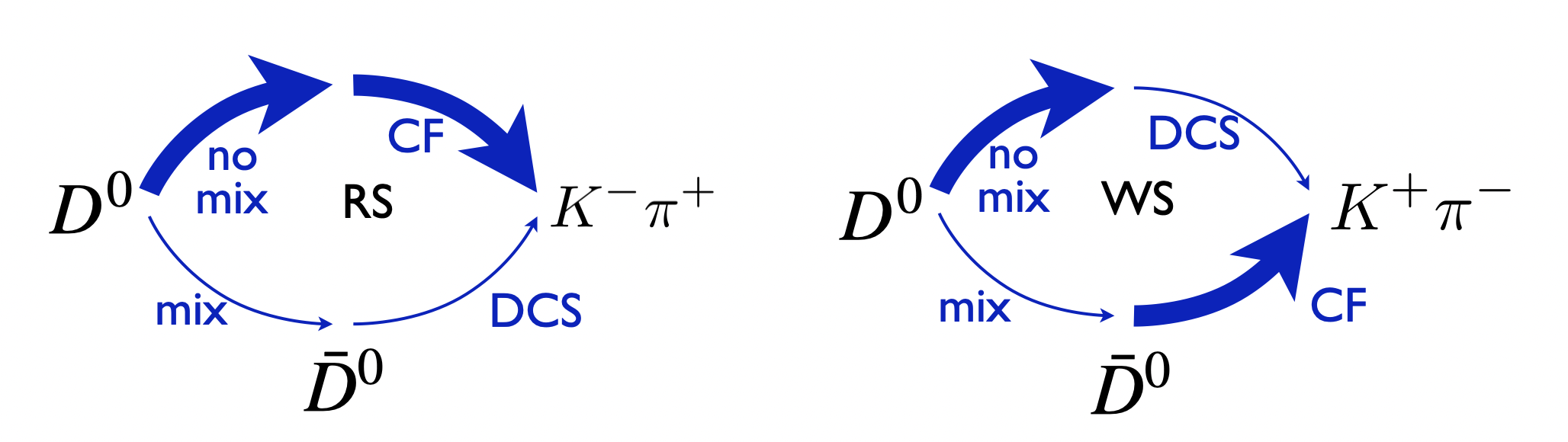}
  \caption{\label{fig:mixing_paths} 
  Schematic visualisation of the interference paths between CF and DCS amplitudes due to $D^0$-$\bar D^0$ mixing, in the RS decay $D^0 \to K^-\pi^+$~(left) and WS decay $D^0 \to K^+\pi^-$~(right).}
\end{figure}

\begin{figure}[t]
 \centering
 \begin{minipage}{0.6\textwidth}
 % \vspace*{1cm}
\begin{overpic}[width=1\linewidth]{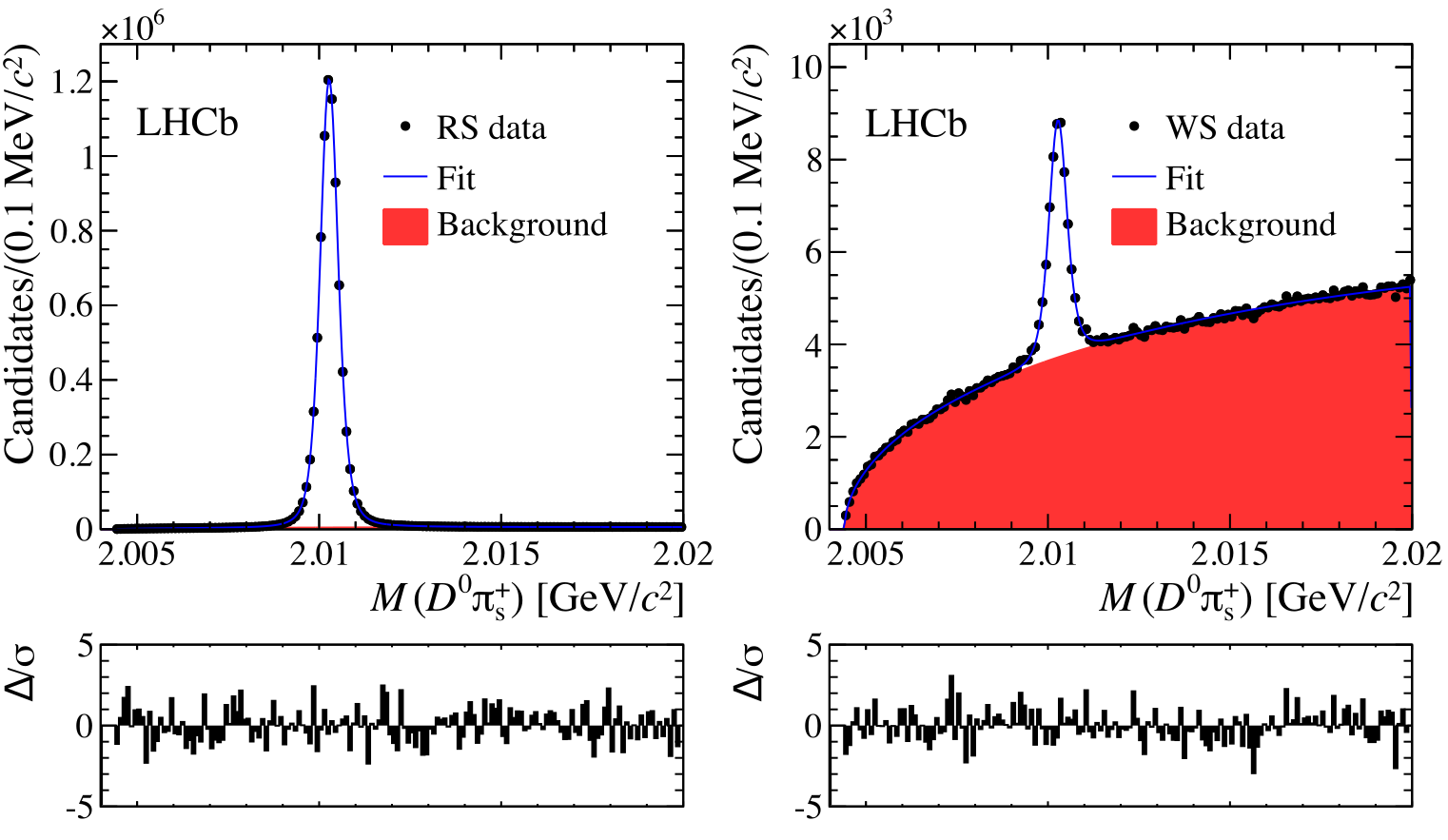} \put(25,130){\includegraphics[width=0.09\linewidth]{lhcb-logo.png}}
  \end{overpic}

  \caption{\label{fig:WSRS} Time-integrated mass distributions for the selected RS (left) and WS (right) candidates with fit projections overlaid. The bottom plots show the normalised residuals between the data points and the fits. Taken from~\cite{LHCb:2012zll}.}
  \end{minipage}
  \qquad
  \begin{minipage}{0.34\textwidth}
\begin{overpic}[width=1\linewidth]{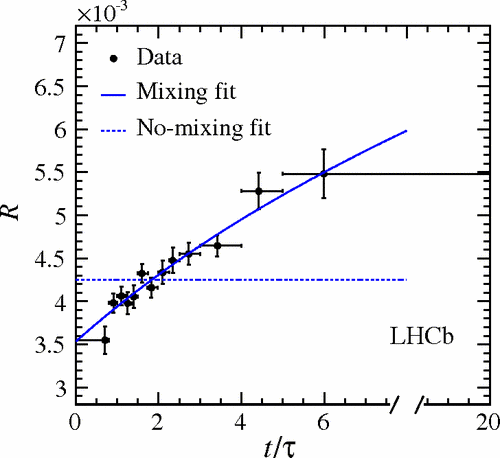} \put(122,30){\includegraphics[width=0.16\linewidth]{lhcb-logo.png}}
 \end{overpic}   \caption{\label{fig:mixing_fit} Decay-time evolution of the ratio, $R(t)$, of WS to RS yields (data points in black) with the projection of the mixing allowed (blue solid line) and no-mixing (blue dashed line) fits overlaid. Taken from~\cite{LHCb:2012zll}.
  }
  \end{minipage}
 \end{figure}

\noindent
{\bf Experimental searches with two-body decays:}
the first observation of $D^0$-$\bar D^0$ meson mixing with a single measurement was obtained by the LHCb collaboration~\cite{LHCb:2012zll} through the lens of wrong-sign (WS) -- right-sign (RS) decays. 
The decays in question were $D^0 \to K^\pm \pi^\mp$, 
see Fig.~\ref{sec5.3:fig1},
where the RS ($K^-\pi^+$) rate is dominated by the CF decay $D^0\rightarrow K^-\pi^+$ and by an extremely suppressed contribution from the DCS decay $\bar{D}^0\rightarrow K^-\pi^+$ following $D^0$-$\bar{D}^0$ mixing; whereas the WS ($K^+\pi^-$) rate receives contribution from the DCS decay $D^0\rightarrow K^+\pi^-$ and the CF decay $\bar{D}^0\rightarrow K^+\pi^-$ following $D^0$-$\bar{D}^0$ mixing. The interplay between CF and DCS modes in the RS-WS decays is sketched in Fig.~\ref{fig:mixing_paths}. Both charm mixing and the Cabibbo suppression contribute to the different yields of the RS-WS candidates, and these differ by roughly~$\mathcal{O}(10^3)$ as it can be seen in Fig.~\ref{fig:WSRS}. \\

\noindent
A measurement of the relative rate between the WS and RS yields, $N( \rm WS)$, $N ( \rm RS)$, in bins of decay time, does not directly give access the mixing parameters $x$ and $y$---(see the preceding chapter)), since, due to final-state interactions and hadronic effects, the observed rates also depend on the relative CP conserving strong phase $\delta$ between the DCS and CF amplitudes. As a result, one can only be sensitive to a linear combination of $x$ and $y$, which depends on the value of $\delta$. 
In fact, in the limit of small mixing parameters, the time-dependent ratio between the observed WS and RS yields, $R(t)$, can be approximated as
\begin{equation}
 R(t) = \frac{N( \rm WS)}{N ( \rm RS)}  \approx R_D + \sqrt{R_D} \, y' \frac{t}{\tau} + \frac{x'^2 + y'^2}{4} \left( \frac{t}{\tau} \right)^2 \,, 
\end{equation}
where $t/\tau$ is the decay time in units of the measured $D^0$-meson lifetime, $R_D$ is the ratio of WS to RS decays at $t= 0$, i.e.\ without contributions from mixing, and $x^\prime$, $y^\prime$ are a combination of the mixing parameters $x$ and $y$, defined as
\begin{equation}
x' = x \cos{\delta} + y \sin{\delta} \,, 
\qquad
y' = -x \sin{\delta} + y \cos{\delta}\,,
\label{eq:strong_phase_offset}
\end{equation}
with $x'^2 + y'^2 = x^2 + y^2$.
In the absence of $D^0$-$\bar D^0$ mixing i.e.\ for $x = 0$ and $y = 0$, it follows that $y' = x' = 0$ and that the ratio $R(t)$ is independent of time and consistent with the Cabbibo suppressed ratio of WS to RS decays, namely $R(t) = R_D$. Conversely, $D^0$-$\bar D^0$ mixing causes this ratio to vary with time, so that the existence of charm mixing can be inferred from a measurement of the decay-time evolution of $R(t)$.  
The time-dependent measurement of the ratio of WS to RS yields performed in~\cite{LHCb:2012zll}, led to the results shown in Fig.~\ref{fig:mixing_fit}. The fit shows that $R(t)$ is not constant in time and presents a clear and beautiful experimental observation of charm mixing, with the no-mixing hypothesis excluded at $9.1\sigma$.\\

\noindent
The WS-RS technique can also be used to search for CPV in the mixing process, by 
comparing $R(t)$ for $D^0$ and $\bar{D}^0$ mesons. The WS-RS ratios for $D^0$ and $\bar{D}^0$ mesons are referred to as $R^+(t)$ and $R^-(t)$, respectively, and defined as
\begin{align}
 R(t)^\pm \approx R_D^\pm + \sqrt{R_D^\pm} \, y'^\pm \frac{t}{\tau} + \frac{x'{^\pm}^2 + y'{^\pm}^2}{4} \left( \frac{t}{\tau} \right)^2 \,.
 \label{eq:wsrstotal}
\end{align}

\noindent
One can then search for signs of CPV by comparing the time-dependent ratio of WS and RS decays. Since $R_D^{+} (R^-_D)$ represents the ratio of $D^0(\bar{D}^0)$ WS-RS decays without mixing, we can fix this term to be insensitive to direct CPV, by substituting $R_D^{\pm} = \sqrt{R_D^+R_D^-}$ into Eq.~\eqref{eq:wsrstotal} which forces into the fit the hypothesis that the $D^0$ and $\bar{D}^0$ are CP symmetric. Similarly, we can test the no-CPV in mixing hypothesis by fixing $y'^- = y'^+ = y'$ and $x'^- = x'^+ = x'$. This decision enforces that the mixing parameters are identical for $D^0$ and $\bar{D}^0$. Recent analyses have been performed with hadronically tagged $D^0 \to K^\pm \pi^\mp$~\cite{LHCb:2024hyb} and semileptonically tagged $D^0 \to K^\pm \pi^\mp$~\cite{LHCb:2025kch} decays. In particular, the results of~\cite{LHCb:2024hyb} have lead to an improvement of the precision compared to the previous best measurement of approximately $60\%$.
However, these measurements still remain consistent with the no-CPV hypothesis---either direct or via mixing.
\\

\noindent
{\bf Experimental searches with multi-body decays:}
It is also possible to carry out WS-RS analysis in multi-body modes.  This was done by Belle, investigating mixing in the WS decays $D^0 \to K^+ \pi^- \pi^0$~\cite{Belle:2005xmv} and $D^0 \to K^+ \pi^- \pi^+ \pi^-$~\cite{Belle:2013nfo}, while a first observation of mixing in the latter mode was made by LHCb in 2016~\cite{LHCb:2016zmn}.
A measurement of $y'$ in $D^0 \to K^+ \pi^- \pi^+ \pi^-$ is of particular interest as it helps to constrain the CPV weak angle $\gamma$ through secondary decays from $B^\pm \to D K^\pm$ ~\cite{LHCb:2022nng}.\\

 \begin{figure}[b]
 \centering
 \includegraphics[scale = 0.13]{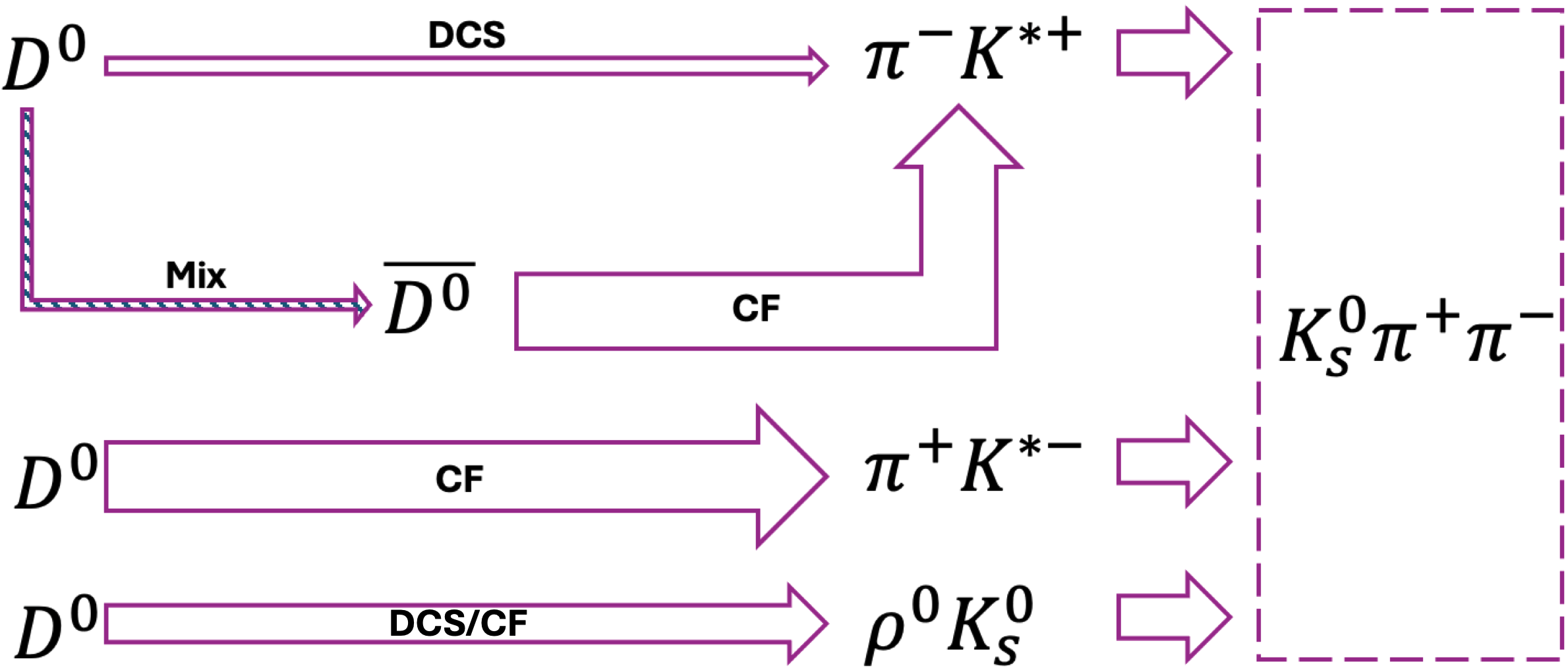}
\caption{Possible decay routes for the mode $D^0\to K^0_S\pi^+\pi^-$. If there is CP violation in $D^0 \to \bar{D}^0$ mixing, this could increase/decrease the ratio of CF to DCS decays.}
\label{fig:decay_roots}
\end{figure}

\noindent
In comparison to two-body analysis, multi-body analysis has the prospect of accessing CPV and mixing parameters directly, rather than as a linear combination. In contrast to Eq.~\eqref{eq:strong_phase_offset}, where parameters $x'$ and $y'$ are a single phase dependent linear combinations of $x$ and $y$, in multi-body final states these can be accessed ``independently" as different regions of the dalitz space will contain different strong phase rotations of the linear combination. The strong phase rotation in these analyses is obtained either from model-dependent analyses, which, may introduce significant additional biases; or via model-independent approaches, using quantum-correlated events at the charm threshold as explained later in the text. Although this makes multi-body decays an extremely powerful tool, these analyses represent a major experimental challenge, often involving a complicated multi-dimensional analysis and systematic study. \\ 

\noindent
The self-conjugate mode $D^0\to K^0_S\pi^+\pi^-$ is one channel where we can extract mixing parameters directly and search for evidence of CPV in charm. This is because both $D^0$ and $\bar{D}^0$ have access to the same final state $K^0_S\pi^+\pi^-$ through CF and DCS decay routes, see Fig.~\ref{fig:decay_roots}. Unlike the WS-RS methodology presented earlier, that can be constrained only up to a global phase, it is possible to access CP-eigenstate processes simultaneously, allowing for absolute strong phase differences across the multibody space to be constrained, cf.\ Fig.~\ref{sec6:fig6}. For this reason, the decay mode $D^0\to K^0_S\pi^+\pi^-$ is often referred to as a ``golden mode''.  
This multi-body final state can be analysed with model-dependent methods such as the amplitude analysis~\cite{Belle:2014Kspipi}, or with model-independent techniques, such as the {\it bin-flip analysis}~\cite{DiCanto:2018tsd} discussed below.\\

\begin{figure}
\centering
 \includegraphics[width=0.45\textwidth]{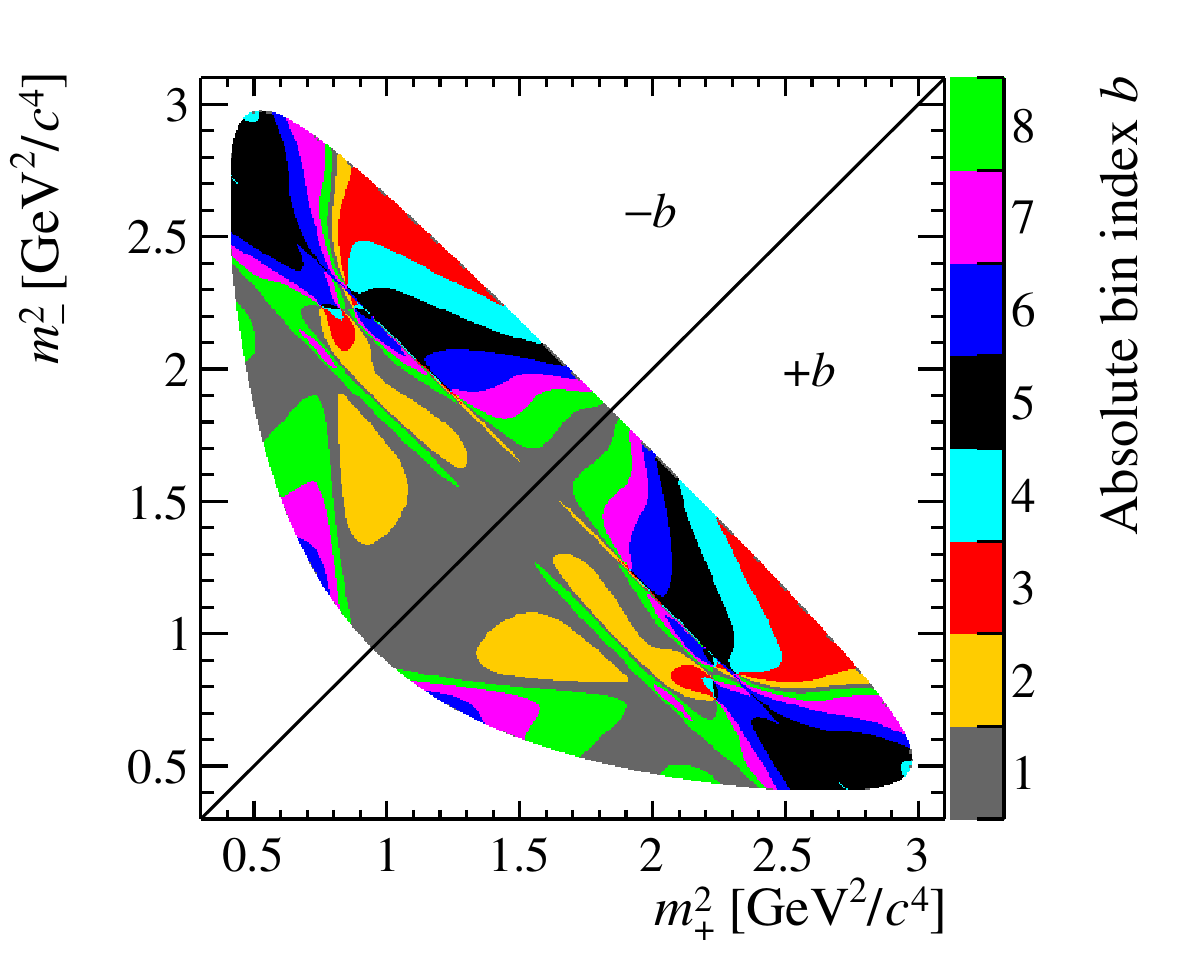}
\caption{``Binning'' of the $D^0\to K^0_S\pi^+\pi^-$ Dalitz plot. Colours indicate the absolute value of the bin index $b$ (each of the 8 colours on the vertical $z$ axis corresponds to one bin in the strong phase difference). $m_+^2$ refers to the invariant mass combination of $K_S \pi^+$ and $m_-^2$ refers to the invariant mass combination of $K_S \pi^-$. The binning is symmetric with respect to the $m_+^2=m_-^2$ axis to be sensitive to the CF and DCS routes described in the text. For each of these bins, the strong phase difference parameters $c_i$ and $s_i$ are measured with quantum correlated charm pairs, see Fig.~\ref{sec6:fig6}. Taken from~\cite{LHCb:2021ykz}.}
\label{fig:binflip}
\end{figure}

\noindent
To experimentally access the CPV structure of $D^0 \to K_S^0 \pi^+ \pi^-$, the mixing and CPV factors need to be properly parametrised. We use the mixing parameters $x$ and $y$;
moreover, CPV in mixing is encoded in a non-zero value of $|p/q|- 1$, while the relative phase $\phi = \arg (\lambda_f)$, 
is sensitive to CPV in the interference between mixing and decay. In CF/ DCS decays, the weak phase contribution to $\lambda_f$ from the decay amplitudes is negligible~\cite{Kagan:2020vri}, hence it is common to approximate $\phi \approx \arg(q/p)$ \footnote{The phase convention adopted here is from Kagan and Silvestrini~\cite{Kagan:2020vri}. $\phi$ is defined with respect to the dominant $\Delta U = 2$ dispersive and absorptive mixing amplitudes. It is referred to in the literature as $\phi_2$.} . A convenient parametrisation reads~\cite{DiCanto:2018tsd}:
\begin{equation}
    \label{eq:xcp}
    x_{\rm CP} = \frac{1}{2}\left[x\left(\abs{\frac{q}{p}} + \abs{\frac{p}{q}}\right) \cos\phi+ y\left(\abs{\frac{q}{p}} - \abs{\frac{p}{q}}\right) \sin\phi\right] \,,
\end{equation}
\begin{equation}
    \Delta x = \frac{1}{2}\left[x\left(\abs{\frac{q}{p}} - \abs{\frac{p}{q}}\right)\cos\phi + y\left(\abs{\frac{q}{p}} + \abs{\frac{p}{q}}\right) \sin\phi\right]\,,
\end{equation}
\begin{equation}
    y_{\rm CP} = \frac{1}{2}\left[y\left(\abs{\frac{q}{p}} + \abs{\frac{p}{q}}\right) \cos\phi- x\left(\abs{\frac{q}{p}} - \abs{\frac{p}{q}}\right) \sin\phi\right]\,,
\end{equation}
\begin{equation}
    \label{eq:deltay}
    \Delta y = \frac{1}{2}\left[y\left(\abs{\frac{q}{p}} - \abs{\frac{p}{q}}\right) \cos\phi- x\left(\abs{\frac{q}{p}} + \abs{\frac{p}{q}}\right) \sin\phi\right]\,.
\end{equation}
From the above equations, we see that in the absence of CPV in mixing and in the interference of mixing and decay, that is for $|q/p| = 1 $ and $\phi = 0$, 
we obtain $\Delta y = \Delta x = 0$, and $x_{CP} = x$, $y_{CP} = y$.
However, CPV in the interference i.e.\ $\phi \neq 0$, or CPV in mixing, $|q/p| \neq 1$, will manifest changes in these values.
With this parametrisation it is possible to use a time-dependent Dalitz plot analysis, coined as a \textbf{bin-flip} analysis~\cite{LHCb:2021ykz} to follow a similar procedure to the WS-RS analysis. Data is partitioned in order to constrain the strong phase at each point in the Dalitz plot, see Fig.~\ref{fig:binflip}. This binning choice is optimised to control the strong phase rotation present each dalitz region balancing between flatness and statistics. The self conjugate properties of $K_s\pi^+\pi^-$ also allow for access to the global phase rotation so the bin-flip parametrisation in Eqs.~\eqref{eq:xcp}-\eqref{eq:deltay} is sensitive to $x$ and $y$ rather than $x'$ and $y'$. 
This partitioning uses strong phase inputs determined from CLEO-c and BESIII~\cite{CLEO:2010iul, BESIII:2020khq}, as discussed further below. The ratio between the upper and lower bins of the Dalitz plot distinguishes the CF from the DCS decays as the charge of the intermediates flip depending on the route, see Fig.~\ref{fig:decay_roots}. This means that the $K^{*+}$ forms in the lower section of Fig.~\ref{fig:binflip} whereas the $K^{*-}$ forms in the upper section. As a function of decay time, the ratio of strong phase bins above and below the meridian gives sensitivity to the variables $x$ and $y$~\cite{LHCb:2021ykz} similar in structure to WS-RS analyses.
\\

\begin{figure}[t]
 \centering
 \begin{minipage}{0.45\textwidth}
 \begin{overpic}[width=1\linewidth]{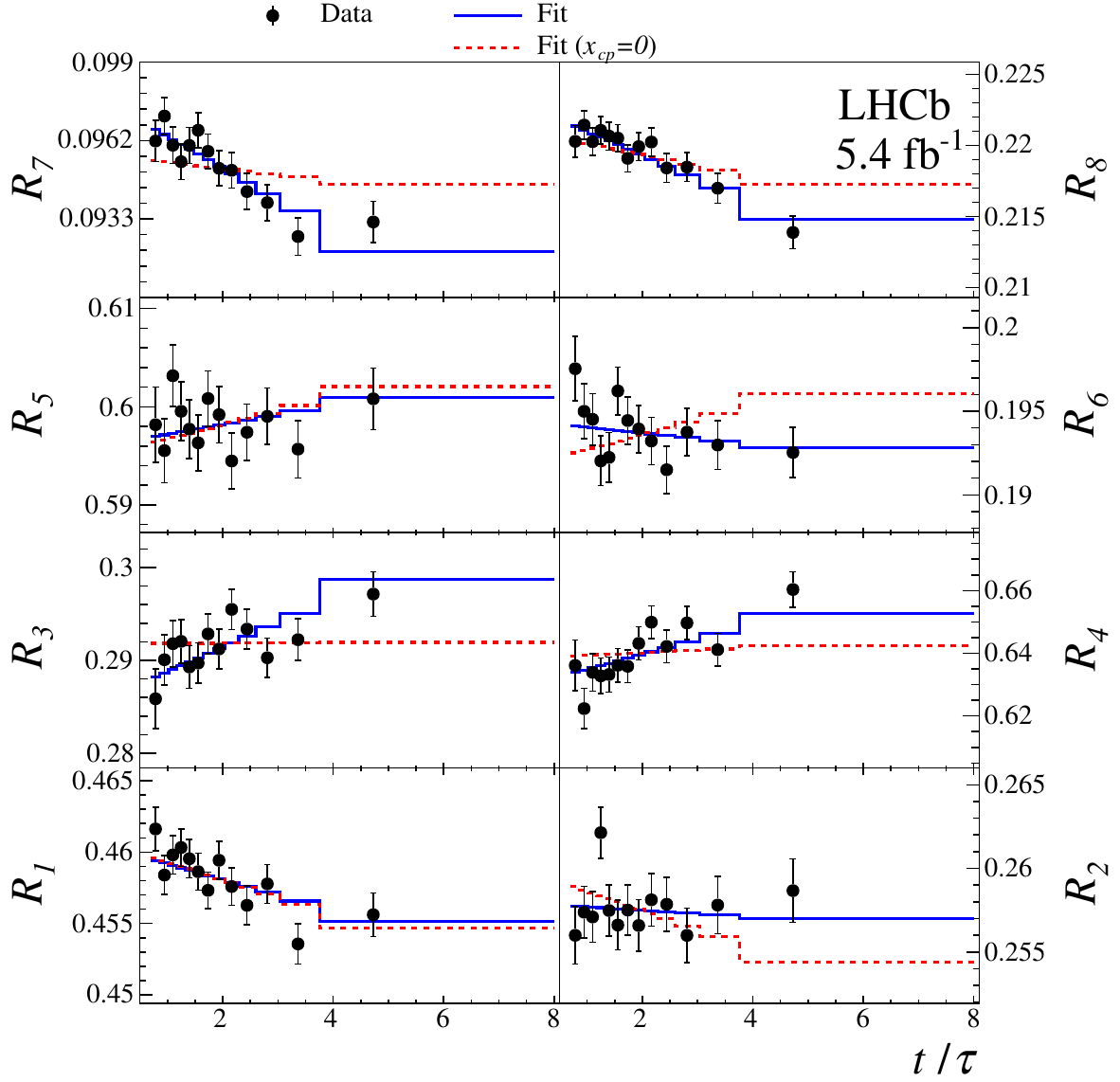} \put(159.5,180){\includegraphics[width=0.10\linewidth]{lhcb-logo.png}}
  \end{overpic} 
    \caption{CP-averaged yield ratios in bins of decay time. Ratios are calculated for bins, $R_i$ from above and below the meridian of Fig.~\ref{fig:binflip}. The red-dotted fit is showing how a non-zero $x_{CP}$ is required to model the mixing effects. Taken from~\cite{LHCb:2021ykz}.}
    \label{bin_flip_bins_ratio}
  \end{minipage}
  \qquad
  \begin{minipage}{0.45\textwidth}
\begin{overpic}[width=1\linewidth]{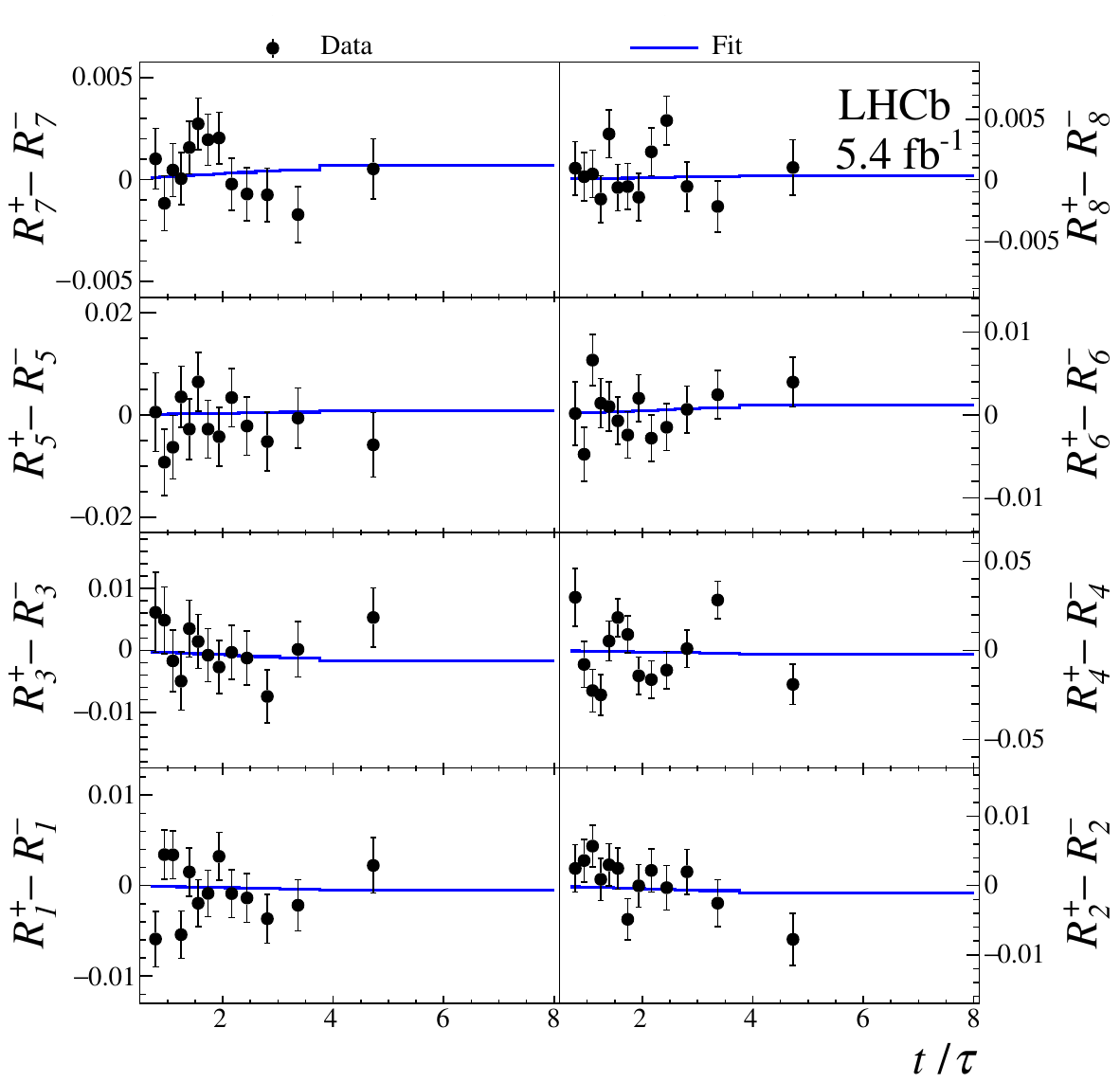} \put(159.5,180){\includegraphics[width=0.10\linewidth]{lhcb-logo.png}}
  \end{overpic} 
    \caption{ Differences of $D^0$ and $\bar{D}^0$ yield ratios in bins of decay time. Ratios are calculated for bins from above and below the meridian of Fig \ref{fig:binflip} for $D^0$, $R^+$, and $\bar{D}^0$, $R^-$. Gradients in this ratio would suggest CP violation. Taken
    from~\cite{LHCb:2021ykz}.}
    \label{bin_flip_bins_ratio_ratio}
  \end{minipage}
 \end{figure}

\noindent
The details of the formalism used in the study are more involved and the keen reader should refer to the original publication~\cite{LHCb:2021ykz}. However, Fig.~\ref{bin_flip_bins_ratio} should show the reader how mixing evolves over this decay, similar to the WS-RS analysis presented earlier in the section, where each plot is a different strong phase rotation. Similarly, Fig.~\ref{bin_flip_bins_ratio_ratio} shows how bins above and below the meridian can be subtracted to search for CPV. This multi-body result presented by LHCb in 2021~\cite{LHCb:2021ykz} was the first non-zero measurement of the mixing parameter $x$ in a single channel. 
The results of the fit, testing the CPV-allowed hypothesis  i.e.\ $|q/p| \neq 1 $ and $\phi \neq 0$, are as follows:
\begin{equation}
\begin{matrix}
x_{\rm CP} = \left(0.397\pm 0.0544\right) \cdot 10^{-2}\,, &
\Delta x = \left(-0.027\pm 0.018\right) \cdot 10^{-2}\,, & 
\\[2mm]
y_{\rm CP} = \left(0.459\pm 0.1471\right) \cdot 10^{-2}\,, &
\Delta y = \left(0.020\pm 0.0383\right) \cdot 10^{-2}\,,  \\
\end{matrix}
\quad
\mbox{(CPV allowed)}\,,
\end{equation}
\vspace*{1mm}
which, using Eqs.~\eqref{eq:xcp} - \eqref{eq:deltay}, lead to
\begin{align}
x = (0.398^{+0.056}_{-0.054}) \cdot 10^{-2}\,,
\quad 
y = (0.46^{+0.15}_{-0.14}) \cdot 10^{-2}\,, \quad
|q/p| = 0.996\pm0.052\,, \quad \phi = \left(-0.056^{+0.047}_{-0.051}\right)\,.
\end{align}
Note that for ease of interpretation, the statistical and systematic errors from the bin-flip analyses have been summed in quadrature~\cite{LHCb:2021ykz}.\\

\noindent
Not only do the values obtained using the bin-flip method dominate the world averages of the parameter $x$, they have helped to define the most precise measure of CPV through mixing and interference in the charm sector. However, since CPV remains still unconfirmed,
the world averages taken from HFLAV~\cite{HeavyFlavorAveragingGroupHFLAV:2024ctg} testing the no-CPV, i.e.\ $|q/p| = 1 $ and $\phi = 0$, and the CPV-allowed, i.e.\ $|q/p| \neq 1 $ and $\phi \neq 0$, hypotheses, respectively, are:
\begin{align}
x = \left(0.4344^{+0.126}_{-0.139} \right)\cdot 10^{-2}, \quad y = \left( 0.646\pm 0.024\right) \cdot 10^{-2} \,, \qquad \mbox{(no CPV)}\,,
\end{align}
\begin{align}
x = \left(0.407\pm{0.044}\right) \cdot 10^{-2}, \quad
y = \left(0.645^{+0.024}_{-0.023}\right) \cdot 10^{-2}, \quad
|q/p| = 0.994^{+0.016}_{-0.015}, \quad \phi = \left( -0.0453\pm 0.0209 \right)\,, \quad \mbox{(CPV allowed)}\,.
\end{align}

\begin{figure}[t]
 \centering
\begin{minipage}{0.43\textwidth}
\vspace*{1cm}
\includegraphics[width=0.95\textwidth]{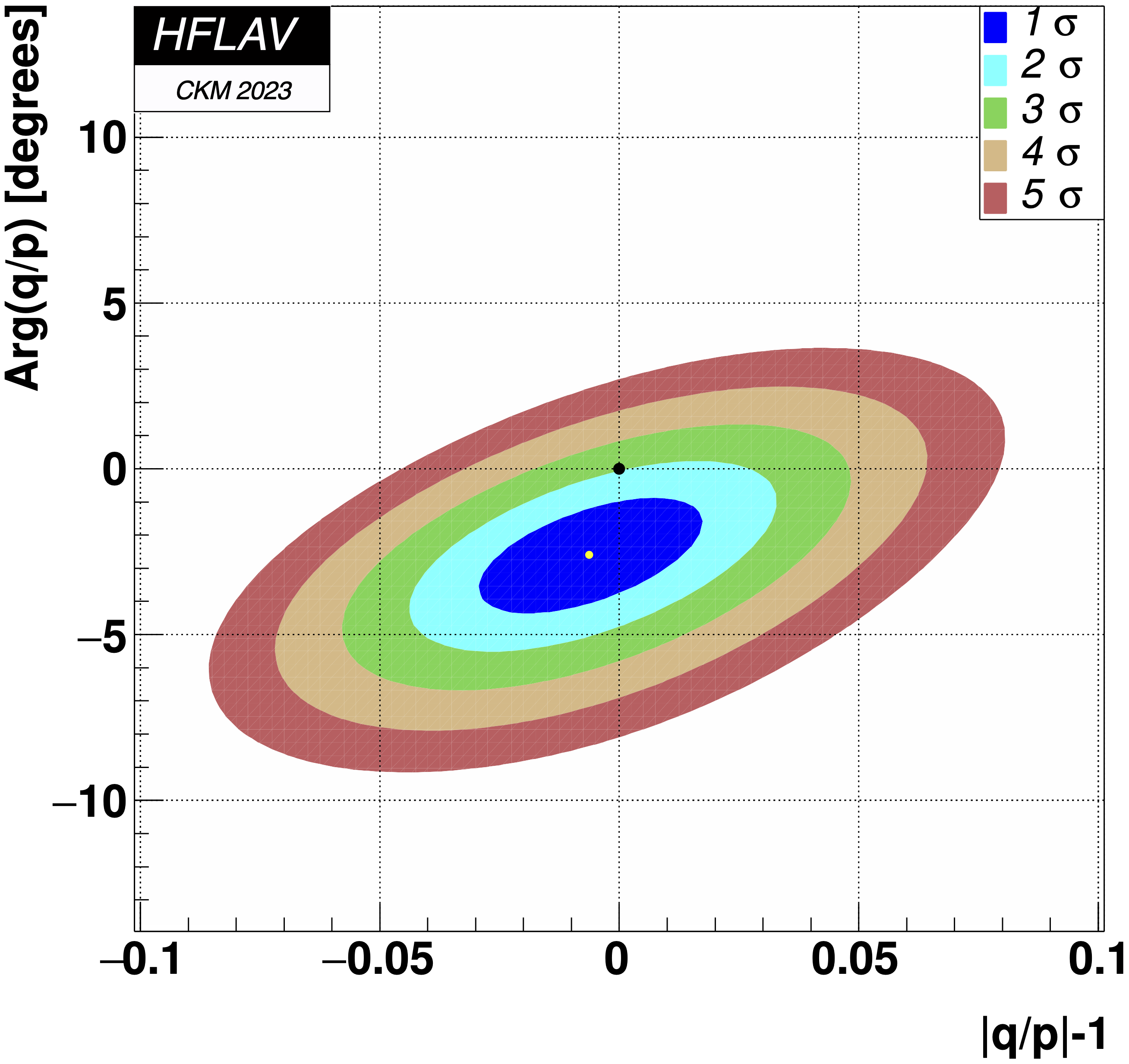}
  \caption{\label{fig:noCPVHFLAV} HFLAV fit of the CPV parameters for charm mixing $|q/p|-1$ and $\phi = \arg(q/p)$. The no-CPV case (0,0) is excluded at $2.1\, \sigma$. Taken from~\cite{HeavyFlavorAveragingGroupHFLAV:2024ctg}.}
  \end{minipage}
  \qquad
  \begin{minipage}{0.45\textwidth}
  \vspace*{1cm}
\includegraphics[width=0.9\textwidth]{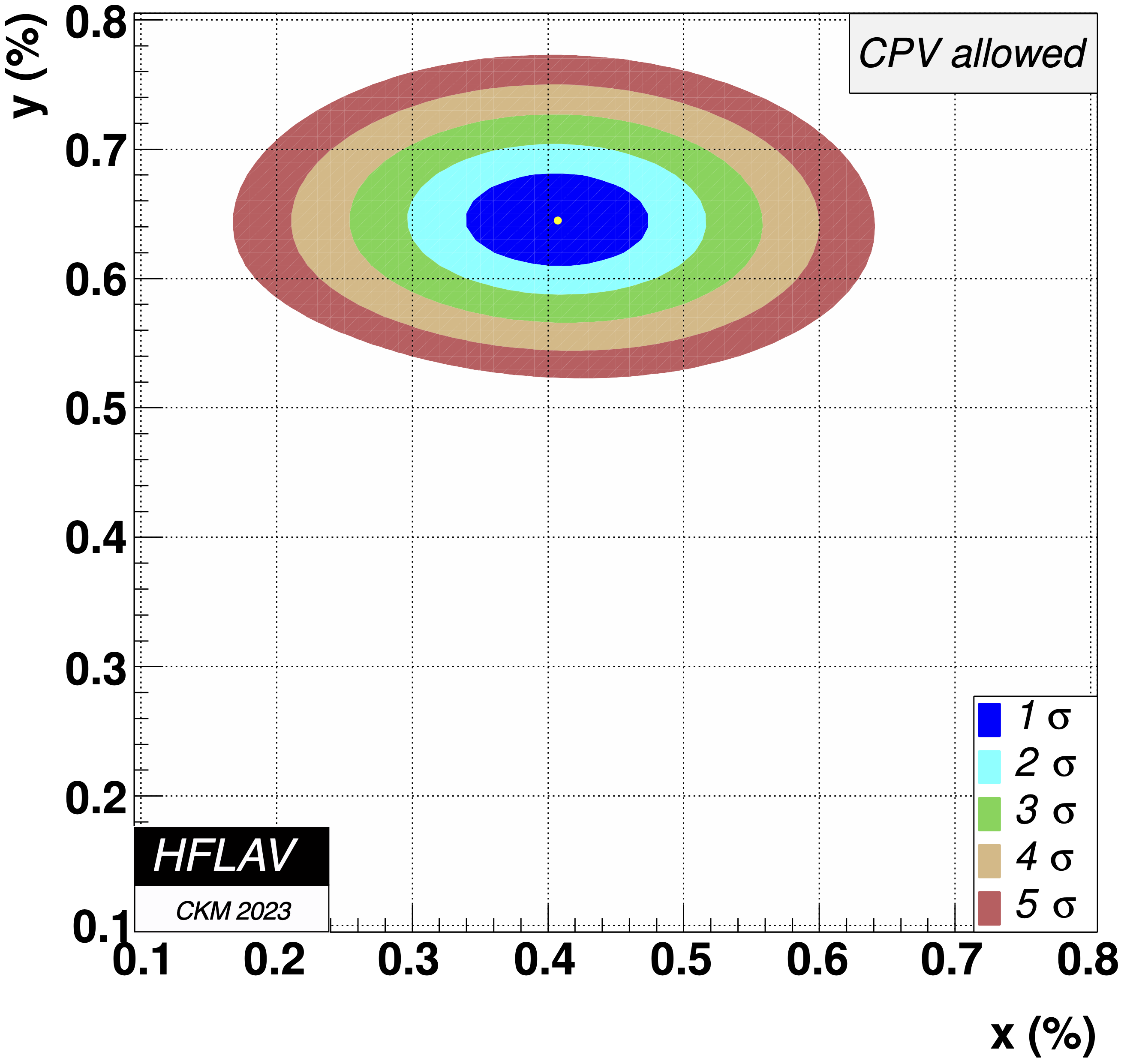}
\caption{HFLAV fit of the mixing parameters $x$ and $y$ in the case of CPV allowed. The no-mixing case (0,0) is excluded at more than $10 \, \sigma$. Taken from~\cite{HeavyFlavorAveragingGroupHFLAV:2024ctg}.}
\label{fig:CPVHFLAV}
  \end{minipage}
 \end{figure}

\begin{figure}
    \centering
   \begin{overpic}[width=0.9\linewidth]{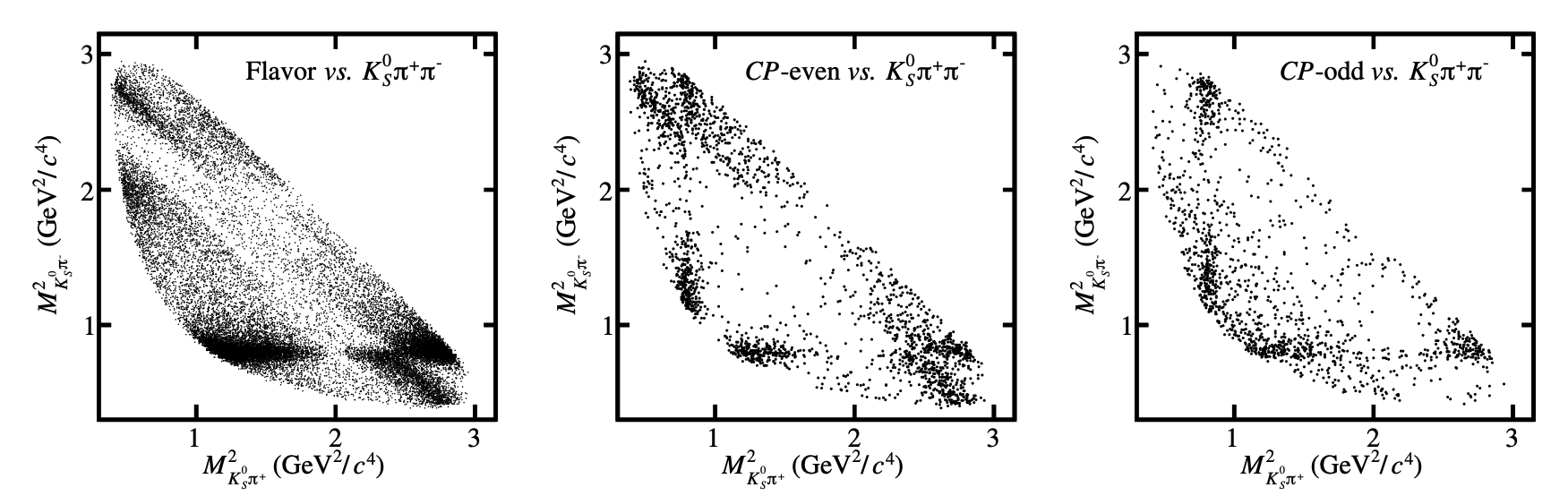} \put(370,95){\includegraphics[width=0.07\linewidth]{bes3-logo.png}}
  \end{overpic} 

    \caption{Comparison of different tags for the $D^0\rightarrow K_S\pi^+\pi^-$ decay at BESIII: flavour tags (left), CP-even tags (middle), CP-odd tags (right). 
    Taken from~\cite{BESIII:2020khq}.}
    \label{sec6:fig6}
\end{figure}

\noindent
The status of charm mixing and CPV, for the CPV-allowed case, is summarised in the HFLAV fits to current experimental measurements shown in Fig.~\ref{fig:noCPVHFLAV}; see also~\cite{UTfit:2022hsi, LHCb:2024yxi} for similar analyses, which also include constraints from current measurements of the CKM angle $\gamma$.  Fig.~\ref{fig:noCPVHFLAV} shows the allowed regions for the CPV parameters $|q/p|-1$ and $\phi \approx \arg(q/p)$, while Fig.~\ref{fig:CPVHFLAV} shows the constraints on the mixing parameters $x$ and $y$. Currently, the no-CPV hypothesis is excluded at $2.1\sigma$ and the no-mixing hypothesis is excluded at more than $10\sigma$. \\

\noindent
Although an exciting step, the measurement reporting the observation of $x$ remains statistically limited, and future analyses will be vital to constrain this further. Unfortunately, $y$ is extremely sensitive to the strong phase partitioning in the bin-flip methodology. In general the mixing-terms for CP-odd yields partially cancel between upper and lower dalitz bins. This increases the uncertainty on $y$ and $\Delta y$ by around a factor of 2. Complementary model-dependent analysis such as time-dependent amplitude analysis~\cite{Belle:2014Kspipi} can offer an opportunity to resolve this. As discussed in Section~\ref{subsec:nonlep} the amplitude analysis models the strong phase, and more importantly, the interference between different states removing the ambiguity caused by this partitioning and cancellation. By defining a model where the evolution of the strong phase is captured as a function of decay time it is possible to extract a less statistically limited determination of $y$. This methodology was applied successfully by Belle in 2014~\cite{Belle:2014Kspipi} where the formalism is outlined in more detail; currently, this measurement remains statistically limited.
\\

\begin{center}
    $\sim \circ \sim \circ$\, {\bf Input from quantum correlated {\boldmath{$D^0$}} mesons for charm mixing
    }\, $\circ \sim \circ \sim$
\end{center}

\noindent
Some hadronic final states can be reached via the decay of both the $D^0$ and the $\bar{D}^0$ as parent particle. Examples of such final states are $K^+K^-$, $ \pi^+\pi^-$, $ K_S^0\pi^+\pi^-$, etc. In general, for these modes, there may be a difference between the magnitude and/or the phase of the decay amplitude for the $D^0$ and the $\bar{D}^0$. For multi-body decays like $ K_S^0\pi^+\pi^-$, the CP-conserving strong phase difference will vary over the phase-space of the decay. 
Knowledge of the strong phase differences in the phase space of $ K_S^0\pi^+\pi^-$ is invaluable input to both direct CPV and mixing measurements done at experiments like 
LHCb and Belle/ Belle II where this information is not directly accessible. 
The idea of using quantum correlated input for $K_S\pi^+\pi^-$ decays was first introduced for measurements of the CKM angle $\gamma$ in~\cite{Giri:2003ty}, and later optimised in \cite{Bondar:2005ki, Bondar:2008hh}.\\

\noindent
For charm mesons produced at the $\psi(3770)$ resonance just above the threshold for production, the kinematics of the decays is well defined, and the processes are relatively background free. 
With this pure source of neutral charm-meson pairs (which are produced either as flavour eigenstates, $D^0 \bar{D}^0$, or CP eigenstates, $D_{\mathrm{CP+}} D_{\mathrm{CP-}}$), we have unique access to information about the strong-phase differences, coherence factors and the CP content of the decays, as briefly described below~\footnote{A novel technique to extract the strong phase differences in $D$-meson decays using quantum correlated $D^0 \bar D^0$ pairs produced at $e^+ e^-$ collisions at energies above the $\psi(3770)$ threshold has been recently proposed in~\cite{BESIII:2025xed, BESIII:2025pod}.}. \\

\noindent
If we have two neutral charm mesons decaying into two-body final states, such as the case of $K^\mp\pi^\pm$ final states, the interference between the amplitudes can be written as:
\begin{equation}
    |\mathcal{A}_1+\mathcal{A}_2|^2= |\mathcal{A}_1|^2+|\mathcal{A}_2|^2+2 |\mathcal{A}_1||\mathcal{A}_2|\cos\delta\,,
\end{equation}
\noindent
where $|\mathcal{A}_{1,2}|$ are the magnitudes of the amplitudes, and $\delta$ is their relative strong phase. 
For multibody decays, such as the case of $K^\mp\pi^\pm\pi^\mp\pi^\pm$ final states, the equation is similar
\begin{equation}
  \int dx  |\mathcal{A}_1+\mathcal{A}_2|^2= {A}_1^2+{A}_2^2+2 R{A}_1{A}_2\cos\delta \,,
\end{equation}
\noindent
where $x$ describes the multi-body phase space, $A_{1,2}$ are the phase-space averaged magnitudes of the amplitudes i.e.\ $A_i^2=\int dx |\mathcal{A}_i|^2$, and $0\leq R \leq 1$ is a coherence factor first introduced in~\cite{Atwood:2003mj}. For a generic final state $f$, one can define the coherence factor $R_f$ and the strong phase difference $\delta_D^f$ as
\begin{equation}
    R_f e^{-i\delta_D^{f}} = \frac{\int {\cal A}^*_{f}(x) {\cal A}_{\bar{f}}(x) dx}{{ A}_{f} { A}_{\bar{f}}}\,,
    \label{sec6.1:eq:coherence}
\end{equation}
where the $\mathcal{A}_f(x)$ are the complex amplitudes at one point of the phase space, $x$. The parameter $R_f$ is usually $<$ 1 for multibody decays. 
The parameters $R_f$ and $\delta_f$ are polar coordinates which can also be presented in a Cartesian basis $(c,s) = (R_f\cos\delta_f, R_f\sin\delta_f)$. If measured in bins $i$ of the phase space, these would be labelled $(c_i, s_i)$.\\
Then, the decay width of a double-tag final state $\psi(3770)\rightarrow DD\rightarrow [F][G]$, where $[F]$ and $[G]$ are generic inclusive final states, is~\cite{Atwood:2003mj}
\begin{equation}
    \Gamma (FG) = \Gamma_0\{A_F^2\ \bar A_G^2 + \bar A_F^2\  A_G^2 - 2 R_F^2R_G^2A_F\bar A_FA_G\bar A_G \cos[\delta(F)-\delta(G)]\}\,,
\end{equation}
where $\Gamma_0=\Gamma (\psi(3770)\rightarrow D^0\bar D^0)$, and $[\delta(F)-\delta(G)]$ is the strong phase difference between the amplitudes of the two final states. \\

\noindent
Strong phase differences can be measured at the CLEO-c and BESIII experiments where two charm mesons are produced coherently using a double tag technique, where an extended list of the possible tag modes includes:\\

\begin{itemize}
\item {\bf Flavour tags:} these can be flavour specific decays, namely processes where the final state can be reached only through a decay of the $D^0$ or $\bar{D}^0$ meson such as the semileptonic decays $K^- e^+\nu_e$. Alternatively, final states such as $K^-\pi^+$, $K^-\pi^+\pi^0$, $K^-\pi^+\pi^-\pi^+$, etc., can be used. 
Such final states are reached mostly via the CF amplitude and receive a small contribution from the DCS amplitude due to $D^0$ mixing.
Because the latter contribution is strongly suppressed,
these hadronic modes can be used as flavour tags. 
\item {\bf CP tags:} these are decays in which the final states are eigenstates of the CP operator. They include {\it CP-odd} states such as $K^+K^-$, $\pi^+\pi^-$, $K_S^0\pi^0\pi^0$, $K_L^0\pi^0$, $K_L^0\omega$, etc., and {\it CP-even} states such as $K_S^0\pi^0$, $K_S^0\eta$, $K_S^0\eta'$, $K_S^0\omega$, $K_L\pi^0\pi^0$, etc.
\item {\bf Mixed CP content:} these are $D$-meson decays in which the final state is not a CP eigenstate and has
mixed CP content. They can be {\it quasi-CP even} states, such as $\pi^-\pi^+\pi^0$~\footnote{This decay is referred to as quasi-CP even because its CP even fraction $F_+$, defined as the rate of CP even signal $D$-meson decays compared to all CP even and odd signal decays, is found to be close to 1: $F_+= 0.9406\pm0.0036\pm0.0021$~\cite{BESIII:2024nnf}.}, and {\it mixed CP} states, such as $K^+K^-\pi^0$, $\pi^+\pi^-\pi^+\pi^-$, $K_S^0\pi^+\pi^-$, $K_L^0\pi^+\pi^-$, etc. The latter two are particularly important for the determination of differences of strong phases in $D^0\rightarrow K_S\pi^+\pi^-$ decays, see e.g.~\cite{BESIII:2020khq, BESIII:2025nsp}.  \\
\end{itemize}

\noindent
In the measurement of the strong-phase differences in the mode $D^0\rightarrow K_S^0\pi^+\pi^-$~\footnote{Note that a similar measurement of the strong-phase differences can be done with the $D^0\rightarrow K_L^0\pi^+\pi^-$ decays.}, the effect of the different tags for $D^0\rightarrow K_S\pi^+\pi^-$ decays can be clearly seen in Fig.~\ref{sec6:fig6}, where the BESIII collaboration published three different Dalitz plots tagged with flavour, CP-odd and CP-even tags~\cite{BESIII:2020khq}. 
To understand how this works in practice, it is easiest to describe 
the effect of the quantum correlation with CP-odd and CP-even states. 
If one side of a quantum-correlated charm pair decays to a CP-odd state, then the other side must be a CP-even state, as the creation of two CP-odd or two CP-even states is not allowed due to the quantum numbers of the parent particle, which are conserved in the strong decay. 
Thus, using the CP-odd state as a tag, only the CP-even contributions will be visible in the Dalitz plot distribution of the signal. 
More specifically, let us take as an example the intermediate contribution $K_S \rho$, which decays to $K_S\pi^+\pi^-$ via $\rho\rightarrow \pi^+\pi^-$ (its location on the Dalitz plot can be seen along the diagonal in Fig.~\ref{sec5.3:fig4_dalitz} (left)). 
As this state is CP-odd, its contribution to the final state $K_S\pi^+\pi^-$, if the latter is used as a signal, will be visible in the Dalitz plot corresponding to a CP-even tag. Conversely, if CP-odd tags are used, the CP-odd contributions in the signal Dalitz plot will be suppressed. These changes in the resonant structure of the Dalitz plot for CP-even and CP-odd tags can be seen comparing the middle and right plots in Fig.~\ref{sec6:fig6}.\\

\begin{figure}[b]
    \centering
\begin{overpic}[width=0.4\linewidth]{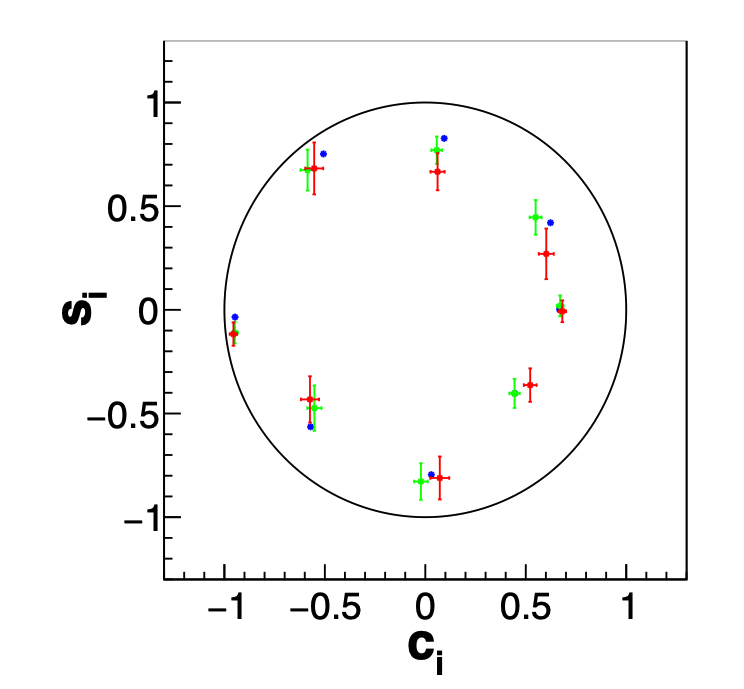} \put(140,145){\includegraphics[width=0.07\linewidth]{bes3-logo.png}}
  \end{overpic}
    \caption{The strong-phase difference parameters $c_i$, $s_i$ measured by the BESIII collaboration for the equal binning scheme shown in Fig.~\ref{fig:binflip}. The data points in red correspond to measurements without any constraints, the points in green correspond to measurements with constraints from model predicted differences between $K_S\pi^+\pi^-$ and $K_L\pi^+\pi^-$ decays, and the blue points correspond to predictions from amplitude models; data and predictions are in good agreement. Taken from~\cite{BESIII:2025nsp}.}
    \label{sec6:fig7}
\end{figure}

\noindent
By
reconstructing both $D$ mesons in the event and from the analysis of the corresponding Dalitz plots, at BESIII it is thus possible to obtain direct sensitivity to the relative strong phases between the $D^0$ and $\bar D^0$ meson decays. The strong-phase differences are commonly split in 8 equal intervals following a model proposed by the BaBar collaboration~\cite{BaBar:2008inr}, and measured in each interval or {\bf bin}. 
These bins can be visualised over the Dalitz plot of the decay. Measurements are made in three different binning schemes: equidistant, where the data in the whole range of the relative strong phase, between $-\pi$ to $\pi$, is divided into discrete bins of equal width; optimal, i.e.\ with maximal sensitivity for measurements of the CKM angle $\gamma$ in a low background environment e.g.\ at the Belle/Belle II experiments; and modified optimal, for a high background environment where the levels of the expected backgrounds at the LHCb experiment are taken into account.  The binning scheme relevant for charm mixing is the equidistant binning scheme, shown in Fig~\ref{fig:binflip}. The measurements of the strong phase differences were first made by the CLEO collaboration~\cite{CLEO:2010iul}, and more recently by the BESIII collaboration~\cite{BESIII:2020khq, BESIII:2025nsp}. 
The strong-phase difference parameters $c_i$ and $s_i$ shown in Fig.~\ref{sec6:fig7}, describe
the average value of the cosine and sine of the strong-phase differences $\cos\Delta\delta_D$ and $\sin\Delta\delta_D$, where $\Delta \delta_D$ indicates the relative strong phase of the $D^0$ and $\bar{D}^0$ amplitudes, over the phase-space bin $i$, respectively. These parameters are defined as~\cite{BESIII:2025nsp}
\begin{equation}
    c_i=\frac{1}{\sqrt{F_iF_{-i}}}\int_i |f_{D^0}(m_+^2,m_-^2)||f_{D^0}(m_-^2,m_+^2)|\cos \left[\Delta\delta_D(m_+^2,m_-^2)\right]dm^2_+dm^2_-\,,
\end{equation}
\begin{equation}
          s_i=\frac{1}{\sqrt{F_iF_{-i}}}\int_i |f_{D^0}(m_+^2,m_-^2)||f_{D^0}(m_-^2,m_+^2)|\sin\left[ \Delta\delta_D(m_+^2,m_-^2)\right]dm^2_+dm^2_- \,,  
\end{equation}
 where $m^2_{\pm}$ is the invariant mass of the $K_S \pi^{\pm}$ system, $f_{D^0}(m_+^2,m_-^2)$ is the amplitude of the $D^0$ decay and $f_{D^0}(m_-^2,m_+^2)$ of the $\bar{D}^0$ decay to $K_S\pi^+\pi^-$, respectively, in correspondence of the points $(m_+^2,m_-^2)$ and $(m_-^2,m_+^2)$ in the Dalitz plot,
$F_i$ is the fraction of events in the $i^{\mathrm{th}}$ bin of the flavour specific decay $D^0\rightarrow K_S\pi^+\pi^-$, where each bin is defined in Fig.~\ref{fig:binflip}. Details on the formalism, the importance of the different tags and the experimental results can be found in ~\cite{CLEO:2010iul, BESIII:2020khq, BESIII:2025nsp}.\\

\noindent
For measurements of the CKM angle $\gamma$ or charm mixing parameters with multibody CF or DSC decays such as $D^0\rightarrow K^-\pi^+\pi^-\pi^+$ and $K\pi\pi^0$, 
the coherence factor, the strong phase difference and the ratio of the suppressed over the leading amplitudes, $r_D^f = {\cal A}_{\bar f}/ {\cal A}_f$, constitute important inputs~\cite{BESIII:2021eud}. Another interesting observable measured at charm at threshold for CP eigenstates is the CP content of decays (mostly relevant for measurements of the CKM angle $\gamma$), defined as $F_+ = N_+/(N_++N_-)$ where $N_+$ and $N_-$ are the decay rates of CP-even and CP-odd charm mesons for the signal mode, and it has been measured for $D^0\rightarrow\pi\pi\pi^0$ and $KK\pi^0$~\cite{BESIII:2024nnf}, $KK\pi\pi$~\cite{BESIII:2022ebh}, $\pi\pi\pi\pi$~\cite{BESIII:2022wqs}. For a more complete overview of the input from quantum correlated charm see e.g.\ \cite{Yabsley:2008sb, Wilkinson:2021tby, Briere:2017pna, Briere:2014ixa, Malde:2016otf}.\\

\noindent
{\bf Theoretical status of charm mixing:}
In the SM the phenomenon of $D^0$-meson mixing is described by box Feynman diagrams (shown in the preceding chapter); the off-shell contributions to these box diagrams, corresponding to internal down, strange and bottom quarks, describe $M_{12}$, while 
$\Gamma_{12}$ is given by the on-shell contributions to the box diagrams with internal down and strange quarks. Since the masses of the down-type quarks are much smaller than $m_W$ - differently from mixing of down-type mesons where $m_t > m_W$ - in the charm sector, both $M_{12}$ and $\Gamma_{12}$ are dominated by long-distance effects, making their theory predictions particularly challenging.\\

\noindent
Starting with the somewhat less involved problem of determining $\Gamma_{12}$, and thus the observable $y$, we can split up the individual contributions due to the down and strange quarks as 
\begin{equation}
    \Gamma_{12}  =  - 
    \left( 
    \lambda_s^2 \Gamma_{12}^{ss}
    + 2 \lambda_s \lambda_d\Gamma_{12}^{sd}
    + \lambda_d^2 \Gamma_{12}^{dd}
    \right) 
    =
    - \lambda_s^2 \left( 
    \Gamma_{12}^{ss} -  2 \Gamma_{12}^{sd}
    +  \Gamma_{12}^{dd}
   \right) 
  + 2 \lambda_s \lambda_b
   \left( 
     \Gamma_{12}^{sd}
    -  \Gamma_{12}^{dd}
   \right) 
 - \lambda_b^2
\Gamma_{12}^{dd}
    \, ,
    \label{sec6.1:eq1}
\end{equation}
where $\lambda_q = V_{cq} V_{uq}^*$ and $\Gamma_{12}^{qq^\prime}$, with $q,q^\prime = \{d,s\}$, denote the individual contributions to the  box diagram with internal $q$ and $q^\prime$ quarks. Note that on the r.h.s.\ of Eq.~\eqref{sec6.1:eq1} the CKM unitarity has been employed to rewrite $\lambda_d  = - \lambda_s - \lambda_b$ in order to make more explicit the  pronounced hierarchies that characterise this observable. In fact, taking into account the numerical values of the CKM elements, see e.g.~\cite{Charles:2004jd}, namely
\begin{equation}
  - \lambda_s^2 =  -4.792 \cdot 10^{-2} - 5.6914 \cdot 10^{-7} i \, ,
  \quad
  2 \lambda_s \lambda_b =   +2.758 \cdot 10^{-5} - 6.129 \cdot 10^{-5} i \, ,
  \quad  
   - \lambda_b^2 =  + 1.563 \cdot 10^{-8}    + 1.764 \cdot  10^{-8} i
      \, ,
\end{equation}
we find that the first term on the r.h.s.\ of Eq.~\eqref{sec6.1:eq1}
has the largest CKM factor, whereas the second and even more the third term are strongly CKM suppressed. 
On the other hand, taking into account that $m_d^2/m_W^2 \approx m_s^2/m_W^2 \ll 1$, we expect the first two terms on the r.h.s.\ of Eq.~\eqref{sec6.1:eq1} to be extremely GIM suppressed. As a result, all three terms on the r.h.s.\ of Eq.~\eqref{sec6.1:eq1} are small and could be of similar size.\\

\noindent
Applying the framework of the HQE, which is well established for the description of inclusive observables in the $b$-sector and it also seems to reproduce the experimental values of the charmed hadron lifetimes, to the theoretical description of $\Gamma_{12}$ yields, similarly to Eq.~\eqref{sec4.1:eq6}, the following expression
\begin{equation}
\Gamma_{12} = 16 \pi^2 \left[\tilde \Gamma_6 \frac{\langle \tilde Q_6 \rangle}{m_c^3}  + \tilde \Gamma_7 \frac{\langle \tilde Q_7 \rangle}{m_c^4}  
 + \ldots \right]\,,
 \qquad 
 {\rm with}
 \qquad
 \tilde \Gamma_d = \tilde \Gamma_d^{(0)} + \left(\frac{\alpha_s}{4\pi}\right) \tilde \Gamma_d^{(1)} + \left( \frac{\alpha_s}{4\pi} \right)^2 \tilde \Gamma_d^{(2)} + \ldots  \,,
 \label{sec6.1:eq2}
\end{equation}
where $\tilde Q_d$ are local $\Delta C = 2$ operators of the type $\tilde Q_d \propto [\bar q c_v][\bar q (i D_\mu) \, \overset{d-6}{\ldots} \, (i D_\nu) c_v] $, and $\tilde \Gamma_d$ are short-distance functions calculable in perturbation theory. Their current status is shown in Table~\ref{sec6.1:tab1}~\footnote{The references in Table~\ref{sec6.1:tab1} mainly concern $B^0_{(s)}$ mixing, but the results obtained can be generalised to $D^0$ mixing with the proper replacements $m_b \to m_c$, $m_c \to m_s$ etc.}. As for the corresponding non-perturbative parameters, determinations of the dimension-six matrix elements from Lattice QCD have been performed in~\cite{Bazavov:2017weg}, while results within QCD sum rules have been obtained in~\cite{Kirk:2017juj}. For the dimension-seven matrix elements, no precise first-principle determinations are available yet. First Lattice QCD results have been published in \cite{Davies:2019gnp} for the case of $B^0_s$-meson mixing, while partial results (only the so-called  condensate contributions) have been determined within the HQET sum rule approach in~\cite{Mannel:2011iqd}.\\
\begin{table}
\centering
\renewcommand{\arraystretch}{1.4}
\begin{tabular}{|c|c|c|}
\hline
$\tilde \Gamma_6^{(2)}$
& $\tilde \Gamma_6^{(1)}$
& $\tilde \Gamma_7^{(0)}$
\\
\hline
\hline
\scriptsize  \it Gerlach, Nierste, Reeck, Stabovenko, Steinhauser~\cite{Gerlach:2022hoj, Gerlach:2022wgb, Reeck:2024iwk, Gerlach:2025tcx}
& 
\scriptsize \it  Beneke, Buchalla, Greub, Lenz, Nierste~\cite{Beneke:1998sy}
& 
\scriptsize \it  Beneke, Buchalla, Dunietz~\cite{Beneke:1996gn}
\\
\scriptsize \it Asatrian, Hovhannisyan, Nierste, Yeghiazaryan~\cite{Asatrian:2017qaz}
& 
\scriptsize \it  Ciuchini, Franco, Lubicz, Mescia, Tarantino~\cite{Ciuchini:2003ww}
&
\scriptsize \it Dighe, H\"urth, Kim, Yoshikawa~\cite{Dighe:2001gc}
\\
& 
\scriptsize \it Lenz, Nierste~\cite{Lenz:2006hd}
& 
\\
\hline
\end{tabular}
\caption{Status of the perturbative contributions $\tilde \Gamma_d$ in the HQE for $D^0$-meson mixing.}
\label{sec6.1:tab1}
\end{table}

\noindent
Theoretical predictions based on Eq.~\eqref{sec6.1:eq2} show that whereas the size of the individual contributions $\Gamma_{12}^{qq^\prime}$ is by some factor larger than the experimental value of $y$, the GIM suppressed combination in Eq.~\eqref{sec6.1:eq1} leads to $y_{\rm HQE} \sim 3.6 \cdot 10^{-7}$~\cite{Lenz:2020efu},
four 
orders of magnitude below the experimental determination of $y$! 
The big question is therefore: how to understand these extreme GIM cancellations? 
More than 30 years ago, in~\cite{Georgi:1992as, Ohl:1992sr} it was suggested that these cancellations might be lifted by higher orders in the HQE, namely by dimension-nine and dimension-twelve contributions. In this case, while the individual contributions to  $\Gamma_{12}^{qq^\prime}$ would become smaller at higher orders - each further order is suppressed by an additional power of $\Lambda_{\rm QCD}/m_c$, in the combination 
$  \lambda_s^2 \Gamma_{12}^{ss}
 + 2 \lambda_s \lambda_d\Gamma_{12}^{sd}
 + \lambda_d^2 \Gamma_{12}^{dd}$, the mass suppression could be
overcompensated by a weaker GIM mechanism. 
This idea has been further investigated in~\cite{Bigi:2000wn, Bobrowski:2012jf, Melic:2024oqj, Dulibic:2025emg} and despite some indications of possible enhancement effects have been observed, a complete picture of higher-dimensional contributions is so far still missing.
Another more recent idea was proposed in~\cite{Lenz:2020efu}. In the latter study, the possibility was investigated that by using an alternative renormalisation scheme to compute $\tilde \Gamma_{12}$, namely by varying the renormalisation scales of the individual contributions describing intermediate $dd$, $sd$, and $ss$ states, separately, the GIM mechanism could be lifted and the theory uncertainties would be enhanced to a level that the experimental value of $y$ could be accommodated~\cite{Lenz:2020efu}.
\\

\noindent
In addition to relying on the inclusive description of $\Gamma_{12}$ based on the HQE,
exclusive methods have also been employed in the literature. In this case, the idea is to directly estimate all the intermediate hadronic states contributing to $\Gamma_{12}$, a task that is clearly very challenging, particularly in the charm sector, see Section~\ref{subsec:nonlep}.
Simplified approaches, in which, e.g.\ only phase-space effects are taken into account without determining the corresponding hadronic matrix elements, yield, however, promising results and the value of $y$ is found to be of the same order as the experimental one~\cite{Falk:2001hx}. Other approaches using dispersive methods have been adopted in~\cite{Li:2022jxc, Xie:2025hzx, Umeeda:2021llf, Li:2020xrz}. Finally, also BSM interpretations of the experimental value of $y$ have been investigated in the literature, see e.g.\ \cite{Golowich:2006gq}.\\

\noindent
The theoretical determination of $M_{12}$ is even more involved and suffers, in principle, from the same type of challenges as for $\Gamma_{12}$. We note that in \cite{Falk:2004wg} a dispersion relation was derived in order to relate $x$ and $y$ and study the relative size between the two, since, being both dominated by long-distance contributions in the SM, naively, no significant hierarchy would be expected. Indeed, the analysis indicated the absence of sizeable differences between $x$ and $y$ in the SM. To conclude, there is still plenty of work to be done in the theoretical description of charm mixing, either within the inclusive or the exclusive approach. In the future, even a direct determination from Lattice QCD calculations might become feasible, following the ideas presented in~\cite{DiCarlo:2025mvt, Hansen:2012tf}. For the above reasons, a robust theory determination of CPV in $D^0$ mixing, which requires the knowledge of both $M_{12}$ and $\Gamma_{12}$, is still missing however, this effect is expected to be small in the SM, see e.g.~\cite{Bobrowski:2010xg}. Experimentally, this phenomenon has not been observed yet, and it provides an opportunity for so-called null tests of the SM.

%%%%%%%%%%%%%%%%%%%%%%%%%%%%%%%%%%%%%%%%%%%%%%%%%%%%%%%%%%%%%%%%%%%%%%%%%%%%%%%%%%%%%%%%%%%%%%%%%%%% %%%%%%%%%%%%%%%%%%%%%%%%%%%%%%%%%%%%%%%%%%%%%%%%%%%%%%%%%%%%%%%%%%%%%%%%%%%%%%%%%%%%%%%%%%%%%%%%
\subsection{CPV in charm hadron decays
}
\label{subsec:CPV}
{\bf Experimental searches with two-body decays:}
The discovery of CPV in the charm sector was announced in 2019 by the LHCb collaboration~\cite{LHCb:2019hro}~\footnote{This important milestone in flavour physics was listed, in the same year, as one of the top 10 breakthroughs of 2019 by the PhysicsWorld~\cite{physicsworld2019}.}~\footnote{A summary of the interesting ``$\Delta A_{\rm CP}$-saga" leading to the discovery of CPV in the charm sector can be found e.g.\ in~\cite{Lenz:2013pwa}, discussing the status after first experimental hints in 2011, and 
in~\cite{Chala:2019fdb}, discussing the status after the final discovery in 2019.}. This observation was performed by measuring the difference between the CP asymmetries in the $D^0 \to K^+ K^-$ and $D^0 \to \pi^+ \pi^-$ decays, a quantity denoted by $\Delta A_{\rm CP}$, namely 
\begin{equation}
\Delta A_{\mathrm{CP}} = A_{\mathrm{CP}}(K^+K^-)-A_{\mathrm{CP}}(\pi^+\pi^-)
\,,
\label{eq:dacp}
\end{equation}
where $A_{\rm CP}(f)$ is the time-integrated CP asymmetry in the decay $D^0 \to f$. 
The time-dependent CP asymmetry 
$A_{\rm CP}(f; t)$, for $f=K^+K^-,~\pi^+\pi^-$, 
is defined as 
\begin{equation}
    A_{\mathrm{CP}}(f; t)=\frac{\Gamma(D^0(t)\rightarrow f)-\Gamma(\bar D^0 (t)\rightarrow f)}{\Gamma(D^0 (t)\rightarrow f)+\Gamma(\bar D^0(t)\rightarrow f)}\,,
\end{equation}
and receives contributions from the direct CP asymmetry, $a_{\mathrm{CP}}^{\mathrm {dir}}$, describing CPV in the decay, see the preceding chapter, but also from the indirect CP asymmetry, $a_{\mathrm{CP}}^{\mathrm {ind}}$, sensitive to CPV in $D^0$-$\bar D^0 $mixing or in the interference between mixing and decay. The effect of the latter, 
will depend on the decay-time distribution. 
The time-integrated asymmetry
\begin{equation}
    A_{\mathrm{CP}}(f)=\frac{\Gamma(D^0\rightarrow f)-\Gamma(\bar D^0 \rightarrow f)}{\Gamma(D^0 \rightarrow f)+\Gamma(\bar D^0\rightarrow f)}\,,
    \label{eq:dacpresult}
\end{equation}
can be thus written as~\cite{Gersabeck:2011xj, HeavyFlavorAveragingGroupHFLAV:2024ctg} 
\begin{equation}
    A_{\mathrm{CP}}(f) \approx a_{\mathrm{CP}}^{\mathrm{dir}}(f)\Bigg(1 +y_{\mathrm{CP}}\frac{\langle t(f) \rangle}{\tau}\Bigg) + a_{\mathrm{CP}}^{\mathrm{ind}}(f)\frac{\langle t(f) \rangle}{\tau} = a_{\mathrm{CP}}^{\mathrm{dir}}(f)-A_\Gamma(f)\frac{\langle t(f) \rangle}{\tau},
    \label{eq:acpag}
\end{equation}
where $a_{\mathrm{CP}}^{\mathrm{dir}}(f)$ is defined as $a_{\rm CP}^{\rm dir}(f) \equiv  
\frac{ |{\cal A}(D \to f)|^2 - | {\cal A}(\bar D \to \bar f)|^2}
     {|{\cal A}(D \to f)|^2 + |{\cal A}(\bar D \to \bar f)|^2}$, $\langle t(f)\rangle$ is the mean decay time of $D^0 \to f$ in the reconstructed sample, $\tau$ is the $D^0$-meson lifetime, and $A_\Gamma(f) = - (a_{\rm CP}^{\rm ind}(f) + y_{\rm CP} a_{\rm CP}^{\rm dir}(f)) \approx - a_{\rm CP}^{\rm ind}(f)$, is defined as the asymmetry between the effective decay times $\hat \tau$ 
of the $\bar D^0$ and $D^0$ decay to the same final state $f$, where $\hat \tau (D^0 \to f) =   \int_0^\infty t \Gamma (D^0 (t)\rightarrow f) dt \Big/\int_0^\infty\,  \Gamma (D^0(t)\rightarrow f) dt$, namely
\begin{equation}
    A_\Gamma(f) \equiv \frac{\hat \tau( \bar D^0 \rightarrow f)- \hat\tau(  D^0 \rightarrow f)}{\hat \tau(\bar D^0 \rightarrow f) + \hat \tau(D^0 \rightarrow f)}\,.
\end{equation}
The observable $A_\Gamma$ is related to  $\Delta Y$ defined as 
\begin{equation}
    \Delta Y_f\equiv\frac{\hat\Gamma(\bar D^0 \rightarrow f)-\hat\Gamma( D^0 \rightarrow f)}{2 \Gamma_D}\,,  
\end{equation} 
where $\hat \Gamma = \hat \tau^{-1}$ by
\begin{equation}
   \Delta Y_f  = {-A_\Gamma(f)}({1+y_{\mathrm{CP}}}) \approx   - A_\Gamma(f) \,, 
\end{equation}
up to small corrections of the order of less than $1\%$
due to $y_{\rm CP} \sim {\cal O} (10^{-3})$, see Section~\ref{subsec:mixing}. Using Eq.~\eqref{eq:acpag}, $\Delta A_{\rm CP}$ in Eq.~\eqref{eq:dacp} becomes
\begin{equation}
    \Delta A_{\mathrm{CP}} \approx \Delta a_{\mathrm{CP}}^{\mathrm {dir}} + \Delta Y\frac{\Delta \langle t \rangle }{\tau},
    \label{eq:dacp-timedep}
\end{equation}
where $\Delta \langle t\rangle$ is the difference of the mean decay time of $D^0 \to K^+ K^-$ and $D^0 \to \pi^+ \pi^-$ in the reconstructed sample. Further details on the formalism can also be found in~\cite{Kagan:2020vri}. Note that in Eq.~\eqref{eq:dacp-timedep} we also dropped the final state dependence on $\Delta Y$ assuming $\Delta Y_{K^+K^-} = \Delta Y_{\pi^+\pi^-} = \Delta Y$. This is valid with corrections expected at $\mathcal{O}(10^{-5})$~\cite{LHCb:2021vmn}, far below current experimental sensitivity, see also the discussion below.\\

\noindent
Experimentally, we can only measure
the so-called {\bf raw asymmetries} for a given decay $D^0\rightarrow f$. In fact, the CP asymmetry $A_{\rm CP}$ cannot be accessed directly, as additional sources of asymmetries also contribute to the measured yields. These are referred to as {\bf nuisance asymmetries} and one of the main experimental challenges is to disentangle them from the CP asymmetry of interest, as their effect is $\sim\mathcal{O}(1\%)$, and the CP asymmetry is at least an order of magnitude smaller. The nuisance asymmetries include detection asymmetries $A_{\rm D}$ and production asymmetries $A_{\rm P}$~\footnote{Note that at $e^+e^-$ colliders, the production asymmetry is from the forward-backward asymmetric
production of charm mesons in $e^+e^-
\rightarrow c\bar c$ events.}. 
The detection asymmetry arises from the different interaction cross sections of final state particles with the material of the detector (kaons, pions) which is momentum dependent~\cite{ParticleDataGroup:2024cfk}, as well as from possible detector asymmetries such as dead channels, from different track reconstruction efficiencies for different sides of the detector, etc.~\footnote{Note that some detector asymmetries at LHCb can be cancelled by swapping the polarity of the magnetic field, however this cancellation is imperfect because of possible movement of the detector, of the imperfect alignment of the detectors, etc. Swapping the magnetic field also has no effect on the asymmetry coming from the material interactions therefore most asymmetry measurements do not rely on swapping the magnetic field.}
The production asymmetry refers to the different production cross sections of particles (e.g.\ $B$ and $D$ mesons) and their antiparticles in proton proton collisions, and there are several different mechanisms contributing to this~\cite{Chaichian:1993rh}, one of which is the combination of the produced heavy quarks ($c,\bar{c}$ or $b,\bar{b}$) with the valence quarks of the protons since $c\bar c$ will be produced in pairs but they will hadronise differently~\footnote{A $c$ quark could couple to valence quarks and form a charmed baryon e.g. $\Lambda_c$, while the excess $\bar c$ quarks could be combined with valence quarks to create more $\bar D^0$ mesons compared to $D^0$ mesons.}.\\
\begin{figure}
    \centering
    \includegraphics[width=0.30\linewidth]{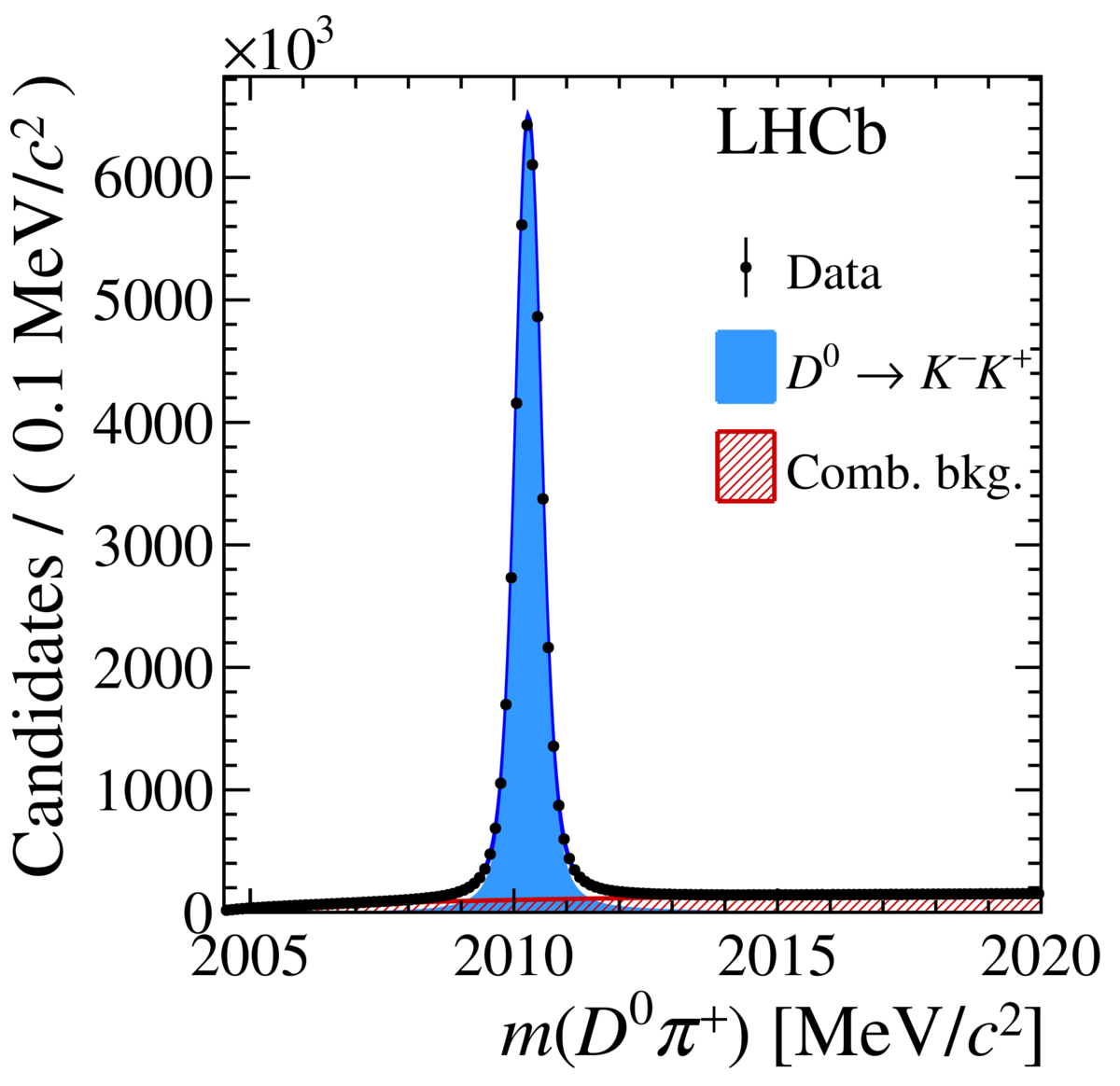}
    \quad 
\begin{overpic}[width=0.3\linewidth]{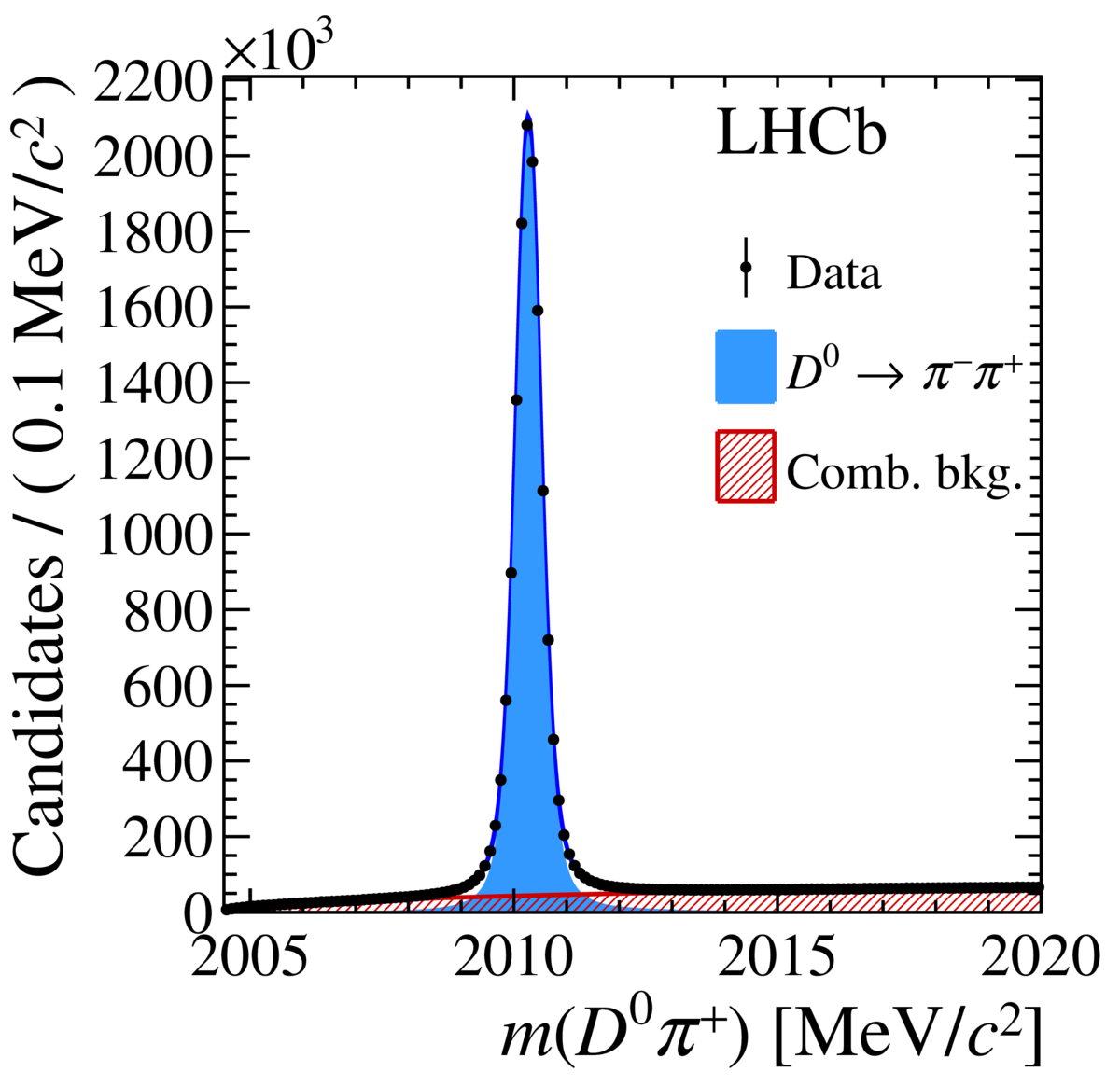} \put(90,108){\includegraphics[width=0.05\linewidth]{lhcb-logo.png}}
  \end{overpic}    \\     
            \includegraphics[width=0.30\linewidth]{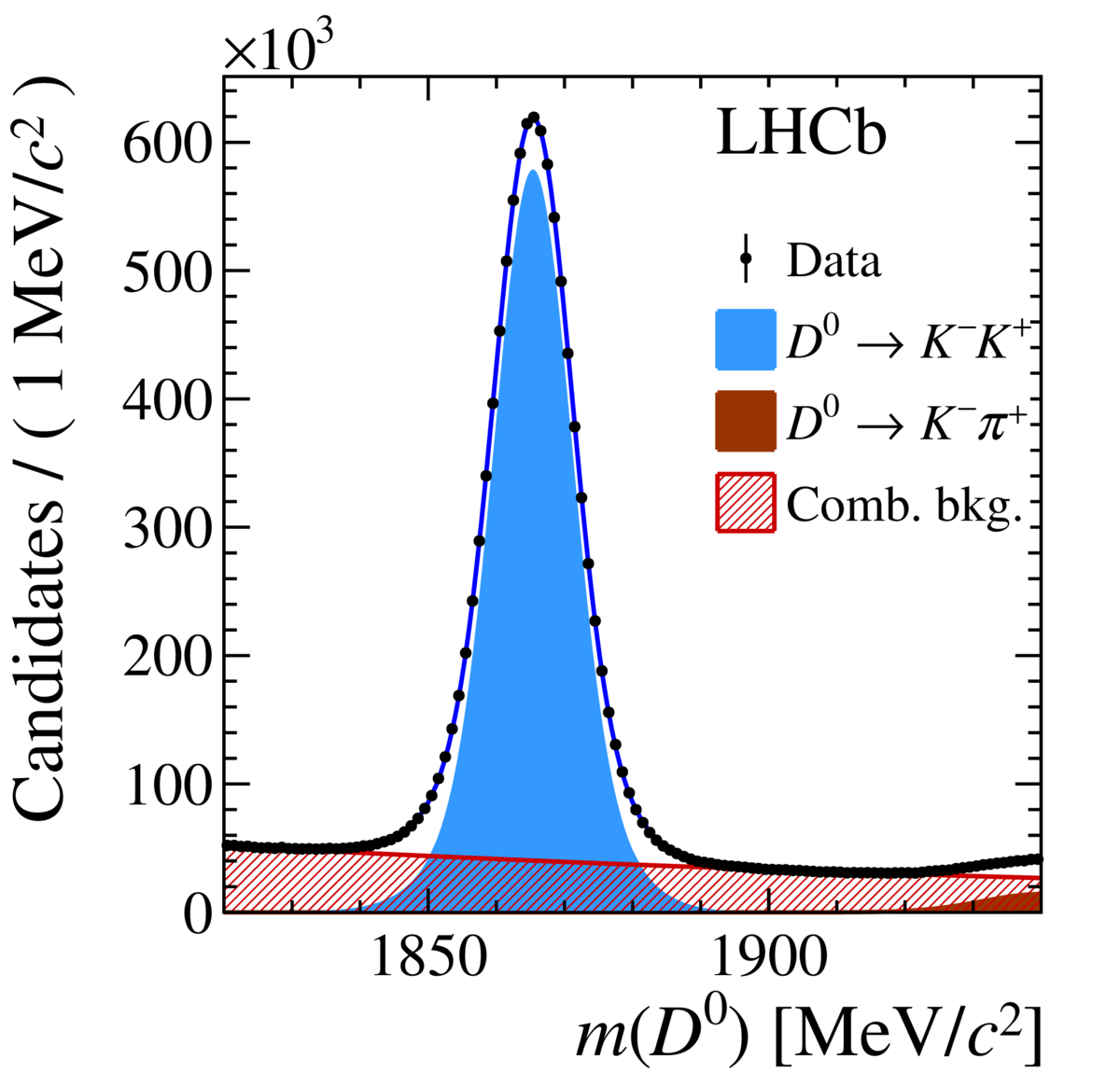}
            \quad
                        \includegraphics[width=0.30\linewidth]{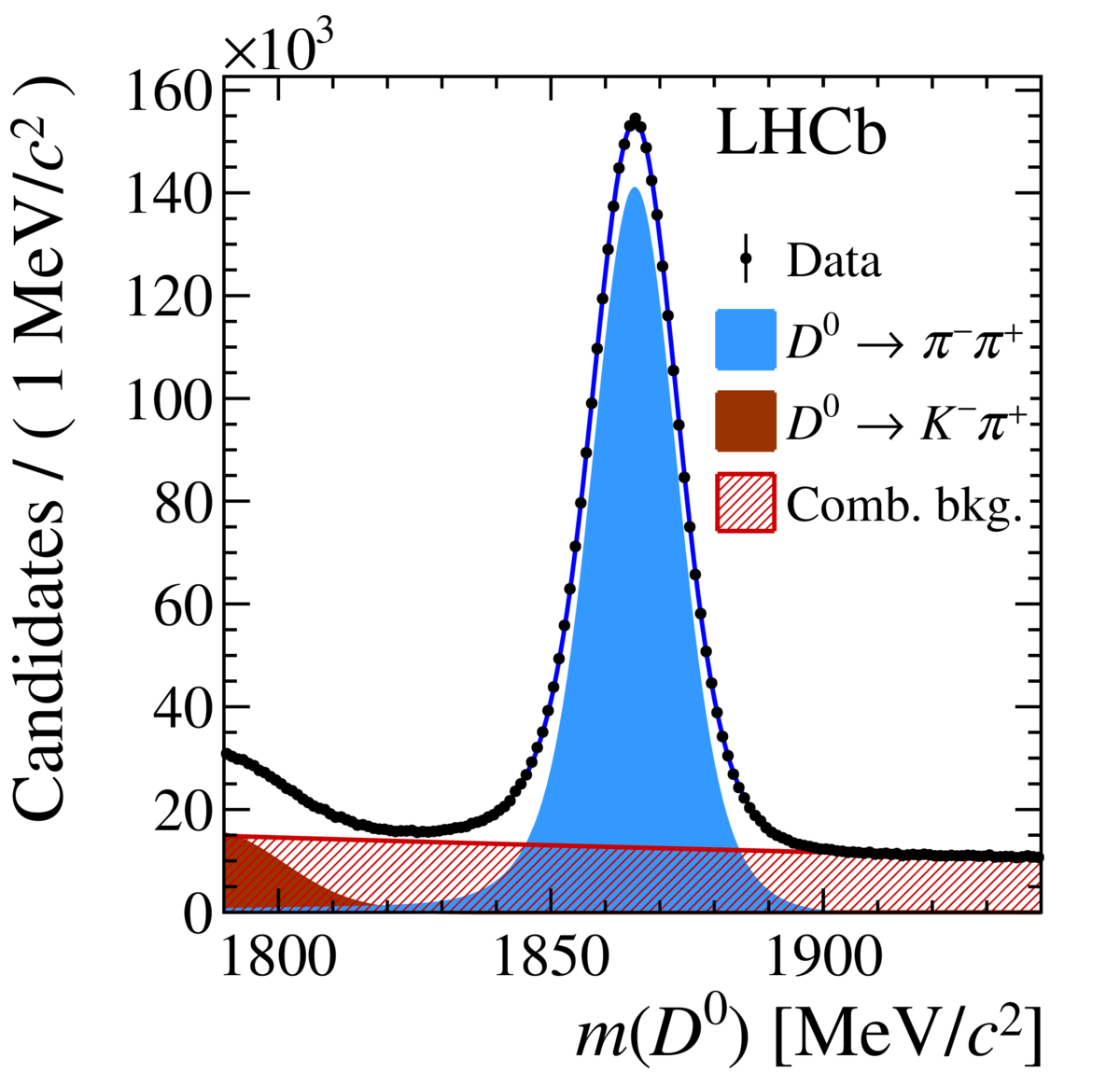}
    \caption{Signal distributions of $D^0\rightarrow K^+K^-$ (left) and  $D^0\to\pi^+\pi^-$ (right) decays. In the two top plots, the distributions of the invariant mass of $D^0\pi$ correspond to promptly produced charm mesons, while in the two bottom plots, the distributions of the $D^0$ invariant mass comes from secondary muon-tagged decays. Taken from~\cite{LHCb:2019hro}.}
    \label{fig:dacp}
\end{figure}

\begin{figure}
    \centering
\begin{overpic}[width=0.45\linewidth]{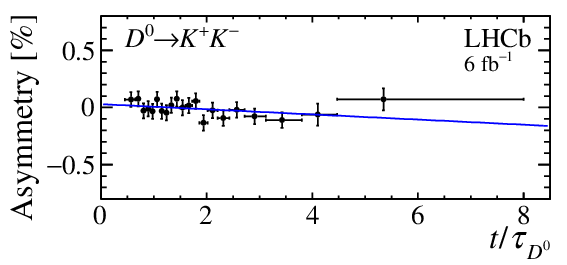} \put(175.5,81){\includegraphics[width=0.05\linewidth]{lhcb-logo.png}}
  \end{overpic}        
\begin{overpic}[width=0.45\linewidth]{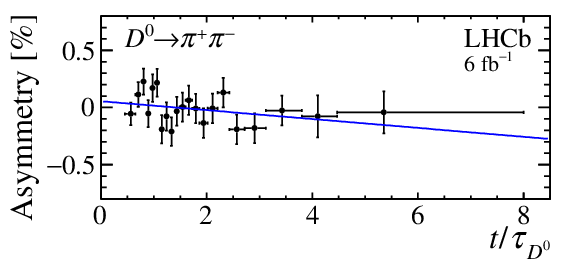} \put(175.5,81){\includegraphics[width=0.05\linewidth]{lhcb-logo.png}}
  \end{overpic}     
    \caption{Decay-time dependent asymmetries of $D^0\rightarrow K^+K^-$ (left) and $D^0\rightarrow \pi^+\pi^-$ (right) decays measured by LHCb. Taken from~\cite{LHCb:2021vmn}. }
    \label{fig:deltay}
\end{figure}

\noindent
Due to the presence of nuisance asymmetries, the raw asymmetries will depend on the specific tagging procedure adopted.
For promptly produced charm mesons, the raw asymmetry reads
\begin{equation}
    A_{\rm raw}^{\pi-{\rm tagged}}(f) \equiv \frac{N(D^{*+}\rightarrow D^0(f) \pi^+) - N(D^{*-}\rightarrow \bar{D}^0 (f) \pi^-)}{N(D^{*+}\rightarrow D^0 (f) \pi^+) + N(D^{*-}\rightarrow \bar{D}^0 (f)\pi^-)}\,,
\end{equation}
where $N(D^{*+}\rightarrow D^0(f) \pi^+)$ is the yield of the neutral $D^0$ mesons promptly tagged  with the charge of the soft pion in the strong decay $D^{*+}\rightarrow D^0 \pi^+$ (similarly for the charge-conjugated mode $D^{*-}\rightarrow \bar{D}^0(f) \pi^-$). The raw asymmetry can be then decomposed as
\begin{equation}
    A_{\rm raw}^{\pi-{\rm tagged}}(f) \approx A_{\rm CP}(f) + A_{\rm P}(D^{*+}) + A_{\rm D}(\pi^+_{tag}) \,,
\label{sec6:eq:nuisance}
\end{equation}
where 
the detection asymmetry
$A_{\rm D}(\pi^+_{tag})$ is related to the reconstruction efficiency of the tagging pion and the production asymmetry $A_{\rm P}(D^{*+})$ is related to the production of $D^{*+}$ mesons in proton proton collision~\footnote{In Eq.~\eqref{sec6:eq:nuisance}, possible detection asymmetries due to the final state particles $K^+K^-$ cancel out: $A_{\rm D}(K^+K^-) = 0$. 
}. 
For $D^0$ mesons  semileptonically tagged via the decay $\bar B \rightarrow D^0 \mu^- \bar \nu_\mu X$, the raw asymmetry reads
\begin{equation}
    A_{\rm raw}^{\mu-{\rm tagged}}(f) \equiv \frac{N(\bar B \rightarrow D^0(f) \mu^- \bar \nu_\mu X) - N( B \rightarrow \bar D^0(f) \mu^+  \nu_\mu X)}{N(\bar B \rightarrow D^0(f) \mu^- \bar \nu_\mu X) + N( B \rightarrow \bar D^0(f) \mu^+  \nu_\mu X)}\,,
\end{equation}
and the corresponding nuisance asymmetries
are given by the production asymmetry of the $B$ meson $A_{\rm P}(B)$ and the detection 
asymmetry related to the tagging of the muon $A_{\rm D}(\mu^-_{tag})$, i.e.\
\begin{equation}
    A_{\rm raw}^{\mu-{\rm tagged}}(f) \approx A_{\rm CP}(f) + A_{\rm P}(B) + A_{\rm D}(\mu^-_{tag}) \,.
\label{sec6:eq:nuisanceSL}
\end{equation}
In the difference of the raw asymmetries the contributions of the nuisance asymmetries cancel to first order, so that $\Delta A_\mathrm{CP}$ 
can be derived from the raw asymmetries measured experimentally~\cite{LHCb:2019hro}, that is 
\begin{equation}
A_{\rm raw}(K^+ K^-) - A_{\rm raw}(\pi^+ \pi^-) \approx A_{\mathrm{CP}}(K^+K^-)-A_{\mathrm{CP}}(\pi^+\pi^-) \equiv \Delta A_{\mathrm{CP}}. 
\label{eq:rawandcpasym}
\end{equation}
The measurement leading to the observation of $\Delta A_{\rm CP}$ in~\cite{LHCb:2019hro} was performed using both prompt and secondary charm decays collected by LHCb. The signal peaks corresponding to the data collected in the Run 2 can be seen in Fig.~\ref{fig:dacp}. The approximate yields are 44 million prompt and 9 million secondary
$D^0 \rightarrow K^+K^-$ decays, and about 14 million prompt and 3 million secondary $D^0 \rightarrow \pi^+ \pi^-$ decays. The results for both samples are, respectively~\cite{LHCb:2019hro} 
\begin{equation}
    \Delta A_{\mathrm{CP}}^{\pi-\mathrm{tagged}}= [-18.2 \pm 3.2 (\mathrm{stat.}) \pm 0.9 (\mathrm{syst.})] \cdot 10^{-4}\,,
\qquad
    \Delta A_{\mathrm{CP}}^{\mu-\mathrm{tagged}}
= [-9 \pm 8 (\mathrm{stat.}) \pm 5 (\mathrm{syst.})] \cdot 10^{-4}\,,
\end{equation}
leading to the combined final result, which includes also previous measurements~\cite{LHCb:2016csn, LHCb:2014kcb}, of 
\begin{equation}
\Delta A_{\mathrm{CP}} = (-15.4 \pm 2.9) \cdot 10^{-4}
\,.
\label{eq:dacpfinal}
\end{equation}
As for the time-dependent observable, $\Delta Y$, related to $A_\Gamma$, the most recent precise measurement comes from the LHCb collaboration~\cite{LHCb:2021vmn} and is determined as the slope of the time-dependent asymmetry measured in bins of decay time as shown in Fig.~\ref{fig:deltay}. 
The LHCb legacy measurements, done with the full dataset (including Run 1 and Run 2 data samples) of $D^0\rightarrow K^+K^-$ and $D^0\rightarrow \pi^+\pi^-$ decays collected by LHCb,  are~\cite{LHCb:2021vmn} 
\begin{equation}
\Delta Y_{K+K-} = (-0.3 \pm1.3 \pm 0.3) \cdot 10^{-4}\,, 
\qquad 
\Delta Y_{\pi^+\pi^-} = (-3.6 \pm 2.4 \pm 0.4) \cdot 10^{-4}\,,
\end{equation}
and their average is
\begin{equation}
\Delta Y = (-1.0 \pm1.1 \pm 0.3) \cdot 10^{-4}\,.
\end{equation}
The results for $\Delta Y_{K^+K^-}$ and $\Delta Y_{\pi^+\pi^-}$ are consistent with zero, and in agreement between each other, indicating that $\Delta Y$ is independent of the mode, at the current level of precision. These are also in line with previous less precise measurements of $A_\Gamma$ by the BaBar~\cite{BaBar:2012bho}, Belle~\cite{Belle:2015etc} and CDF~\cite{CDF:2014wyb} experiments.\\

\begin{figure}
    \centering
    \includegraphics[width=0.5\linewidth]{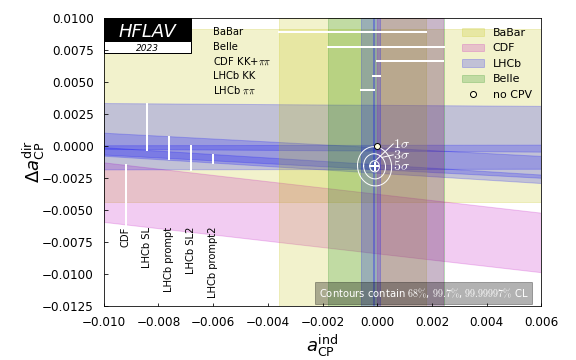}
    \caption{HFLAV plot showing the current status of the direct CPV, $\Delta a^{\rm dir}_{\mathrm{CP}}$ (along the vertical axis), and indirect CPV, $ a^{\rm ind}_{\mathrm{CP}}$ (along the horizontal axis), in charm decays, obtained combining measurements from LHCb (blue), Belle (green), BaBar (yellow) and CDF (pink) experiments, where the individual measurements are presented by white bars. Note that as there are several results from the LHCb collaboration, the darker blue regions are simply a result of the overlap of the individual blue bands. The global fit result is indicated by a cross, and the ellipses correspond to the 1$\sigma$, 3$\sigma$ and 5$\sigma$ regions. The no-CPV point is denoted with a white dot. Taken from~\cite{HeavyFlavorAveragingGroupHFLAV:2024ctg}. 
    }
    \label{fig:HFLAVfit}
\end{figure}

\noindent
The interplay between direct and indirect CPV for the two-body $D^0\rightarrow K^+K^-$ and $D^0\rightarrow \pi^+\pi^-$ decays can be seen in Fig.~\ref{fig:HFLAVfit}, where the inputs from time-integrated measurements are displayed on the $y$ axis, and time-dependent measurements are shown on the $x$ axis. The slight slope of the blue~(LHCb) and pink~(CDF) bands corresponding to measurements of $\Delta A_\mathrm{CP}$ comes from the last term in Eq.~\eqref{eq:dacp-timedep}.
The HFLAV fit~\cite{HeavyFlavorAveragingGroupHFLAV:2024ctg} gives
\begin{equation}
  a^{\rm ind}_{\mathrm{CP}} = (-0.010 \pm 0.012)\%\,, 
  \qquad 
\Delta a^{\rm dir}_{\mathrm{CP}} = (-0.159 \pm0.029)\%\,.
\label{sec6.2:eqDeltaacp}
\end{equation}

\noindent
Although the measurement of $\Delta A_{\mathrm{\rm CP}}$ is a great achievement, the question of the origin of CPV still remains, for which it would be beneficial to study individual CP asymmetries. In the context of sub-percent precision for $ \Delta A_{\rm CP}$ measurements, 
all other sources of asymmetry in 
Eqs.~\eqref{sec6:eq:nuisance}, \eqref{sec6:eq:nuisanceSL} need to be properly controlled. This can be achieved
through a careful cancellation with control channels, as it has been done in~\cite{LHCb:2014kcb, LHCb:2016nxk, LHCb:2022lry}, with \cite{LHCb:2022lry} being the most precise measurement. 
Typically, one selects CF modes as control channels as, within the SM, we do not expect any CPV effects in the latter, and therefore we are only left with the production asymmetry associated with the parent particle, and the detection asymmetries related to the final-state particles. 
For the measurement of $A_{\rm CP}(K^+K^-)$, this cancellation is outlined below 
using the following CF modes: the promptly tagged decay $D^0\rightarrow K^-\pi^+$ via $D^{*+}\rightarrow D^0(\rightarrow K^-\pi^+)\pi^+_{tag}$,  
and the decays $D^+\rightarrow K^-\pi^+\pi^+$ and $D^+\rightarrow \bar K^0\pi^+$~\footnote{Note that there are alternative CF control samples which can be used. Also, at LHCb, the decay mode $D^+\rightarrow \bar K^0\pi^+$ is reconstructed with two charged pions, and is dominated by the $K_S$ state.}. 
The corresponding measured raw asymmetries read:  
\begin{eqnarray}
        {\textcolor{blue}{D^{+*}\rightarrow D^0(K^-\pi^+)\pi^+ :~~ }}  A_{\rm raw}(K\pi) & = & {\textcolor{blue}{A_{\rm P}(D^{*+}) + A_{\rm D}(\pi^+_{tag}) + A_{\rm D}(\pi^+) - A_D(K^+)}} \,,  \label{eq:controlasymmetries1} \\
             \textcolor{red}{D^+ \rightarrow K^- \pi^+ \pi^+ :~~} A_{\rm raw}(K\pi\pi) &=& \textcolor{red}{ A_{\rm P}(D^+) - A_{\rm D}(K^+) + A_{\rm D}(\pi^+_1) + A_{\rm D}(\pi^+_2)}\,,  \label{eq:controlasymmetries2}\\
           \textcolor{violet}{D^+ \rightarrow \bar{K}^0\pi^+ :~~} A_{\rm raw}(\bar K^0\pi) &=& \textcolor{violet}{ A_{\rm P}(D^+) + A_{\rm D}(\pi^+) + A(\bar{K}^0)}\,,
    \label{eq:controlasymmetries3}
\end{eqnarray}
where the nuisance asymmetries for each decay mode have been colour coded to make more transparent how their combinations can cancel in the signal mode.  Here, $A(\bar{K}^0)$ is the neutral kaon asymmetry including effects of CPV, mixing, and regeneration after interacting with the material of the detector~\footnote{Reference~\cite{Yu:2017oky} suggests that in future measurements with large data samples, the effect of CP violation from the interference of the $D^0$ decay into $K^0$ and $\bar{K}^0$, and neutral-kaon mixing may also become relevant}. \
In the CF decay $D^{+*}\rightarrow D^0(K^-\pi^+)\pi^+$ we have the same production asymmetry associated with the $D^{*+}$, and the same detection asymmetry of the soft pion as in the signal modes used for the $\Delta A_{\mathrm{CP}}$ measurement. However, in contrast to the symmetric signal modes $D^0\rightarrow K^+K^-$ or $D^0\rightarrow \pi^+\pi^-$, the detection asymmetry of the final state $K\pi$ does not cancel, and this is why we need both $D^+$ modes: the $K\pi$ asymmetry of the $D^{+*}\rightarrow D^0(K^-\pi^+)\pi^+$ mode will cancel if we subtract the $K\pi$ asymmetry measured with two of the final state particles in $D^+ \rightarrow K^- \pi^+ \pi^+$; then the remaining production asymmetry and the detection asymmetry of the remaining $\pi^+$ will be cancelled with the measurement of the asymmetries in the $D^+ \rightarrow \bar{K}^0\pi^+$ mode. 
This is can be achieved in the following way, namely starting with Eq.~\eqref{sec6:eq:nuisance} for $f = K^+ K^-$, subtracting Eq.~\eqref{eq:controlasymmetries1}, adding Eq.~\eqref{eq:controlasymmetries2}, and then subtracting Eq.~\eqref{eq:controlasymmetries3}, leading to
 \begin{eqnarray}
     &A_{\rm raw}(K^+K^-)-A_{\rm raw}(K\pi)+A_{\rm raw}(K\pi\pi)-A_{\rm raw}(\bar K^0\pi) = 
     \nonumber \\
    & =  {A}_{\rm CP}(K^+ K^-) + [A_{\rm P}(D^{*+}) - \textcolor{blue}{A_{\rm P}(D^{*+})}] + [A_{\rm D}(\pi^+_{tag}) - \textcolor{blue}{A_{\rm D}(\pi^+_{tag})}] + 
    \label{sec6.2:nuiscanc}\\
 &   [\textcolor{blue}{A_{\rm D}(K^+)} - \textcolor{red}{A_{\rm D}(K^+)}] + [\textcolor{red}{A_{\rm D}(\pi^+_1)} - \textcolor{blue}{A_{\rm D}(\pi^+)}] + 
    [\textcolor{red}{A_{\rm P}(D^+)} - \textcolor{violet}{A_{\rm P}(D^+)}] + 
    \nonumber \\
  &  [\textcolor{red}{A_{\rm D}(\pi^+_2)} - \textcolor{violet}{A_{\rm D}(\pi^+)}] - \textcolor{violet}{A(\bar{K}^0)}. \,
  \nonumber 
\end{eqnarray}
\noindent
One condition for this method to be successful is that the kinematic distributions of the particles whose asymmetries we want to cancel match, and this can be achieved through a re-weighting of the distributions. All terms in the square brackets of Eq.~\eqref{sec6.2:nuiscanc} cancel, and re-arranging for $A_{\rm CP}(K^+ K^-)$ yields
\begin{eqnarray}
    {A}_{\mathrm {CP}}(K^+ K^-) = A_{\rm raw}(K^+K^-)-A_{\rm raw}(K\pi)+A_{\rm raw}(K\pi\pi)-A_{\rm raw}(\bar K^0\pi) + \textcolor{violet}{A(\bar{K}^0)}\,,
\end{eqnarray}
where now all the asymmetries, apart from $A(\bar{K}^0)$, refer to the raw asymmetries in the respective decay mode, and these can be measured directly in the experiment. As the effects of CPV, mixing, and regeneration of neutral kaons are very well known, the asymmetry $A(\bar{K}^0)$ can be obtained as detailed in~\cite{LHCb:2014kcb, Fetscher:1996fa}. Using this method $A_{\mathrm {CP}}(K^+K^-)$ has been measured to be~\cite{LHCb:2022lry}
\begin{equation}
A_{\mathrm {CP}}(K^+K^-) = [6.8 \pm 5.4 \text{ (stat)} \pm 1.6 \text{ (syst)}] \cdot 10^{-4}\,. \\
\end{equation}
\begin{figure}[t]
\centering
\begin{overpic}[width=0.45\linewidth]{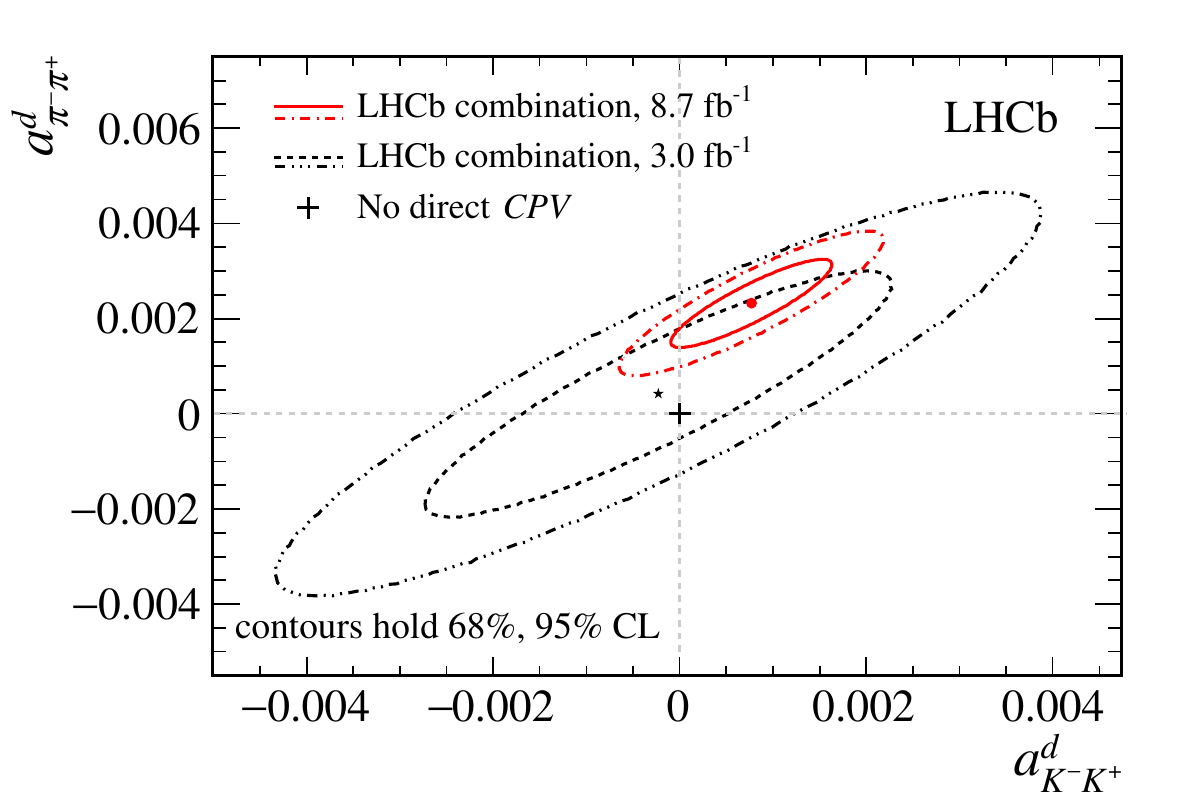} \put(170,116){\includegraphics[width=0.05\linewidth]{lhcb-logo.png}}
  \end{overpic}

\caption{
Central values and two-dimensional confidence regions in the ($a^d_{K^-K^+}, a^d_{\pi^-\pi^+}$) plane for the Run 1 measurement using 3~fb$^{-1}$ (in black), and the legacy LHCb measurement using the full Run 1 and Run 2 samples corresponding to about 8.7~fb$^{-1}$ (in red). The no-direct CPV point is at (0,0). Note that here $a_{f}^d = a_{\rm CP}^{\rm dir}(f)$. 
Taken from~\cite{LHCb:2022lry}.
}\label{fig:Contor}
\end{figure}
\noindent
Note that this measurement has been performed with kaons as the yields are higher than for the $\pi^+\pi^-$ channel. Assuming that time-dependent CPV is final state independent, $\Delta Y_{K^+K^-} = \Delta Y_{\pi^+ \pi^-}$, the values of direct CPV in the two SCS decay modes have been obtained from the combination of the current LHCb average for $\Delta A_{\rm CP}$ together with the measurement of $A_{\rm CP}(K^+K^-)$. These values respectively are~\cite{LHCb:2022lry}~\footnote{A recent combination of measurements of charm-mixing parameters performed by the LHCb collaboration~\cite{LHCb:2024yxi} yields $a_{\rm CP}^{\rm dir}(K^+K^-) = (6^{+6}_{-5}) \cdot 10^{-4} $ and $a_{\rm CP}^{\rm dir}(\pi^+\pi^-) = (22\pm 6) \cdot 10^{-4} $.}
\begin{equation}
a^{\rm dir}_{\rm CP}(K^+K^-) = (7.7 \pm 5.7) \cdot 10^{-4}\,, 
\qquad
a^{\rm dir}_{\rm CP}(\pi^+\pi^-) = (23.2 \pm 6.1) \cdot 10^{-4}\,,
\label{sec6.2:eq0}
\end{equation}
providing the first evidence for CPV in a specific $D^ 0$ decay i.e.\ in $D^0\rightarrow\pi^+\pi^-$. This result is most interesting when plotted as in Fig.~\ref{fig:Contor}, which reveals a departure from the strict U-spin symmetric limit $a^{\rm dir}_{\rm CP}(K^+K^-) + a^{\rm dir}_{\rm CP}(\pi^+\pi^-) = 0$ of $2.7\sigma$. We note however, that this tension is reduced if also the U-spin breaking in the corresponding branching fractions is taken into account, see~\cite{PajeroatCharm}. \\

\noindent
Further study into $\Delta A_{\rm CP}$ and direct CP-violating quantities will be exciting for the future of charm physics. As of now, no other experiment has the statistical power to measure a non-zero $\Delta A_{\mathrm{CP}}$. 
Therefore, it is important to re-measure this observable with future samples, and in other two- or multi-body decays. A related decay mode
is the $D^0 \rightarrow K^0_S K^0_S$ channel for which the most recent combination of Belle/ Belle II results gives $A_\mathrm{CP} = (-0.6 \pm 1.1 \pm 0.1) \%$~\cite{Belle-II:2025ezw}. Very recently, the LCHb collaboration reported $A_\mathrm{CP} = (1.86 \pm 1.04 \pm 0.41) \%$~\cite{LHCb:2025ezf}, which represents the most precise single-experiment determination of this quantity to date. 
Previous measurements include the LHCb result in~\cite{LHCb:2021rdn}, the Belle/ Belle II measurements in~\cite{Belle:2024vho} and most impressively, although with a slightly lower precision, the CMS result in~\cite{CMS:2024hsv}. 
Sum rules provide additional tests of the SM~\cite{Grossman:2012eb, HeavyFlavorAveragingGroupHFLAV:2024ctg} and can relate the decay rates of final states with isospin partners e.g. two-body decays such as $D^0\rightarrow K^+K^-$, $D^0\rightarrow \pi^+\pi^-$, $D^0\rightarrow \pi^0\pi^0$ or $D^+\rightarrow \pi^+\pi^0$, $D^0\rightarrow \pi^+\pi^-$, $D^0\rightarrow \pi^0\pi^0$. In the SM, $A_\mathrm{CP} (D^+\rightarrow\pi^+\pi^0)$ is expected to vanish, as it does not receive the contribution of two interfering amplitudes with different strong and weak phases.
The most precise measurement of the CP asymmetry in $D^+\rightarrow\pi^+\pi^0$ comes from the 
Belle II experiment $A_\mathrm{\rm CP}=(-1.8 \pm 0.9 \pm 0.1)\% $~\cite{Belle-II:2025wsy}. Belle II recently released a new measurement of the CP asymmetry in $D^0\rightarrow \pi^0\pi^0$, $A_\mathrm{CP}=(-0.30\pm0.72\pm0.20)\%$~\cite{Belle-II:2025rmf} slightly less precise than their previous measurement of $A_\mathrm{CP}=(-0.03\pm0.64\pm0.10)\%$~\cite{Belle:2014evd}. \\

\noindent
Time-integrated asymmetries $A_\mathrm{CP}$ have been measured in about 60 modes~\cite{HeavyFlavorAveragingGroupHFLAV:2024ctg}, all compatible with CP conservation at the current precision. Some of the more precise measurements are shown in Table~\ref{tab:acp}. 
Measurements of CP asymmetries in decays with $K_L$ in the final state, e.g.\ $D^0 \rightarrow K_LX, ~~X~=~\omega,~\phi, ~\eta, ~\eta^\prime$~\cite{BESIII:2022xhe}  have been reported by BESIII, however due to the double tag technique needed to infer the presence of the $K_L$ (by the amount of the missing energy), the yields are limited, and the uncertainties on the CP asymmetries in such decay modes are of the order of a few percent. 
Asymmetries with sub-percent precision have been measured for several multi-body decay modes as well e.g.\ in  
$D^+ \rightarrow K_S\pi^+\pi^0$~\cite{CLEO:2013rjc}, $D^+ \rightarrow K^+K^-
\pi^+$~\cite{BaBar:2012ize}, 
$D^0 \rightarrow \pi^+\pi^-
\pi^0 $~\cite{BaBar:2008xzl}, $D^0 \rightarrow K_S\pi^+\pi^- $~\cite{CDF:2012lpc}, $D^0 \rightarrow K^+K^-
\pi^+\pi^-$~\cite{Belle:2018pcz}, however, there are special techniques for searches of CPV in multi-body decays which offer a greater sensitivity, and they are discussed further below. The first observation of CPV in baryon decays ($\Lambda_b^0
\rightarrow pK^-\pi^+\pi^-$) was reported by the LHCb collaboration in 2025~\cite{LHCb:2025ray}. The CPV searches in the charm baryon sector are limited, a couple of examples are the searches in SCS decays $\Lambda^+_
c \rightarrow \Lambda K^+$~\cite{Belle:2022uod}, and in the difference of CP asymmetries in the SCS decays $\Lambda^+_c
 \rightarrow p K^+K^-$ and
$\Lambda^+_
c \rightarrow p\pi^+\pi^-$~\cite{LHCb:2017hwf}, also in the rare decays $\Lambda^+_
c \rightarrow p\mu^+\mu^-$~\cite{LHCb:2025bfy}. These are all compatible with CP symmetry conservation.\\

\begin{table}
    \centering
    \renewcommand{\arraystretch}{2}
    \begin{tabular}{|c||c|c|}
    \hline
        {\it Decay mode} & $a^\mathrm{dir}_\mathrm{CP}$ &  {\it Reference}\\
        \hline
        \hline
$D^0\rightarrow K^-K^+$ & $(+7.7 \pm 5.7 ) \cdot 10^{-4}$   & LHCb~\cite{LHCb:2022lry}\\
\hline
$D^0\rightarrow \pi^-\pi^+$& $(+23.2 \pm 6.1)\cdot 10^{-4}$& LHCb~\cite{LHCb:2022lry}\\
\hline\hline
   % \hline
        {\it Decay mode} & $A_\mathrm{CP}$ &  {\it Reference}\\
        \hline
        \hline
        $D^0 \rightarrow \pi^0\pi^0$ & $(-0.03\pm0.64\pm0.10)\%$  & Belle~\cite{Belle:2014evd}\\
                 \hline
         $D^0 \rightarrow K^0_
S K^0_
S$           & $(-0.6 \pm1.1 \pm0.1)\%$ & Belle/ Belle II~\cite{Belle-II:2025ezw}\\
        \hline
       $D^0 \rightarrow K_S\pi^0$  & $(-0.21 \pm 0.16 \pm 0.07)\%$ & Belle~\cite{Belle:2014evd}\\
       \hline
         $D^0 \rightarrow K_S\eta$         & $(+0.54 \pm 0.51 \pm 0.16)\%$ & Belle~\cite{Belle:2011npc}\\

\hline
$D^0 \rightarrow K_S\eta^\prime$ & $(+0.98 \pm 0.67 \pm 0.14)\%$ & Belle~\cite{Belle:2011npc}\\
        
        \hline
        \hline$D^+ \rightarrow \pi^+\pi^0$ & $(-1.3 \pm 0.9 \pm 0.6)\%$ & LHCb~\cite{LHCb:2021rou}\\
        \hline
        $D^+ \rightarrow \pi^+\eta$  & $(+0.13 \pm 0.50 \pm 0.18)\%$ & LHCb~\cite{LHCb:2022pxf}\\
        \hline
        $D^+ \rightarrow \pi^+\eta^\prime$ & $(+0.43 \pm 0.17 \pm 0.10)\%$ & LHCb~\cite{LHCb:2022pxf}\\
        \hline
       $D^+ \rightarrow \phi \pi^+$  & $(+0.003 \pm 0.040 \pm 0.029)\%$ & LHCb~\cite{LHCb:2019dwr}\\
         \hline
        $D^+ \rightarrow (K^0/\bar K^0)K^+$ & $(-0.004 \pm 0.061 \pm 0.045)\%$ & LHCb~\cite{LHCb:2019dwr}\\

\hline
\hline
 $D^+
_s \rightarrow \pi^+\eta^\prime$& $(-0.04 \pm 0.11 \pm 0.09)\%$& LHCb~\cite{LHCb:2022pxf}\\
\hline
$D^+
_s \rightarrow(K^0/\bar K^0)\pi^+ $& $(+0.16 \pm 0.17 \pm 0.05)\%$& LHCb~\cite{LHCb:2019dwr}\\
\hline

    \end{tabular}
    \caption{An overview of some of the most precise measurements of $A_\mathrm{CP}$ with about a percent or sub-percent precision, all of which are consistent with CP symmetry conservation, apart from the direct CP asymmetry measurement in $D^0\rightarrow \pi^-\pi^+$ which constitutes the first evidence of CPV in a single charm decay mode. The most precise individual asymmetries are presented in the top two rows. Extended list of results can be found in ~\cite{HeavyFlavorAveragingGroupHFLAV:2024ctg}.}
    \label{tab:acp}
\end{table}

\noindent
{\bf Experimental searches with multi-body decays:} Searching for direct CPV in multi-body decays, 
as already described in the case of CPV in mixing in Section~\ref{subsec:mixing},
builds off the study of the corresponding Dalitz space. Before discussing the available model-independent methods, 
it is important to touch on the application of model-dependent amplitude analysis for the study of direct CPV in multi-body channels. With a suitable amplitude analysis it is possible to investigate local CPV effects through asymmetries in the magnitudes, $a_r$, and phases, $\phi_r$ of each individual resonance, cf. Eq.~\eqref{sec5.3:eqAA}. 
For more details, the reader should refer to the wealth of model-dependent results for multi-body modes obtained by e.g. CLEO \cite{CLEO:2012beo, dArgent:2017gzv}, LHCb \cite{LHCb:2018mzv} and BaBar~\cite{BaBar:2012ize}, etc..\\

\begin{figure}[t]
 \centering
\begin{minipage}{0.44\textwidth}
\vspace*{1cm}

\begin{overpic}[width=1\linewidth]{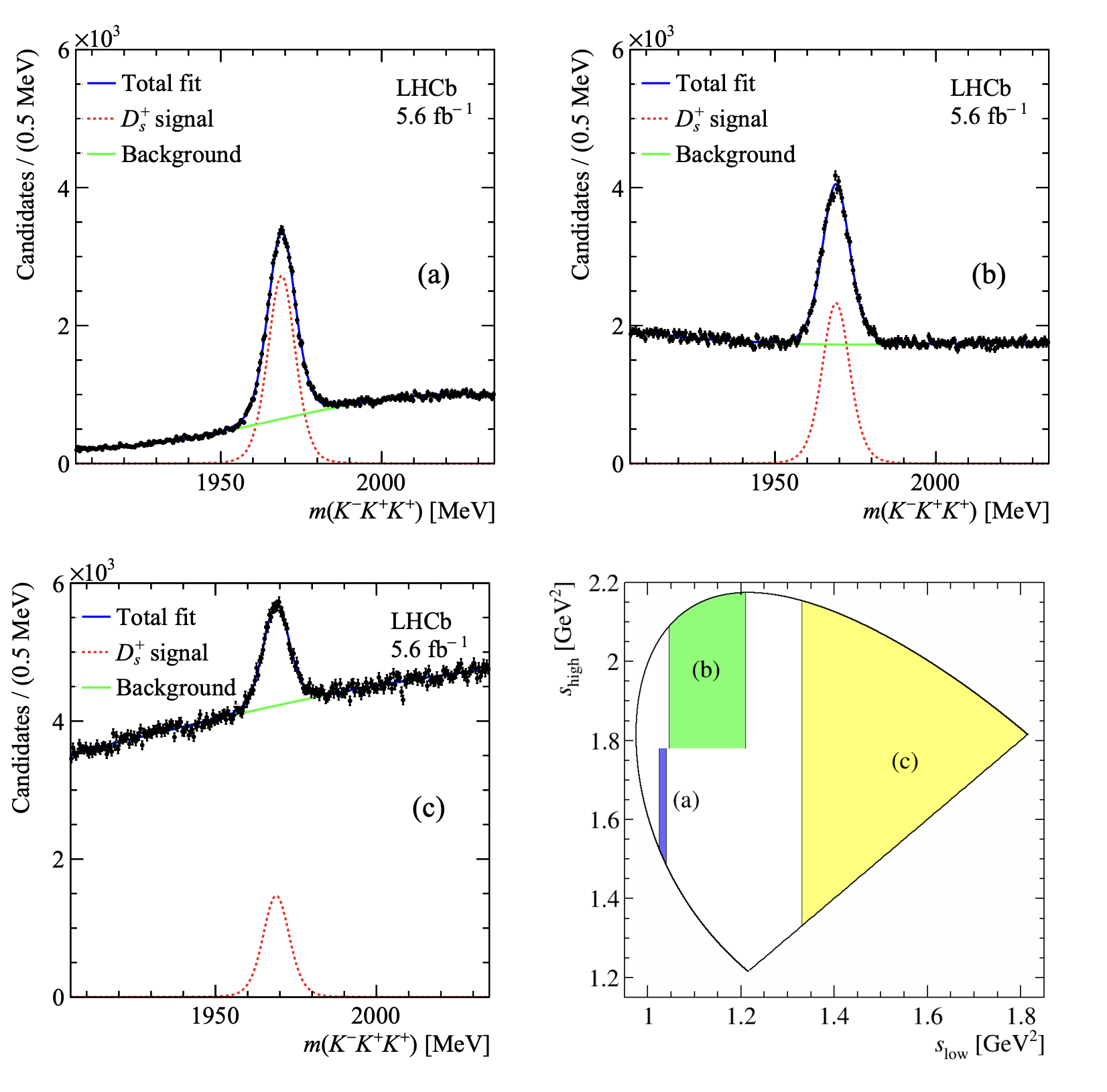} \put(178,186){\includegraphics[width=0.06\linewidth]{lhcb-logo.png}}
  \end{overpic} 

\caption{ Mass distributions for the $D^+_s$
candidates in three representative Dalitz plot bins
(a,b,c), defined in the lower right subfigure. Fitting these distributions greatly reduces the error on $N^i$ when using the Miranda method. The Dalitz description here is $s_{low}$ for the invariant mass squared of the lowest momenta $K^+K-$ combination, and $s_{high}$ for the invariant mass squared of the highest momenta $K^+K-$ combination. Taken from \cite{LHCb:2023qne}.}
\label{sec5.3:fig_miranda}
\end{minipage}
  \qquad \qquad
  \begin{minipage}{0.44\textwidth}
  \vspace*{1cm}

\begin{overpic}[width=1\linewidth]{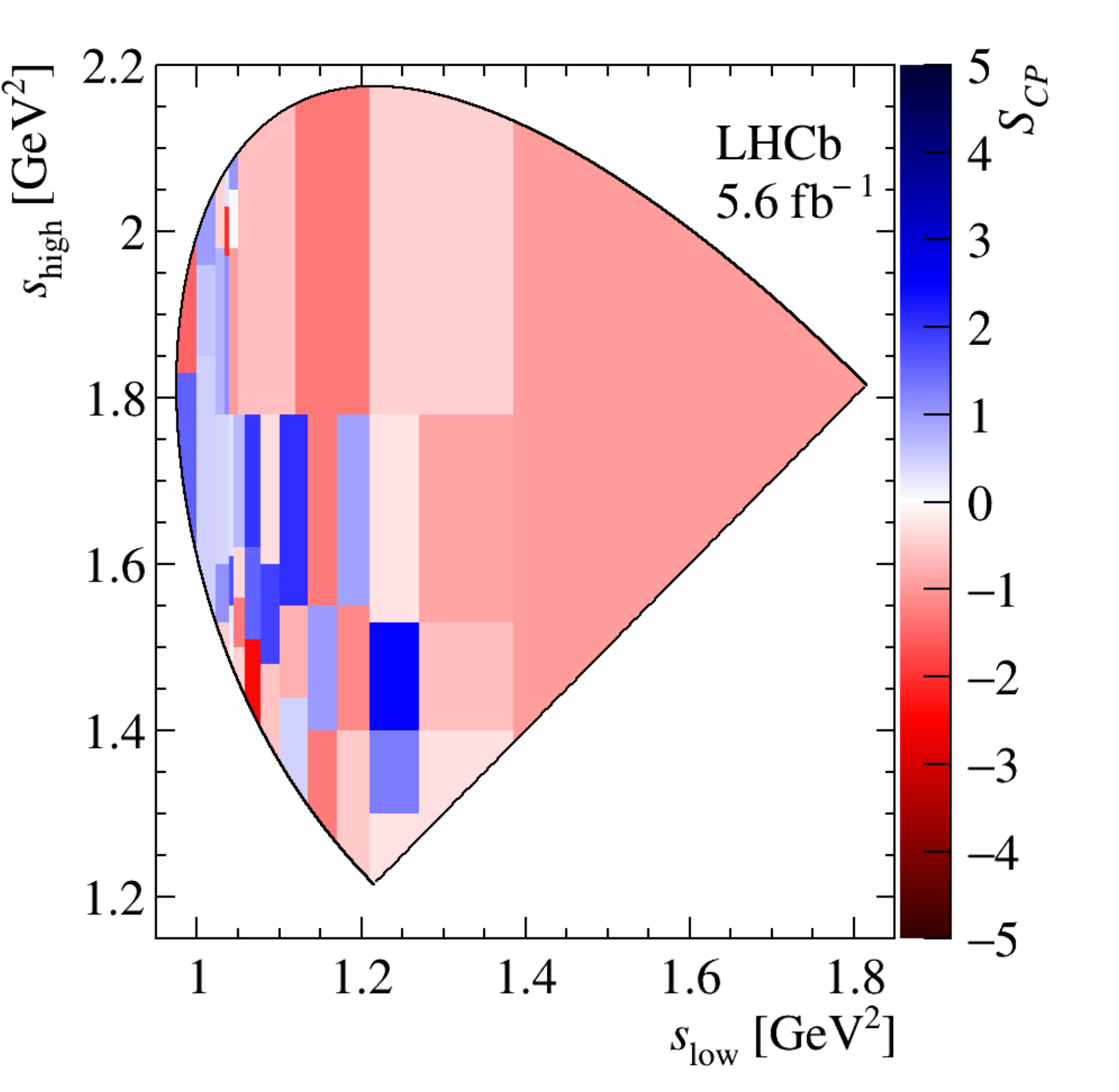} \put(132,172){\includegraphics[width=0.12\linewidth]{lhcb-logo.png}}
  \end{overpic} 
\caption{Asymmetry distributions for the $D^+_s$
candidates. One can see how the asymmetry can vary across the space although the global asymmetry remains consistent with zero. The Dalitz plot description here is $s_{low}$ for the invariant mass squared of the lowest momenta $K^+K-$ combination, and $s_{high}$ for the invariant mass squared of the highest momenta $K^+K-$ combination. Taken from \cite{LHCb:2023qne}.}
\label{sec5.3:fig_miranda_b}
  \end{minipage}
 \end{figure}

\noindent
Moving onto model-independent methods, one such method that aims to analyse the significance of local CP asymmetries in the Dalitz space is the \textbf{Miranda method}~\cite{Bediaga:2009tr}. The original Miranda method computes the difference between matter and anti-matter candidates in each bin of the Dalitz plot, before computing their significance. The local CP observable $\mathcal{S}_{\rm CP}$ is defined as
\begin{equation}
\mathcal{S}_{CP}^i = \frac{N^i-\alpha \bar{N^i}}
{\sqrt{\alpha(\delta^2_{N^i}+\delta^2_{\bar{N^i}})}} \ , \hskip .5cm  
{\rm with} ~\alpha=\frac{\sum_i N^i}{\sum_i \bar{N^i}}\ ,
\label{eq:scp1}
\end{equation}
\noindent
where $N^i$ and $\bar{N^i}$ represent the number of signal candidates in the $i^{th}$ bin of the Dalitz plot, and $\delta_{N^i}$, $\delta_{\bar{N^i}}$ their corresponding statistical uncertainties. The factor $\alpha$ normalises the total candidates between the compared samples. To explore CPV, a  $\chi^2$ test is then conducted, with $\chi^2=\sum_i(\mathcal{S}_{\rm CP}^i)^2$ --- with the number of degrees of freedom being the number of bins minus one. In the absence of CP violating effects we would expect this sum to follow a normal distribution as the local effects average over the whole kinematic region. Although this robust method has been used in a number of analyses \cite{LHCb:2011nqf, LHCb:2013qzm, LHCb:2013byb, LHCb:2020zkk}, we point out how the main issue 
when using this method is extracting a good estimate of $N^i$ for each bin in high-background measurements. In the case of high-background measurements this estimates of $N^{i}$ can become biased. However, in 2023 LHCb extended the method carrying out a 3-body mass fit in each bin, see Fig.~\ref{sec5.3:fig_miranda} extracting a more unbiased estimate of $N^i$~\cite{LHCb:2023qne}. Although this extension continued to support the no-CPV hypothesis, see Fig.~\ref{sec5.3:fig_miranda_b}, it was a meaningful extension to multi-body methodologies.  \\

\begin{figure}[b]
{\centering
\includegraphics[width=0.43\textwidth]{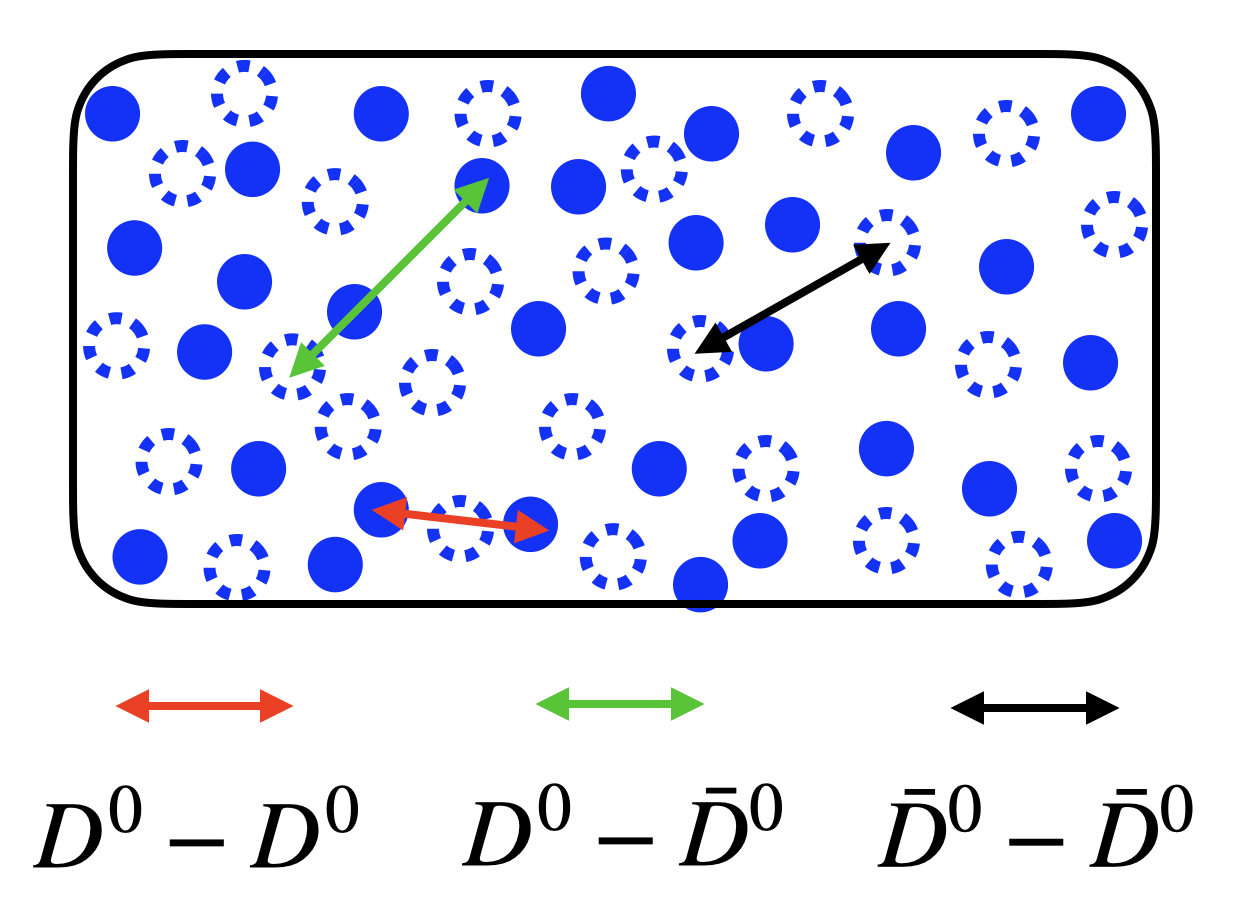}
\caption{Visualisation of the energy test method. 
The blue dots represent a sample $n$ formed of $D^0$ mesons and the white dots a sample $\bar{n}$ formed of $\bar D^0$ mesons. Searches of CPV among the samples 
are performed comparing the following three 
differences  
$D^0 - D^0$ (red), which evaluates the density of the $D^0$ sample, $\bar D^0 - \bar D^0$ (black), which evaluates the density of the $\bar D^0$ sample and $D^0 - \bar D^0$ (green), that evaluates differences in density between these samples. These three contributions correspond to the three terms in Eq.~\eqref{sec6.2:eq29}.}
\label{sec5.3:fig4}
}
\end{figure}

\noindent
An alternate method is the \textbf{energy test (ET)}~\cite{ET_1, ET_2, Parkes_2017}. The ET defines a test statistic to compare the relative density between two Dalitz spaces at every point in the space. This is defined as
\begin{equation}
    T \equiv \frac{1}{2n(n-1)} \sum_{i,j \neq i}^{n} \psi_{ij} + \frac{1}{2\bar{n}(\bar{n}-1)} \sum_{i,j \neq i}^{\bar{n}} \psi_{ij} - \frac{1}{n\bar{n}} \sum_{i,j}^{n, \bar{n}} \psi_{ij} \,,
    \label{sec6.2:eq29}
\end{equation}
where the first term corresponds to the sum of a weighted distance $\psi_{ij}$ between pairs of events, ${(i,j)}$ in the sample $n$. The second term calculates the same for the sample $\bar{n}$ and the final sum represents the weighted distance between events in each sample. The normalisation terms remove any global asymmetries. The different sums can be visualised simply in Fig.~\ref{sec5.3:fig4}. In the case in which both samples are identical - or near to, the $T$ value will fluctuate around zero. As the difference increases between the samples the value of $T$ will also rise. In order to make any inference about the significance of this $T$ value, it needs to be converted to a $p$-value. To do this, under the hypothesis of CP symmetry, we take permutations of our test sample with randomly assigned tags - where tags refer to the flavour of the $D$ meson. The $p$-value for the no-CP-violation hypothesis is obtained as the fraction of permutation $T$ values greater than the observed $T$ value. It is an important distinction that we extract a $p$-value for the no CP-violation hypothesis as we cannot run permutations for the inverse. The challenge of calculating these permutations for large datasets is circumvented by~\cite{Barter:2018xbc} applying a scaled version of these permutations\\

\noindent
LHCb has demonstrated the strength of ET with searches in $D^0 \to \pi^+ \pi^- \pi^0$, $D^0 \to \pi^+ \pi^- \pi^+ \pi^-$ and most recently ${D}^0\to {K}_S^0{K}^{\pm }{\pi}^{\mp}$~\footnote{The decays of $D^0\rightarrow K_S K^{*0}$, intermediate contributions to the phase space of $D^0\rightarrow K_SK^\pm\pi^\mp$ decays, have been identified as possible discovery channels~\cite{Nierste:2017cua}. } yielding results consistent with CP conservation \cite{LHCb_2023_ET, LHCb:2016qbq, ET_2024_K}. The analysis of decays with more particles such as $D^0 \to \pi^+\pi^-\pi^+\pi^-$ provide increased computational challenges with the increased dimensionality. The LHCb analysis on the Run I dataset was consistent with CP conservation~\cite{ET_4body_2017}. There are also experiential cases to implement complementary statistical methods such as the \textit{Earth Movers Distance}~\cite{2023_Earth_Movers}. Any hints of CPV from these statistical methods with larger datasets will provide a smoking gun for multi-body CPV in charm physics.\\

\noindent
Finally we can use multi-body decays to search for T violation, where T is the symmetry under time reversal, and by extension CPV, as a violation of the CP symmetry is equivalent to a violation of the T symmetry, through the use of \textbf{T-odd triple products}~\cite{Valencia:1988it}. If we only measure momenta, for a non trivial triple product we require 4 particles. If we consider a decay $\vec{P} \to \vec{p}_{h_1} + \vec{p}_{h_2} + \vec{p}_{h_3} + \vec{p}_{h_4}$ measured in the frame to the decaying particle, 
we can construct two decay planes made up of $\{ \vec{p}_{h_1}, \vec{p}_{h_2}\}$ and $\{\vec{p}_{h_3}, \vec{p}_{h_4}\}$. These planes intersect at the straight line given by $\vec{p}_{h_1} + \vec{p}_{h_2} = -\vec{p}_{h_3} - \vec{p}_{h_4}$. We then define a unit vector for $\vec{p}_{h_1} + \vec{p}_{h_2}$ as $\hat{z}$ and two vectors normal to the previously defined planes---$\{ \vec{p}_{h_1}, \vec{p}_{h_2}\}$ and $\{\vec{p}_{h_3}, \vec{p}_{h_4}\}$---and this vector. We notate these vectors as $\hat{n}_{12}, \hat{n}_{34}$. Thus we can describe the angle $\omega$ between these vectors, which gives us
\begin{equation}
\hat{n}_{12} \cdot \hat{n}_{34} = \cos{\omega}\,, \qquad \hat{n}_{12} \times \hat{n}_{34} = \sin{\omega}\,  \hat{z}\,,
\end{equation}
leading to a T-odd triple product of the form
\begin{equation}
(\hat{n}_{12} \times \hat{n}_{34}) \cdot \hat{z} = \sin{\omega} \,.
\end{equation}
However, from an experimental perspective, momentum conservation allows us to define the above triple product as
\begin{equation}
C_{\rm T} \equiv \vec{p}_{h_1} \cdot \left( \vec{p}_{h_2} \times \vec{p}_{h_3}\right)\,, 
\end{equation}
up to a normalisation term omitted for clarity. In essence, these triple products define a volume between vectors that is sensitive to the different resonant formation between the decay products. The CP analogous observable being given as ${\bar{C}_{\rm T}}$.
Taking this into account, we can generate asymmetries of the form
\begin{equation}
A_{\rm T-odd} = \frac{N(C_{\rm T} > 0) 
- N(C_{\rm T} < 0)}{N(C_{\rm T} > 0) + N(C_{\rm T} < 0)}\,,
\qquad
\bar{A}_{\rm T-odd} = \frac{\bar{N}(\bar{C}_{\rm T} < 0) - \bar{N}(\bar{C}_{\rm T} > 0)}{\bar{N}(\bar{C}_{\rm T} < 0) + \bar{N}(\bar{C}_{\rm T} > 0)}\,,
\end{equation}
with $N$, $\bar{N}$ are the yields of $D_{(s)}$ and $\bar{D}_{(s)}$ respectively. This asymmetry is expected to be non-zero, being sensitive to final state interactions. However, these effects should cancel between the $D_{(s)}$ and $\bar{D}_{(s)}$ so we can be sensitive to CPV in T-odd is using
\begin{equation}
a_{\rm CP}^{\rm T-odd} = \frac{1}{2} \left( A_{\rm T-odd} - \bar{A}_{\rm T-odd} \right) \,.
\end{equation}
By construction this method cancels strong phase effects, meaning that any non-zero result would indicate T violation, and by extension CP, as $A_{\rm T-odd}$ and 
$\bar{A}_{\rm T-odd}$ are conjugate states. 
This has been used to explore CP and T violating effects in the CF $D^+ \to K^+K^-K^0_S\pi^+$ and 
$D^+_s \to K^+K^0_S\pi^+\pi^-$ 
alongside the CS 
$D^+_s \to K^+K^-K^0_S \pi^+$ decays~\cite{Belle_2023_t_odd, BABAR_2011_t_odd}. 
The experimental measurements, summarised by HFLAV~\cite{HFLAV:2012imy}, currently remain consistent with zero, see Table~\ref{sec5.3:tab3}.
However, with the projected luminosity collection of both Belle II and LHCb this methodology might well provide insight through multi-body channels.
\\

\begin{table}
\renewcommand*{\arraystretch}{2}
\centering
\begin{tabular}[t]{|c|c||c|c|}
 \hline
 \it Mode & $a^{\rm T-odd}_{\rm CP}$ & \it Mode & $a^{\rm T-odd}_{\rm CP}$ \\
 \hline
 \hline
 $D^0 \rightarrow K^+ K^- \pi^+ \pi^-$ 
 & $+0.0029 \pm 0.0022$
 & $D^0 \rightarrow K_S^0 \pi^+ \pi^- \pi^0$ & $-0.00028 \pm 0.00138^{+0.00023}_{-0.00076}$\\
 \hline
 $D^0 \rightarrow K_S^0 K_S^0 \pi^+ \pi^-$ & $-0.0195 \pm 0.0142^{+0.0014}_{-0.0012}$
 &
 $D^+ \rightarrow K_S^0 K^+ \pi^+ \pi^-$ & $-0.0027 \pm 0.0070$\\
 \hline
 $D^+ \rightarrow K^+ K^- K_S^0 \pi^+$ & $-0.0334 \pm 0.0266 \pm 0.0035$
 &
 $D^+ \rightarrow K^+ K^- \pi^+ \pi^0$ &
 $+0.026 \pm 0.066 \pm 0.013$\\
 \hline
 $D^+ \rightarrow K^+ \pi^- \pi^+ \pi^0$ & 
 $-0.013 \pm 0.042 \pm 0.001$
 &
$D^+ \rightarrow K^- \pi^+ \pi^+ \pi^0$ & $+0.002 \pm 0.015 \pm 0.008$\\
\hline
$D_s^+ \rightarrow K_S^0 K^+ \pi^+ \pi^-$ & $-0.0087 \pm 0.0055$
&
$D_s^+ \rightarrow K^- K^+ \pi^+ \pi^0$ & $(+2.2 \pm 3.3 \pm 4.3) \cdot 10^{-3}$\\
\hline
\end{tabular}
    \caption{The current status of T-odd triple product measurements as summarised by HFLAV~\cite{HFLAV:2012imy}.}
\label{sec5.3:tab3}
\end{table}

\begin{figure}
    \centering
    \begin{subfigure}[t]{1\textwidth}
        \centering
        \includegraphics[scale = 0.4]{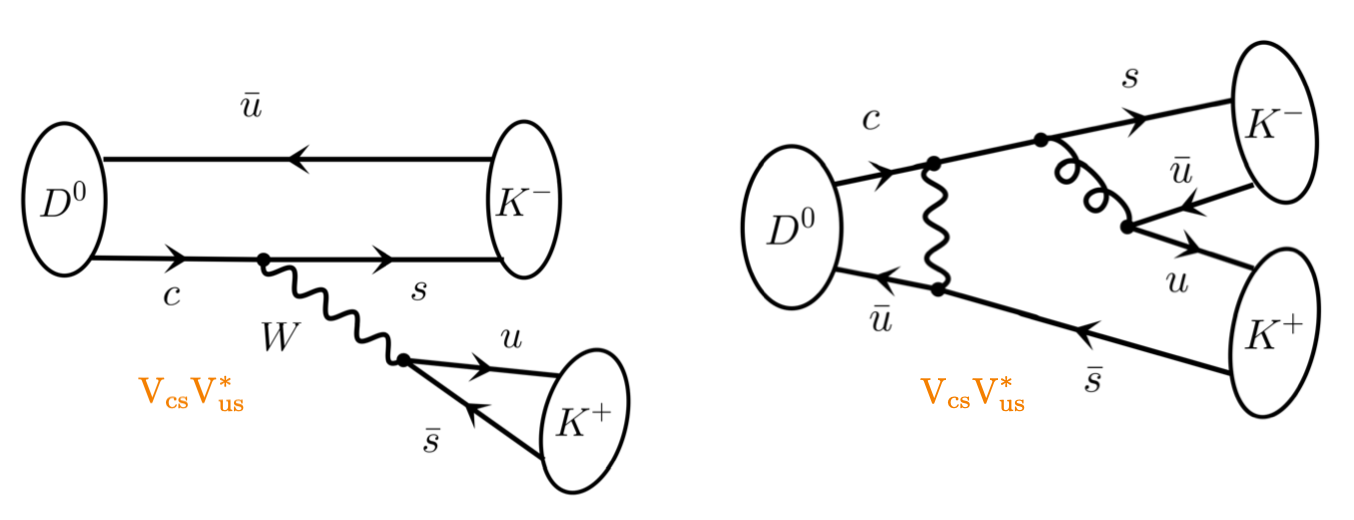}
        \caption{}
    \end{subfigure}
    \begin{subfigure}[t]{1\textwidth}
        \centering
        \includegraphics[scale = 0.4]{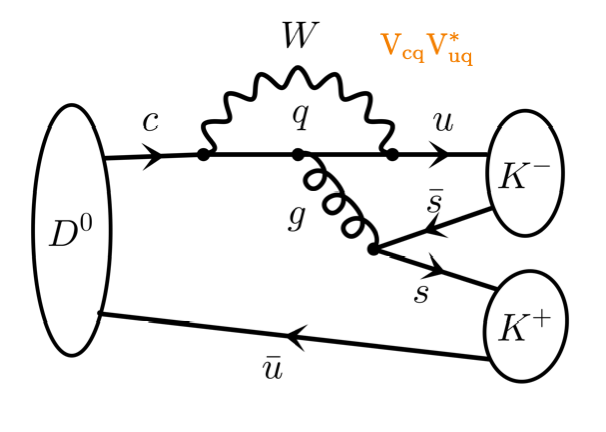}
        \caption{}
    \end{subfigure}
    \caption{(a) Examples of tree-level (left) and exchange (right) topologies contributing to $A_{\rm T}$ in Eq.~\eqref{sec6.2:eq1}. (b) Example of penguin topology with internal $q = d,s,b $ quark pair contributing to $A_{\rm P}^q$ in Eq.~\eqref{sec6.2:eq1}. Analogous diagrams with $s \to d$ contribute to $D^0 \to \pi^+ \pi^-$.}
    \label{sec6.2:fig1}
\end{figure}

\noindent
{\bf Theoretical status of charm CPV:}
Due to the very small imaginary part of the CKM matrix elements entering charm decays, CPV in the charm sector is expected to be a negligible effect. 
Direct CPV can only arise in decays which receive contributions from two different amplitudes with different strong and weak phases. For this reason, the SCS decays $D^0 \to \pi^+ \pi^-$  and $D^0 \to K^+ K^-$ are ideal candidates for searches of charm CPV, 
since, in the SM, they proceed via different hadronic amplitudes with different weak and strong phases, such as the tree-level and penguin topologies shown in Fig.~\ref{sec5.3:fig1}. Specifically, making the dependence on the CKM matrix elements explicit, the amplitude for e.g.\ the decay $D^0 \to K^+ K^-$ can be expressed as

\begin{equation}
        {\cal A} (D^0 \to K^+ K^-)  =  
\lambda_d (A_{\rm T} + A^d_{\rm P})
+ \lambda_s  A^s_{\rm P}
+ \lambda_b  A^b_{\rm P}
        \, ,
     \label{sec6.2:eq1}        
    \end{equation}
where $A_{\rm T}$ denotes the {\it tree amplitude}, which includes e.g.\ the tree-level and the exchange topologies shown in Fig.~\ref{sec6.2:fig1}~(a), while $A_{\rm P}^q$ is the {\it penguin amplitude} which includes the penguin topology with an internal $q$-quark pair, as shown in Fig.~\ref{sec6.2:fig1}~(b). Note that additional diagrams that may contribute in the SM can be also recast in the form of Eq.~\eqref{sec6.2:eq1}.  
Using the CKM unitarity, Eq.~\eqref{sec6.2:eq1} can be written as
\begin{equation}
        {\cal A} (D^0 \to K^+ K^-)  =  
\lambda_s {\cal A}_{KK}
\left[1 - \frac{ \lambda_b}{\lambda_s}  \frac{{\cal P}_{KK}}{{\cal A}_{KK}}
\right]
        \, ,
     \label{sec6.2:eq2}        
    \end{equation}
    with ${\cal A}_{KK}$ being the {\it CKM leading} amplitude, as it multiplies the dominant CKM factor $\lambda_s$ and ${\cal P}_{KK}$ being the {\it CKM sub-leading} amplitude, highly suppressed by the value of $\lambda_b$. Moreover, ${\cal A}_{KK}$ receive contributions from tree-level, exchange, and penguin topologies, whereas ${\cal P}_{KK}$ receive contributions from penguin topologies, see e.g.~\cite{Lenz:2023rlq}~\footnote{Additional contributions such as the ``penguin-annihilation" topologies also contribute to ${\cal A}_{KK}$ and ${\cal P}_{KK}$, these, however, are expected to be further suppressed by additional gluon corrections. For a comprehensive list of the possible topologies contributing to two-body $D^0$-meson decays, see e.g.~\cite{Muller:2015lua}.}.
Taking into account that $|\lambda_b/\lambda_s |\ll 1$, the corresponding expression for the branching ratio and the direct CP asymmetry to good approximation becomes 
\begin{equation}
B (D^0 \to K^+ K^-) \simeq {\cal N}_{KK} |\lambda_s |^2 |{\cal{A}}_{KK}|^2 \,,
\qquad 
a_{\rm CP}^{\rm dir} (D^0 \to K^+ K^-)  
\simeq
- 2  \left|   \frac{ \lambda_b}{\lambda_s}   \right| \sin \gamma 
\left| \frac{{\cal P}_{KK}}{{\cal A}_{KK}} \right|
 \sin \phi_{KK} 
     \, ,
     \label{sec6.2:eq3}
    \end{equation}
where $\gamma$ denotes the CKM angle in the unitarity triangle and $\phi_{KK}$ denotes the relative strong phase between the CKM sub-leading and CKM leading amplitudes. Notice that the expressions in Eq.~\eqref{sec6.2:eq3} can be easily generalised to the case of $D^0 \to \pi^+ \pi^-$ by replacing $\lambda_s \to \lambda_d$, $KK \to \pi\pi$ and $\sin \gamma \to - \sin \gamma$. From Eq.~\eqref{sec6.2:eq3} we see that, owing to $|\lambda_s| \approx |\lambda_d|$, in the limit of exact U-spin symmetry, the direct CP asymmetries for the $K^+ K^-$ and $\pi^+ \pi^-$ final states would be equal in magnitude and opposite in sign, leading to the sum rule $a_{\rm CP}^{\rm dir} (D^0 \to \pi^+ \pi^-) + a_{\rm CP}^{\rm dir} (D^0 \to K^+ K^-) = 0 $. However, the latter is violated by the experimental data in Eq.~\eqref{sec6.2:eq0} at the 2.7$\sigma$ level~\footnote{As noted above, this tension is reduced if also the U-spin breaking in the corresponding branching fractions is taken into account, see~\cite{PajeroatCharm}.}. This has been referred to as {\it U-spin anomaly}~\cite{Schacht:2022kuj} and its possible implications have been discussed in~\cite{Iguro:2024uuw, Bause:2022jes, Schacht:2022kuj}.\\

\noindent
Combining the expressions in Eq.~\eqref{sec6.2:eq3}, we arrive at the following form for $\Delta a_{\rm CP}^{\rm dir}$:
\begin{equation}
\Delta a_{\rm CP}^{\rm dir} \simeq - 2  \left|   \frac{ \lambda_b}{\lambda_s}   \right| \sin \gamma 
\left( \left| \frac{{\cal P}_{KK}}{{\cal A}_{KK}} \right|
 \sin \phi_{KK} + \left| \frac{{\cal P}_{\pi\pi}}{{\cal A}_{\pi\pi}} \right|
 \sin \phi_{\pi\pi}  \right)\,.
 \label{sec6.2:eq4}
\end{equation}
From the known values of the CKM matrix elements and assuming $|{\cal P}/{\cal A} | \approx 0.1$, as suggested by perturbative QCD estimates~\cite{Chala:2019fdb}, the expected size of the direct CP asymmetry in these channels and thus of $\Delta a_{\rm CP}^{\rm dir}$ is of the order of ${\cal O}(10^{-4})$, the latter being about a factor 10 smaller than the experimental determination in Eq.~\eqref{sec6.2:eqDeltaacp}\footnote{This estimate implicitly assumes large strong phases, i.e.\ $\sin \phi \sim {\cal O} (1)$. Smaller values of the strong phase would further decrease the expected size of $\Delta  a_{\rm CP}^{\rm dir}$.}.
An estimate of the penguin amplitudes ${\cal P}_{KK}$, ${\cal P}_{\pi\pi}$, using LCSRs was derived in~\cite{Khodjamirian:2017zdu}. In the latter reference, the value of the leading CKM amplitudes, needed to determine $\Delta a_{\rm CP}^{\rm dir}$, was extracted from the precise experimental data on the branching ratios~\cite{ParticleDataGroup:2024cfk}, cf.\ Eq.~\eqref{sec6.2:eq3}, yielding an upper bound of $|\Delta a_{\rm CP}^{\rm dir}|_{\rm SM}< 2.3 \cdot 10^{-4}$, in line with naive estimates and a factor of 6 below the experimental value. This result was confirmed in \cite{Lenz:2023rlq}, where the framework of LCSRs was employed to also estimate the value of the leading CKM amplitudes, without recurring to their experimental determinations. The conclusions obtained within LCSRs also agree with the results of~\cite{Pich:2023kim}, where the size of final state interactions was studied with the use of DRs. The experimental value in Eq.~\eqref{sec6.2:eqDeltaacp} has thus lead to several BSM analyses, see e.g.~\cite{Chala:2019fdb, Dery:2019ysp, Calibbi:2019bay}. \\
The ratio of penguin over tree-level amplitudes in $D \to \pi \pi $ decays, e.g.\ $D^0 \to \pi^+ \pi^-$, $D^0 \to \pi^0 \pi^0$, has been investigated in~\cite{Gavrilova:2023fzy} in the isospin limit and using the experimental values of branching ratios and CP asymmetries; moreover the possibility that $\Delta a_{\rm CP}^{\rm dir}|_{\rm exp}$ could be accommodated by large non-perturbative effects and thus be consistent with the SM has been advanced in \cite{Grossman:2019xcj}, based on the use of U-spin relations; in~\cite{Bediaga:2022sxw}, from a study of final state interactions; in \cite{Schacht:2021jaz}, based on an analysis of rescattering effects with potential large contributions due to nearby resonances such as the $f_0(1710)$ and $f_0(1790)$; and in \cite{Li:2019hho, Cheng:2019ggx, Wang:2020gmn}, based on the use of topological amplitude decompositions and the SU(3)$_F$ symmetry. However, no sign of a large enhancement due to nearby resonances was found in~\cite{Pich:2023kim} and in the latter work also some inconsistencies in the approach of~\cite{Bediaga:2022sxw} were pointed out. Finally, it is important to stress that analyses such as~\cite{Cheng:2019ggx} often have to recur to strong assumptions on the size of the hadronic matrix elements. Hence, currently, an exhaustive theoretical interpretation of the discovery of CPV in charm decays has not yet been reached.\\

\noindent
To conclude, other interesting channels for the study of CPV in charm are $D^0 \to K_S  K_S$, and $D^0 \to  K_S K^{0 *}$, $D^0 \to K_S \bar K^{0 *}$, namely SCS $D^0$ decays into two pseudoscalar or a vector and pseudoscalar neutral kaons final states. In the SM, these modes proceed only via exchange and loop-suppressed penguin-annihilation topologies and due to the interference of the $c \to d \bar d u$ and $c \to s \bar s u$ tree-level transitions, non-vanishing CPV effects are possible even if only the exchange topology is considered, that is, no penguin contributions are needed for non-zero direct CPV, contrary to the case of modes such as $D^0 \to K^+ K^-$ and $D^0 \to \pi^+ \pi^-$ discussed above. For $D^0 \to K_S K_S$, the branching ratio is strongly suppressed in the SU(3)$_F$-symmetric limit, as discussed, for example, in \cite{Brod:2011re, Atwood:2012ac, Hiller:2012xm, Nierste:2015zra}. Consequently, the size of SU(3)$_F$-breaking effects plays a crucial role not only in determining the decay rate but also in controlling the magnitude of CP violation. In \cite{Nierste:2015zra}, it was argued that CP-violating effects in this channel could be enhanced up to the percent level, while a recent analysis \cite{Fleischer:2025zhl} derived a lower bound at the per-mille level. The pseudoscalar vector modes also proceed only via exchange and penguin-annihilation topologies, however, due to the presence of distinguishable particles in the final state, the corresponding decay amplitude is not suppressed in the limit of SU(3)$_F$ symmetry~\cite{Nierste:2017cua}. Also in this channel, CPV effects could be enhanced~\cite{Nierste:2017cua}, in addition, experimentally, CPV searches would benefit from the higher efficiency due to the prompt decay $K^{0*} \to K^+ \pi^-$.

%%%%%%%%%%%%%%%%%%%%%%%%%%%%%%%%%%%%%%%%%%%%%%%%%%%%%%%%%%%%%%%%%%%%%%%%%%%%%%%%%%%%%%%%%%%%%%%%%%%%%%%%%%%%%%%%%%%%%%%%%%%%%%%%
\section{Conclusion and outlook}
\label{sec:conclusions}
Charm physics is a very active field of research with important milestones expected in the future. In this chapter we have presented a broad and pedagogical discussion of the physics of the charm quark, and provided an introduction to the principal experimental and theoretical methods used to study it, as well as an overview of the current status for fundamental observables like lifetimes of charmed hadrons, rare decays, CPV and mixing. The charm quark, with its peculiarities, represents a unique system to test the SM, and plays a key role in studying the interplay between the strong and weak interactions which govern its hadronic decays. In recent years, numerous achievements have been made in investigating the properties of the charm sector of the SM. \\

\noindent
On the experimental side, important breakthrough results have been obtained, such as the discovery of CPV in $D^0$ decays and the significant improvement in the precision of the mixing parameters $x$ and $y$, firmly establishing charm meson mixing. Moreover, searches for CPV in other decay modes are actively pursued, and many new particles
containing hidden or open charm have been discovered.
Limits on the branching ratios of numerous rare charm decays have been improved by orders of magnitude and rigorous tests of the SM through studies of null tests, i.e.\ observables that vanish in the SM, like CP asymmetries and angular observables have been performed.
Our knowledge of the lifetime of charm
mesons and baryons has also been improved, and a wealth of studies of charm decays with multi-body final states have provided information
on the intermediate resonances in such decays, as well as on their dynamics.\\

\noindent
Precise theoretical predictions for charm observables are notoriously challenging; however, important theory advances have also been made. Several studies of the validity of the HQE for charm hadron lifetimes have been performed, showing that this framework is capable of describing the observed experimental pattern, although with large uncertainties. Moreover, the framework of LCSRs, well established in the bottom sector, has been applied to the analysis of hadronic effects in charm meson decays and in particular to the study of CPV in $D^0 \to K^+ K^-$, $D^0 \to \pi^+ \pi^-$, leading to promising results, even if also with large uncertainties. These findings, with a higher precision could e.g.\ imply that the measured value of CPV in the charm system, $\Delta A_{\rm CP}$ could have a BSM origin. 
In addition, other methods have been used to study the non-perturbative dynamics in SCS two-body charm meson decays. For instance, dispersion relations have been employed to investigate final state interactions and estimate the size of CPV in these modes. Moreover, sum rules based on the approximate SU(3)$_F$ symmetry of QCD have been derived to obtain relations between CP asymmetries in different channels, to be tested experimentally. Great progress has also been made in understanding the QCD nature of exotic states.
Finally, numerous studies of rare charm decays have been performed, and several null tests observables have been proposed in order to test the SM. Such observables could serve as a smoking gun of NP if any experimental signal were observed.\\ 
 
\noindent
Charm physics has a bright future. Many of the available theoretical approaches
can be systematically improved with current technology, i.e.\ by including higher order
perturbative and power corrections.
Other theoretical methods, such as lattice QCD also have a great potential to shed further light in the non-perturbative dynamics of charmed hadron decays, and recent progress was made in developing new tools to be applied to the description of charm mixing.
Future more precise experimental data will be of crucial help in answering the still open questions on e.g.\ the nature of the observed value of CPV and of $D^0$-meson mixing. In fact, LHCb, with its upgraded detector and the planned Upgrade II, and Belle II~\cite{LHCb:2018roe} can investigate CP-violating effects in other decay channels than $D^0 \to K^+ K^-, \pi^+ \pi^-$ as well as in $D^0$-meson mixing and a planned Super Tau Charm Facility~\cite{Achasov:2023gey} could contribute with further precise data.
Moreover, studies from BESIII~\cite{BESIII:2020nme} and Belle II~\cite{Aihara:2024zds} of inclusive semileptonic charm hadron decays could provide important information on the size of the still poorly known non-perturbative parameters of the HQE in charm.
More precise tests of LFU can also be performed at these experiments. Furthermore, particularly important is the role of Belle II for decays with neutral particles in the final state.
A future $e^+ e^-$ collider running as an ultimate $Z$ factory will open exciting possibilities for charm physics~\cite{Monteil:2021ith}, for example stringent tests of the SM via FCNC processes and further searches of CPV in charm decays could provide indications for physics beyond the SM.  
All these experiments will continue their searches of not yet observed conventional and exotic hadrons containing charm, as well as contribute to improve our knowledge of charm baryon decays. Maybe the charm quark will in the end provide the ultimate clue to explain our existence.

\begin{ack}[Acknowledgments]
 We thank
 Alex Gilman for useful discussions on leptonic and semileptonic charm decays results, Patrick Koppenburg for providing Fig.~\ref{sec3.4:fig4}, Marco Gersabeck and Dominik Mitzel for helpful discussions, Claus Grupen, Jack Jenkins, Alexander Khodjamirian, Ali Mohamed and Eleftheria Solomonidi for valuable comments on the draft and Aleksey Rusov for providing some of the plots used in this review. 
Moreover, we are grateful to Tommaso Pajero for his careful reading of the document and his helpful feedback. 
 All Feynman diagrams have been drawn using the software tool FeynGame~\cite{Harlander:2024qbn}. D.~Friday acknowledges funding support from STFC, UK. E.~Gersabeck acknowledges funding support from the Royal Society, UK.
\end{ack}

%%%%%%%%%%%%%%%%%%%%%%%%%%%%%%%%%%%%%%%%%%%%
%%%%%%%%%%%%%%%%%%%%%%%%%%%%%%%%%%%%%%%%%

\bibliographystyle{unsrt}

\bibliography{reference}

\end{document}